\documentclass[a4paper,12pt,notitlepage]{article}

\usepackage{color}
\usepackage{amsmath,amsthm,amssymb,amscd}
\usepackage{graphicx}
\usepackage{cite}
\usepackage{psfrag}
\usepackage{pstricks}
\usepackage{pst-node}
\usepackage{pst-plot}
\usepackage{url}
\usepackage[dvips]{hyperref}

\newcommand{\bb}[1]{{\mathbb{#1}}}

\newcommand{\bC}{\ensuremath{\mathbb{C}}}

\newcommand{\bF}{\ensuremath{\mathbb{F}}}

\newcommand{\bP}{\ensuremath{\mathbb{P}}}

\newcommand{\bR}{\ensuremath{\mathbb{R}}}

\newcommand{\bT}{\ensuremath{\mathbb{T}}}

\newcommand{\bZ}{\ensuremath{\mathbb{Z}}}

\newcommand{\be}{\ensuremath{\mathbf{e}}}

\newcommand{\scA}{\ensuremath{\mathcal{A}}}
\newcommand{\scB}{\ensuremath{\mathcal{B}}}
\newcommand{\scC}{\ensuremath{\mathcal{C}}}

\newcommand{\scE}{\ensuremath{\mathcal{E}}}

\newcommand{\scM}{\ensuremath{\mathcal{M}}}
\newcommand{\scN}{\ensuremath{\mathcal{N}}}

\newcommand{\scS}{\ensuremath{\mathcal{S}}}

\newcommand{\scW}{\ensuremath{\mathcal{W}}}

\newcommand{\Ob}{\mathfrak{Ob}}
\newcommand{\Homfrak}{\mathfrak{Hom}}

\newcommand{\bCx}{\bC^{\times}}
\newcommand{\Hom}{\operatorname{Hom}}

\newcommand{\coh}{\operatorname{coh}}

\newcommand{\id}{\mathrm{id}}
\newcommand{\vspan}{\operatorname{span}}

\def\tr{\mathop{\rm tr}\nolimits}
\def\Tr{\mathop{\rm Tr}\nolimits}
\def\mod{\mathop{\rm mod}\nolimits}

\def\Log{\mathop{\rm Log}\nolimits}

\newcommand{\Fuk}{\ensuremath{\mathop{\mathfrak{Fuk}}}}
\newcommand{\dirFuk}{\ensuremath{\mathop{\mathfrak{Fuk}^{\to}}}}
\newcommand{\m}{\mathfrak{m}}

\newcommand{\Ext}{\operatorname{Ext}}

\newcommand{\Arg}{\operatorname{Arg}}

\theoremstyle{definition}
\numberwithin{equation}{section}
\newtheorem{thm}{Theorem}[section]
\newtheorem{theorem}{Theorem}[section] 

\newtheorem{exa}[thm]{Example}
\newtheorem{prop}[thm]{Proposition}

\newtheorem{conj}[thm]{Conjecture}

\theoremstyle{plain}
\newtheorem{Rule}{Rule} % \rule already defined
\renewcommand{\qed}{\hfill \hbox{\rule[-2pt]{3pt}{6pt}} \par}

\newcommand{\beq}{\begin{equation}}
\newcommand{\eeq}{\end{equation}}
\newcommand{\beqa}{\begin{eqnarray}}
\newcommand{\eeqa}{\end{eqnarray}}

\newcommand{\drawsquare}[2]{\hbox{%
\rule{#2pt}{#1pt}\hskip-#2pt%  left vertical
\rule{#1pt}{#2pt}\hskip-#1pt%  lower horizontal
\rule[#1pt]{#1pt}{#2pt}}\rule[#1pt]{#2pt}{#2pt}\hskip-#2pt%  upper horizontal
\rule{#2pt}{#1pt}}% right vertical

\newcommand{\fund}{{\raisebox{-.5pt}{\drawsquare{6.5}{0.4}}}}
\newcommand{\fundbar}{{\overline{\raisebox{-.5pt}{\drawsquare{6.5}{0.4}}}}}

\newcommand{\symm}{{\raisebox{-.5pt}{\drawsquare{6.5}{0.4}}\hskip-0.4pt%
        \raisebox{-.5pt}{\drawsquare{6.5}{0.4}}}}%  symmetric second rank

\newcommand{\asymm}{{\raisebox{-3.5pt}{\drawsquare{6.5}{0.4}}\hskip-6.9pt%
        \raisebox{3pt}{\drawsquare{6.5}{0.4}}}}%  antisymmetric second rank

\newcommand{\rap}[2]
{\setbox1=\hbox{#1}%
\setbox2=\hbox to\wd1{\hss #2\hss}%
\mbox{\rlap{\box1}\box2}}

\newcommand{\sla}[1]{\rap{$#1$}{/}}
\def\tr{\mathop{\rm tr}\nolimits}
\def\mod{\mathop{\rm mod}\nolimits}
\def\notni{\mathop{\sla\ni}\nolimits}

\newcommand{\ol}{\overline}
\newcommand{\wt}{\widetilde}

\newcommand{\B}{{{\cal C}^*}}

\newcommand{\balpha}{\mbox{\boldmath $\alpha$}}
\newcommand{\bbeta}{\mbox{\boldmath $\beta$}}

\newcommand{\com}[1]{}

\begin{document}

\begin{titlepage}

%% Set the number of the title with 0
\setcounter{page}{0}

%% change the footnote symbol
\renewcommand{\thefootnote}{\fnsymbol{footnote}}

\thispagestyle{empty}
\begin{flushright}
UT-08-05\\
March, 2008
\end{flushright}

\vskip 1.5 cm

\begin{center}
\noindent{\LARGE{Brane Tilings and Their Applications}  \\\vspace{0.5cm}
}
\vskip 1.5cm
\noindent{\large{Masahito Yamazaki}\footnote{E-mail: yamazaki(at)hep-th.phys.s.u-tokyo.ac.jp}}\\ 
\vspace{1cm}
\noindent{\small{\textit{Department of Physics, University of 
Tokyo, Tokyo 113-0033, Japan}}}
%\small{\textit{Hongo 7-3-1, Bunkyo-ku, Tokyo 113-0033, Japan}}}
\end{center}

\vspace{1cm}

\begin{abstract}
We review recent developments in the theory of brane tilings and four-dimensional $\mathcal{N}=1$ supersymmetric quiver gauge theories. This review consists of two parts. In part I, we describe foundations of brane tilings, emphasizing the physical interpretation of brane tilings as fivebrane systems. 
In part II, 
we discuss application of brane tilings to
AdS/CFT correspondence and homological mirror symmetry. More topics, such
as orientifold of brane tilings, phenomenological model building, similarities with BPS solitons in supersymmetric gauge theories, are also briefly discussed.
This paper is 
a revised version of the author's master's thesis submitted to Department of Physics, Faculty of Science, the University of Tokyo on January 2008, and is 
based on his several papers and some works in progress \cite{UY1,UY2,UY3,UY4,IIKY,IKY,FNOSY}.
\end{abstract}

\end{titlepage}

\tableofcontents
\newpage

\renewcommand{\thefootnote}{\arabic{footnote}}

%%%%%%%%%%%%%%%%%%%%%%%%%%%%%%%%%%%%%%%%%%%%%%%%%%%%%%%%%%%%%%%%%%%%%%%%%%%%%%%
%%%%%%%%%%%%%%%%%%%%%%%%%%%%%%%%%%%%%%%%%%%%%%%%%%%%%%%%%%%%%%%%%%%%%%%%%%%%%%
\section{Introduction}\label{intro.sec}

Free fermions are omnipresent
in string theory.
They are so important because they are ``exactly solvable''. 

Dimer models are yet another statistical mechanical models which are exactly solvable on arbitrary two-dimensional lattices, as discovered independently in the early 60s' by Kasteleyn \cite{Kasteleyn} and Temperley and Fisher \cite{TemperleyFisher}. Such models have been studied in statistical mechanics since long ago (see e.g. \cite{Fowler} for early discussions), and they appear in many areas of  science, such as statistical mechanics, condensed matter physics, chemistry and biochemistry (see \cite{Kenyon} for an introduction to the dimer model). 

\begin{figure}[htbp]
\centering{\includegraphics[scale=0.4]{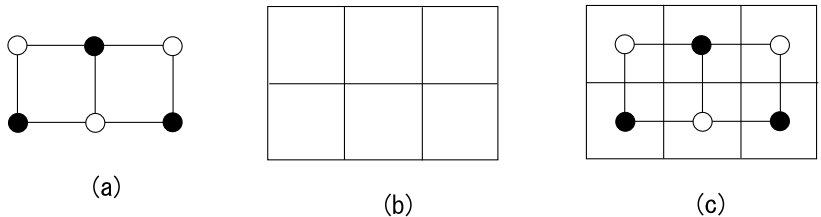}}
\caption{Example of a bipartite graph is shown in (a), which has three black/white vertices and seven edges. In this example, you can superimpose (a) with $2\times 3$ boxes (b), to obtain (c).
}
\label{bipartite}
\end{figure}

\begin{figure}[htbp]
\centering{\includegraphics[scale=0.4]{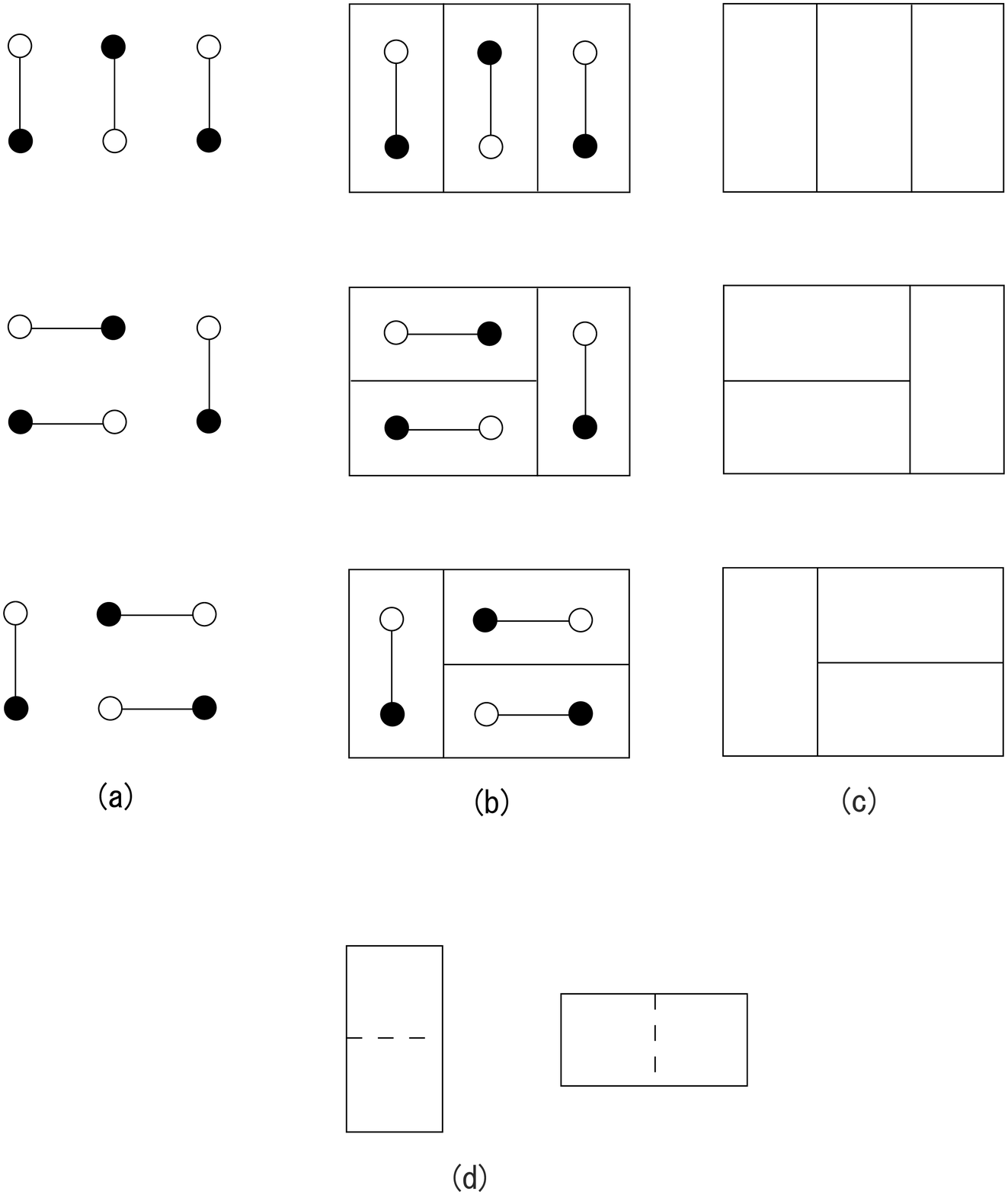}}
\caption{Here we show an example of perfect matchings on the bipartite graph of Figure \ref{bipartite}. The problem is to count the number of perfect matchings, and you can easily verify that the answer is three, as shown in (b). Alternatively, you can see this problem as the tiling problem of the dual graph (in this case, $2\times 3$ boxes) by `dominoes'. as shown in (b) and (c): we consider the tiling of $2\times 3$ regions ( Figure \ref{bipartite} (b)) using two types of dominoes (d) as basic constituents.}
\label{PM}
\end{figure}

Let us explain what dimer models are. Dimer models are defined on \textit{bipartite graphs}. A bipartite graph is a graph consisting of vertices which are colored either black or white and edges connecting vertices of different colors (see Figure \ref{bipartite} for example).
On this bipartite graph, we consider \textit{perfect matchings}. Here perfect matching refers to a matching, or a subset of edges without common vertices, of the graph that touches all vertices exactly once (see Figure \ref{PM} for example)\footnote{Strictly speaking, in some literature of dimer models, perfect matchings simply refers to subset of edges without common vertices (not necessarily touching all vertices). In this paper, however, we always use the term perfect matching as defined in the main text.}.

In statistical mechanics, we consider partition function, which is the weighted sum of perfect matchings:
\beq
Z=\sum_{\textrm{perfect matching}} (\textrm{weight of perfect matching}), \label{partition}
\eeq
where the choice of weight depends on the specific problem we want to consider. For example, the Kasteleyn matrix we discuss in \S\ref{another.subsec} is an example of such a partition function. As an another example, if the weight for perfect matchings are all equal to one, then the partition function is simply given by the total number of perfect matchings. Knowing the possible number of perfect matchings on a bipartite graph is an interesting combinatorial problem.

For example, in the case of the bipartite graph in Figure \ref{bipartite}, it is easy to see that we have three perfect matchings (Figure \ref{PM}). 
In this example, you can think of perfect matchings as a tiling of the dual graph (in this case, $2\times 3$ boxes) by `dominoes' (Figure \ref{PM}). You can consider similar problem by changing the shape of dominoes or the whole region. This type of counting problem has a long history, but it is still continues to be an interesting problem in combinatorics\cite{ArdilaStanley}.

%It is also easy to generalize the result to the case of $2\times N$ boxes (previously $N=3$), and the answer is $N$th Fibonacci number.

As the graph becomes more and more complicated, the problem of counting perfect matchings of a bipartite graph becomes more and more difficult, but Kasteleyn has used the technique of Kasteleyn matrix to solve the problem for arbitrary bipartite graphs on $\bT^2$, as will be explained in \S\ref{another.subsec}.

So far the story is purely combinatorial in nature, but one fact makes this more interesting.
It is known since long ago that dimers are related to, and in some sense equivalent to, free fermions (see \cite{Dijkgraaf:2007yr} for recent discussions). In fact, dimers were used to exactly solve the two-dimensional Ising model\cite{Kasteleyn2,Fisher,FisherJMP,McCoyWu}, which is well-known to be related to free fermions: there exists a one-to-one mapping between Ising model on a lattice and dimer model on another lattice\cite{FisherJMP}.

Since dimer model is similar to (and in a sense generalization of) free fermions, and since we know that free fermions are everywhere in string theory, it is natural to ask whether dimers have their role to play in string theory. 

\begin{figure}
\centering{\includegraphics[scale=0.6]{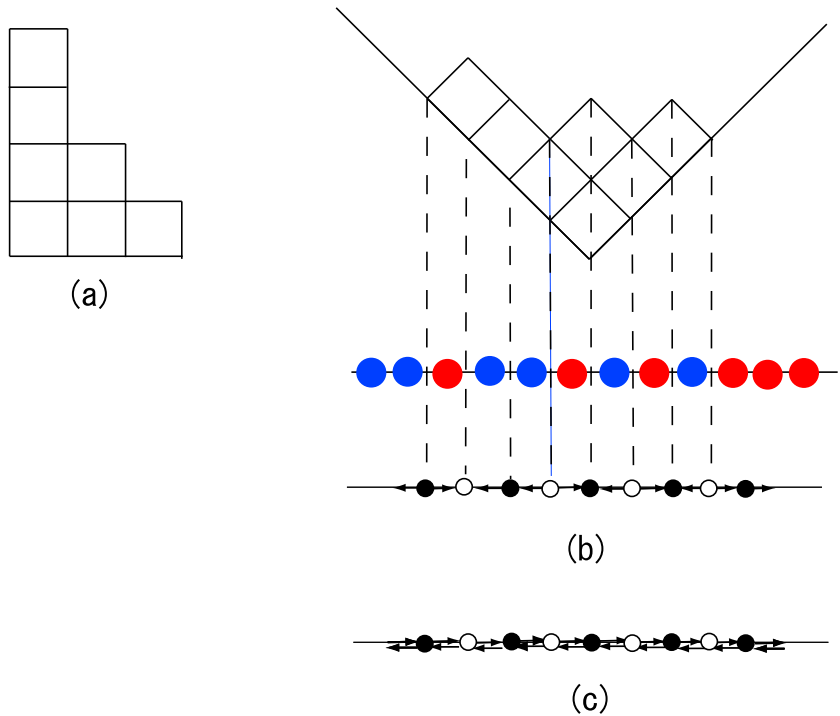}}
\caption{Young diagram (a) is also represented as Maya diagram (b). Maya diagram, in our language, is in one-to-one correspondence with the choice of subset of edges of the bipartite graph shown in (c). In this sense enumeration of Young diagrams is equivalent to enumeration of such subsets of edges of bipartite graph  (c), which is essentially one-dimensional version of the dimer model.}
\label{2dYoung}
\end{figure}

In fact, the answer is yes. Although usually unnoticed, one-dimensional version of dimer models has already appeared in many contexts, in the form of Young diagrams. Young diagrams are used in the representation theory of Lie algebras and symmetric groups, and appear in many places in physics, such as large $N$ limit of two-dimensional QCD\cite{Migdal:1975zg, Gross:1993hu,Cordes:1994fc}. In the literature, we have another method to represent Young diagram, by using Maya diagram. For a Young diagram as in Figure \ref{2dYoung} (a), rotate it by 45 degrees as in Figure \ref{2dYoung} (b). Then, if you go along the shape of the Young diagram from the far left to the right, we are going either downwards or upwards. Place a blue (resp. red) ball each time we go downwards (resp. upwards)\footnote{Usually, we use white and black colors in Maya diagram. Here, we choose to use blue and red in order to avoid confusion with black and white vertices of bipartite graph shown in Figure \ref{2dYoung} (c).}. The diagram so obtained is the so-called Maya diagram (Figure \ref{2dYoung} (b)).

This is usual story, but in fact you can look at Young diagram and Maya diagram  as a one-dimensional dimer problem. Consider a bipartite graph as shown in Figure \ref{2dYoung} (c). Then the Maya diagram is in one-to-one correspondence with the choice of subset of edges of the bipartite graph, such that one and only one is chosen from two edges connecting the same set of vertices. Namely, red (resp. blue) node of the Maya diagram corresponds to right-going (resp. left-going) edge of the bipartite graph. Although this subset itself is not a perfect matching, it is the one-dimensional analogue of a perfect matching in the two-dimensional dimer model.

\begin{figure}[htbp]
\centering{\includegraphics[scale=0.45]{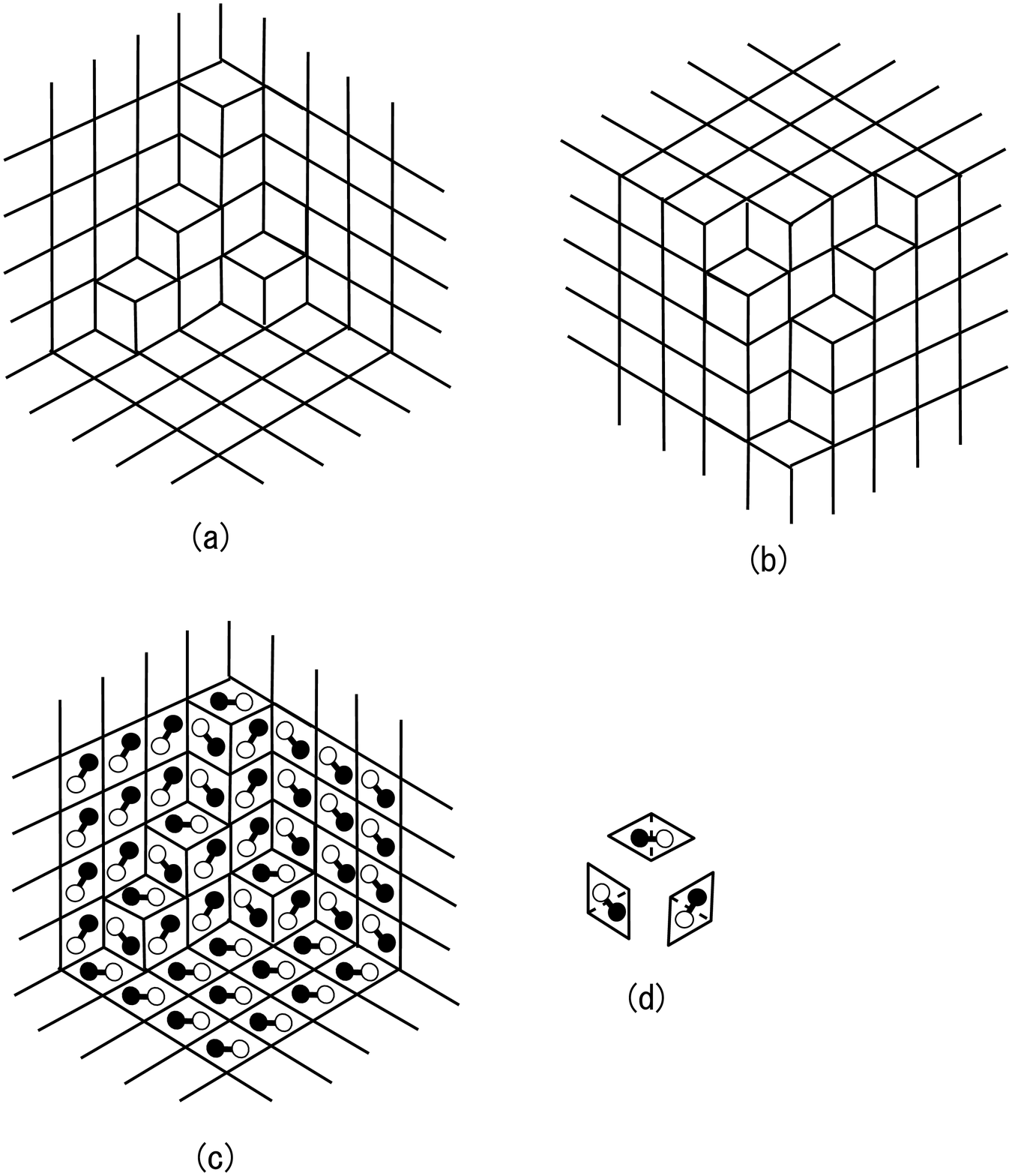}}
\caption{This figure shows an example of three-dimensional version of Young diagram (a). If you rotate (a) by 180 degrees, we have (b), which looks like melting of a crystal. By projecting this figure onto two-dimensions, we have a perfect matching of a bipartite graph defined on honeycomb bipartite graph (c), or equivalently tiling of plane using three types of rhombi shown in (d) (this is an analogue of ``domino tiling'' in Figure \ref{PM}). This one-to-one correspondence between three-dimensional Young diagram and perfect matching in dimer model is a higher-dimensional generalization of more familiar correspondence shown in Figure \ref{2dYoung}. The interesting fact is that this type of three-dimensional Young diagram appears in string theory, in the ``melting crystal'' picture of \cite{Okounkov:2003sp}.}
\label{3dYoung}
\end{figure}

In this example, only one-dimensional dimer appears.
One might thus be motivated to generalize this discussion to two-dimensional dimer models. In fact, it is not difficult to do this. Three-dimensional Young diagram is in one-to-one correspondence with perfect matchings on a honeycomb lattice, as shown in Figure \ref{3dYoung}. This means that, instead of two-dimensional dimer models, we can search for three-dimensional Young diagrams in string theory.

Now the interesting fact is that these three-dimensional Young diagrams have already appeared in string theory. In \cite{Okounkov:2003sp}, three-dimensional Young diagram, or ``crystal melting'', is proposed as a statistical mechanical model of topological A-model on non-compact toric Calabi-Yau. By geometric engineering\cite{Katz:1996fh,Katz:1996th}, this is also related to Nekrasov's partition functions\cite{Nekrasov:2002qd} of four-dimensional $\mathcal{N}=2$ supersymmetric gauge theories (and their lift to five dimensions).

\begin{figure}[htbp]
\begin{center}
\includegraphics{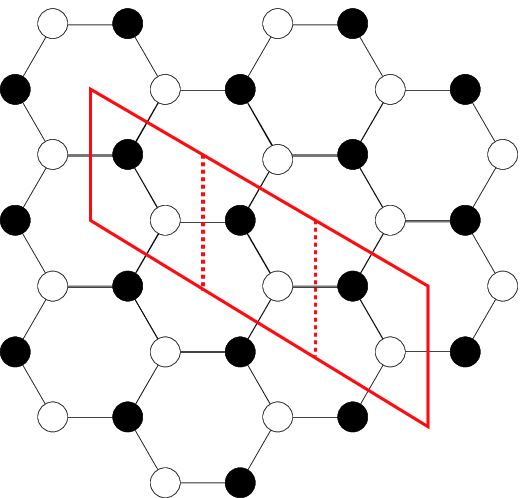}
\includegraphics{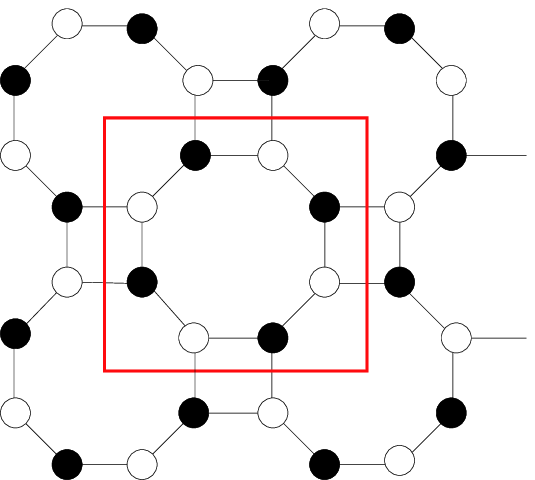}
\end{center}
\caption{Example of brane tilings, or bipartite graphs on $\bT^2$. The region represents fundamental region of torus. It suffices to write graphs only in the fundamental region, but sometimes it is convenient to write the graph as a periodic tiling of two-dimensional plane, which is the reason for the name ``brane tiling''. The left figure is the bipartite graph for $\bZ_3$-orbifold of $\mathbb{C}^3$, and the right for canonical bundle over $\bP^1\times \bP^1$, which is denoted $K_{\bP^1\times \bP^1}$.}
\label{BTfirst}
\end{figure}

Dimers also appear in seemingly different (although as we will see in \S\ref{D6.subsec} they are at least indirectly related by chain of dualities) context in string theory: dimers appear in the context of four-dimensional $\mathcal{N}=1$ supersymmetric quiver gauge theories as well, and that is the main topic of this paper. The difference with the case of topological A-model is that here dimer models are always written on two-dimensional torus $\mathbb{T}^2$ (see Figure \ref{BTfirst} for example). In this case, bipartite graphs are called \textit{brane tilings}\cite{Hanany:2005ve,Franco:2005rj,Franco:2005sm}. They have turned out to be quite powerful tools to study $\mathcal{N}=1$ supersymmetric (often superconformal) quiver gauge theories and their relation with toric Calabi-Yau manifolds.

Soon after brane tilings were discovered, they were applied to AdS/CFT duality for Sasaki-Einstein manifolds\cite{Butti:2005sw,Franco:2005rj,Franco:2005sm} and had a great success.
 Fortunately, since 2004, we have explicit metrics of infinite families of Sasaki-Einstein manifolds, which are called $Y^{p,q}$\cite{Gauntlett:2004yd} or $L^{a,b,c}$\cite{Cvetic1,Cvetic2,Martelli:2005wy}, and this has motivated an upsurge of interest in this field, 
 culminating for example in the construction of gauge theory duals of $Y^{p,q}$ and $L^{a,b,c}$ Sasaki-Einstein manifolds\cite{Franco:2005rj,Franco:2005sm,Butti:2005sw} and in the proof\cite{Butti:2005vn,Butti:2005ps} of the equivalence of $a$-maximization\cite{Intriligator:2003jj} and Z-minimization\cite{Martelli:2005tp,Martelli:2006yb} (see \S\ref{AdSCFT.sec}).

Despite these many successes, it is fair to say that, for some time, we still did not understand properly the physical significance of these bipartite graphs. It was already stated in the original references that the bipartite graphs represent systems of D5-branes and NS5-branes, but concrete picture was still lacking. For example, we have hexagons in Figure \ref{BTfirst}, but does this indeed mean that D-branes take hexagonal face?\footnote{For impatient readers, we comment here that the answer is no. The shape of fivebranes are represented by fivebrane diagrams, rather than bipartite graphs themselves. See the discussion in \S\ref{D5NS5.sec}.}

Important development was later made in \cite{Feng:2005gw}, where physical interpretations of brane tilings from mirror D6-brane picture was clarified. Later, the understanding from fivebrane systems is developed by our group\cite{Imamura:2006db,Imamura:2006ie,IIKY,IKY}, and we can now confidently say that brane tilings undoubtedly represent physical brane systems.

We would like to stress again that although a brane tiling (a bipartite graph) is certainly interesting and useful, the underlying D-brane picture (which is represented by fivebrane diagrams) should be much more powerful and have much wider implications. By the word ``brane tiling'' in the title of this paper, we have in mind not only the bipartite graphs, but the whole theory of physical fivebrane systems represented by bipartite graphs.

Since almost three years have passed since the first proposal, and since many aspects of brane tilings we can confidently talk about, it is time now to collect and review known facts, place them in a unified perspective, clarify connection with various topics, and prepare for what will come next. This review is intended as a modest step toward this ambitious goal. 

The structure of the rest of this paper is as follows. 

This paper is divided into two parts, part I and part II. 
In part I, we describe foundations of the theory of brane tilings. 
Although we have emphasized in Introduction the aspects of brane tilings as a statistical mechanical model, we would like to emphasize physical D-brane picture in following sections. 
We choose this way of presentation because we believe it is the best way of clarifying the physical interpretation of many technicalities of brane tilings. In fact, the author used to have many complaints from his colleagues that theory of brane tilings contains too many intricate `rules' without proper physical explanation, which make them reluctant to do research on brane tilings. But we would like to stress that this is no longer the case. Of course, we do not still understand all the aspects of brane tilings clearly enough. However, we are now certain that we have real physical, string-theoretic understanding of basic aspects of brane tilings themselves, and this review tries to tell you about these recent exciting developments.

We begin, as a warm-up, with a quick introduction to quiver gauge theories and their D3-brane realization in \S\ref{D3.sec}. Next in \S\ref{D5NS5.sec}, we describe basics of brane tilings. There we have tried to emphasize the physical interpretation of brane tilings as fivebrane systems, since that will become crucial later in some parts of this paper.
Only after these explanations we are going to describe in \S\ref{more.sec} more detailed aspects of brane tilings, such as inclusion of fractional branes and flavor branes, analysis of BPS conditions, and Seiberg duality.

In part II, we move onto applications of brane tilings.

As a first application, in \S\ref{AdSCFT.sec} we describe application to $AdS_5/CFT_4$ correspondence in the case of $\mathcal{N}=1$ SUSY and Sasaki-Einstein manifold. This topic is interesting in its own right, so we have also included some slightly detailed topics which are not directly related to brane tilings. The main goal of this section is to check the holographic relation between central charges of quiver gauge theories and volumes of Sasaki-Einstein manifolds, and brane tiling tells us the precise relation between Sasaki-Einstein manifolds and quiver gauge theories.

Mirror symmetry is another field where brane tilings play an important role (\S\ref{HMS.sec}). In particular, we explain an intriguing fact that brane tilings are used to give rigorous mathematical proofs of homological mirror symmetry conjecture, as shown in \cite{UY1,UY2,UY3,UY4}. Brief introduction to mathematical machineries, such as derived categories, is also included.

Brane tilings have many more applications, some of which are briefly summarized in \S\ref{others.sec}. They include application to phenomenological model building, soliton junctions in supersymmetric gauge theories, counting of gauge invariant operators, and the possible higher-dimensional extension of brane tilings.

Finally, in \S\ref{final.sec} we summarize, and discuss many questions which are still open as of this writing.

\begin{figure}[htbp]
\centering{\includegraphics[scale=0.6]{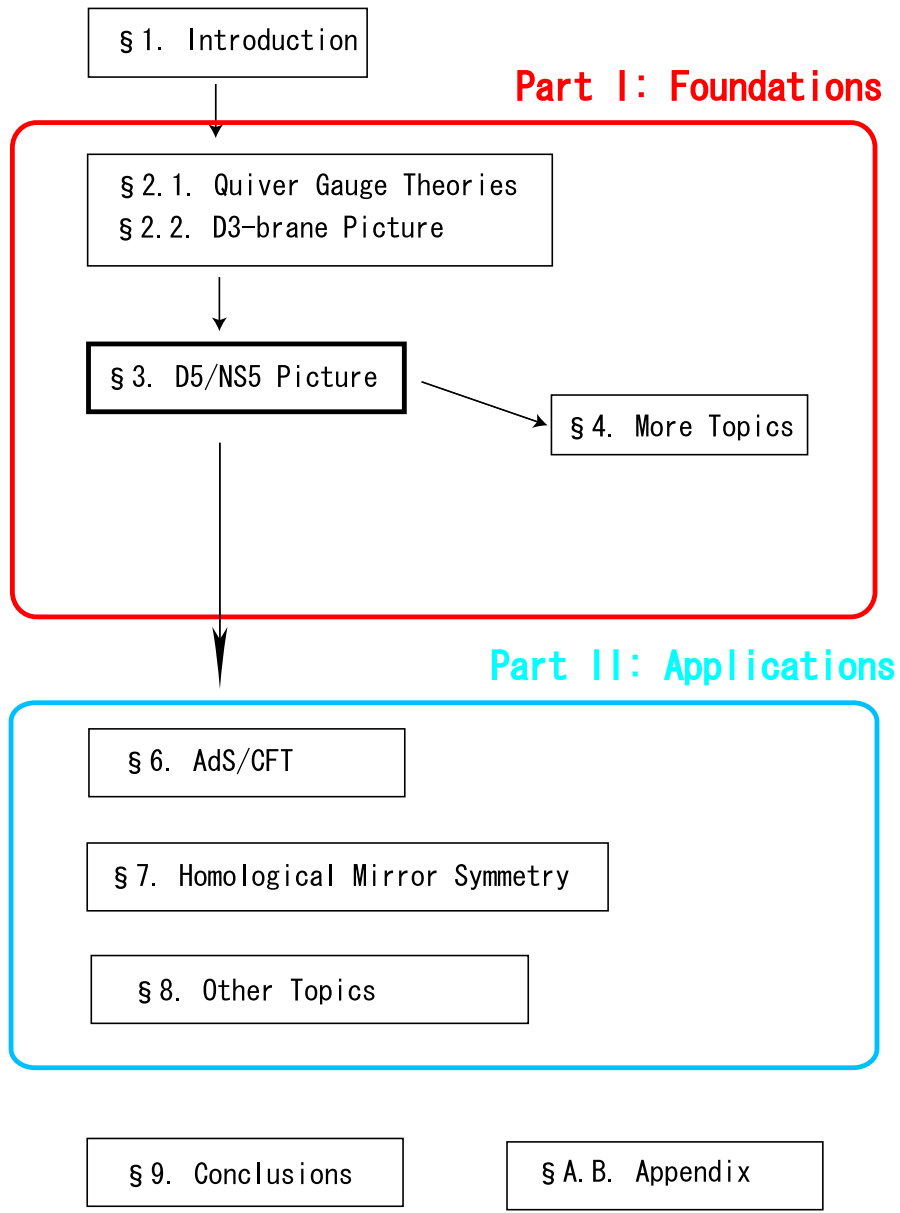}}
\caption{The dependence of sections in this paper. }
\label{flowchart}
\end{figure}

For readers' convenience, the dependence of sections is shown in Figure \ref{flowchart}. Note that sections in part II are mostly independent, and readers can choose according to their tastes. 

%\paragraph{Suggestions to the reader}

This review is mainly intended for people in the string theory community, but some parts of it should be of interest to researchers in other areas, such as statistical mechanics and mathematics. The interested reader should also consult a nice review by Kennaway\cite{Kennaway:2007tq}, which gives a concise summary of the subject.

Throughout this review we have tried to emphasize connection with various topics. We have chosen to do so in the hope of conveying you the richness of the subject. We hope by going through this review, the reader can find many unexpected connection with various topics. At the same time, however, this makes this review slightly long with many references. Thus we have made some suggestions below to readers of certain backgrounds, although they should not be taken too seriously. In any case, We recommend \S\ref{D5NS5.sec}, since that is the most important section in this paper. %One should also consult \S\ref{D3.sec} as needed.

\begin{itemize}
\item Readers interested in physical interpretations or computational methods of brane tilings are strongly encouraged to read \S\ref{D5NS5.sec}. If you have more time, you can pick up some materials in \S\ref{more.sec}. 
\item Readers interested in $AdS_5/CFT_4$ correspondence for Sasaki-Einstein manifolds should read \S\ref{AdSCFT.sec}. This section can be read almost independently. You should also consult some materials in \S\ref{D5NS5.sec} as needed.
\item Readers wanting to use brane tilings in string-phenomenological model building should read \S\ref{D5NS5.sec}, discussion of fractional branes and flavor branes in \S\ref{fractional.subsec} and \S\ref{flavor.subsec}, %\S\ref{orientifold.sec} 
and relevant part of \S\ref{orientifold.subsec}, in combination some materials from the paper \cite{IKY}.
\item Readers (including mathematicians) interested in homological mirror symmetry are encouraged to read \S\ref{D6.subsec} and \S\ref{HMS.sec}, after skimming through \S\ref{D5NS5.sec}, in combination with original works \cite{UY1,UY2,UY3}. 
\end{itemize}

Although we have tried to cover many materials in this field, certainly we have not covered all the topics. We would like to apologize in advance for any important omissions. 

%% arXiv %% 
Finally, this paper is based on the author's Master's thesis submitted in January 2008 to Department of Physics, Faculty of Science, the University of Tokyo, and was defended in February 2008. Most of this paper is based on previous literature in this field, and in particular the author's papers and some works in progress \cite{UY1,UY2,UY3,UY4,IIKY,IKY,FNOSY}. However, we tried in this review to clarify points which are not explicitly stated, and some of the explanations we believe are new.

%% thesis %%
%% Finally, many part of this thesis is based on the author's previous papers and some works in progress \cite{UY1,UY2,UY3,UY4,IIKY,IKY,FNOSY}. The contents of this paper is based on these papers and many more papers in references, but we hope to clarify many points which are not explicitly stated, and we believe that some of the explanations are new.

\newpage

%%%%%%%%%%%%%%%%%%%%%%%% Part I %%%%%%%%%%%%%%%%%%%%%%%%%%%%%%%%%%%
\part{Foundations of brane tilings}\label{part1}
%%%%%%%%%%%%%%%%%%%%%%%%%%%%%%%%%%%%%%%%%%%%%%%%%%%%%%%%%%%%%%%%%%

%\section{Quiver gauge theories and their D3-brane realization}
\section{Quiver gauge theories from toric Calabi-Yau cone} \label{D3.sec}

In this section, we give a quick overview of $\mathcal{N}=1$ superconformal quiver gauge theories and their string theory realizations using D3-branes and Calabi-Yaus. Readers already familiar with earlier developments in quiver gauge theories can skip \S\ref{quiver.subsec} and refer back to it as needed.

\subsection{Preliminaries on quiver gauge theories} \label{quiver.subsec}
\subsubsection{Quiver diagram and quiver gauge theories} \label{quiverdiagram.subsubsec}
\begin{figure}[htbp]
\begin{center}
\begin{tabular}{cc}
\includegraphics[scale=0.38]{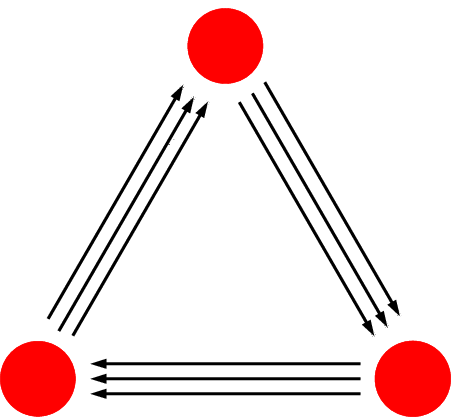} &
\includegraphics[scale=0.13]{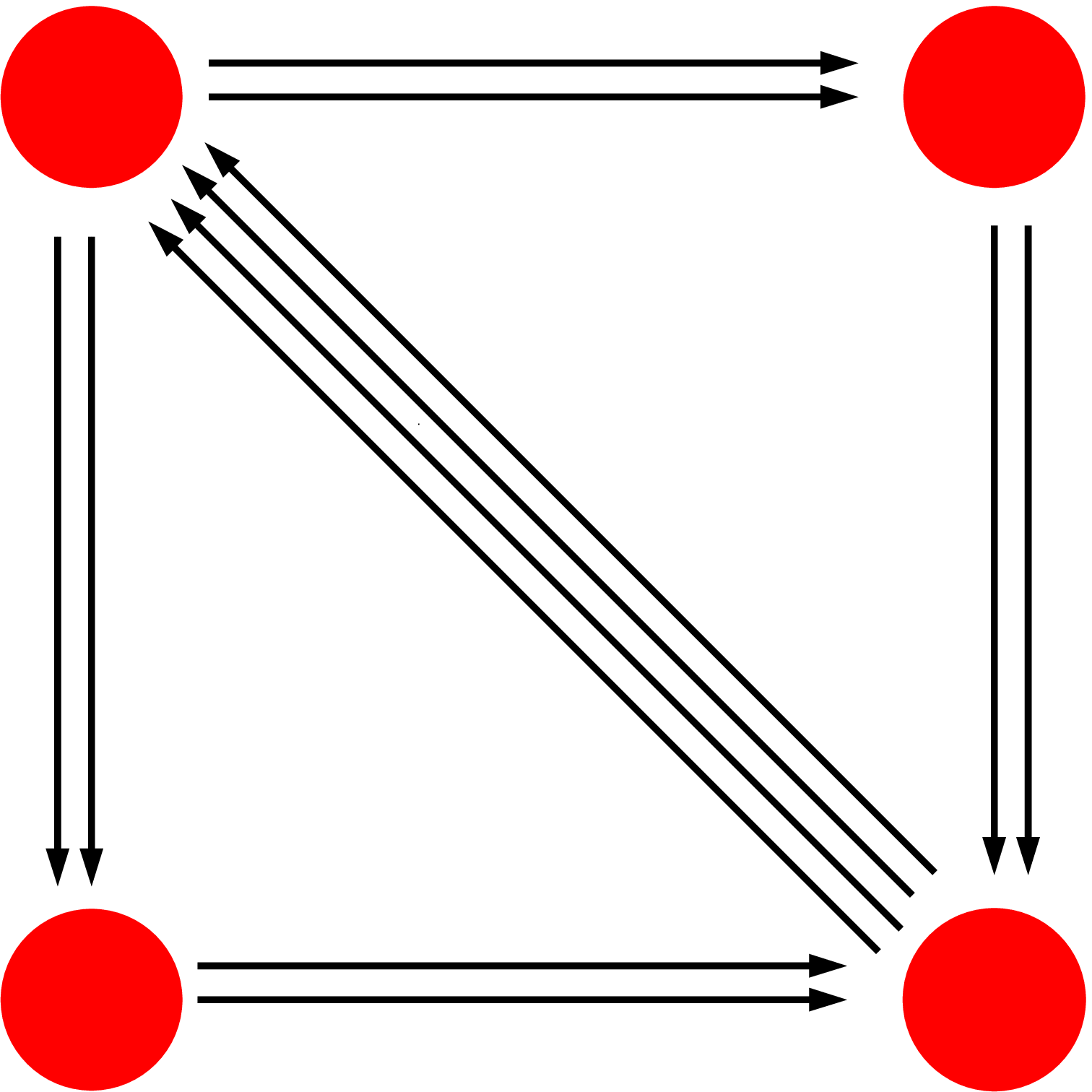}
\end{tabular}
\end{center}
\caption{Examples of quiver diagram. As will be explained below, the quiver on the left correspond to a toric Calabi-Yau cone $\bb{C}^3/\bb{Z}_3$, and the one on the right corresponds to $K_{\bb{P}^1\times\bb{P}^1}$. These are the quiver diagrams corresponding to  bipartite graphs shown in Figure \ref{BTfirst}.}
\label{quivereg}
\end{figure}

The term `quiver' was first introduced into mathematics by a mathematician Gabriel in the early 70's\cite{Gabriel} (actually, instead of `quiver' he used the German word `K\"{o}cher'.). 
Although dictionary meaning of quiver is ``a portable case for holding arrows'', in physics and mathematics quiver (or quiver diagram) simply means an ``oriented graph''. By an oriented graph we mean a collection of vertices (nodes) and oriented edges (links, arrows)\footnote{In this paper, we make no distinction between the words `vertex' and `node'. The same applies to the words `edge' and `link'.}. Some examples of quiver diagrams are shown in Figure \ref{quivereg}.

Note in some cases some arrows might start from one vertex and end at the same vertex. Also, in general we have multiple arrows going from one vertex to another\footnote{It is said that oriented graphs we consider here are called `quivers'  because multiple arrows as in Figure \ref{quivereg} look similar to arrows in stored in a case.}, as shown in Figure \ref{quivereg}. 

\begin{figure}[htbp]
\begin{center}
\centering{\includegraphics[scale=0.5]{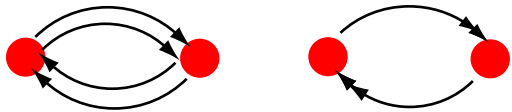}}
\caption{Two different ways of representing the same quiver with multiple arrows. This is the quiver corresponding to conifold, as will be explained below.}
\label{quiverarrows}
\end{center}
\end{figure}

When we have multiple arrows between two vertices, two different representations are used as in Figure \ref{quiverarrows}, but they both represent the same quiver.

So far, a quiver is simply a graph, but then what is the real meaning of this graph in physics? To a physicist, choosing a quiver means to specify a gauge theory (quiver gauge theory). Let us explain how to construct a gauge theory from a quiver diagram.

\begin{itemize}
\item First, to each vertex we associate a gauge group $U(N)$. We do not bother about the differences between $U(N)$ and $SU(N)$ for the most of this paper . The rank $N$ of the gauge group can differ from one vertex to another, as we will discuss in \S\ref{fractional.subsec}, but for the moment we simply take all to them to be $N$.

Also, although it is also possible to extend the construction to other gauge groups, e.g. $SO$ or $Sp$, we restrict attention to $U(N)$ case for the moment. $SO$ and $Sp$ gauge groups appear in orientifolded brane tilings\cite{Franco:2007ii,IKY}. See also \S\ref{orientifold.subsec} for brief discussion.

\item Second, to each oriented edge, we assign a bifundamental field. This field transforms as fundamental with respect to the gauge group at the startpoint of the edge, and anti-fundamental with respect to the gauge group at the endpoint (Figure \ref{bifundex}). When we have multiple edges, we have precisely as many bifundamentals as the number of edges.

\begin{figure}[htbp]
\begin{center}
\centering{\includegraphics[scale=0.5]{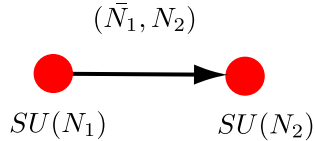}}
\caption{An arrow in a quiver diagram represents a bifundamental field, which transforms as $(\bar{N_1},N_2)$ under $SU(N_1) \times SU(N_2)$ gauge groups.}
\label{bifundex}
\end{center}
\end{figure}

For example, as a special case, when the startpoint and endpoint of an arrow coincide, the bifundamental field is an adjoint representation of the gauge group. In this sense, bifundamentals are generalization of adjoint representations\footnote{In orbifold case, bifundamental fields actually come from adjoint scalars in $\mathcal{N}=4$ theories.}. We can also consider inclusion of fundamentals and antifundamentals as will be explained in \S\ref{flavor.subsec}, and anti-symmetric and symmetric representations in orientifolded brane tilings \ref{orientifold.subsec}.
\end{itemize}

So far, we have not specified spacetime dimension, but throughout this review, all quiver gauge theories are defined in four spacetime dimensions. Also, in this review we only consider theories with at least $\mathcal{N}=1$ supersymmetry (in some examples we have $\mathcal{N}=4$ or $\mathcal{N}=2$ supersymmetry). In other words, we are going to consider four-dimensional $\mathcal{N}=1$ supersymmetric quiver gauge theories.

In order to actually specify $\mathcal{N}=1$ theory, quiver diagram is not enough and we need to know superpotential (we simply assume that the K\"ahler potential is canonical). We thus need quiver diagram and superpotential to actually specify four-dimensional $\mathcal{N}=1$ supersymmetric quiver gauge theories. We are going to return to superpotentials later in \S\ref{quiver.subsubsec}, when we discuss the relation between fivebrane diagrams and quiver gauge theories.

\subsubsection{Cancellation of gauge anomalies}\label{anomaly.subsubsec}
The important thing to notice is that quiver gauge theories are in general chiral and we have non-trivial condition for cancellation of anomalies. In order to write down  this condition, we label vertices of quiver diagram by $a,b,\ldots $ and edges by $I,J,\ldots$. For an edge $I$ and one of its endpoint $a$, define $\sigma(I,a)$ by
\beq
\sigma(I,a)=
\begin{cases}
+1 & \textrm{ (when $a$ is an endpoint of $I$), } \\
-1 & \textrm{ (when $a$ is a startpoint of $I$).} 
\end{cases}
\label{sigma}
\eeq

Suppose we have gauge group $SU(N_a)$ at each vertex $a$. Then, using the definition as in \eqref{sigma}, the gauge anomaly cancellation\footnote{For $SU(2)$ case, we do not have ordinary gauge anomaly, but we should care about global anomaly (Witten's anomaly) \cite{Witten:1982fp}. 
} with respect to gauge group at vertex $a$ is represented by
\beq
\sum_{\genfrac{}{}{0pt}{}{I}{I=(a,b)}} \sigma(I,a) N_b=0, \label{anomcancel}
\eeq 
where the summation is over all edges $I$ one of whose endpoints is $a$, and the other endpoint is denoted $b$. The rank of the gauge group at vertex $b$ is denoted $N_b$. For the discussion in \S\ref{fractional.subsec}, let us rewrite \eqref{anomcancel} in another form. Define $\sigma(a,b)$ for two gauge groups $a, b$ (connected by an edge $I$) by
\beq
\sigma(a,b)=\sigma(I,a). \label{sigma2}
\eeq
In other words, for two vertices $a, b$ of the quiver digram connected by an edge $I$, $\sigma(a,b)=+1$ if the bifundamental field $I$ transforms as $(N_a, \bar{N_b})$ under $SU(N_a)\times SU(N_b)$, and $\sigma(a,b)=-1$ if $I$ transforms as $(\bar{N_a}, N_b)$.
Then \eqref{anomcancel} is rewritten as
\beq
\sum_{b\in a} \sigma(b,a) N_b=0, \label{anomcancel2}
\eeq
where $b\in a$ means that $b$ is adjacent to $a$. Also, in general there are several edges connecting $a$ and $b$, and in that case the summation over all such edges is implicitly taken in the formula \eqref{anomcancel2}.

Let us explain the origin of this formula.
An arrow which starts at edge $a$ and ends at edge $b$ corresponds to a bifundamental transforming as $(\overline{N_a},N_b)$ under $SU(N_a)\times SU(N_b)$. From the viewpoint of $SU(N_a)$ gauge group, this bifundamental looks like a set of $N_b$ anti-fundamentals, which contributes $-N_b=\sigma(I,a) N_b$ to gauge anomaly. An arrow with opposite orientation then contributes with opposite sign and is again given by $\sigma(I,a) N_b$. The condition for vanishing of the total gauge anomaly is therefore given by \eqref{anomcancel}.

\begin{figure}[htbp]
\centering{\includegraphics[scale=0.4]{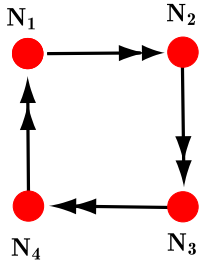}}
\caption{The quiver diagram corresponding to canonical bundle over $\bP^1 \times \bP^1$. In this case, possible anomaly restricts ranks to be $(N_1,N_2,N_3,N_4)=(N,N+M,N,N+M)$. We will see a meaning of this anomaly-free rank assignments from the viewpoint of fivebranes systems in \S\ref{fractional.subsec}.}
\label{F0quiverI}
\end{figure}

As an example, let us take the example of the quiver shown in Figure \ref{F0quiverI}. In this example, \eqref{anomcancel} simply says
\beq
N_1=N_3,\ \ N_2=N_4. \label{N13N24}
\eeq
Thus we have $(N_1,N_2,N_3,N_4)=(N,N+M,N,N+M)$, where $N$ is a non-negative integer and $M$ is an integer with $N+M \ge 0$. Here appearance of $M$ means we can assign different ranks to difference vertices, and this corresponds to fractional branes in the Calabi-Yau setup explained in \S\ref{D3.subsec}. The meaning of such anomaly-free rank assignments from fivebrane viewpoint will be discussed in \S\ref{fractional.subsec}.

Consider the special case where $N_a$ are the same for all vertices. Then the condition \eqref{anomcancel} simply means that, for each vertex $a$, the number of incoming and outgoing arrows are the same. This condition is satisfied for the quiver in Figure \ref{F0quiverI}. We will see in \S\ref{strong.subsec} that this condition is automatically satisfied for all quiver gauge theories realized by brane tilings.

%%%%%%%%%%%%%%%%%%%%%%%%%%%%%%%%%%%%%%%%%%%%%%%%%%%%%%%%%%%%%%%%%%%%%%%%%%%%%%%

%\subsection{D3-branes probing non-compact singular Calabi-Yau} 
\subsection{Quiver gauge theories from D3-branes and Calabi-Yau}\label{D3.subsec}

\subsubsection{Orbifold singularities}\label{orbifold.subsubsec}
Perhaps the reader might wonder at this point why we want to study such complicated quiver gauge theories. Certainly, we have many motivations, as we amply discuss in this review.  Here we give one important motivation: quiver gauge theories appear quite naturally in string theory compactifications.

\begin{figure}[htbp]
\centering{\includegraphics[scale=0.19]{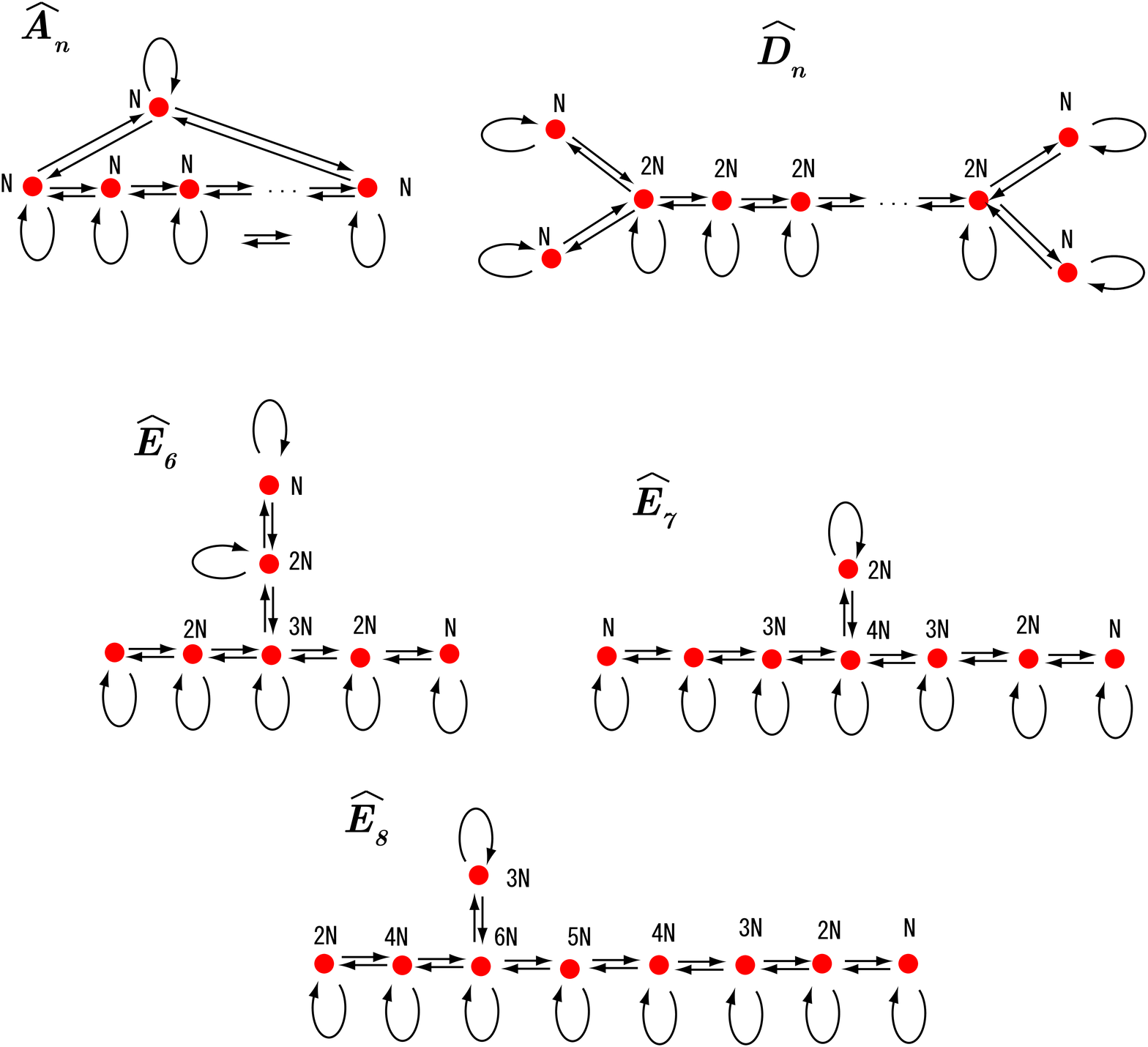}}
\caption{Extended Dynkin diagrams for semisimple simply-laced Lie-algebras. This classification is in one-to-one correspondence with the classification of discrete subgroups of $SU(2)$ shown in Table \ref{SU2}. }
\label{Dynkin}
\end{figure}

Historically speaking, this is actually the original motivation to study quiver gauge theories. In 1994 celebrated work, Douglas and Moore\cite{Douglas:1996sw} has discovered that, by probing ALE spaces by D-branes, we have four-dimensional $\mathcal{N}=2$ superconformal quiver gauge theories on the probe D-branes. More specifically, let $\Gamma$ be a discrete subgroup of $SU(2)$. Such discrete subgroup is classified completely by Felix Klein\cite{Klein} (see Table \ref{SU2}). Interestingly enough, this classification coincides with the classification of semisimple simply-laced Lie algebras (Figure \ref{Dynkin}).

\begin{table}[htbp]
\caption{Classification of discrete subgroups of $SU(2)$. More precisely, the subgroups shown below is that of $SO(3)$, and corresponding lift to $SU(2)$ is called with `binary' in front, such as ``binary tetrahedral group''. This classification is in one-to-one correspondence with the classification of Dynkin diagrams, or semisimple Lie algebras, as shown in Figure \ref{Dynkin}.}
\begin{center}
\begin{tabular}{|c||c|}
\hline
$A_n$ & cyclic \\
\hline
$D_n$ & dihedral \\
\hline
$E_6$ & tetrahedral \\
\hline
$E_7$ & octahedral \\
\hline
$E_8$ & icosahedral \\
\hline
\end{tabular}
\end{center}
\label{SU2}
\end{table}

Consider the orbifold space $\bC^2/\Gamma$, whose resolutions are called Asymptotically Locally Euclidean (ALE) spaces. Then the question is what kind of gauge theory we have on the stack of $N$ D-branes. Then answer\cite{Douglas:1996sw,Johnson:1996py,Lawrence:1998ja} is that the theory is quiver gauge theory, with the quiver diagram given as follows.

Consider discrete subgroup $\Gamma$ of $SU(2)$, and denote the set of all irreducible representations by $\{ \rho_i \}$. Since $\Gamma$ is a subgroup of $SU(2)$ (which acts on $\bC^2$), we can consider standard two-dimensional representation denoted by $\rho_r$. Then decompose the tensor product into irreducible representations:
\beq
\displaystyle \rho_r\bigotimes \rho_i =\bigoplus_j a^{ji}\rho_j   .
\label{rhodecomp}
\eeq
From this decomposition we obtain a quiver diagram as follows. First, prepare a node for each irreducible representation $\rho_i$. Next, write $a^{ij}$ arrows starting from the node $\rho_i$ and ending at $\rho_j$. The supersymmetric quiver gauge theory specified by such quiver diagram is the theory we have on the D-branes probing $\bC^2/\Gamma$.

We have another way to write quiver diagram from $\Gamma$. The quotient space $\bC/\Gamma$ has a rational double point at the origin. Take a minimal resolution of this space, and we have exceptional divisors. Then we have a graph as a dual of the configuration of exceptional divisors.

%\begin{figure}[htbp]
%\centering{\includegraphics[scale=0.45]{dual.eps}}
%\caption{The configuration of exceptional divisors of minimal resolution of quotient space $\bC^3/\Gamma$ is shown in (a), in the case $\Gamma$ is  The dual graph of (a) is exactly the corresponding Dynkin diagram (b).}
%\label{dual.eps}
%\end{figure}

Interestingly enough, the quiver diagram obtained from $\Gamma$ by these two different methods coincide, and moreover is exactly the same with the extended Dynkin diagram of ADE type, shown in Figure \ref{Dynkin}. This is famous McKay correspondence\cite{McKay} (see \cite{Slodowy,Reid} for reviews) \footnote{In \S\ref{quiver.subsubsec}, we explain one more method to obtain quiver diagram from $\Gamma$.}. Thus McKay correspondence has a natural role to play in string theory!

Moreover, if we consider the moduli space of these $\mathcal{N}=2$ quiver gauge theories, the moduli space coincides with the ALE space itself, which is the elegant physical realization of the hyperK\"ahler construction of ALE spaces\cite{Kronheimer} in mathematics.

We can generalize this discussion to subgroup $\Gamma$ of $SU(3)$. The classification of such $\Gamma$ was done by Blichfeldt in 1917\cite{Blichfeldt}, although in this case classification is much more subtle. Similar to $SU(2)$ case, we consider the orbifold $\bC^3/\Gamma$. Such orbifolds leave $\mathcal{N}=1$ supersymmetry, and the corresponding gauge theory is the $\mathcal{N}=1$ supersymmetric quiver gauge theory.
 
\begin{figure}[htbp]
\centering{\includegraphics[scale=0.25]{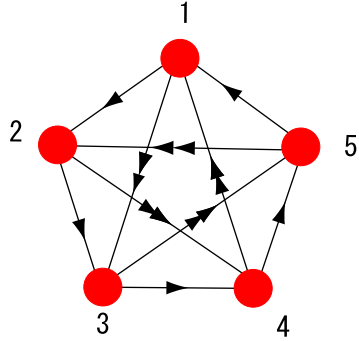}}
\caption{Example of McKay quiver for $(1,2,2)/5$. We will recover this result in \S\ref{strong.subsec}.}
\label{McKayeg}
\end{figure}

Again from the rules \eqref{rhodecomp}, we can write down the corresponding quiver diagram by decomposition of representation, which is a trivial generalization\footnote{The only complication in $SU(3)$ case is that we also need superpotential, since SUSY is broken down to $\mathcal{N}=1$. For Abelian orbifolds, superpotential is obtained by brane tiling techniques, as we will explain in later sections.} of the $SU(2)$ case. For examples of non-Abelian orbifolds, see \cite{Hanany:1998sd,He:2002fp}. Here we only consider Abelian orbifolds. As a simple example, we show in Figure\ref{McKayeg} the McKay quiver corresponding to $\bZ_5$ orbifold of $\bC^3$, with the $\bZ_5$ action given by $(z_1,z_2,z_3)\to (\omega z_1,\omega^2 z_2, \omega^2 z_3)$ (here $\omega^5=1$). Such a discrete subgroup is sometimes denoted $(1,2,2)/5$. 

\begin{figure}[htbp]
\centering{\includegraphics[scale=0.25]{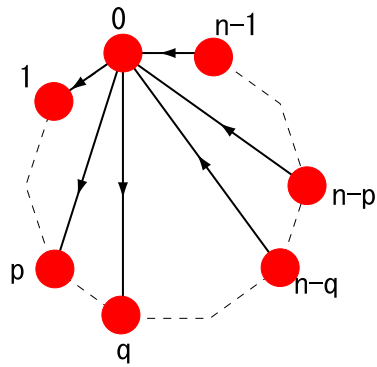}}
\caption{Example of McKay quiver for $(1,p,q)/(1+p+q)$. This reduces to Figure \ref{McKayeg} if we set $p=q=2$.}
\label{McKayeg2}
\end{figure}

More generally, if we consider the orbifold $(1,p,q)/(1+p+q)=\bZ_{1+p+q}$, we have the quiver shown in Figure \ref{McKayeg2}. These McKay quivers will appear again in \S\ref{quiver.subsubsec}, in the discussion of brane tilings.

%%%%%%%%%%%%%%%%%%%%%
\subsubsection{Toric Calabi-Yau singularities} \label{toric.subsubsec}

The orbifold examples we have been discussing so far are interesting, but we want to consider more general case. Of course, if we consider arbitrary Calabi-Yau, we are completely at a loss, since for non-orbifold case we no longer have the rules as in \eqref{rhodecomp}. Our strategy here is to concentrate on toric Calabi-Yau three-folds (although we will refer to non-toric case as well in some parts of this review). 

For reviews of toric geometry, see excellent books\cite{Fulton,Oda}, and the physics paper\cite{Leung:1997tw}. Here we remind the reader that toric Calabi-Yau manifold is a  special class of Calabi-Yau manifold which is specified by a convex polytope, the so-called toric diagram (or more precisely, fan). In other words, everything we need to know about toric Calabi-Yau are contained in the toric diagram, at least in principle. Of course, these are special class of Calabi-Yaus. For example. when $\Gamma\subset SU(2)$ is a DE-type discrete subgroup, $\bC^2/\Gamma\times \bC$ is not toric. It should be stressed, however, that toric Calabi-Yaus contains infinitely many examples of Calabi-Yaus and they constitute an important subclass of Calabi-Yau manifolds.

\begin{figure}[htbp]
\begin{center}
\includegraphics[scale=0.4]{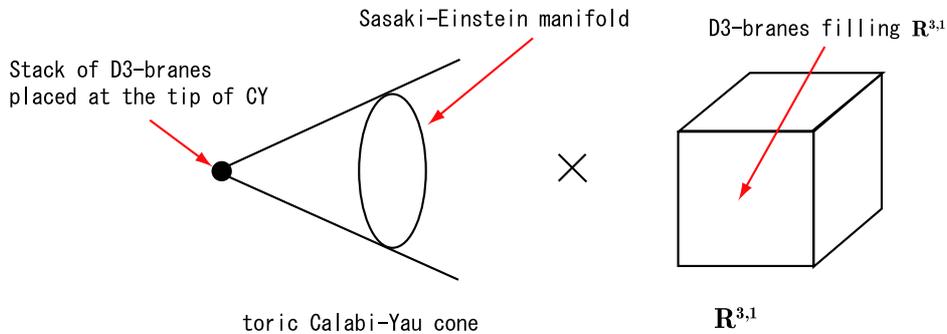}
\end{center}
\caption{We probe the tip of non-compact singular Calabi-Yau by a stack of $N$ D3-branes.}
\label{cone.eps}
\end{figure}

We now prepare non-compact Calabi-Yau 3-fold which have cone-type singularity. Then we probe this geometry by D3-branes (Figure \ref{cone.eps}), neglecting the possible back-reaction onto geometry. As shown in the Figure, D3-branes are placed transverse to Calabi-Yau.
Also, by a cone we mean a Calabi-Yau 3-fold, which is obtained from some five-dimensional manifold $(S,g_S)$ by the metric cone construction\footnote{Such manifolds, i.e., manifolds whose metric cone is Calabi-Yau, are called Sasaki-Einstein manifolds and going to play an important role in AdS/CFT, as we will see in section \S\ref{AdSCFT.sec}.}. That is, the metric of Calabi-Yau $(C(S),g_{C(S)})$ can be expressed in the form 
\begin{equation}
ds_{C(S)}^2=dr^2+r^2 ds^2.
\end{equation}
The manifold $(S,g_S)$ is called toric Sasaki-Einstein manifolds, which is the topic of \S\ref{AdSCFT.sec}.

\begin{figure}[htbp]
\centering{\includegraphics[scale=0.40]{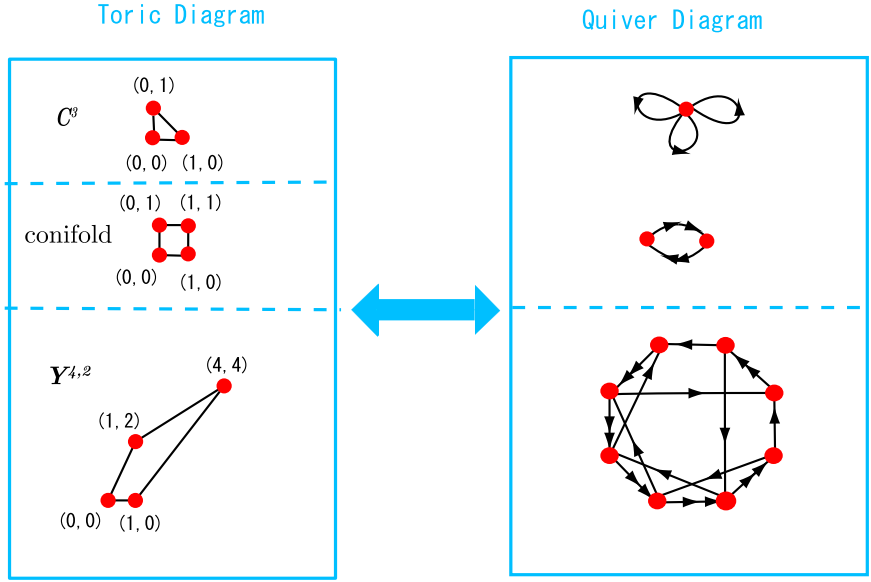}}
\caption{Toric diagram versus quiver diagram. From above, the example of $\bC^3$, conifold and $Y^{4,2}$ is shown. Here $Y^{4,2}$ is a Sasaki-Einstein manifold whose explicit metric is known, which will be discussed in \S\ref{explicit.subsubsec}. Their relation is a relation between geometry and gauge theory, or in AdS/CFT language gravity and gauge theory.}
\label{theproblem}
\end{figure}

Since the Calabi-Yau is non-compact, and the volume of Calabi-Yau is infinite, gravity decouples and we have gauge theory on these probe D-branes. The moduli space of this gauge theory should coincide with the toric Calabi-Yau cone. The general belief since 90's was that the gauge theory is a quiver gauge theory, at least for toric Calabi-Yaus.
But in this general setting, it is not obvious  at all which Calabi-Yau gives which quiver gauge theory. 
Since toric Calabi-Yaus are specified by the toric diagram and since quiver gauge theories are specified by the quiver diagram, the problem is how to translate the geometry (toric diagram) into gauge theory (quiver diagram) (Figure \ref{theproblem}).

\begin{figure}[htbp]
\centering{\includegraphics[scale=0.45]{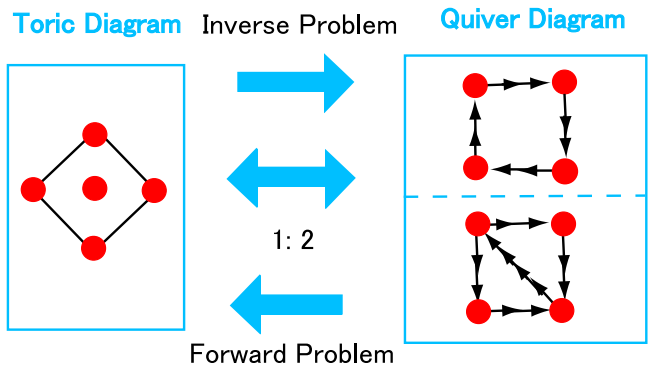}}
\caption{The problem to obtain toric diagram from quiver diagram is called the forward problem, and the other direction called the inverse problem. This Figure shows an example of $K_{\bP^1\times \bP^1}$, and it is known in the literature that at least two different quiver diagrams correspond to the same toric diagram. In general, the correspondence between toric diagram and quiver diagram is one-to-many.}
\label{manytoone}
\end{figure}

In the literature, the problem of obtaining toric diagram from quiver diagram is called the forward problem, and other direction, namely the problem of obtaining quiver diagram from toric diagram, is called the inverse problem. The reason for these names is that in general several different quivers correspond to the same toric diagram, and the correspondence between toric diagrams and quiver diagrams are in general one-to-many (see Figure \ref{manytoone}). This phenomena is dubbed ``toric duality'', and it is believed\cite{Beasley:2001zp} that this is related to the Seiberg duality\cite{Seiberg:1994pq}. We will discuss this point more in detail in \S\ref{Seiberg.subsec}.

If you look at this correspondence from a slightly different viewpoint, you will see that it is essentially the setup of AdS/CFT correspondence. If you consider the back-reaction of D-branes onto geometry, we have AdS near-horizon geometry\footnote{However, when we consider other ingredients, such as fractional branes, the gauge theory is not necessarily conformal. See \S\ref{fractional.subsec} for discussion on fractional branes.}. AdS/CFT in this case states that type IIB string theory on $AdS_5\times S$ is equivalent to four-dimensional $\mathcal{N}=1$ superconformal quiver gauge theories. This will be studied in detail in \S\ref{AdSCFT.sec}.

So, the problem is to find the precise relation between toric diagrams and quiver diagrams. Before the discovery of brane tilings, we have no efficient way to answer this question. Although general algorithm already existed\cite{Feng:2000mi} using partial resolutions, computation often becomes unpractical 
as the toric diagram becomes more and more complicated. Also, that method does not tell you about the superpotential of $\mathcal{N}=1$ theories.

All these problems are solved by the method of brane tilings, as we will see. 
Brane tiling gives us ``fast forward algorithm'' and ``fast inverse algorithms'' which solve these problems. But before describing the details of such algorithms, we should explain the physical meaning of brane tilings.

\subsection{Summary}
In this section, we prepared some basic knowledge needed to understand the rest of this paper. In \S\ref{quiver.subsec}, we briefly reviewed quiver diagram and quiver gauge theories. The condition for gauge anomaly cancellation was also discussed. In \S\ref{D3.subsec}, we described stringy realization of quiver gauge theories from non-compact Calabi-Yaus probed by D3-branes. For the orbifold case, the story is well-known, but for more general toric Calabi-Yaus, we do not know the precise relation between toric diagram and quiver diagram. As we will see in the next section, fivebrane configurations represented by brane tilings give a clear-cut answer to this problem.

%%%%%%%%%%%%%%%%%%%%%%%%%%%%%%%%%%%%%%%%%%%%%%%%%%%%%%%%%%%%%%%%%%%%%%%%%%%
%\section{Construction of quiver gauge theories by D5-branes and NS5-brane}
\section{Brane tiling as a fivebrane system}\label{D5NS5.sec}

In this section we explain configuration of D5-branes and NS5-brane which is represented by a bipartite graph. This fivebrane system provides string theory realization of a four-dimensional $\mathcal{N}=1$ superconformal quiver gauge theory. Along the way we will see a clear physical interpretation of brane tilings.
The explanation of this section is an expanded version of part of \S 1 of our paper\cite{IIKY}.

\subsection{The strong coupling limit} \label{strong.subsec}

\subsubsection{Step-by-step construction of fivebrane systems} \label{step-by-step.subsubsec}
The goal of this subsection is to construct a fivebrane system which represents  a $\mathcal{N}=1$ superconformal quiver gauge theory. 
The final fivebrane system is slightly complicated, so we proceed step by step.

As a first step, we consider type IIB string theory and a stack of $N$ D5-branes.  Then, as is well-known, we have $U(N)$ (or $SU(N)$) gauge theory on the D5-branes. 
Here we do not bother the difference of $U(N)$ and $SU(N)$  because $U(1)$ factor decouples in IR \footnote{However, these $U(1)$s play crucial roles in the discussion of baryonic $U(1)$ global symmetries in \S\ref{versus.subsec}.}. 

Since we want to have four-dimensional gauge theories, two of six directions of D5-branes are redundant. We therefore choose to compactify two spatial directions on two-dimensional torus $\bb{T}^2$ with radius $R$. Out of ten-dimensional spacetime coordinates $x^0, x^1,\ldots x^9$ ($x^0$ being time coordinate and others spatial coordinate) we take $x^5$ and $x^7$ to be directions of ${\mathbb T}^2$. 

We now have $N$ D5-branes wrapping $\mathbb{T}^2$. If $R$ is small and Kaluza-Klein modes decouple, we have four-dimensional $\mathcal{N}=4$ supersymmetric Yang-Mills theory. But this is certainly not what we want! We want gauge theory with multiple gauge groups, and we want to reduce supersymmetry down to $\mathcal{N}=1$.

\begin{figure}[htbp]
\begin{center}
\includegraphics[scale=0.4]{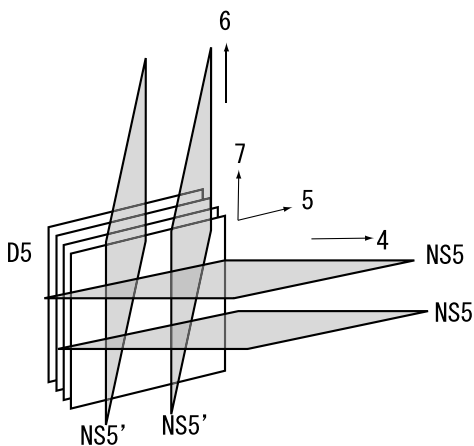} 

\caption{The worldvolume of D5-branes are divided by NS5-branes into several different regions. The two directions of D5-branes shown in the figure are $\bb{T}^2$ directions ($x^5$ and $x^7$), and NS5-branes intersect the D5-brane with 1-cycles on $\bb{T}^2$. This example corresponds to the case of the conifold, as we will see. Although in this figure two NS5-branes seemingly intersect at 1-cycle, that intersection is an artifact of visualizing four-dimensional figure in three dimensions. The two NS5-branes span in 45 and 67 directions, respectively.}
\label{NS5divide}
\end{center}
\end{figure}

\begin{figure}[htbp]
\begin{center}
\includegraphics[scale=0.4]{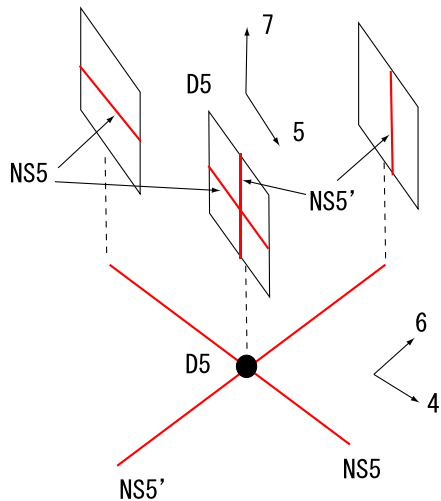} 

\caption{We have semi-infinite cylinders of NS5-branes. The D5-branes shown in Figure \ref{NS5divide} is a point in 46-directions shown here, and is located at the intersection point of NS5 cylinders, that is, at the tip of the Calabi-Yau cone in the T-dual Calabi-Yau setup. The lines shown in 46 directions are identical to the web diagram.}
\label{NS5junction}
\end{center}
\end{figure}

For those purposes we use another ingredient, NS5-brane.
 In order to obtain multiple gauge groups, we divide the D5 worldvolume using semi-infinite cylinder of NS5-branes, as shown in Figure \ref{NS5divide}, \ref{NS5junction} and Table \ref{conifoldtbl}.

Although originally the same D5-brane, each D5-brane region is now separated by semi-infinite cylinder of NS5-branes, and each region has its own gauge group $SU(N)$, and we have multiple gauge groups, as desired. Also, since the introduction of NS5-brane in two different directions leaves quarter of original $\mathcal{N}=4$ supersymmetry, we are left with $\mathcal{N}=1$ supersymmetry. Thus we have succeeded in obtaining both $\scN=1$ SUSY and multiple gauge groups.

\begin{table}[htbp]
\caption{Brane configuration corresponding to Figure \ref{NS5divide} and \ref{NS5junction}. This is the conifold case, as will be explained below. The two directions (57) are compactified on $\bb{T}^2$. Compare with Figure \ref{NS5divide} and Figure \ref{NS5junction}. We have two semi-infinite NS5-cylinders extending in 45-directions. We also have two in 67-directions, which are denoted by NS5'. 
Due to the introduction of NS5-brane in two directions, supersymmetry is broken to $\mathcal{N}=1$. In general, the brane configuration takes more general form shown in Table \ref{config_fst.tbl}.}
\label{conifoldtbl}
\begin{center}
\begin{tabular}{ccccccccccc}
\hline\hline
&0&1&2&3&4&5&6&7&8&9\\
\hline
D5 & $\circ$ & $\circ$ & $\circ$ & $\circ$ & & $\circ$ & & $\circ$ & &\\
NS5 & $\circ$ & $\circ$ & $\circ$ & $\circ$ &$\circ$& $\circ$ & & & & \\
NS5' & $\circ$ & $\circ$ & $\circ$ & $\circ$ & &  &$\circ$&$\circ$&  & \\
\hline
\end{tabular}
\end{center}
\label{conifoldconfig}
\end{table}

\begin{figure}[htbp]
\begin{center}
\centering{\includegraphics[scale=0.4]{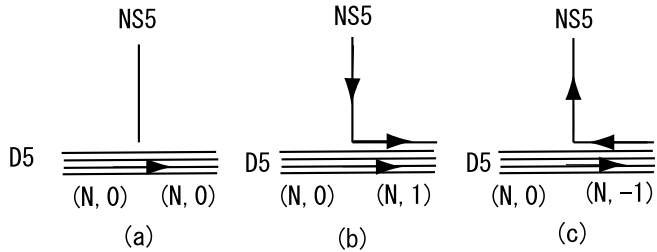}}
\caption{Junction of $N$ coincident D5 and NS5-brane. Due to the conservation of NS5-charge, junction (a) is not allowed and we must have (b) or (c), depending on the orientation of NS5-brane. In case (b), we have bound states of $N$ D5-brane and $1$ NS5-branes, or $(N,1)$-brane. In case (c), we also have bound states of $N$ D5-branes and $1$ NS5-brane, but since orientations of D5 and NS5 are opposite, we have $(N,-1)$-brane.}
\label{junction}
\end{center}
\end{figure}

But this is not the end of the story. In our brane configuration, we have junctions of D5-branes and NS5-brane. As first pointed out in \cite{Imamura:2006db}, in order for the conservation of NS5-charge, the junction should be either (b) or (c) of Figure \ref{junction}, depending on the orientation of NS5.

\begin{figure}[htbp]
\begin{center}
\centering{\includegraphics[scale=0.5]{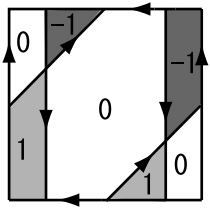}}
\caption{In this case, each region of $\bb{T}^2$ divided by cycles of NS5-branes is either $(N,0)$, $(N,1)$ or $(N,-1)$-brane. Numbers shown represent NS5-charges. $(N,+1)$-branes (resp. $(N,-1)$-branes) are shaded light (resp. dark) gray. In this case no $(N,k)$-brane with $|k|\ge 2$ appears.}
\label{division1}
\end{center}
\end{figure}

\begin{figure}[htbp]
\begin{center}
\centering{\includegraphics[scale=0.8]{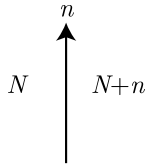}}
\caption{The convention of sign of NS5-charge. The arrow represents a cycle of NS5-brane, When we cross an arrow,
 NS5-charge increase or decreases by one, depending on orientation of the arrow.}
\label{inflow2}
\end{center}
\end{figure}

From this fact, we see that some regions of $\bb{T}^2$ becomes bound states of $N$ D5-branes and $1$ NS5-brane. In general, bound states of $N$ D5-branes and $k$ NS5-branes are called $(N,k)$-branes. In the case $k<0$, $(N,k)$-brane is a bound state of $N$ D5 and $|k|$ NS5, with D5 and NS5 having opposite orientations. In this language, we should expect at least $(N,1)$-branes and $(N,-1)$-branes to appear. The example of such a division of $\bT^2$ is shown in Figure \ref{division1}. We call such diagrams {\em fivebrane diagrams} \cite{IKY}, since they represent the structure of fivebrane systems. In the fivebrane diagram, we have used the convention of NS5-charge as shown in Figure \ref{inflow2}.

\begin{figure}[htbp]
\begin{center}
\centering{\includegraphics[scale=0.5]{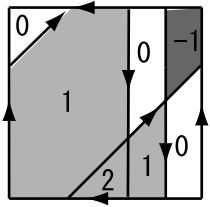}}
\caption{By changing the position of one 1-cycle from Figure \ref{division1}, we obtain this figure. In this case, one of the regions of $\bb{T}^2$ divided by cycles of NS5-branes are $(N,2)$-brane.}
\label{division2}
\end{center}
\end{figure}

Interestingly enough, in general $(N,k)$ branes with $|k|\ge 2$ appear as in Figure \ref{division2}. In this review, we do not discuss these cases for the most part, since we are not quite sure what kind theory we have on these branes\footnote{Presumably that theory should be some deformation of ordinary quiver gauge theories, since by changing the position of cycles of NS5-branes we have a transition between the two theories.}; at least, they are not conventional quiver gauge theories. (see, however, \S\ref{strong.subsubsec}, particularly Figure \ref{spp-phase-2.eps} for discussion of such brane configurations.).

\begin{figure}[htbp]
\begin{center}
\centering{\includegraphics[scale=0.45]{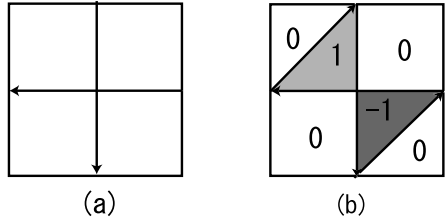}}
\caption{The configuration like (a) is prohibited by \eqref{pq0}, whereas (b) is possible.}
\label{consistentNS5}
\end{center}
\end{figure}

Now we know some consistency conditions are imposed on the NS5-cycles. Since 57-directions are torus, the NS5-charge should be the same after we go around an arbitrary cycle of $\bb{T}^2$. More formally, let $\balpha_{\mu}$ be the cycle of NS5-brane. Here $\mu$ runs from $1$ to $d$, where $d$ is the number of cycles of NS5-brane. Let us denote the winding number of cycle $\balpha_{\mu}$ by $(p_{\mu},q_{\mu})$. Here $p_{\mu}$ and $q_{\mu}$ refers to the winding number in $\alpha$ and $\beta$-cycles, respectively. Then we have

\beq
\sum_{\mu} p_{\mu}=\sum_{\mu} q_{\mu}=0,\label{pq0}
\eeq
or
\beq
\sum_{\mu=1}^d \balpha_{\mu}=0 \mbox{ in } H_1(\bT^2,\bZ). \label{pq0-2}
\eeq 
As an example, the configuration as in Figure \ref{consistentNS5} (a) is not allowed, but (b) is a consistent configuration.

%Some might wonder about the conservation of D5-brane charge, but that is trivia%lly satisfied.

\begin{figure}[htbp]
\begin{center}
\centering{\includegraphics[scale=0.45]{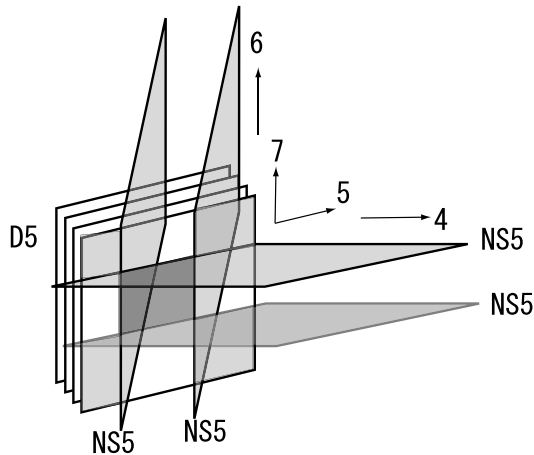}}
\caption{All the NS5-brane now join into a single NS5-brane, although seemingly we have many NS5-branes. In the region shaded dark gray, the NS5-brane has opposite orientation from others. Like Figure \ref{NS5divide}, the cycles of NS5-branes do not intersect on a line, which is an artifact of writing figure in three-dimensions. }
\label{NS5patch}
\end{center}
\end{figure}

Now we should look at NS5-branes slightly differently from previous discussions. In our previous discussions, we have simply put several NS5-branes orthogonal to D5. But now, we also have NS5-branes parallel to D5-branes, as shown in Figure \ref{junction}. These seemingly different NS5-branes join together and we have a single NS5-brane! (see Figure \ref{NS5patch} for example) As in (b) and (c) of Figure \ref{junction}, one should also remember that the orientation of the NS5-brane is opposite in some regions.

Of course, it should be kept in mind that that the shape of NS5-brane is in some sense singular. For example, as can be seen from Figure \ref{junction}, NS5-brane bends 90 degrees and change its direction suddenly at junctions. As we will explain later in \ref{note.subsec}, this is simply because we are considering some limit (the strong coupling limit) in which the brane configuration simplifies dramatically. The shape of NS5-brane is smooth for general string coupling constant, and in the weak coupling limit even becomes a holomorphic curve (see \S\ref{weak.subsec}).

Since now we have single NS5-brane, the final brane configuration is, if you write NS5-worldvolume schematically as $\bb{R}^{3,1}\times \Sigma$, the one shown in Table \ref{config_fst.tbl}.
\begin{table}[htbp]
\caption{The brane configuration corresponding to brane tilings. The directions 5 and 7 are compactified, and $\Sigma$ is a two-dimensional surface in 4567 space. Note that all the semi-infinite cylinders of NS5-branes as in Table \ref{conifoldtbl} now merge into a single NS5-brane. We have so far used the example of the conifold, but the surface $\Sigma$ takes more general form, as we will explain.}
\label{config_fst.tbl}
\begin{center}
\begin{tabular}{ccccc|cccc|cc}
\hline\hline
& 0 & 1 & 2 & 3 & 4 & 5 & 6 & 7 & 8 & 9 \\
\hline
NS5 & $\circ$ & $\circ$ & $\circ$ & $\circ$ &\multicolumn{4}{c|}{$\Sigma$} & & \\
D5  & $\circ$ & $\circ$ & $\circ$ & $\circ$ && $\circ$ && $\circ$ & &\\
\hline
\end{tabular}
\end{center}
\end{table}
We should still remember that $\Sigma$ is not an arbitrary two-dimensional surface in 4567-space. It intersects $\bb{T}^2$ with 1-cycle. That is, one of its two directions is in the compact 5,7-directions, and the other one is in non-compact 4,6-directions.

\subsubsection{Relation with D3-brane picture} \label{D3.subsubsec}
The Tables \ref{conifoldtbl} and \ref{config_fst.tbl} are the key to understanding the relation with  D3-brane setup discussed in \S\ref{D3.subsec}. Since 5,7-directions are simply $\bb{T}^2$, we can take T-duality along these directions. Then D5-branes are turned into D3-branes, and NS5-branes are now turned into geometry, actually a Calabi-Yau manifold (As we mentioned above, one of 5,7-directions are orthogonal to NS5, so we are taking T-duality perpendicular to NS5, which turns NS5 into geometry). The net result is shown in Figure \ref{config_2nd.tbl}.
You will immediately notice that this is precisely the setup of the D3-brane probing Calabi-Yau which we already discussed in \S\ref{D3.subsec}.
\begin{table}[htbp]
\caption{The brane configuration obtained by taking T-duality in 5,7-directions of the fivebrane systems shown in Table \ref{config_fst.tbl}. D5-branes are turned into D3-branes, and NS5-brane is turned into a toric Calabi-Yau manifold. The system obtained by this way is precisely Calabi-Yau setup with D3-branes, which we already discussed in \S\ref{D3.subsec}.}
\label{config_2nd.tbl}
\begin{center}
\begin{tabular}{ccccc|cccccc}
\hline\hline
& 0 & 1 & 2 & 3 & 4 & 5 & 6 & 7 & 8 & 9 \\
\hline
CY3 &  &  &  &  & $\circ$ &$\circ$ &$\circ$ &$\circ$ &$\circ$ &$\circ$ \\
D3  & $\circ$ & $\circ$ & $\circ$ & $\circ$ &&  &&  & &\\
\hline
\end{tabular}
\end{center}
\end{table}

At this point you can understand why we started from D5-branes, with two spatial directions compactified. From D3-brane picture, the torus $\bb{T}^2$ are subtorus of the $U(1)^3$-isometry of toric Calabi-Yau, and we have taken T-duality along that $\bb{T}^2$ to turn Calabi-Yau into NS5-brane.

The important point here is that by taking T-duality, Calabi-Yau geometry is turned into NS5-branes. Although NS5-brane contains essentially the same information as Calabi-Yau, by taking appropriate limit NS5-branes become flat and we are left with brane configurations in {\em flat} spacetimes! This is the beauty of Hanany-Witten\cite{Hanany:1996ie} type construction (see \cite{Giveon:1998sr} for extensive review).

We should perhaps stress here that this is not the only way to use NS5-brane. For example, even before the discovery of brane tiling, we have many literature on realizing chiral gauge theories using type IIA theory with NS5-brane and D4-branes \cite{Elitzur:1997fh,Witten:1997ep,Uranga:1998vf}. 
The type IIA picture, although simple, is limited to orbifolds and generalized conifolds (and their orientifolds) and as we will see, D5/NS5-configurations we consider here are by far the most powerful and applies to arbitrary toric Calabi-Yau 3-fold.

\paragraph{A short reminder of toric geometry}
We still haven't discuss how to specify NS5-cycles $\balpha_{\mu}$ on torus. This choice should reflect the choice of the toric diagram. In order to see the connection, let us spend some time recalling some basic facts in toric geometry. %and on Buscher's rule\cite{Buscher:1987qj,Buscher:1987sk}.

\begin{figure}[htbp]
\centerline{\includegraphics{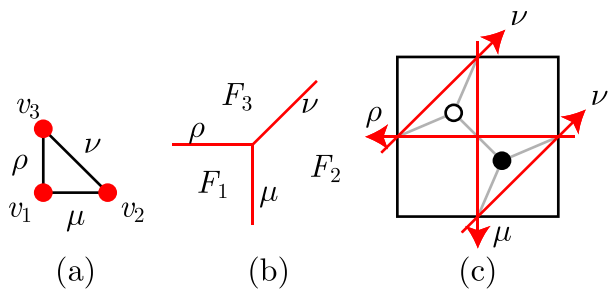}}
\caption{Some diagrams for $\bC^3$. Shown here are toric diagram (a), web-diagram (b) and fivebrane diagram with bipartite graph (c). The significance of the bipartite graph, and its connection with the fivebrane diagram, will be discussed below.}
\label{c320.eps}
\end{figure}

Let us start from a toric Calabi-Yau cone $\cal M$ described by a
toric diagram.
In this paper we only consider three-dimensional Calabi-Yau manifolds.
Let $v_i\in \Gamma=\bZ^3$ be the set of lattice points in
the toric diagram.
By $SL(3,\bZ)$ transformation, we can take the coordinate
system in which the components of $v_i$ are given by
\begin{equation}
v_i=(p_i,q_i,1).\label{vi}
\end{equation}
The toric diagram is
ordinary represented as
a two-dimensional diagram by using the first two components of these vectors.
An example of $\bC^3$ case is shown in Figure \ref{c320.eps} (a).

We define\footnote{The reader should be careful when comparing with literature, because $\mathcal{C}^*$ is sometimes also written as $\mathcal{C}$. Our notation is in accord with the fact that $\mathcal{C}^*$ is in the dual of Lie algebra of torus.} the dual cone $\B$ as the set of vectors $w\in \mathbb{R}^3$
satisfying
\begin{equation}
v_i\cdot w\geq0\quad
\forall i.
\label{dualcone}
\end{equation}
When we consider resolutions of the toric Calabi-Yau,
K\"ahler parameters come to the right hand side of this inequality.
In this paper we will not discuss such resolutions
and the right hand side is always zero.
In such a case
we do not have to use all $v_i$
to define $\B$ by (\ref{dualcone}),
and we only need $v_\alpha$ corresponding to the corners of
the toric diagram. Here symbols $\alpha,\beta \ldots$ is used to denote lattice points in the corners of toric diagram, and $i,j \ldots$ denotes all the lattice points in the toric diagram (including its boundary). We further assume that the label $\alpha$ increase one by one as we go around the perimeter of toric diagram in counterclockwise manner.
The boundary of the dual cone $\partial\B$ consists of flat faces called facets.
Each facet corresponds to each vector $v_{\alpha}$,%in (\ref{vi}),
 and is defined as the set of points satisfying $v_\alpha\cdot w=0$ and $v_{\beta} \cdot w \ge 0 ~~(\forall \beta\ne \alpha)$.
We denote the facet corresponding to $v_\alpha$ by $F_\alpha$.
The structure of the base manifold $\B$ is conveniently expressed as
a planar diagram by projecting the facets onto a two-dimensional
plane by simply neglecting the third coordinate.
It is called a web-diagram.
Figure \ref{c320.eps} (b) is an example of web-diagram for $\bC^3$.
The lines in the web-diagram represent the edges of the
base manifold $\B$.

We can regard the Calabi-Yau manifold as
the $\bT^3$-fibration over the dual cone $\B$ (\ref{dualcone}), although strictly speaking some cycles of $\bT^3$ shrinks on facets as we will explain.
Let $(\phi_1,\phi_2,\phi_3)$ be the coordinates in the toric
fiber.
We choose the period of each coordinate to be $2\pi$.
We can regard $\Gamma$ as the lattice associated with
the toric fiber $\bT^3$.
Namely, we can associate points in $\Gamma$ with cycles in
$\bT^3$.
By this identification, we can regard an arbitrary non-vanishing vector $v\in\Gamma$
as a generator of $U(1)$ isometry
of the $\bT^3$.
%We denote the symmetry generated by $v$ by $U(1)[v]$.
%Two flavor symmetries which do not rotate the supercharges
%are $U(1)[(1,0,0)]$ and $U(1)[(0,1,0)]$,
%and R-symmetry is $U(1)[(a_1,a_2,1)]$.
%When $a_1$ and $a_2$ are appropriately chosen
%this gives the R-symmetry in the superconformal algebra.

On a facet $F_\alpha$ the cycle specified by $v_\alpha$ in $\bT^3$ fiber
shrinks and the fiber becomes $\bT^2$.
In order to parameterize the $\bT^2$ fiber on each facet,
the following coordinate change is convenient\footnote{Strictly speaking, we should write ${\theta_i}^{\alpha}$ since $\theta_i$ depends on facet $\alpha$. In this review, we do not show the $\alpha$-dependence explicitly for notational convenience.}.
\begin{equation}
(\phi_1,\phi_2,\phi_3)=\theta_1(1,0,0)+\theta_2(0,1,0)+\theta_3(p_i,q_i,1).
\end{equation}
This is equivalent to
\begin{equation}
\theta_1=\phi_1-p_{\alpha}\phi_3,\quad
\theta_2=\phi_2-q_{\alpha}\phi_3,\quad
\theta_3=\phi_3. \label{thetadef}
\end{equation}
On the facet, the $\theta_3$-cycle shrinks and
$(\theta_1,\theta_2)$ is a pair of good coordinates on $\bT^2$.
%The third angular coordinate $\theta_3$ is identified
%with the argument of the complex coordinate $x^8+ix^9$
%in the fivebrane system.

%%%%%%%%
\paragraph{T-duality and Buscher's rule}
After a brief review of toric geometry, we now want to apply T-duality along the two-cycle. Basically, T-duality is taken along $\phi_1,\phi_2$-directions. As we have seen, however, these are in general not good coordinates since we have shrinking cycles. 
%and we have to deal with possible Dirac string-like singularities. 
Instead, we choose to use coordinates $\theta_1,\theta_2$ for each facet $\alpha$ \footnote{Thus we are taking T-duality in each patch of the manifold. This is somehow similar to the recent work of \cite{Ishii:2007ex}.}.

\begin{figure}[htbp]
\centering{\includegraphics[scale=1]{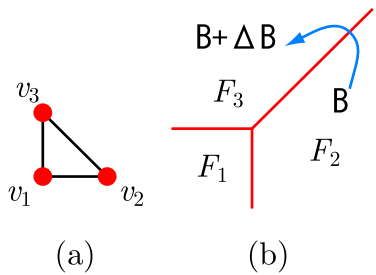}}
\caption{The value of NS-NS B-field jumps as in \eqref{Bjump}. This means that we have NS5-brane in the intersection of two facets, or in the direction of the normals to toric diagram.}
\label{Bfieldjump}
\end{figure}

Recall that T-duality exchanges momentum and winding. This means corresponding gauge fields, i.e. metric and B-field, should also be exchanged. In the original Calabi-Yau picture, we have no NS-NS B-field, but the metric is not flat. After T-duality, we have a non-trivial B-field, which is the source of NS5-brane. This is represented in general by Buscher's rule explained in Appendix \S\ref{Buscher.sec}. 
After applying  the Buscher's rule \eqref{Buscher2} twice, we have 
non-trivial B-field
\beq
B=v_1\wedge (d\theta^1+v_1) +v_2 \wedge (d\theta^2+v_1),
\eeq
where $v_1$ and $v_2$ are gauge fields in $\theta_1$ and $\theta_2$-directions, respectively.
The important thing is that this B-field is dependent on the choice of the facet $\alpha$. Since $\theta$ is related to $\phi$ as in \eqref{thetadef}, the value of $B$ jumps as we go from one facet $\alpha$ to another $\alpha+1$:
\beq
B\to B-(\Delta p \  v_1+\Delta q \ v_2) \wedge d\theta_3, \label{Bjump}
\eeq
where $\Delta p=p_{\alpha+1}-p_{\alpha}, \Delta q=q_{\alpha+1}-q_{\alpha}$.
This signifies the presence of NS5-brane in the intersection of $F_{\alpha+1}$ with $F_{\alpha}$ (See Figure \ref{Bfieldjump}). \footnote{In principle, we can compute the jump of NS-charge by integrating the difference \eqref{Bjump} over $S^3$. This is difficult in practice, since we do not know the explicit form of the metric. } Consequently, semi-infinite cylinder of NS5-branes should be extending in the direction of the primitive normals to the toric diagram, as in Figure \ref{NS5junction}. To put in other terms, the web diagram coincides with junction of semi-infinite cylinders of NS5-branes.

\begin{figure}[htbp]
\centering{\includegraphics[scale=1]{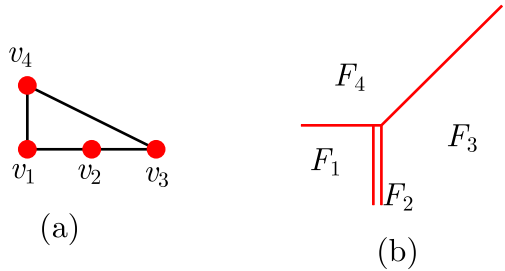}}
\caption{The orbifold $\bC\times \bC^2/\bZ_2$, whose toric diagram is shown in (a), is turned by T-duality into NS5 junctions as shown in (b).}
\label{AnSing.eps}
\end{figure}

This is familiar in the simple example shown in Figure \ref{AnSing.eps}. In this case, $\bC\times \bC^2/\bZ_2$ is turned into two parallel NS5-branes. More generally, it is well-known that T-duality maps $A_n$ singularity to $n$ parallel NS5-branes, and you can directly verify this fact using explicit metric. 

Our discussion for general toric Calabi-Yau is a generalization of this well-known fact, although it is often difficult to verify this fact directly using explicit form of the metric. You might think this is a pity, but at the same time it is good, since it is often extremely difficult to find explicit metrics and we do not need to know them for most of our purposes.

We have understood our brane configuration in 46-directions, but what about 57-directions?
Since we are using smeared solution in SUGRA (see Appendix \ref{Buscher.sec}), Buscher rule does not tell us anything about $\bT^2$-directions we want to know. But BPS conditions dictates $\Sigma$ is Lagrangian, namely the K\"ahler form in 4567-space
\beq
\omega=dx_4\wedge dx_5+dx_6\wedge dx_7
\eeq
should vanish on NS5-brane.
This means that when a NS5-brane span $(p,q)$-directions in 46-directions, it should be extending in $(q,-p)$-direction in 57-directions. The net result is that we have 1-cycle of NS5-brane on $\bT^2$ with winding $(q,-p)$ for each primitive normal $(p,q)$ of the toric diagram. 
By rotating 90 degrees for convenience, NS5-cycle is turned into 1-cycle with winding $(p,q)$, namely the same winding number with the primitive normal of the toric diagram. We are going to use this convention throughout the rest of the paper.

Summarizing, shrinking cycles of winding number $(p,q)$ are mapped by T-duality  (and 90 degrees rotation in 57-space) to NS5-brane wrapping cycles of winding number $(p,q)$. This means in particular that previously defined $d$, which is the number of cycles of NS5-branes, is equal to number of lattice points on the boundary of the toric diagram. Note also that the consistency conditions discussed previously in \eqref{pq0} is trivially satisfied (it amounts to the condition that if we go around the perimeter of toric diagram, then we are back to the same place). By this way we now understand the relation with toric Calabi-Yau and NS5-cycles.

\begin{figure}[htbp]
\begin{center}
\centering{\includegraphics[scale=0.5]{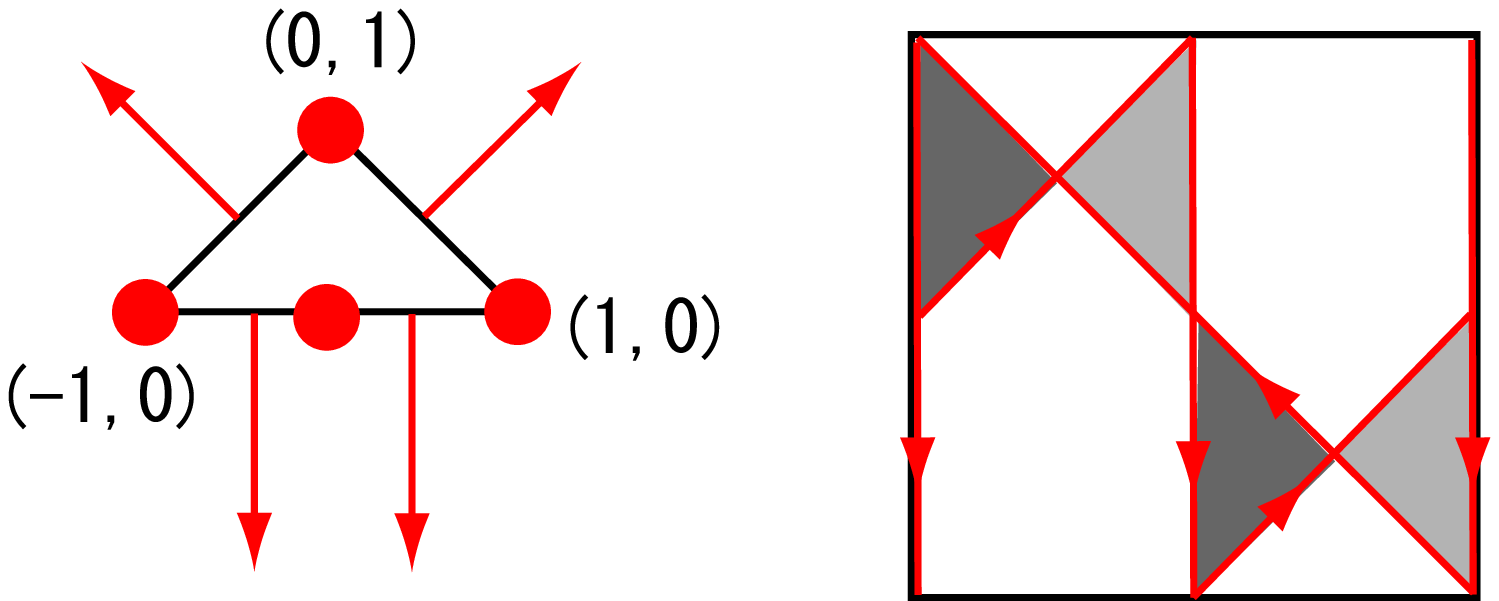}}
\caption{Examples of correspondence between NS5 cycles and toric diagram. Here we show the example of $\bC^3/\bZ_2$ ($\bZ_2\subset SU(2)$) which is previously discussed in Figure \ref{Dynkin}.}
\label{SU2eg}
\end{center}
\end{figure}

\begin{figure}[htbp]
\begin{center}
\centering{\includegraphics[scale=0.5]{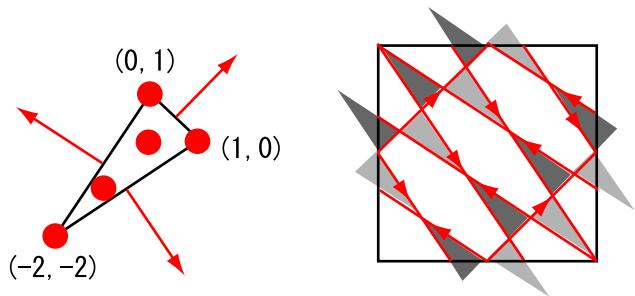}}
\caption{Examples of correspondence between NS5 cycles and toric diagram. Here we show the example of $(1,2,2)/5$, which is previously discussed in Figure \ref{McKayeg}.}
\label{SU3eg}
\end{center}
\end{figure}

Examples are shown for $\bC^2/\bZ_2\times \bC$ (Figure \ref{SU2eg}) and $\bC^3/\bZ_5$ with $\bZ_5=(1,2,2)/5$ (Figure \ref{SU3eg}). These orbifold examples are discussed previously in Calabi-Yau setups in \S\ref{D3.subsec}.

\subsubsection{Relation with quiver gauge theories} \label{quiver.subsubsec}
Up to now we have discussed fivebrane systems and their relation with toric Calabi-Yau manifolds. We now discuss the precise relation between fivebrane systems and quiver gauge theories.

\paragraph{From fivebrane systems to quiver gauge theories}
The question we want to ask is what kind of gauge theories we have on fivebrane systems. 

As already said, each $(N,0)$-brane region corresponds to a $SU(N)$ gauge group. How about $(N,1)$-branes and $(N,-1)$-branes? Do we have $SU(N)$ gauge groups also on these branes in the low-energy effective field theory? In fact, the answer is no. We have only $U(1)$ on $(N,\pm 1)$-branes, as can be seen by applying $SL(2,\bZ)$-transformation. This shows that only regions of $(N,0)$-brane corresponds to vertices of the quiver diagram.

Some readers might worry about the $U(1)$ gauge group that we have not taken account of. For example, $(N,\pm 1)$-branes couple to $(N,\pm 1)$-branes
\footnote{In general, only $(p,q)$-strings can end on $(p,q)$-branes. You can see this fact by applying $SL(2,\bZ)$-duality to D-brane, which couples to the fundamental string.} and we have seemingly a $U(1)$ gauge field. However, as analyzed in \cite{Imamura:2006db}, that $U(1)$ gauge field is fixed by the boundary conditions at junctions and therefore not dynamical. That $U(1)$ is only a global symmetry, and is identified with one of the global anomalies of $\scN=1$ superconformal quiver gauge theories \cite{Imamura:2006ie}.

\begin{figure}[htbp]
\begin{center}
\centering{\includegraphics[scale=0.4]{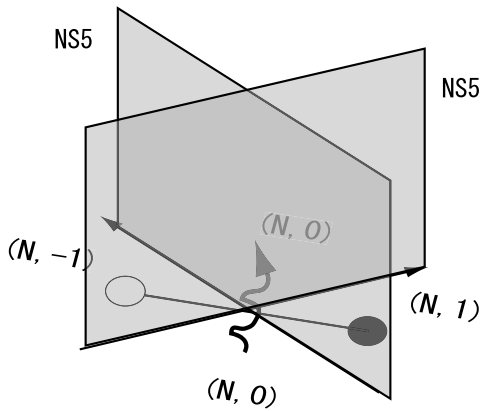}}
\caption{The wavy arrow represents an open string, or a bifundamental field in the quiver gauge theory. The open string connecting two $(N,0)$-branes have to go through NS5-branes, and the presence of NS5-brane allows strings only in one direction, thus making chiral theories. The black/white vertices and the edge connecting them represent part of bipartite graphs, which will be commented on in a moment.}
\label{chiral}
\end{center}
\end{figure}

Next, for each intersection point of $(N,0)$-branes, we have massless open strings, which corresponds to bifundamental fields in quiver gauge theories. Due to the presence of NS5-brane, only strings in one direction is allowed and the theory becomes chiral (Figure \ref{chiral}). 

You can understand this fact as follows. We explain in the conifold example for simplicity. In the conifold example shown in Table \ref{conifoldconfig}, we have two types of NS5-branes, NS5 and NS5'. If we remove NS5', we have four-dimensional $\scN=2$ superconformal quiver gauge theory. In particular, we have $\mathcal{N}=2$ matter hypermultiplets. The scalar degrees of freedom of this multiplet is four (when counted as real) and this corresponds to the move of D5-brane in 6789 space. Now put NS5' back, and supersymmetry is broken down to $\scN=1$. In this process, $\scN=2$ hypermultiplet is broken down into two $\scN=1$ chiral multiplets. However, due to the presence of NS5', D5-brane can no longer move freely in 67 directions and this means that only one of two chiral multiplets remains. This explains that we have a chiral gauge theory. This feature, that the gauge theory is chiral, is extremely important because our world is described by chiral gauge theories but it is often not easy to realize chiral gauge theories from string theory. 

\begin{figure}[htbp]
\begin{center}
\centering{\includegraphics[scale=0.4]{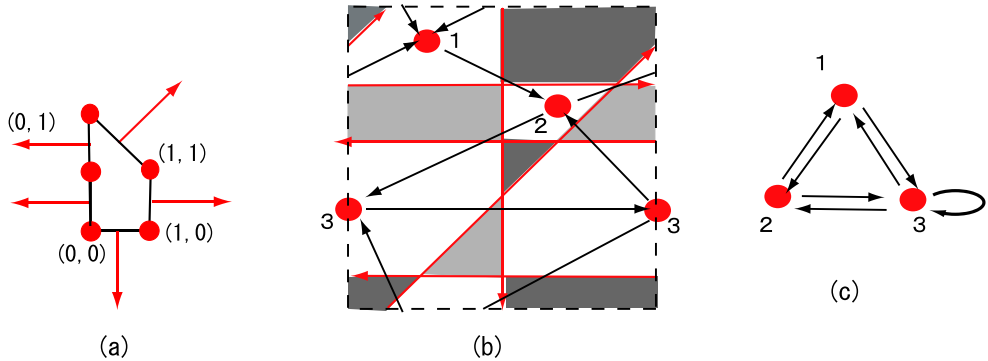}}
\caption{Example of quiver diagrams read off from fivebrane systems. The toric diagram (a) corresponds to the so-called Suspended Pinched Point (SPP). From the toric diagram, we have a fivebrane diagram as in (b) (this is the same fivebrane diagram as in Figure \ref{division1}, although corresponding toric diagram is rotated and turned upside down.). We have a quiver diagram on $\bT^2$ (b), which reduces to the usual quiver diagram (c) if we forget that it is written on torus. Note sometimes $(N,0)$-brane can have intersection with itself, and in that case we have adjoint field. In this example and the following, the fundamental region of torus is sometimes represented by dotted line, so that the figure does not become too messy.}

\label{quiverread}
\end{center}
\end{figure}

\begin{figure}[htbp]
\begin{center}
\centering{\includegraphics[scale=0.4]{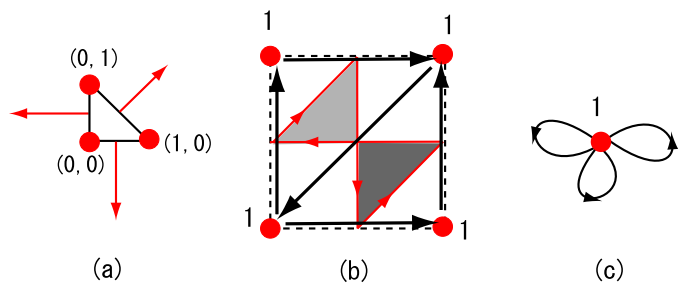}}
\caption{The $\bC^3$ example. The toric diagram (a), the fivebrane diagram (b), and the quiver diagram (c). In this example, the quiver diagram has only one node with three adjoint fields, and the gauge theory is familiar $\scN=4$ supersymmetric Yang-Mills. }
\label{C3cycles.eps}
\end{center}
\end{figure}

\begin{figure}[htbp]
\begin{center}
\centering{\includegraphics[scale=0.4]{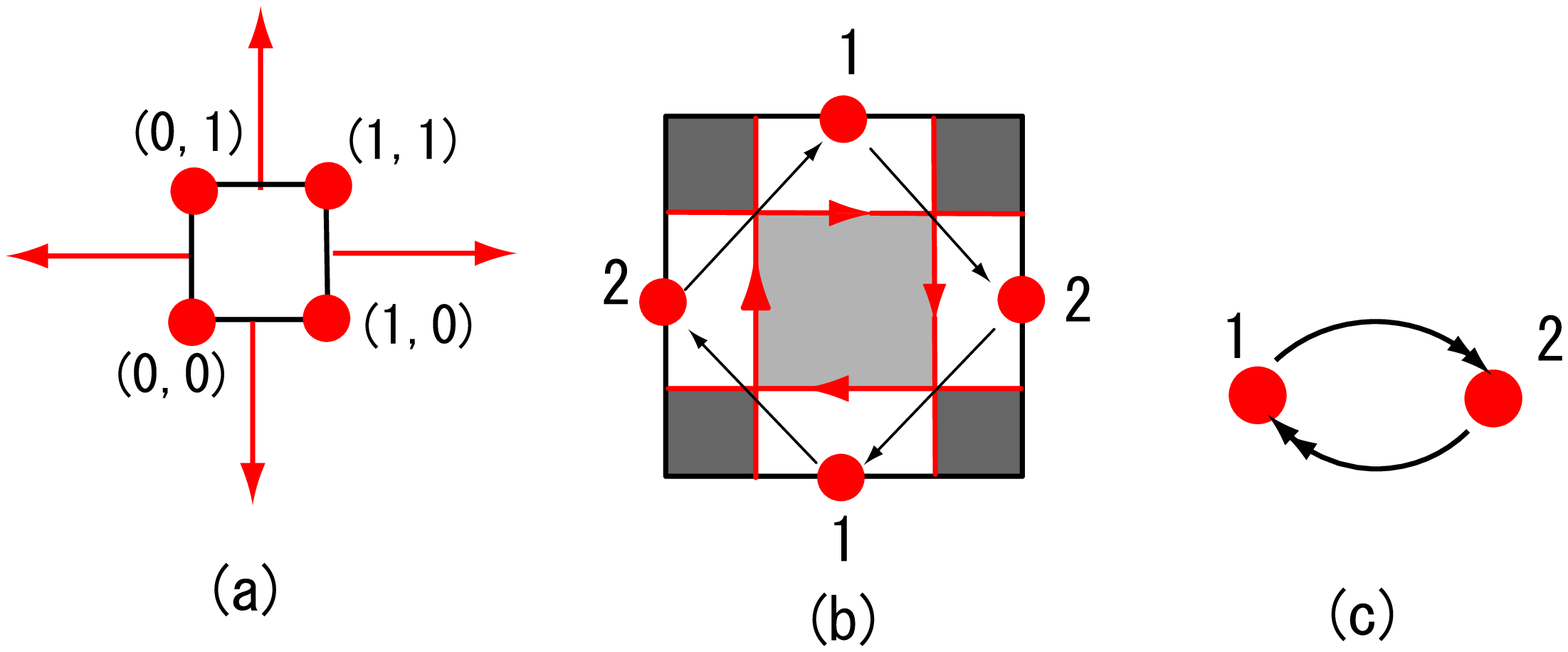}}
\caption{The conifold example. The corresponding quiver appeared in a famous work by Klebanov and Witten \cite{Klebanov:1998hh} (see \S\ref{AdSCFT.sec}).}
\label{conifoldcycles.eps}
\end{center}
\end{figure}

Now, from above considerations it is clear that we have quiver gauge theories on D5-branes. Each $(N,0)$-region corresponds to a vertex of the quiver diagram, and we need an arrow of the quiver diagram for each intersection point of $(N,0)$-brane. An example of the quiver diagram read off from fivebrane diagram is shown in Figure \ref{quiverread}. More examples, corresponding to $\bC^3$ and the conifold, are also shown in Figure \ref{C3cycles.eps} and Figure \ref{conifoldcycles.eps}. You can also verify that bipartite graphs in Figure \ref{BTfirst} give quiver diagrams as in Figure \ref{quivereg}.

\begin{figure}
\centering{\includegraphics[scale=0.4]{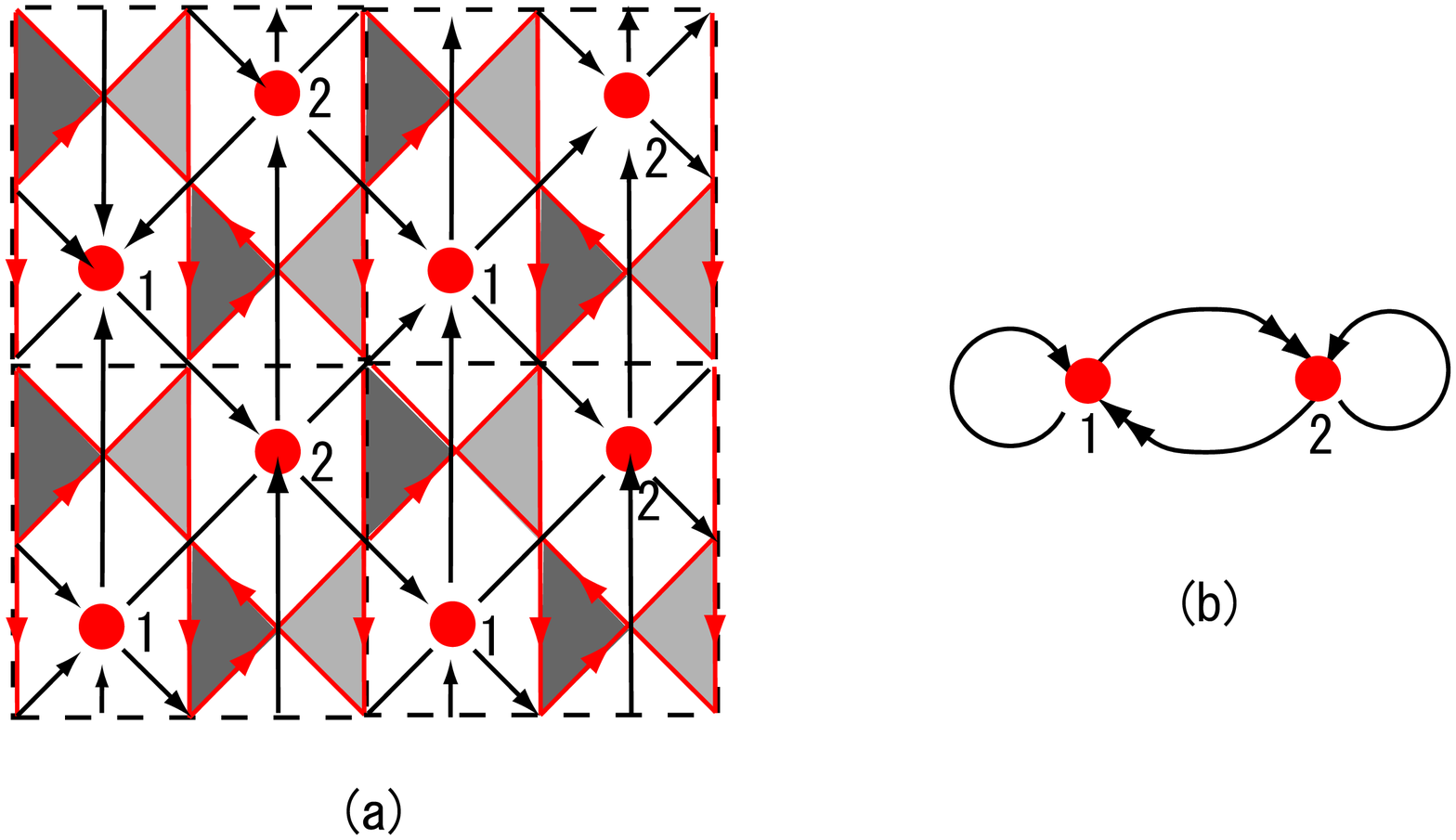}}
\caption{The quiver diagram obtained from fivebrane diagram, in the case of $\bC^2/\bZ_2\times \bC$ (Figure \ref{SU2eg}). As expected, this reproduces the $A_2$ extended Dynkin diagram shown in Figure \ref{Dynkin}. The red lines represent cycles of NS5-branes, while black oriented lines represent arrows of the quiver diagram. The quiver is similar to the quiver for $\bC^2/\bZ_2\times \bC$ shown in Figure \ref{conifoldcycles.eps}. The only difference is the existence of adjoint field at each node.}
\label{SU2egquiver}
\end{figure}

\begin{figure}
\centering{\includegraphics[scale=0.4]{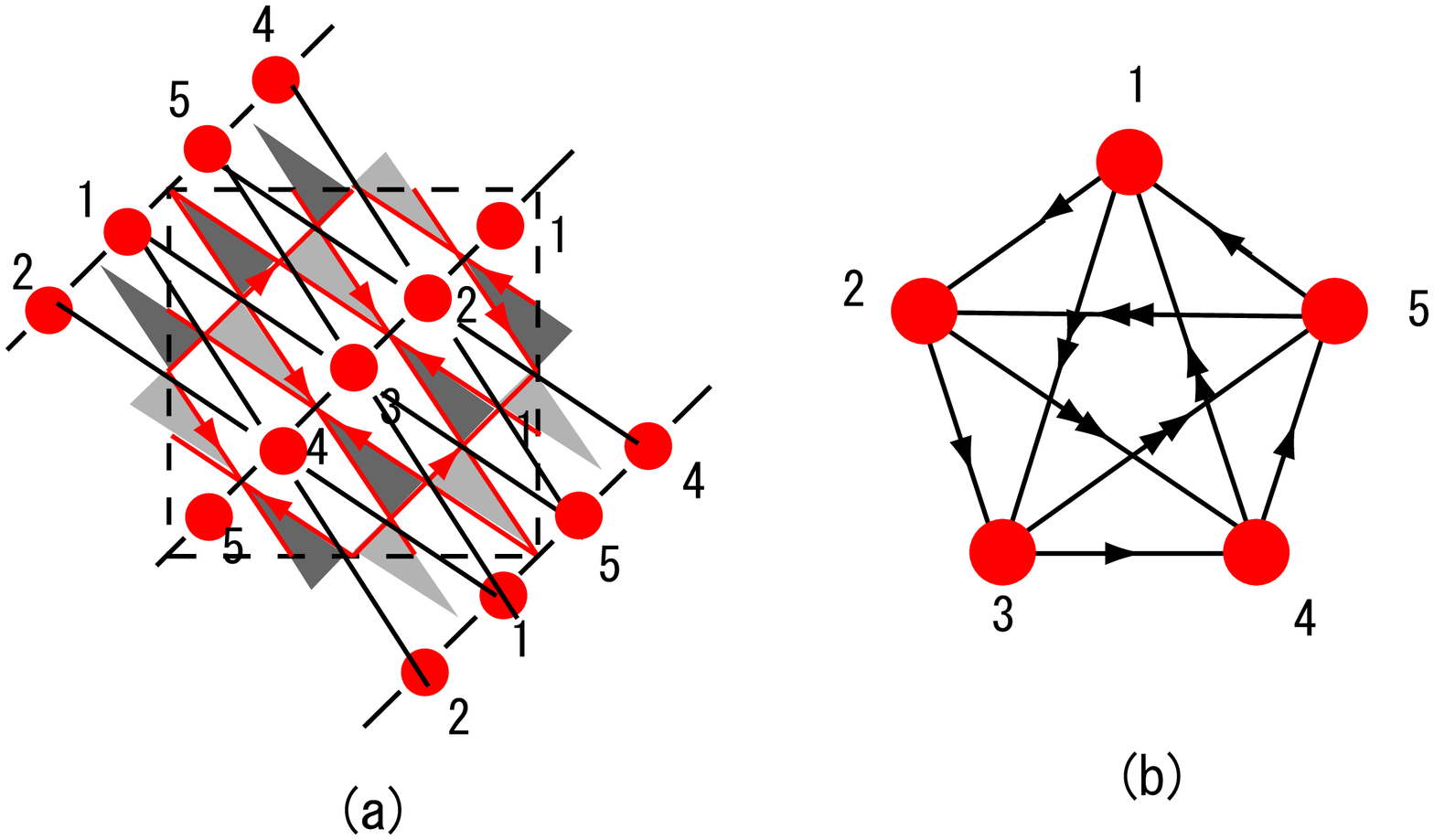}}
\caption{The quiver diagram obtained from the fivebrane diagram in the case of $\bC^3/\bZ_5$, with $\bZ_5 = (1,2,2)/5$ (Figure \ref{SU3eg}). As expected, this reproduces the $(1,2,2)/5$ McKay quiver shown in Figure \ref{McKayeg}. The square region surrounded by dotted lines represent fundamental region of torus.
}
\label{SU3egquiver}
\end{figure}

We also show in Figure \ref{SU2egquiver} and \ref{SU3egquiver} the quiver diagrams corresponding to fivebrane diagrams in Figure \ref{SU2eg} and \ref{SU3eg},
Note that these are exactly the same as the quivers shown in Figure \ref{Dynkin} and \ref{McKayeg}, respectively. More generally, we can prove 
\begin{thm}
Let $\Gamma$ be an Abelian discrete subgroup of $SU(3)$, and let $\Delta$ be a toric diagram whose corresponding Calabi-Yau is the orbifold $\bC^3/\Gamma$. Then the quiver (and superpotential)\footnote{Mathematically speaking, superpotential (or its F-term relations) is an ideal in the path algebra of quiver. See \S\ref{pathalg.subsubsec}.} obtained from the method discussed here coincides with the McKay quiver of $\Gamma$. 
\qed
\end{thm}
Recall that McKay quiver is obtained from $\Gamma$ by decomposing tensor product of representations (see \eqref{rhodecomp}). The proof of this statement is combinatorial in nature and can be found in \cite{UY1}. In \S\ref{orbifold.subsubsec}, we have explained two different methods to obtain quiver diagram from $\Gamma$. We now have another, and these three methods all give the same answer.

\bigskip

\begin{figure}[htbp]
\begin{center}
\centering{\includegraphics[scale=0.4]{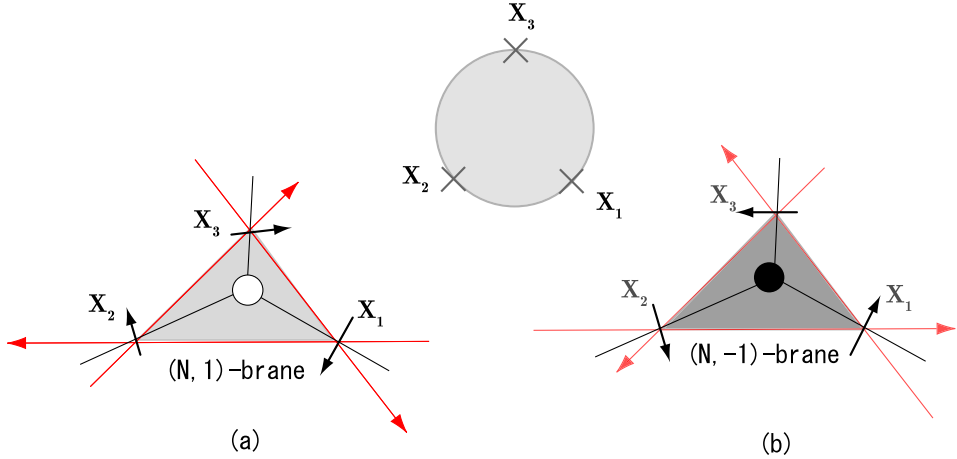}}
\caption{We have disc amplitude of string interactions, for each face of $(N,\pm 1)$-brane. The arrows, labeled $X_1, X_2$ and $X_3$, represent bifundamental fields. The order of operators inside the trace depends on the orientation of NS5-brane. For example, (a) with $(N,1)$-brane contributes operators of the form
 $\tr(X_1 X_2 X_3)$ to the superpotential, and $(N,-1)$-brane as in (b) contributes $\tr (X_1 X_3 X_2)$ term.}
\label{discamp}
\end{center}
\end{figure}

We have now verified our original claim that fivebrane systems realize four-dimensional $\mathcal{N}=1$ supersymmetric quiver gauge theories, but fivebrane systems knows more than that. For example, we can read off superpotentials. For each region of $(N,\pm 1)$-brane, we can span a disk and thus we have tree-level disk amplitude of string theory interactions, which in turn means that we have such a term in superpotential, as in Figure \ref{discamp}.

More formally, for each face of $(N,\pm 1)$-brane (which are labeled by $k$), we have superpotential terms like 
\begin{equation}
h_k \tr \prod_{I\in k} \Phi_I \label{hkterm},
\end{equation}
where $I\in k$ means that bifundamental $I$ is around the face $k$, and the product inside the trace is taken in clockwise (or counterclockwise, depending on the sign of NS5-charge) manner.
Also, $h_k$ is a superpotential coupling (Yukawa coupling), and meaning of this parameter will be explored in full detail in \S\ref{marginal.subsec}.

The expression of superpotential is now given by 
\begin{equation}
\sum_k h_k \tr \prod_{I\in k} \Phi_I. \label{hktermsum}
\end{equation}
 Here we simply take them to be

\beq
h_k=
\begin{cases}
+1 & (k: (N,1)-\textrm{brane}) \\
-1 & (k: (N,-1)-\textrm{brane})
\end{cases}
. \label{hkpm1}
\eeq

The meaning of this choice will become clear in \S\ref{marginal.subsec}. Here we simply comment that it is chosen so that the theory becomes conformal and the corresponding geometry being toric Calabi-Yau cone.

Let me give you several examples of superpotentials. In $\bC^3$ example of Figure \ref{C3cycles.eps}, we have

\beq
W=\tr (XYZ-XZY)=\tr (X[Y,Z]), 
\eeq
which is the well-known superpotential of $\mathcal{N}=4$ super Yang-Mills. We can instead consider the superpotential

\beq
W=\tr (e^{\sqrt{-1}\gamma}XYZ-e^{-\sqrt{-1}\gamma}XZY).
\eeq

This deformation is the so-called $\beta$-deformation of $\mathcal{N}=4$ gauge theory. It preserves $\scN=1$ supersymmetry, but the corresponding geometry is no longer toric Calabi-Yau.

For Figure \ref{conifoldcycles.eps}, we have
\beq
W= \tr (A_1 B_1 A_2 B_2)-\tr(A_1 B_2 A_2 B_1). \label{conifoldW}
\eeq
This is again the famous superpotential corresponding to the conifold\cite{Klebanov:1998hh}.

\begin{figure}[htbp]
\begin{center}
\centering{\includegraphics[scale=0.4]{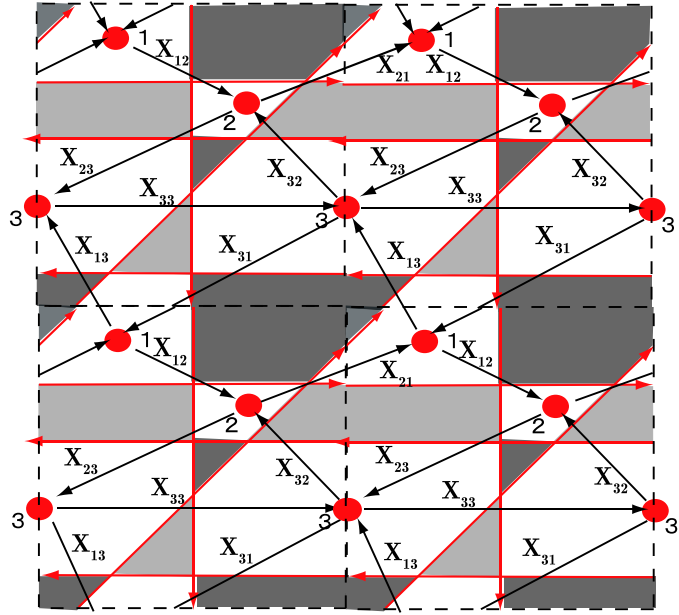}}
\caption{The planar quiver for SPP shown again. The fundamental region of Figure \ref{quiverread} is enlarged, so that it becomes easier to read off superpotentials. $X_{ij}$ refers to a bifundamental, or an arrow of the quiver starting at vertex $i$ and ending at $j$.}
\label{superpotentialeg}
\end{center}
\end{figure}

As a more complicated example, consider the case of SPP shown in Figure \ref{quiverread}. In Figure \ref{superpotentialeg} we have enlarged the fundamental region, so that superpotential is easily read off. The result is given by

\beq
W=\tr(
X_{21}X_{12}X_{23}X_{32}-X_{23}X_{33}X_{32}+X_{33}X_{31}X_{13}-X_{31}X_{12}X_{21}X_{13}
). \label{sppW}
\eeq

If you look at several examples of superpotential, then you will notice that each bifundamental field $X_I$ is contained exactly twice in the superpotential, one in the term with plus sign and one minus sign. The superpotentials are therefore of the special kind. From fivebrane viewpoint, this follows because each bifundamental is located at the intersection of two $(N,0)$-branes and one $(N,1)$ and one $(N,-1)$-brane (see Figure \ref{chiral}).

In the language of Calabi-Yau geometry, this corresponds to the condition that the Calabi-Yau is toric (we sometimes say that superpotential $W$ satisfies toric condition). Suppose the superpotential takes the form
\beq
W=+\tr (X_I X_J\ldots)-\tr (X_I X_K\ldots) +(\textrm{terms independent of } X_I),\label{toricW}
\eeq
then the F-term equation $\frac{\partial W}{\partial X_I}=0$ takes the form
\beq
X_J\ldots=X_K\ldots ,
\eeq
i.e. (monomial)=(monomial). Since the vacuum moduli of quiver gauge theory corresponds to Calabi-Yau, this relation becomes part of defining equations of Calabi-Yau. Since toric geometry allow only relations of (monomial)=(monomial) form, this ensures that the Calabi-Yau is toric. This means that as long as we use brane tilings, we can only describe toric Calabi-Yaus, or their deformations (such as $\beta$-deformation).

\paragraph{The connection with bipartite graphs}

Perhaps the reader might be wondering at this point what has become of the bipartite graph we first encountered in Introduction (see Figure \ref{BTfirst}). So far, we have not mentioned bipartite graphs up to this point. In fact, in many cases we can simply use fivebrane systems and can forget about bipartite graphs. It is true, at the same time, that bipartite graphs are quite useful in some cases (see more on this in \S\ref{another.subsec} and \S\ref{GLSM.subsec}, for example), and they have interesting connections with various topics, as emphasized in Introduction. Moreover, most of the literature use bipartite graphs. We therefore explain the connection between fivebrane diagrams and bipartite graphs. 

\begin{figure}[htbp]
\centering\includegraphics[scale=0.4]{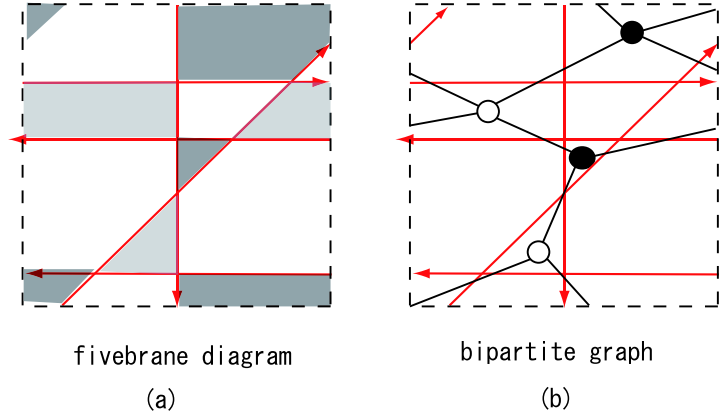}
\caption{The relation between fivebrane diagram (left) and bipartite graph (right). On the light (resp. gray) region, or region of $(N,1)$-brane (resp. $(N,-1)$-brane), we place white (resp. black) vertices. Each edge of bipartite graph corresponds to an intersection point of $(N,1)$-brane and $(N,-1)$-brane. }
\label{bipartiteconnection}
\end{figure}

The procedure to obtain bipartite graphs from fivebrane diagrams is as follows. First, place white (resp. black) vertex to each $(N,1)$-brane (resp.$(N,-1)$-brane)\footnote{You can choose a different convention and reverse the assignment of black and white. The convention here is in accordance with \cite{IIKY}. It also depends on the convention of the sign of the NS-charge.}. Second, we connect black and white vertices whenever the polygon around one vertex has an intersection point with polygon around another vertex. 
See Figure \ref{bipartiteconnection} for example.  Note by construction, this graph is automatically bipartite (the intersection always looks like Figure \ref{chiral}).

\begin{figure}[htbp]
\centering\includegraphics[scale=0.4]{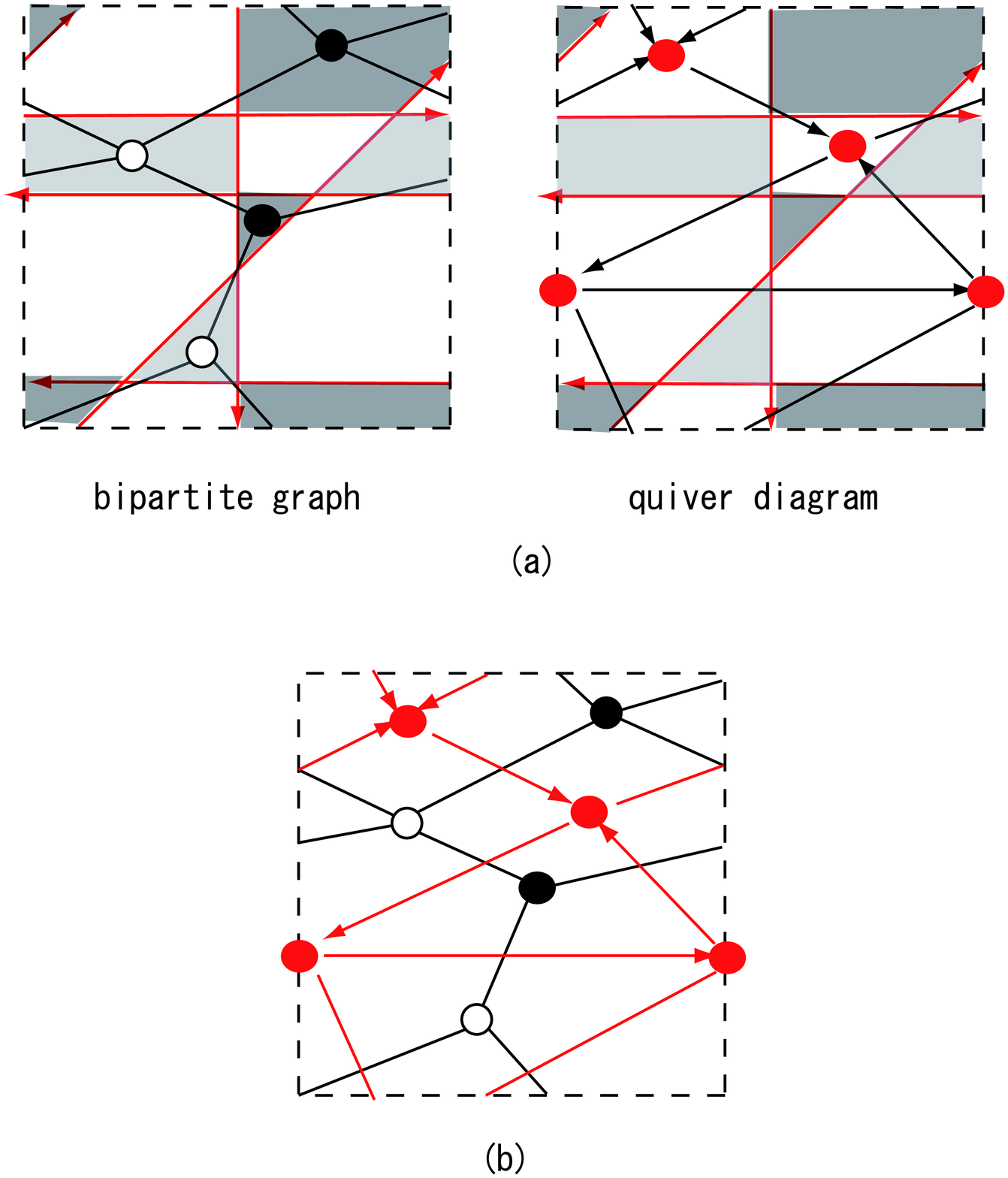}
\caption{In (a), we show both the bipartite graph (left) and the quiver diagram (right), together with fivebrane diagram. Combining these as in (b), we see that the dual of the bipartite graph is the quiver diagram. The orientation of an arrow of the quiver diagram corresponds to the color of vertices in the bipartite graph.}
\label{dualisquiver}
\end{figure}

Note this also explains the relation between bipartite graphs and quiver diagrams: if we write quiver diagram on torus (which is sometimes called periodic quiver or planar quiver), the quiver diagram is the dual of the bipartite graph (Figure \ref{dualisquiver}).

Summarizing, we now have the relation between bipartite graphs and quiver diagrams as in Table \ref{relationtable}.

\begin{table}[htbp]
\caption{The relation between fivebrane system, the bipartite graph, and the quiver diagram. Face of the quiver is written in parentheses since they appears only when we write them on $\bT^2$.}

\begin{center}
\begin{tabular}{|c|c|c|c|} 
\hline
fivebrane & bipartite graph & quiver diagram & quiver gauge theory \\
\hline
 \hline
 $(N,1)$-brane & white vertex & (face)  &  superpotential term  \\
(light gray)&  & & ($+$ sign)\\
 \hline
 $(N,-1)$-brane & black vertex & (face) &superpotential term \\
(dark gray) & & & ($-$ sign) \\
 \hline
 $(N,0)$-brane & face &  vertex & gauge group \\
 \hline
open string & edge &  edge & bifundamental \\
 \hline
 
\end{tabular}
\label{relationtable}
\end{center}
\end{table}

Note also that quiver gauge theories obtained in this way automatically satisfies the anomaly cancellation condition discussed in \S\ref{quiver.subsec} (see \eqref{anomcancel}). Because of the bipartiteness of the graph, the face of the bipartite graph is always a polygon with $2n$ (even) number of edges. This means that the corresponding vertex of quiver has $n$ outgoing arrows and $n$ incoming arrows, which ensures the anomaly cancellation condition \eqref{anomcancel} in the case of $N_a=N$ for all  $a$.

We can consider more general rank assignments as discussed in \S\ref{anomaly.subsubsec}. This is also possible in our fivebrane setup, as long as they are consistent with D5-charge and NS5-charge conservation. In fact, we will verify in \S\ref{fractional.subsec} that charge conservation implies gauge anomaly cancellation.

Now let us make a remark before closing up this paragraph.
Our explanation makes clear that not all bipartite graphs are realized in the strong coupling limit. The only relevant bipartite graphs are those which can be obtained from the division of $\bT^2$ by straight lines (recall NS5-branes should be straight in order to preserve $\mathcal{N}=1$ supersymmetry).

For example, in the case of del Pezzo 3, we have at least four bipartite graphs  as will be shown in Figure \ref{dP3phases} in page \pageref{dP3phases}. However, by using straight NS5 cycles, we can obtain only one bipartite graph. This suggests that although we have several different bipartite graphs in the weak coupling limit, we have only one (or at least smaller number of) bipartite graphs in the strong coupling limit. In other words, by changing string coupling constant, several different `phases' of quiver merge into one. The physical significance of this fact, however, is not clear as of this writing.

\paragraph{Zig-zag path and fivebrane diagram} %\label{HV.subsec}

In previous section we explained that bipartite graphs can be obtained from fivebrane diagrams. Conversely, we can obtain fivebrane diagram from bipartite graphs.

\begin{figure}[htbp]
\centering\includegraphics[scale=0.4]{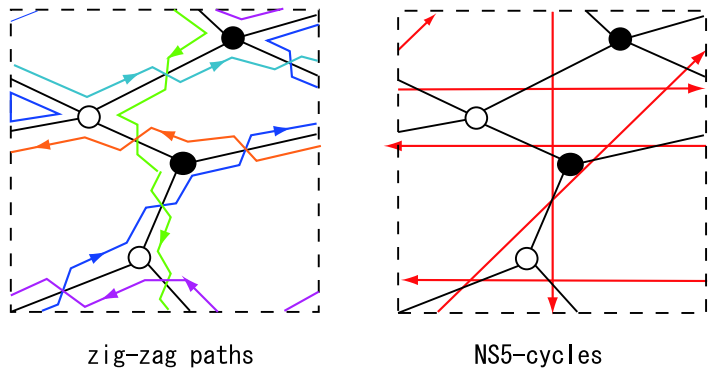}
\caption{Zig-zag paths of bipartite graphs. Comparing with Figure \ref{bipartiteconnection}, you will immediately notice that zig-zag paths correspond one-to-one with cycles of NS5-brane.}
\label{zigzageg}
\end{figure}

To explain that, let us define zig-zag path\footnote{This is also called rhombus loops, or ``train tracks'' in earlier mathematics literature\cite{KenyonSchlenker}.}. Let us start from one edge and go in either direction you like along the edge of the graph. Then you will come across a vertex. Turn maximally left at white vertex and maximally right at black vertex, and proceed further along an edge of the bipartite graph. Since the bipartite graph is finite, if you continue to do this, you will be back to the original position and we have a closed path.  This is a zig-zag path. The example of zig-zag paths are shown in Figure \ref{zigzageg}.

Comparing with Figure \ref{bipartiteconnection}, you will immediately notice that zig-zag paths are in one-to-one correspondence with the cycles of NS5-brane, and the winding number of zig-zag path and that of the corresponding NS5-cycle are the same.
Since winding number of NS5 cycles are given by primitive normals to the toric diagram, we can read off the toric diagram from bipartite graphs. 

Historically speaking, this relationship between zig-zag path and the toric diagram is conjectured first by Hanany and Vegh\cite{Hanany:2005ss}. This conjecture was based on several examples, and was partly motivated from similar considerations in mathematics literature\cite{Kenyon,KenyonSchlenker}. In our explanation, this is no longer a conjecture and the correctness of their algorithm is guaranteed automatically by construction. In \cite{Hanany:2005ss}, the method to write fivebrane diagram from toric diagram as explained above is called a ``fast-inverse algorithm'', because it is an inverse algorithm in the sense of Figure \ref{manytoone}.

\begin{figure}[htbp]
\centering\includegraphics[scale=0.5]{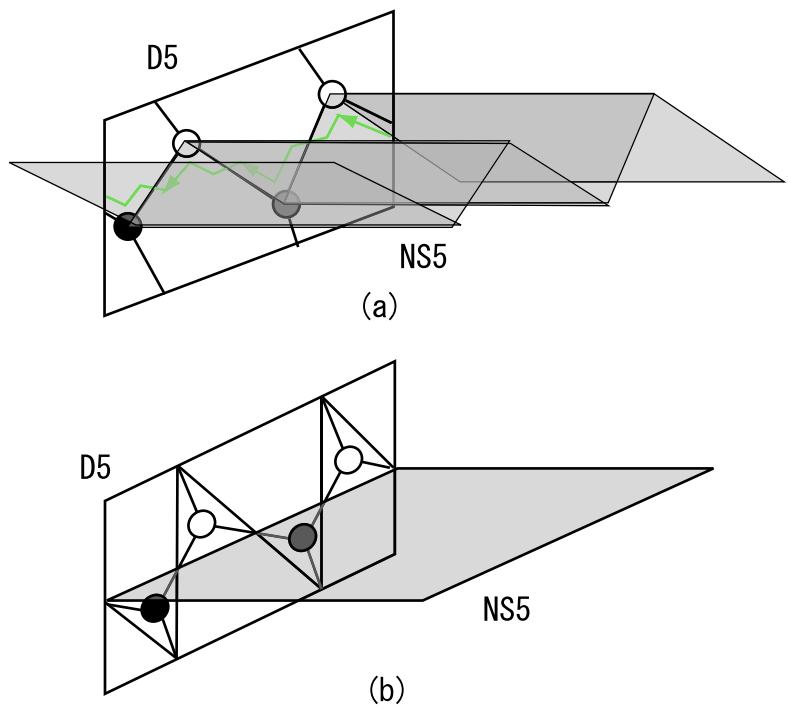}
\caption{As far as topological information is concerned, zig-zag paths of bipartite graphs can be directly identified with cycles of NS5-branes (a). This is deformation of the real brane configuration (b).}
\label{zigzagNS5}
\end{figure}

Sometimes, it is often cumbersome to rewrite fivebrane diagram from bipartite graph and vice versa. As far as topological information is concerned, we can directly identify zig-zag path as NS5-brane, as shown in Figure \ref{zigzagNS5} (a). We should keep in mind, however, that the real NS5-brane cycle is not the zig-zag path, but rather deformation of it (Figure \ref{zigzagNS5} (b)). 

%%%%%%%%%%%%%%%%
\subsubsection{Conformality, R-charges and isoradial embeddings} \label{R-charge.subsubsec}

In Introduction, we explained that our brane configuration are related by AdS/CFT correspondence to some conformal field theory (more on this in \S\ref{AdSCFT.sec}). In the case of $\bC^3$ (Figure \ref{C3cycles.eps}), we have $\scN=4$ SYM, which is conformal. But what about other cases, for example, the case of the conifold (Figure \ref{conifoldcycles.eps})? 

%As we will see in \S\ref{AdSCFT.sec}, 
As clarified first in \cite{Klebanov:1998hh}, the real meaning of quiver diagram and the superpotential, and the corresponding conformal field theory, is as follows. Start with a quiver gauge theory with the matter contents given by the quiver diagram. The theory flows to an IR fixed point by renormalization group flow, where the theory becomes strongly coupled and conformal. The theory we want to consider is the such conformal field theory perturbed by the superpotential.

If the theory in UV has a global $U(1)_R$-symmetry, that should correspond to $U(1)_R$-symmetry of superconformal algebra in IR \footnote{When we have non-anomalous $U(1)$ global symmetries, the problem becomes more complicated because of possible mixing between these $U(1)$s. We will explain in \S\ref{a-max.subsec} how to compute R-charges in this more general setting.}. The problem is whether we can understand R-charges of superconformal $U(1)_R$ from the viewpoint of fivebrane systems.
In \cite{Hanany:2005ss}, graphical representation of R-charge $U(1)_R$ of superconformal algebra is proposed. In their work, they propose to use bipartite graphs of special kinds, which admit  ``isoradial embedding''. Isoradial embedding is an embedding of the bipartite graph onto the plane, where the nodes of each face are on a  circle of unit radius, with the edges of the tiling being straight lines.

\begin{figure}[htbp]
 \centering{\includegraphics[scale=0.45]{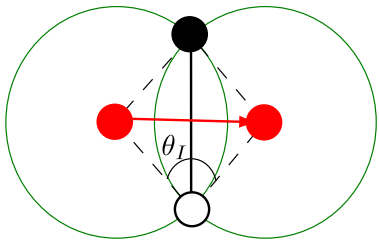}}
 \caption{The proposal to realize R-charges on bipartite graphs which admit isoradial embedding. The angle $\theta_I$ shown in this figure is the R-charge of the field divided by $\pi$. The black and white nodes represent nodes of the bipartite graph, and the edge connecting them is the edge of the bipartite graph. The red nodes denotes center of the circle, and corresponds to nodes of the quiver diagram. The red arrow is an edge of the quiver diagram.}
 \label{thetaI}
 \end{figure}

 As an example, the bipartite graph of left figure in Figure \ref{BTfirst} can be isoradially embedded, but the right figure in Figure \ref{BTfirst} does not admit any isoradial embedding. The concept of isoradial embedding is known in mathematics since long ago\cite{Duffin,Mercat,Kenyon}, but their physical meaning is first proposed in \cite{Hanany:2005ss}. Their proposal is that, for isoradially embedded bipartite graphs, the R-charge satisfies the condition
\beq
R_I=\pi \theta_I \label{Rtheta}
\eeq
 where $\theta_I$  is given by the angle shown in Figure \ref{thetaI}.
 
So far this is still a conjecture. As a consistency check, we invoke the following relations satisfied for angles:

\beq
\sum_{I\in k}\theta_I =2 \pi ,  \ \
\sum_{I\in a}(\pi-\theta_I) =2 \pi \label{thetacond},
\eeq
or rewritten in terms of \eqref{Rtheta},
\beq
\sum_{I\in k}R_I=2 , \ \
 \sum_{I\in a}(1-R_I)=2 \label{Rcond},
\eeq
where $R_I$ denotes R-charge for an edge $I$.
These are actually the conditions imposed on R-charges. The first condition shows that R-charge of the superpotential is 2.   The second equation comes from the vanishing of the NSVZ $\beta$-function. The NSVZ exact $\beta$-function\cite{Novikov:1983uc} is given by
\beq
\beta_a=\frac{d}{d \ln \mu}\frac{1}{g_a^2}=
\frac{N}{1-g_a^2 N/ 8\pi^2} \left[
3-\frac{1}{2} \sum_{I\in a}(1-\gamma_I)
\right], \label{NSVZ}
\eeq
where $\gamma_I$ is the anomalous dimension of the field $\Phi_I$, and the summation is taken over all fields $I$ which has $a$ as one of its endpoints. The dimension $\Delta_I$ of operator $\Phi_I$ is represented by 
R-charge as $\Delta_I=1+\frac{1}{2}\gamma_I$, and from superconformal algebra we have $\Delta_I=\frac{3}{2}R_I$. Thus \eqref{NSVZ} can be rewritten as 
\beq
\beta_a=\frac{N}{1-g_a^2 N/ 8\pi^2} \frac{3}{2} \left[
2- \sum_{I\in a}(1-R_I)
\right]. \label{NSVZ2}
\eeq
This shows that the second equation in \eqref{Rcond} is equivalent to the condition of vanishing of $\beta$-function, or the condition of conformality.

This is an interesting observation. If this is correct, 
changing the R-charge corresponds to changing the angles of cycles of NS5-branes, and $a$-maximization (see later discussion in \S\ref{a-max.subsec}) suggests that there is some preferred brane configuration over others. See also \cite{Kenyon} for further discussion on R-charges from mirror symmetry viewpoint.

As the same time, we should perhaps warn the readers that the correctness or the real physical meaning of the proposal is not clarified yet, to the best of the author's knowledge. How we can understand this from fivebrane systems, for example, is far from trivial. 
Indeed, from our viewpoint, brane tilings are fivebrane systems, and the angle formed by two NS5-cycles  are simply given by that formed by two primitive normals to the toric diagram, at least in the strong coupling limit. In slightly complicated examples, we can easily check that this does not coincide with R-charge of superconformal $U(1)_R$ (see examples of \S\ref{a-max.subsec}). 
We should also point out that the gauge theory described by fivebrane diagrams are in UV, whereas the conformal field theory is in IR, and thus it is far from trivial how to understand conformality from fivebrane viewpoint.

\subsection{A note on string coupling}\label{note.subsec}

So far, we have discussed how to construct fivebrane systems representing four-dimensional superconformal quiver gauge theories. The fivebrane system is a complicated system of D5-branes and NS5-brane, and in some sense singular because NS5-branes suddenly change their direction perpendicularly at the intersection with D5-branes (Figure \ref{junction}, \ref{NS5patch}).

The source for all this happening is that we have taken the limit $g_{str}\to \infty$ in the explanation of previous section. In the strong coupling limit, since
\beq
T_{D5}\sim \frac{1}{l_s^6}\frac{1}{g_{\textrm{str}}}, ~~
T_{NS5}\sim \frac{1}{l_s^6}\frac{1}{g_{\textrm{str}}^2},
\eeq
we have $T_{D5} \gg T_{NS5}$ and thus D5-branes become flat and NS5-branes are perpendicular to them. This is precisely the situation we have seen previously (Figure \ref{NS5patch}). 
 
 Readers might think this strong coupling limit is a terrible limit, because in that limit we cannot confidently say what kind of theories we have on the D-branes! We would like to stress, however, that this limit should {\em  not} be taken as physical strong coupling limit. In principle, and in simple examples, we can explain everything in the decoupling limit $g_{str}\to 0$. Here we first choose to use this strong coupling limit simply because it is a certain limit in which brane configuration simplifies considerably and is directly related to toric data. 

Let us make the limit more precise. As defined above, let $R$ be a radius of torus. We take the limit $R\to 0$ because we want to decouple KK modes. At the same time, we are taking the limit $g_{str} \to \infty$. Furthermore, we want to keep the combination $R^2/{(g_{\rm str} l_s^2)}$ finite since that quantity corresponds to (some order one coefficient times) $1/g^2$, with $g$ the coupling constant of gauge theory. For proper discussion of this point, see \eqref{oong} of \S\ref{marginal.subsec}. The rough reason is that $1/g^2$
 comes from dimensional reduction of DBI action of D5-brane, which is proportional both to $1/g_{\rm str}$ and to $R^2$ (area of $\bT^2$). The factor $1/l_s^2$ is needed for dimensional reasons.

Summarizing, what we mean by ``the strong coupling limit'' is 
\beq
g_{str}\to \infty, l_s\to 0, R\to 0,\ \ \textrm{ with } \frac{R^2}{g_{\rm str} l_s^2} \textrm{ kept fixed }.
\eeq
Note that in this limit, $T_{D5}$ and $T_{NS5}$ become infinity, and thus back-reaction to brane configuration can safely be ignored.

%%%%%%%%%%%%%%%%%%%%%%%%%%%%%%%%%%%%%%%%%%
\subsection{The weak coupling limit} \label{weak.subsec}

Next we consider the another limit, the weak coupling limit. More precisely speaking, the limit is 
\beq
g_{str}\to 0, l_s\to 0, R\to 0,\ \ \textrm{ with } \frac{R^2}{g_{\rm str} l_s^2} \textrm{ kept fixed }. \label{weakcouplinglimit}
\eeq
In this limit, tension of NS5-brane is much larger than that of D5-brane and thus the NS5 surface $\Sigma$ (see Figure \ref{configre.tbl}) becomes holomorphic curve (holomorphic curve is always a special Lagrangian submanifold). We next discuss the shape of this holomorphic curve.

\begin{table}[htbp]
\caption{The brane configuration corresponding to brane tilings is shown again. In the weak coupling limit, the surface $\Sigma$, which is a two-dimensional surface in 4567 space, becomes a holomorphic surface with respect to variables $x,y$ in \eqref{xydef}.}
\label{configre.tbl}
\begin{center}
\begin{tabular}{ccccc|cccc|cc}
\hline\hline
& 0 & 1 & 2 & 3 & 4 & 5 & 6 & 7 & 8 & 9 \\
\hline
NS5 & $\circ$ & $\circ$ & $\circ$ & $\circ$ &\multicolumn{4}{c|}{$\Sigma$} & & \\
D5  & $\circ$ & $\circ$ & $\circ$ & $\circ$ && $\circ$ && $\circ$ & &\\
\hline
\end{tabular}
\end{center}
\end{table}

\subsubsection{The shape of NS5-brane} \label{NS5shape.subsubsec}
Let us forget about D5-branes for the moment and concentrate on NS5-brane.
Actually, one can write down an explicit expression for $\Sigma$. It is the zero locus of the Newton polynomial of the toric diagram $\Delta$ ($\subset \bR^2$):

\begin{equation}
P(x,y)=\sum_{(k,l)\in \Delta\cap \bZ^2}c_{k,l}x^ky^l.
\label{newton}
\end{equation}

Here, if we define complex variables
\begin{equation}
s=x^4+ix^5,\quad
t=x^6+ix^7, \label{stdef}
\end{equation}
$x$ and $y$ are given by 
\begin{equation}
x=e^{2\pi s},\quad y=e^{2\pi t}. \label{xydef}
\end{equation}
If we restore $R$ (radius of $\bT^2$), then \eqref{stdef} should be replaced by 
\begin{equation}
s=\frac{x^4+ix^5}{R},\quad
t=\frac{x^6+ix^7}{R}. \label{stdefnew}
\end{equation}
In our following discussions, we set $R=1$ for simplicity and thus the period of $x_5$ and $x_7$ is normalized to be $1$. 

It is not difficult to check that the zero locus of the Newton polynomial \eqref{newton} preserves $\scN=2$ SUSY. The fact that the shape of NS5-brane is given by Newton polynomial is discussed first in the context of mirror symmetry\cite{Hori:2000kt,Hori:2000ck} (as we will explain in \S\ref{D6.subsec}, the fivebrane system we are discussing now is related by T-duality to mirror Calabi-Yau). Another way of understanding this fact is to use holomorphy and asymptotic behavior, as will be explained in a moment.

In \eqref{newton} the coefficients $c_{k,l}$ are arbitrary as long as generic. The meaning of these coefficients will be studied in detail in \S\ref{BPS.subsec} and \S\ref{marginal.subsec}. There we identify (part of) degrees of freedom of changing coefficients with the those of geometrical deformations preserving $\mathcal{N}=1$ symmetry in fivebrane systems, or equivalently with those of exactly marginal deformations in $\scN=1$ superconformal quiver gauge theories. In Calabi-Yau language, they correspond to deformation of K\"ahler structure in toric Calabi-Yau side, and thus to the deformation of complex structure in mirror D6-brane side (see \S\ref{D6.subsec}). In fact, the mirror manifold $\scW$ of the toric Calabi-Yau $\scM$ is given by (see \S\ref{D6.subsec}) the equation
\beq
P(x,y)=uv,
\eeq
where $u,v \in \bC^{\times}$.
and you can see from this that $c_{k,l}$ corresponds to complex structure deformations.

The one-dimensional complex manifold given by $P(x,y)=0$ is a Riemann surface (with punctures), with genus $g$ given by the number of internal lattice points of the toric diagram, and $d$ is equal to the number of points on the boundary of the toric diagram. In particular, the $d$ here coincides with the previously introduced $d$ in \S\ref{strong.subsec}, or the number of NS5 cycles.

%\beq
%\{P(x,y)=0 \} \sim \Sigma_{g,d}, \ \ \textrm{with }g=N_{\rm int}.
%\eeq

As an example of this fact, let us give you a simple example of the conifold. In this case
\begin{equation}
P(x,y)=c_{(0,0)}+c_{(1,0)}x+c_{(0,1)}y+c_{(1,1)}xy . \label{conifoldeq}
\end{equation}
Seemingly we have four complex parameters, but if we use the freedom of overall rescaling of $x$ and $y$ (recall from \eqref{xydef} this corresponds to the shift of origin in 46-directions), we have only one parameter left and we have
\begin{equation}
P(x,y)=c+x+y+xy.
\label{conifoldeq2}
\end{equation}

\begin{figure}[htbp]
\centering{\includegraphics[scale=0.5]{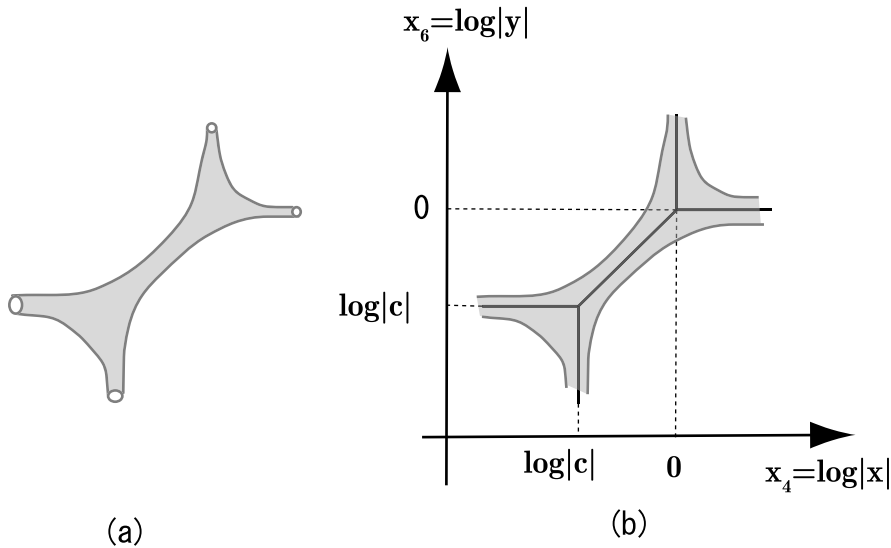}}
\caption{The shape of NS5-brane (a) and its amoeba (b) for the conifold, whose Newton polynomial is shown in \eqref{conifoldeq2}.}
\label{conifolddeform}
\end{figure}

In this case, $P(x,y)=0$ can be solved with respect to $y$ as $y=-\frac{x+c}{1+x}$ and we have complex plane and the genus is 0. The puncture is at $(x,y)=(0,-c),(-c,0),(\infty,-1),(-1,\infty)$, and we have four punctures. This coincides with the above explanation. See Figure \ref{conifolddeform} for a picture of the Riemann surface and its projection onto 46-plane.

From Figure \ref{conifolddeform}, we see that $|c|$ represent resolution parameter. When we consider singular Calabi-Yau, this is zero, and we are left with a single real parameter $\textrm{arg}(c)$. The meaning of this parameter will be discussed in \S\ref{BPS.subsec} and \S\ref{marginal.subsec}.

In the conifold case, it is useful to consider the projection onto $46$-plane, as in Figure \ref{conifolddeform}. In general, $\Sigma$ is a two-dimensional surface in four-dimensional space and is difficult to visualize. It is thus natural to consider projections. The projection of $\Sigma$ (shape of NS5-brane) onto $4$ and $6$ directions (the D5 directions) is actually known in mathematics literature since long ago and is called {\em amoeba}\cite{GKZ}. It is originally defined in the study of the monodromy of the so-called GKZ-hypergeometric functions\cite{GKZ1,GKZ2,GKZ3}. It also appears in many different contexts, such as real algebraic geometry, tropical geometry and recently in the instanton counting\cite{Nakatsu} and certain soliton systems in supersymmetric gauge theories\cite{FNOSY}. See \cite{PassareRullgard_Amoebas,Mikhalkin_Amoebas} for review on amoebae.

More formally, let $P(x,y)$ be a Laurent polynomial of two variables\footnote{This definition has obvious generalization to $n$-variable case, but we concentrate on two-variable case here.}. Then the amoeba $\mathcal{A}(P)$ of $P(x,y)$ is given by the image of $P^{-1}(0)$ by the map $\Log$:
\beq
\begin{array}{ccccc}
 \Log & : & (\bC^{\times})^2 & \to & \bR ^2 \\
 & & \rotatebox{90}{$\in$} & & \rotatebox{90}{$\in$} \\
 & & (x, y) & \mapsto &
  {\displaystyle (\log|x|, \log|y|)}.
\end{array}
\eeq
That is,
\begin{equation}
\begin{array}{ccccc}
 \mathcal{A}(P)\equiv\Log(P^{-1}(0)).
\end{array}
\end{equation}
Note since we have exponential in the definition of $x,y$ \eqref{xydef}, this map $\Log$ is simply a projection onto $x^4, x^6$-plane.

\begin{figure}[htbp]
\begin{center}
\centering{\includegraphics[scale=0.7]{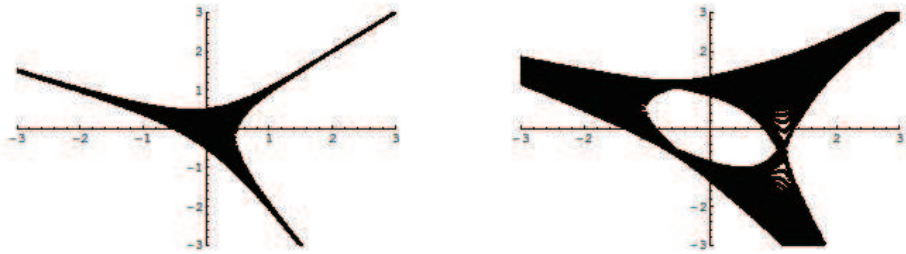}}
\caption{Examples of amoebae are shown. The left figure is for $W=x+y+\frac{1}{xy}$, and the right for $W=x+y-3+\frac{1}{xy}$. By changing coefficients, the shape of amoeba changes in a subtle way, but the 'tentacles' of amoeba is always the same.}
\label{amoebaeg}
\end{center}
\end{figure}

The example of amoeba is shown in Figure \ref{amoebaeg}.
You can see from these examples that the shape of amoeba changes in a subtle way if we change the coefficients of the Newton polynomial of convex polytope $\Delta$.
The `tentacles' of amoeba, however, always have the same slope as that of the normals to $\Delta$.

We can understand this fact as follows. Since we want to consider the behavior at infinity, we consider $(x,y)=(r^n u, r^m v)$ with $r\to\infty$ with $u,v\in \bCx$. Then the equation $P(x,y)=0$ becomes 

\begin{equation}
P(x,y)=\sum_{(i,j)\in \Delta}c_{i,j}r^{ni+mj}u^i v^j,
\label{newtonnew}
\end{equation}
and the leading term $P_{(n,m)}$ becomes
\begin{equation}
P_{(n,m)}(x,y)=\sum_{(i,j)\in \Delta,\ ni+mj\textrm{ takes largest value}}c_{i,j}r^{ni+mj}u^i v^j.
\label{leading}
\end{equation}

This means that the term $x^k y^l$ with the largest value of $nk+ml$ becomes dominant in the $r\to\infty$ limit. For general choice of $(n,m)$, only single term is dominant and we have no solution to $P_{(n,m)}(x,y)=0$ in $(\bCx)^2$. When $(n,m)$ is orthogonal to an edge of the toric diagram, however, we have several terms with the same value of $nk+ml$ and we can have non-trivial solution to $P_{(n,m)}(x,y)=0$. This explains why spines extend in the direction of normals to the toric diagram $\Delta$.

Although we have checked this fact explicitly, physically speaking this is to be expected, since in the strong coupling limit (\S\ref{strong.subsec}), we have provided independent argument for the existence of semi-infinite cylinder of NS5-branes for each normal to the toric diagram (see Figure \ref{Bfieldjump}). Here we have verified this fact again in the weak coupling limit.

If you turn this argument around, you can see that this fact explains the form of the holomorphic function given in \eqref{newton}. That is, we want to have semi-infinite cylinders in the weak coupling limit, as we have studied in the strong coupling limit. Then we have to choose $P(x,y)$  such that for each primitive normal $(n_{\mu}, m_{\mu})$ to the toric diagram ($\mu=1,2,\ldots d$), $P(x,y)$ coincides with $P_{(n_{\mu},m_{\mu})}$ in the direction $(x,y)=(r^{n_{\mu}} u, r^{m_{\mu}} v)$.  The form of $P(x,y)$ given in \eqref{newton} is the most general form of such polynomials. You can also understand from these facts that the coefficients $c_{k,l}$ in \eqref{newton} are left as arbitrary parameters.

Finally, we briefly comment on the limit $R\to 0$, where $R$ is the radius of $\bT^2$. Recall we take this limit both in the strong and the weak coupling limit, in order to decouple KK modes of $\bT^2$. In this limit, amoeba degenerates into `spines', or the so-called $(p,q)$-web or the web diagram in the physics literature \footnote{We note that although we have exactly the same graphs as old works\cite{AharonyHK}, the physical meaning is slightly different, because there $(p,q)$-web represents $(p,q)$-brane, whereas in our case it simply represents NS5-brane wrapping $(p,q)$ cycles of $\mathbb{T}^2$. These two viewpoints are related by chain of dualities.}. 
In mathematics literature, this limit $R\to 0$ is called ``tropical limit'', `dequantization' or `ultradiscretization' depending context. The resulting geometry is called ``tropical geometry''\footnote{You should not take the meaning of this name too seriously. According to \cite{SpeyerSturmfels}, the name `tropical' was coined by a French mathematician Jean-Eric Pin \cite{Pin}, in honor of their Brazilian colleague Imere Simon \cite{Simon}.}, which is an active area of research in mathematics (see \cite{Mikhalkin,FirstSteps} for reviews). Tropical geometry seems to play no fundamental role in the discussion of brane tilings and quiver gauge theories (at least so far), but they are useful in other situations, such as in application to soliton junctions (\cite{FNOSY}; see \S\ref{soliton.sec} for brief discussion.).

\paragraph{Coamoebae}

The story so far is more or less well-known, but we get something new if we consider another projection onto two-dimensional torus specified by $x^5$ and $x^7$ directions. This is known as coamoeba, and is first defined by Passare and Tsikh (unpublished). It is rediscovered later by physicists\cite{Feng:2005gw} and is renamed alga.

More formally, the coamoeba $\tilde{\mathcal{A}}(P)$ of a Laurent polynomial $P(x,y)$ is the image of $P^{-1}(0)$ by the following map $\Arg$: 
\begin{equation}
\begin{array}{ccccc}
 \Arg & : & (\bCx)^2 & \to & (\bR / \bZ )^2 \\
 & & \rotatebox{90}{$\in$} & & \rotatebox{90}{$\in$} \\
 & & (x, y) & \mapsto &
  {\displaystyle \frac{1}{2 \pi}(\arg(x), \arg(y))}.
\end{array}
\end{equation}
That is,
\beq
\tilde{\mathcal{A}}(P)\equiv\Arg(P^{-1}(0)).
\eeq

\begin{figure}[htbp]
\begin{center}
\centering{\includegraphics[scale=0.7]{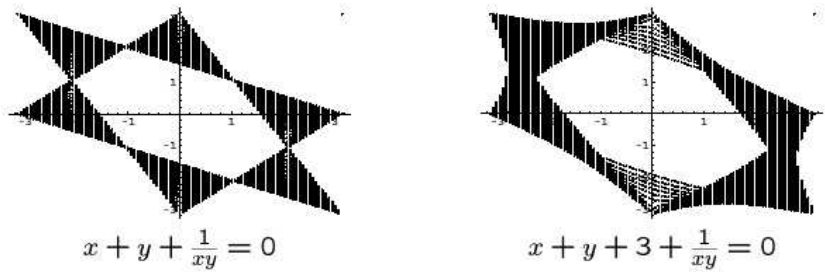}}
\caption{Examples of coamoebae are shown. By changing coefficients, the shape of amoeba changes in a subtle way, but the ``asymptotic boundary'' of coamoeba is always the same, and corresponds to the primitive normals of toric diagram.}
\label{coamoebaeg}
\end{center}
\end{figure}

\begin{figure}[htbp]
\begin{center}
\centering{\includegraphics[scale=0.7]{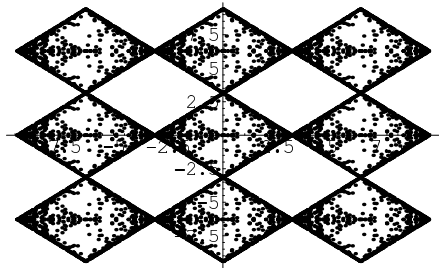}}
\caption{Coamoeba for $K_{\bP^1\times \bP^1}$, drawn using Monte-Carlo. Figure taken from \cite{Feng:2005gw}.}
\label{coamoebaegFeng}
\end{center}
\end{figure}

\begin{figure}[htbp]
\centering{\includegraphics[scale=0.4]{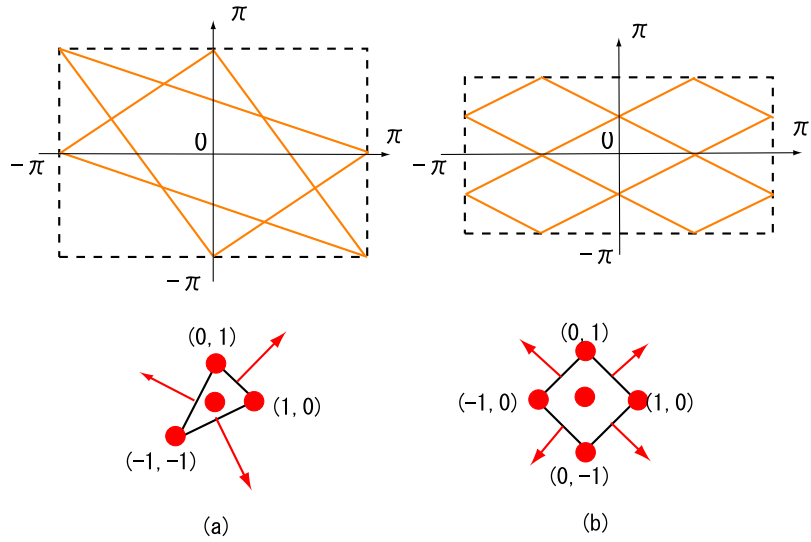}}
\caption{The asymptotic boundary of coamoeba in the case of $\bC^3/\bZ_3$ (a) and $K_{\bP^1\times \bP^1}$ (b). Their coamoebae are shown in Figure \ref{coamoebaeg} and \ref{coamoebaegFeng}, respectively. Their toric diagrams are also shown below, and the reader will recognize that the slope of asymptotic boundaries coincide with that of normals to toric diagram.}
\label{asymptotic}
\end{figure}

The examples of coamoebae are shown in Figure \ref{coamoebaeg}, and Figure \ref{coamoebaegFeng}. %, \ref{coamoebaegFeng2}.
As you can see in these examples, the shape of the coamoeba changes again in a subtle way, depending on coefficients.
But if you see more closely, you will notice that they have several asymptotic lines, and their slope is again given by the slope of the toric diagram (Figure \ref{asymptotic}).

We can understand the appearance of these asymptotic lines by an analysis similar to that in the case of amoeba \cite{UY3}.
Let us now carry that out.
From the analysis of amoeba case, we only need to consider the limit $(x,y)=(r^{n_{\mu}} u, r^{m_{\mu}} v)$ with $r\to\infty$, where $(n_{\mu}, m_{\mu}) \in \bZ^2$
is the primitive outward normal vector of and edge $\mu$ of $\Delta$.
Let $l_{\mu}$ be the integer such that
the defining equation for the edge $\mu$ is
\beq
 n_{\mu} i + m_{\mu} j = l_{\mu}.
\eeq

Then, in the $r \to \infty$ limit, we have
\beq
 P(r^{n_{\mu}} u, r^{m_{\mu}} v)
  = r^{l_{\mu}} P_{\mu} (u, v) + O(r^{l_{\mu}-1}),
\eeq
with the leading term $P_e(u,v)$ given by
\beq
 P_{\mu} (x, y)
  = \sum_{n_{\mu} i + m_{\mu} j = l_{\mu}} c_{ij} x^i y^j.
\eeq

Now assume for simplicity
\footnote{A comment for those interested in homological mirror symmetry. In our discussion of homological mirror symmetry in \S\ref{HMS.sec}, this assumption is not important since the category of A-branes, the directed Fukaya category $D^b\dirFuk P$, does not depend on the choice of
sufficiently general $P$. A-brane category its is defined by symplectic geometry, and thus does not depend on complex structure deformation.} that for an edge $e$,
the leading term $P_e(x,y)$ is a binomial
\beq
 P_{\mu}(x, y)
  = c_1 x^{i_1} y^{j_1}
     + c_2 x^{i_2} y^{j_2}, \label{Ptwoterms}
\eeq
where $c_1, c_2 \in \bC$ and
$(i_1, j_1), (i_2, j_2) \in \bZ^2$.

Put $\alpha_i = \arg(c_i)$ for $i = 1, 2$ and
\beq
 (x, y) =
   (r^{n_{\mu}} |c_2| \be(\theta), r^{_{\mu}} |c_1| \be(\phi)),
\eeq
where $\be(\theta)\equiv \exp(2\pi \sqrt{-1} \theta)$.
Then the leading behavior of $W$ as $r \to \infty$
is given by
\begin{equation} \label{eq:coamoeba_boundary}
 r^{l_{\mu}} W_e(\be(\theta), \be(\phi))
  =  r^{l_{\mu}} |c_1 c_2| \{ \be(\alpha_1 + i_1 \theta + j_1 \phi)
                + \be(\alpha_2 + i_2 \theta + j_2 \phi) \}.
\end{equation}
Hence the coamoeba of $W^{-1}(0)$ asymptotes in this limit
to the line on the torus $\bT^2$:
\beq
 (\alpha_2 - \alpha_1) + (i_2 - i_1) \theta + (j_2 - j_1) \phi
  + \frac{1}{2} = 0 \  \mod \bZ. \label{asymbound}
\eeq
This line will be called
an {\em asymptotic boundary} of the coamoeba of $W^{-1}(0)$. The slope of this line is given by $\left(j_2-j_1,-(i_2-i_1)\right)$, which is identical to $(n_{\mu},m_{\mu})$ (recall we are now using \eqref{Ptwoterms}). This shows that asymptotic boundaries we defined just now coincide with asymptotic lines shown in Figure \ref{asymptotic}, and this is what we originally expected.

As in amoeba case, the appearance of asymptotic boundaries is consistent with analysis in the strong coupling limit. In the strong coupling limit, we have cycles of NS5-brane with winding number $(n(e),m(e))$ for each primitive normal $(n(e),m(e))$ to toric Calabi-Yau, and this corresponds to asymptotic boundaries of coamoeba.
However, we should notice at the same time that in the weak coupling limit, the story is not that simple as in the strong coupling limit. For example, although asymptotic boundary coincides with the real boundary of coamoeba for triangle and parallelogram toric diagram by suitable choice of coefficients\cite{UY1,UY2}, this is not the case in general toric diagram.

\begin{figure}[htbp]
\begin{center}
\centering{\includegraphics[scale=0.5]{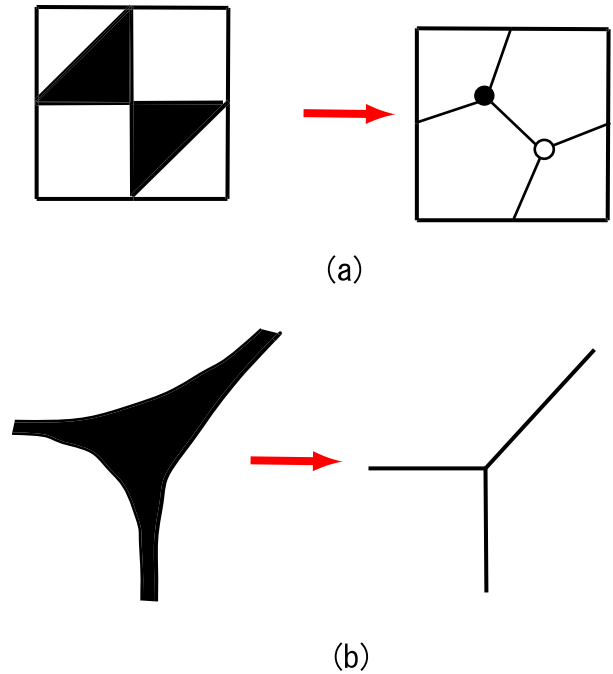}}
\caption{Reducing the fivebrane diagram (coamoeba) to the bipartite graph (a) is an analogue of reducing the shape of NS5-brane (amoeba) to the web-diagrams (b).}
\label{coamoebatropical}
\end{center} 
\end{figure}

Since this coamoeba story is almost parallel to amoeba case, it is natural to ask the question whether we have an analogue of  ``tropical limit'' ($R\to 0$) for coamoeba. In a sense, the answer is yes and the resulting object is precisely the bipartite graph, although the meaning of this limit is different from $R\to 0$. Our motto is that bipartite graphs are ``tropical analogue'' of fivebrane diagrams (Figure \ref{coamoebatropical}).

\subsubsection{Inclusion of D5-branes} \label{D5cycle.subsubsec}
\begin{figure}[htbp]
\centering{\includegraphics[scale=0.5]{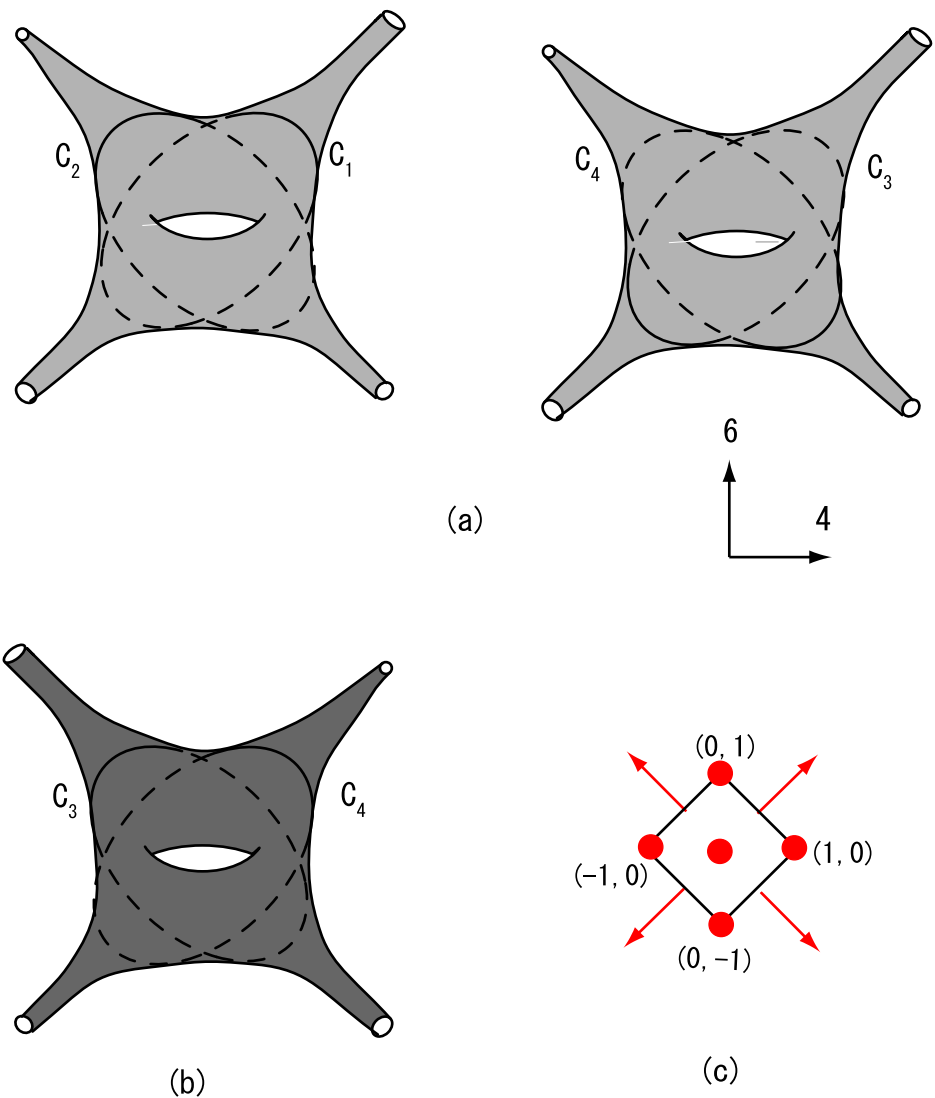}}
\caption{The fivebrane configuration in the weak coupling limit. The NS5 surface $\Sigma$ is a torus with four punctures. The intersections $C_a$ of $\Sigma$ with D5-branes $D_a$ are shown in (a). The reversed figure is also shown in (b). The toric diagram is shown (c), which has four lattice point on its boundary and one internal lattice point. This corresponds to the fact that $\Sigma$ has genus 1 and 4 punctures.}
\label{F0cycles.eps}
\end{figure}

After spending much time explaining the geometry of NS5-surface $\Sigma$, we now go back to Table \ref{configre.tbl}. In the Table, we also have D5-branes, which we denote by $D_a$ ($a=1,2,\ldots N_G$, where $N_G$ is the number of gauge groups of the quiver gauge theory, or the number of nodes of quiver diagram.). These D5-branes have intersections with NS5-brane along one-cycle $C_a$ of $\Sigma$, which is a boundary of the disc $D_a$. An example of $K_{\bP^1\times \bP^1}$ is given in Figure \ref{F0cycles.eps}.

In \S\ref{step-by-step.subsubsec}, we have the condition \eqref{pq0-2} for the conservation of NS5-charge. In the weak coupling limit, we have a condition for D5-charge conservation, which amounts to
\beq
\sum_a N_a C_a=0 \textrm{ in } H_1(\bar{\Sigma},\bZ), \label{ND0}
\eeq
where $N_a$ stands for the number of D5-branes on the disc $D_a$, or the rank of the corresponding gauge group. Also, $\bar{\Sigma}$ stands for the compactification of surface $\Sigma$. Since $\Sigma$ is a non-compact surface, you might think that RR-charge can escape to infinity. But since cylinders of NS5-branes cannot be sources or sinks of RR-charge, we can use $\bar{\Sigma}$ to write down condition for RR-charge conservation, as in \eqref{ND0}. 

In most of our previous discussions, $N_a$s are all common and set to $N$. 
In this case, \eqref{ND0} simplifies to 
\beq
\sum_a C_a=0 \textrm{ in } H_1(\bar{\Sigma},\bZ). \label{sumD0}
\eeq
Note that this is similar to \eqref{pq0-2}, which reads
\beq
\sum_{\mu} \balpha_{\mu}=0 \textrm{ in } H_1(\bT^2,\bZ)
\eeq

In the example of Figure \ref{F0cycles.eps}, the Riemann surface $\Sigma$ is a torus with four punctures, and we have four cycles of D5-branes. The number of D5-brane cycles, which is four in this case, is the same with the number of nodes of the quiver diagram shown in Figure \ref{F0quiverI}.

Now let us check that we have $\scN=1$ supersymmetric quiver gauge theories on the D5-branes. We have already verified this fact in the strong coupling limit in \S\ref{quiver.subsubsec}, but we can provide an independent argument in the weak coupling limit. One important difference from the strong coupling analysis is that we have cycles of D5-branes on NS5-branes in the weak coupling limit, whereas we have cycles of NS5-branes on D5-branes in the strong coupling limit. In other words, we have some sort of intersecting D5-brane models in the weak coupling limit. In the usual discussion of intersecting D-brane models, we have intersecting D-brane in some manifold, say torus, but in our discussion the geometry itself is flat and we have NS5-brane instead. As we will explain in \S\ref{D6.subsec}, by taking T-duality, we can go to more conventional intersecting D6-branes.

The discussion is almost similar to the strong coupling limit.
First, for each disc $D_a$ of D5-branes, we have a $SU(N_a)$ gauge group. In other words, $N_a$ D5-branes are wrapping cycle $D_a$. Second, for each intersection point of D5-branes, we have a massless open string, or a bifundamental field. The presence of NS5-brane makes the theory chiral, precisely as in the strong coupling case (Figure \ref{chiral}). Third, if we can span a disc, then we have a term in the superpotential, again exactly as in the case of strong coupling limit (Figure \ref{discamp}).

\begin{figure}[htbp]
\centering{\includegraphics[scale=0.45]{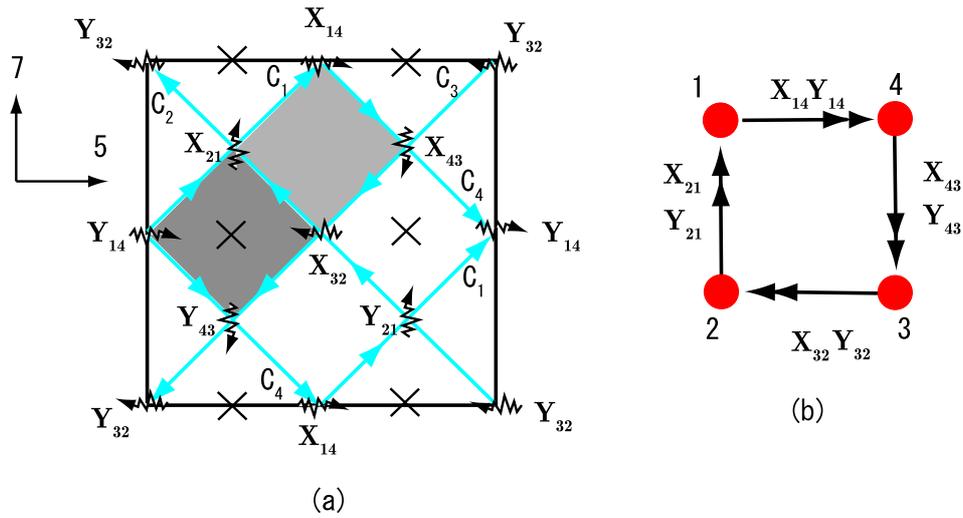}}
\caption{Another representation of Figure \ref{F0cycles.eps}. The large black square in (a) represents NS5-brane. Be careful with the difference with previous figures. In the discussion of \S\ref{strong.subsec}, the torus represents D5-brane, whereas in this figure the torus represents the shape of NS5-brane $\Sigma$. The four crosses represent punctures. The four blue cycles $C_a$ represents the cycles of D5-branes, and the black wary arrows, placed at intersection points of D5 cycles, represents bifundamental field. The light gray square contributes to a term in the superpotential, whereas the dark gray square does not because of the existence of puncture inside it. The corresponding quiver diagram is shown in (b).}
\label{F0Sigma}
\end{figure}

We now again take $K_{\bP^1\times \bP^1}$ as an example. The torus $\Sigma$ of Figure \ref{F0cycles.eps} can also be represented as a fundamental region of square with opposite edges identified, as in Figure \ref{F0Sigma} (a). In the Figure, the cycles of D5-branes are represented as blue arrows, and punctures of Riemann surface is represented as four crosses. The four cycles $C_a (a=1,2,\ldots,4)$ have 8 intersection points, which corresponds to 8 bifundamental fields in the quiver diagram (b). Concerning the superpotential, we have, for example, a term like $\tr (X_{14}X_{43}X_{32}X_{21})$ since we have a disc amplitude spanning region colored light gray in (a). However, we do not have a term like $\tr(Y_{14} Y_{43}X_{32}X_{21})$ coming from disc spanning region colored dark gray, since we have a puncture in that region. Therefore, we have four terms of in the superpotential, not eight.

Finally, we comment on one difficulty in the weak coupling descriptions, since some readers might be wondering at this point why we did not discuss the weak coupling limit at first. Anyway, the weak coupling limit (the decoupling limit) is actually the limit we are interested in.

The reason for first presenting strong coupling limit is that, in general, it is a non-trivial problem to obtain $C_a$ and $D_a$. Given a toric data, we do not know even the homology class of $C_a$. Of course, since D5-branes are special Lagrangian submanifolds in 4567-space, in principle we should be able to find them explicitly for arbitrary toric diagram, but that is difficult in practice. 
%\footnote{The author would like to thank Reiko Miyaoka for correspondence on this point.}. 
In the next subsection we explain how brane tiling bypasses this problem, and gives the information about D5-brane cycles $C_a$.

\subsection{Untwisting}\label{untwist.subsec}
So far, we have explained the strong coupling limit and the weak coupling limit
 separately. Each has its own its advantages and disadvantages. In the strong coupling limit, the relation with toric diagram is simple and the brane configuration is simply represented by fivebrane diagrams, but the gauge theory interpretation is not necessarily clear. In the weak coupling limit (or the decoupling limit), the gauge theory interpretation is clear but brane configuration is slightly complicated because we have to use a holomorphic curve to represent the shape of NS5-brane.

Actually, we can relate the strong coupling limit to the weak coupling limit, using `untwisting' operation\cite{Feng:2005gw}. Physically speaking, this corresponds to changing the string coupling constant from infinity to zero. 
The reader might worry at this point, since we do not know the precise shape of branes in general string coupling constant. However, untwisting is a topological operation, and we do not have to consider the real shapes of branes in order to know the homology class of $C_a$.

\begin{figure}[htbp]
\centerline{\includegraphics[scale=0.8]{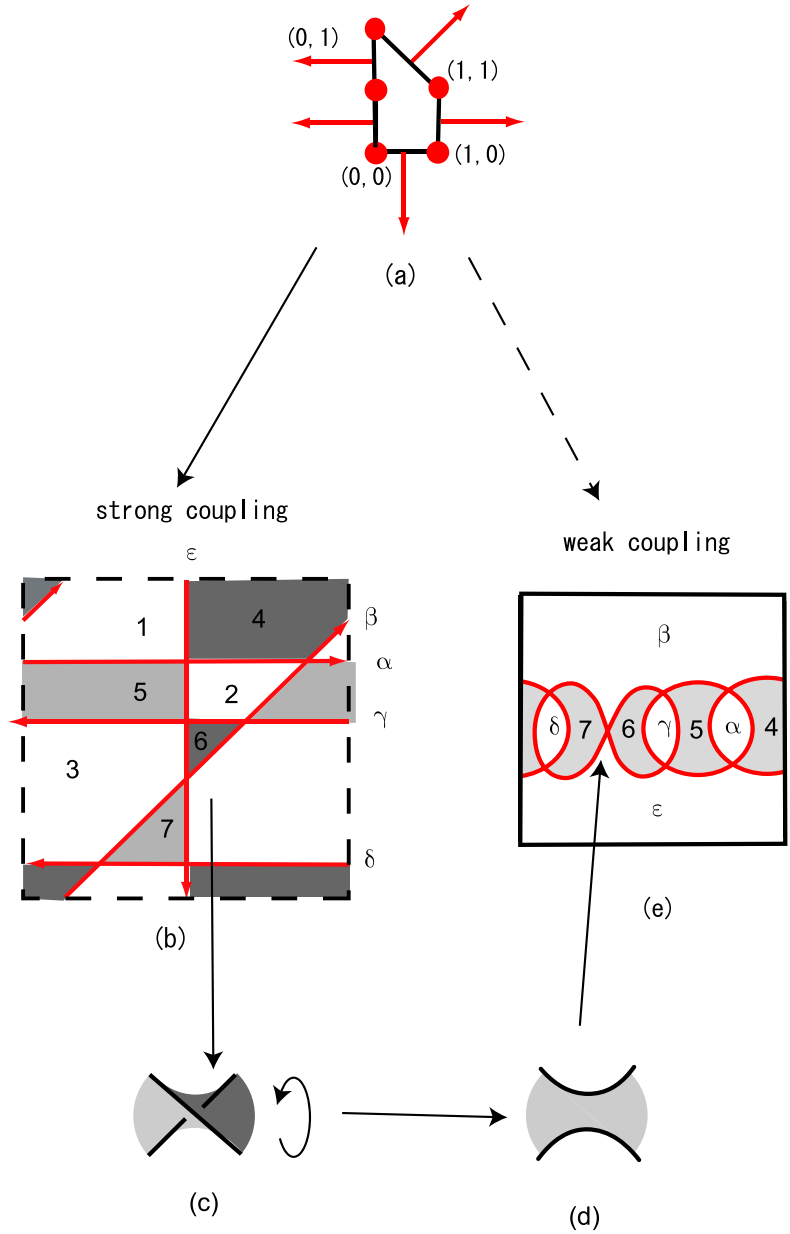}}
\caption{The untwisting operation of the SPP tiling, whose toric diagram is shown in (a). As discussed previously in Figure \ref{quiverread}, the fivebrane diagram is (b). Locally, untwisting corresponds to replacing (c) by (d), and the resulting surface is (e), where the two vertical lines on the left and right sides are identified. Through the untwisting, $(N,-1)$ faces are turned over and become $(-N,1)$
faces. It follows that all shaded faces in (e) have NS5 charge $+1$.}
\label{untwistspp.eps}
\end{figure}

Let us first forget about the D5-branes and concentrate on the NS5-brane.
Namely, we remove the D5-branes, and we regard the white faces
in Figure \ref{untwistspp.eps}(a) as holes.
This makes the torus an NS5-brane composed of pieces connected at
the corners.
In Figure \ref{untwistspp.eps} the pieces of
NS5-branes
are represented as shaded faces.
Because of the bipartite property of the system, any two shaded faces contacting each other
have opposite NS5 charges.
This means that the NS5-brane changes the orientation at the contact points
of faces.
In other words, the NS5-brane is `twisted' at the intersections of cycles.

In order to obtain the surface $\Sigma$ (the shape of $\Sigma$) in the weak coupling limit, we gradually take the string coupling constant smaller and smaller, so that the tension of NS5-brane becomes stronger and stronger as compared with D5-branes. Then , $(N,-1)$ faces with opposite NS5 orientations are turned over so that they become $(-N,1)$ faces, as shown in Figure~\ref{untwistspp.eps}(b).
By shrinking the holes to punctures, we finally obtain the surface $\Sigma$. \footnote{For mathematically inclined readers, we remark that this procedure is somehow similar to the construction of Seifert surface in knot theory (see \cite{Kennaway:2007tq} for discussions).}
In the SPP case depicted in Figure~\ref{untwistspp.eps},
we obtain a genus $0$ surface (i.e. a sphere) with the five
punctures corresponding to five cycles in the original bipartite graph.

We can check the consistency of this procedure. First, since winding cycles of $\bT^2$ (zig-zag path) are mapped into puncture of $\Sigma$, the surface $\Sigma$ should have $d$ punctures. But we have verified previously in \S\ref{weak.subsec} that the number of punctures of the Newton polynomial \eqref{newton} is exactly $d$. You can also verify that the untwisted surface is genus $g$, where $g$ is given by the number of lattice points, as in \S\ref{weak.subsec}. This is easily carried out by the explicit computation of Euler number\cite{Feng:2005gw}.

Now we take D5-branes into account. D5-branes intersects NS5-brane with 1-cycle, and in the strong coupling limit, that 1-cycle is simply given by edges going around the face of $(N,0)$-brane (recall each region of $(N,0)$-brane corresponds to a stack of $N$ D5-branes). After the untwist, that 1-cycle is now mapped to winding cycles $C_a$ of $\Sigma$, and each D5-brane becomes a disc $D_a$ bounding $C_a$ (here we follow the same notation as in \S\ref{D5cycle.subsubsec}). We can therefore read off the intersection cycles of NS5-brane and D5-branes, and the result should be the Figure \ref{F0cycles.eps} and \ref{F0Sigma}, for example. In the discussion of the weak coupling limit, one of the difficulties is that we do not have the information regarding such cycles in general. By untwisting, however, we can directly obtain the homology class of the cycle $C_a$. This operation of untwisting will play crucial roles in, for example, \S\ref{marginal.subsec} and \S\ref{HMS.sec}.

If we summarize such correspondence between the strong and weak coupling limit, the result is
\begin{itemize}
\item The zig-zag path, or the winding cycle of $\bT^2$ in bipartite graph, is turned into punctures of $\Sigma$ in the weak coupling limit. Physically, this corresponds to semi-infinite cylinder of NS5-brane.
\item The face of bipartite graph on $\bT^2$ in the strong coupling limit is mapped to a disc $D_a$ bounding 1-cycle $C_a$ of $\Sigma$. Physically, this corresponds to a stack of $N$ D5-branes.
\end{itemize}

\subsection{Summary}
In this section, we explained fivebrane setup which realizes $\scN=1$ supersymmetric quiver gauge theories. The fivebrane system is complicated in general string coupling constant, but simplifies in two limits (\S\ref{note.subsec}). As we have seen, each limit has its own advantages and disadvantages.

The first limit is the  strong coupling limit (\S\ref{strong.subsec}). This strong coupling limit is not a physical strong coupling limit, and we take this limit for simplicity of the analysis. In this limit, brane configurations simplifies dramatically, and has direct connection with toric Calabi-Yau geometry. D5-branes become flat, and semi-infinite cylinders of NS5-branes are attached to D5-branes. Unfortunately, the relation with gauge theory is not clear in this limit. We have emphasized fivebrane viewpoint represented by fivebrane diagrams, but these diagrams are equivalent to more conventional bipartite graphs.

Another limit is the weak coupling limit or the decoupling limit (\S\ref{weak.subsec}). In this limit, the connection with gauge theory is clear, but fivebrane configuration is more complicated. The shape of NS5-brane is represented by a holomorphic curve, and its intersection with D5-brane is a 1-cycle on the surface. The connection with toric Calabi-Yau is not direct as in the strong coupling limit, although we can easily read off corresponding quiver gauge theory from brane configuration. 

The two descriptions are related by the procedure of untwisting (\S\ref{untwist.subsec}). By untwisting, junctions of semi-infinite NS5-brane cylinders in the strong coupling limit is turned into a smooth holomorphic surface in the weak coupling limit, and regions of D5-branes are turned into 1-cycles on the holomorphic surface. By this procedure, fivebrane configuration in the weak coupling limit is directly connected with Calabi-Yau geometry.

%%%%%%%%%%%%%%%%%%%%%%%%%%%%%%%%%%%%%%%%%%%%%%%%%%%%%%%%%%%%%%%%
\section{More on brane tilings}\label{more.sec}

In the previous section, we have discussed basic aspects of brane tilings. We now turn to more detailed discussions. The inclusion of fractional branes and flavor branes is discussed in \S\ref{fractional.subsec} and in \S\ref{flavor.subsec}. In \S\ref{BPS.subsec}, we analyze the BPS conditions of fivebrane systems, and relate that analysis to exactly marginal deformations of quiver gauge theories in \S\ref{marginal.subsec}. These two subsections are linked, and the reader interested in \S\ref{marginal.subsec} should read \S\ref{BPS.subsec} in order to understand the real significance of the discussion. More combinatorial aspects, such as Kasteleyn matrix (\S\ref{another.subsec}) and perfect matchings (\S\ref{GLSM.subsec}) are also discussed. They are related, respectively, to the fast forward algorithm and solving F-term conditions. The discussion of Seiberg duality (\S\ref{Seiberg.subsec}) is also contained. Finally, we close this section in \S\ref{D6.subsec} with a brief discussion of the relation with mirror Calabi-Yau setup. Most of the discussions in each subsection is independent, and the only exception is \S\ref{BPS.subsec} and \S\ref{marginal.subsec}.

\subsection{Fractional branes} \label{fractional.subsec}
Most of the discussions so far deals with the case where the rank of gauge groups are all equal to the same number $N$, but we can also consider more general rank assignments. In Calabi-Yau setup, this corresponds to the introduction of fractional branes. 

As we discussed in \S\ref{quiver.subsec}, anomaly cancellation conditions are imposed on the rank assignments. In \cite{Imamura:2006ub}, it is shown that anomaly cancellation conditions are derived from the conditions of RR-charge conservation.

As an example, let us consider the case of $K_{\bP^1\times \bP^1}$, shown in Figure \ref{F0cycles.eps} and \ref{F0Sigma}. The condition of RR-charge \eqref{ND0} reads (since $N_1=\balpha+\bbeta$ in $H_1(\overline{\Sigma},\bZ)=H_1(\bT^2,\bZ)\bZ \balpha+\bZ\bbeta$ and so forth \footnote{$\balpha, \bbeta$ denotes $\alpha$ and $\beta$-cycles of $\bT^2$, or its homology class. We use bold letters throughout this review in order to reserve $\alpha, \beta$ for labels of corners of the toric diagram.})
\beq
N_1(1,1)+N_2(-1,1)+N_3(-1,-1)+N_4(1,-1)=0 ,
\eeq
or
\beq
N_1=N_3,\ N_2=N_4 \label{D13D24}.
\eeq
Now if you compare this with \eqref{N13N24} in \S\ref{quiver.subsec}, you will immediately recognize that these two conditions are the same. 

To give a more general argument, we note that the gauge anomaly cancellation condition \eqref{anomcancel2} (which we reproduce here for the convenience of the reader)
\beq
\sum_b \sigma(b,a)N_b=0,
\eeq
can be rewritten as 
\beq
\sum_b \langle C_b, C_a \rangle N_b=\langle \sum_b N_b C_b, C_a \rangle=0, \label{sigma3}
\eeq
where $\langle *,*\rangle$ denotes the intersection number of two cycles. This is because we have a bifundamental field for each intersection point of D5-brane cycles. Also, the sign of the intersection number corresponds to the chirality of bifundamental field, and this determines the sign of $\sigma(b,a)$ (recall the definition of $\sigma(b,a)$ in \eqref{sigma} and \eqref{sigma2}).
Clearly, the final expression \eqref{sigma3} follows from RR-charge conservation condition \eqref{ND0}.

\begin{figure}[htbp]
\centering{\includegraphics[scale=0.45]{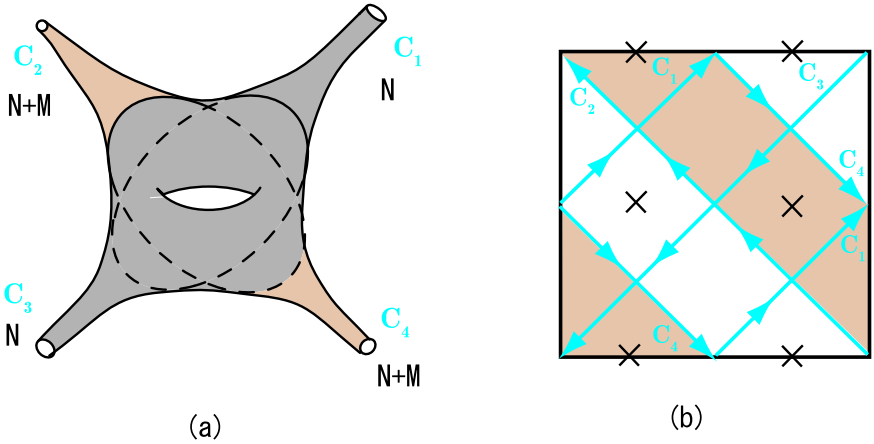}}
\caption{An example for fivebrane system with fractional branes. We show the case of $K_{\bP^1\times \bP^1}$. See Figure \ref{F0cycles.eps}, \ref{F0Sigma} for figures without fractional branes. Here we consider the rank assignments $(N_1,N_2,N_3,N_4)=(N,N+M,N,N+M)$ for all D5-brane cycles $C_1, C_2,\ldots C_4$. As shown in (a), of four semi-infinite cylinders, two of them, colored light brown, becomes $(M,1)$-branes due to RR-charge conservation. More accurate figure is (b), where again the region colored light gray is $(M,1)$-brane, whereas other region is the usual NS5-brane.}
\label{fractional.eps}
\end{figure}

Now one difference arises from our previous discussions. The difference of D5-brane charge in two neighboring regions should be compensated by the D5-charge of NS5 cylinders, and this turns each asymptotic part of the NS5-brane into D5-NS5 bound state, or $(p,1)$-brane (Figure \ref{fractional.eps}). The tension of $(p, 1)$-branes is given by
\beq
T_{(p,\pm 1)}\sim \frac{1}{\sqrt{p^2 g_{\mbox{str}}^2 + g_{\mbox{str}}^4}}\frac{1}{l_{s}^6},
\eeq
which is comparable to that of the D5-brane in the strong coupling limit. This means we cannot use the strong coupling limit argument as in \S\ref{strong.subsec}. We should stress, however, that this is necessarily not a problem. We can safely consider the decoupling limit, and everything is fine.

Finally, we mention that inclusion of fractional brane breaks conformal invariance and quiver gauge theories enjoy the phenomena of the so-called duality cascade . These theories are interesting, because show have QCD-like behavior such as confinement and chiral symmetry breaking, as exemplified in the seminal work of Klebanov and Strassler\cite{Klebanov:2000hb}. See \cite{Franco:2005zu,Bertolini:2005di,Berenstein:2005xa,Dymarsky:2005xt,Argurio:2006ew,Brini:2006ej,Evslin:2007au} for some of related literatures.

%%%%%%%%%%%%%%%%%%%%%%%%%%%%%%%%%%%%%%%%%%%%%%%%%%%%%%%%%
\subsection{Flavor branes} \label{flavor.subsec}
%%%%%%%%%%%%%%%%%%%%%%%%%%%%%%%%%%%%%%%%%%%%%%%%%%%%%%%%%
Another important ingredient in brane tilings is the flavor brane, which give fields
belonging to the fundamental and the anti-fundamental representations, or quarks and anti-quarks. In this paper, we only discuss flavor D5-branes in 46-directions. By T-duality transformation
they are transformed into D7-branes wrapped on toric divisors in the
toric Calabi-Yau geometry.
In \cite{Franco:2006es} non-compact D7-branes wrapped on divisors in
toric Calabi-Yau cones are investigated, and the matter contents
supplemented by the flavor branes are proposed. We are going to give a more general argument, following Section 5 of our paper \cite{IKY}.

Let us begin with the Calabi-Yau setup. A flavor D7-brane wrapped on a divisor in the Calabi-Yau $3$-fold is
represented as a curve on the NS5-brane worldvolume,
connecting two punctures. Since such a D7-brane runs to infinity, its volume is infinite and thus the gauge field on the D7-brane is no longer dynamical. This means such D7-branes are indeed ``flavor branes''.

In fivebrane diagrams, the two punctures
are represented as two 1-cycles (recall the untwisting rule in \S\ref{untwist.subsec}),
and the curve corresponds to an intersection of these two 1-cycles.
This intersection point is the flavor D5-brane worldvolume
projected onto the 57-plane.

\begin{figure}[htbp]
\centerline{\includegraphics{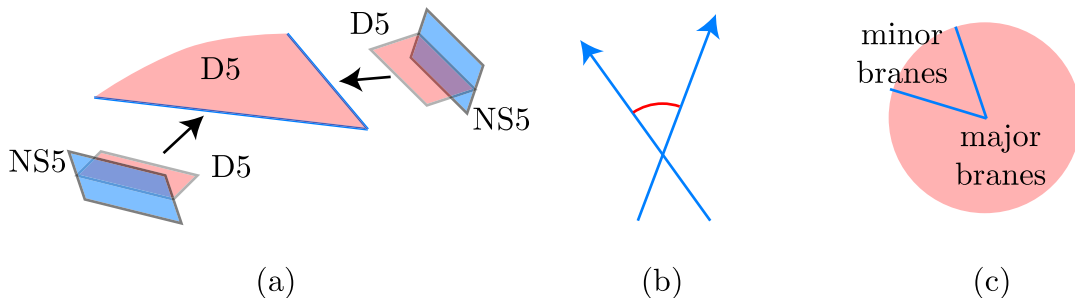}}
\caption{(a) shows a flavor brane stretched between two legs
in a web-diagram.
In the internal 57-space, the D5-brane is attached on the NS5-branes.
In the corresponding fivebrane diagram (b) we use an arc to represent the flavor brane.
As is shown in (c) there are two possible flavor branes associated a pair of legs in a web-diagram.
We call them minor and major branes.
}
\label{flavor.eps}
\end{figure}

On the web-diagram, a flavor brane is represented as
a fan between two external legs corresponding to the
two zig-zag paths
(Figure \ref{flavor.eps} (a)).
When we specify two legs, there are always two
fans defined by these legs.
One has minor central angle and the other has
major central angle.
We call these two possible flavor branes for a given pair of external legs
minor branes and major branes
(Figure \ref{flavor.eps} (c)).
In order to distinguish these two types of flavor branes,
we represent flavor branes as arcs at the intersection of cycles
(\ref{flavor.eps} (b)).
Arcs are drawn in the angle corresponding to the fans
on the web-diagram.
We can define these angles because the directions of the cycles
are the same as those of external legs in the web-diagram.
We are now going to treat each of these two cases in detail.

\paragraph{Minor flavor branes}
In \cite{Franco:2006es}
the following superpotential
is proposed for quarks $q$ and $\wt q$ emerging from
the introduction of flavor branes placed on an intersection $I$:
\begin{equation}
W=\wt q\Phi_Iq.
\label{QPQ}
\end{equation}
This superpotential corresponds only to
minor flavor branes as
will be confirmed in the following.

If we assume that quark fields are supplied from the
D3-D7 strings in the Calabi-Yau perspective, the fundamental fields must become
massless when D3-branes coincide with the D7-branes.
Hence massless loci of fundamental fields in the moduli space
should be identified with the worldvolume of the D7-branes.

When the quark mass term is given by (\ref{QPQ}),
the massless locus is given by $\Phi_I=0$.
(Following the usual procedure to obtain Calabi-Yau geometry,
we here treat all the gauge groups as $U(1)$).
In order to determine the corresponding divisor in the moduli space,
we should solve the F-term conditions imposed on bi-fundamental fields.
As will be explained in \S\ref{GLSM.subsec}, the solution is given by \cite{Franco:2006gc}
\begin{equation}
\Phi_I=\prod_{D\ni I}\rho_{D},
\label{ppr}
\end{equation}
where $\rho_{D}$ are complex fields
defined for each perfect matching $D$ \footnote{We use $D, D', \ldots$ to denote perfect matchings, whereas we use $\alpha, \beta$ to denote the corners of the toric diagram, or corresponding perfect matching.},
and $D\ni I$ means that the product is taken over all
the perfect matchings which include the edge $I$.
By this relation we can describe the moduli space
of quiver gauge theory as the moduli space of
gauged linear sigma model (GLSM) with the fields $\rho_{D}$.
The equation (\ref{ppr}) means that
the massless locus is given by the union
of loci defined by $\rho_{D}=0$.
Because we are interested in divisors,
we do not take care of subspace of moduli space with dimension
less than $2$.
We only focus on the complement of the submanifold
corresponding to the legs and the center of the
web-diagram.
We can show that
in this submanifold
GLSM fields $\rho_{D}$ which do not correspond to
corners of the toric diagram do not vanish.
This allow us to forget about such fields and we have only to
take care of fields $\rho_\alpha$, which corresponds to corners in
the toric diagram.
The following theorem can be proved:
\begin{theorem}\label{divisor}
A divisor $F_\alpha$ is given by $\rho_\alpha=0$.
\end{theorem}
The proof is given in Appendix A.1 of \cite{IKY}.
With the theorem \ref{divisor} we obtain
\begin{equation}
\mbox{massless locus}=\bigcup_{\alpha\ni I}F_{\alpha},
\end{equation}
and the theorem \ref{minor} means that this is precisely the
worldvolume of the minor branes associated with the edge $I$.

\begin{figure}[htbp]
\centerline{\includegraphics{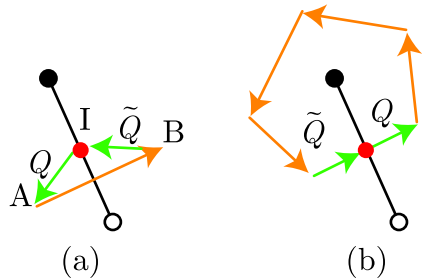}}
\caption{Closed paths representing quark mass terms for minor flavor branes (a) and major flavor branes (b) are shown.}
\label{fundmass0.eps}
\end{figure}

\paragraph{Major flavor branes}
In order to obtain the worldvolume of major flavor branes,
we need different quark mass terms from (\ref{QPQ}).
Let us assume the following form of quark mass term:
\begin{equation}
W=\wt Q{\cal O}Q,
\label{WQOQ}
\end{equation}
where ${\cal O}$ is composite operator made of bi-fundamental fields.
We denote quark fields provided by major flavor brane by $Q$ and $\wt Q$
while we write $q$ and $\wt q$ the quark fields for minor branes.
Now we can prove the theorem\cite{IKY}
\begin{theorem}\label{minor}
Let $I$ be an edge in a bipartite graph,
and $\{F_\alpha,F_\beta,\ldots,F_\gamma\}$ be the set of
facets whose associated perfect matchings include the edge $I$.
Then, facets in the set
$\{F_\alpha,F_\beta,\ldots,F_\gamma\}$ form
one continuous region
in the web-diagram,
and the central angle of the region is always a minor angle.
\end{theorem}
This theorem means that 
the worldvolume of major flavor branes associated with the
edge $I$ is given by
\begin{equation}
\mbox{major branes}=\bigcup_{\alpha\notni I}F_{\alpha},
\end{equation}
where $D \notni I$ means that the product is taken over all
the perfect matchings which do not include the edge $I$.
By the theorem \ref{divisor}
this is given by ${\cal O}=0$ with the operator ${\cal O}$ defined by
\begin{equation}
{\cal O}=\prod_{D\notni I}\rho_{D}.
\end{equation}
In order to write the superpotential (\ref{WQOQ}),
we need to rewrite the operator ${\cal O}$
in terms of bi-fundamental fields in the gauge theory.
It is easy to see that
\begin{equation}
{\cal O}=\prod_{J\in k, J\neq I}\Phi_J,
\label{opp}
\end{equation}
where $k$ is one of two endpoints of the edge $I$,
and $J\in k$ means edges sharing the vertex $k$.
Namely, the product in 
(\ref{opp}) is taken over all edges ending on $k$ but $I$.
When we regard this as the operator in the gauge theory with
non-Abelian gauge groups, the constituent fields should
be ordered so that the color indices
of adjacent fields match.
The operator ${\cal O}$ are graphically represented as the
path consisting of orange arrows in Figure \ref{fundmass0.eps} (b).
In the definition of the operator ${\cal O}$
there are two choices of the endpoint of the edge $I$.
Let ${\cal O}_B$ and ${\cal O}_W$ be the two operators
obtained by choosing black and white endpoints of $I$, respectively.
Because the superpotential of bi-fundamental fields
includes
\begin{equation}
W=\tr(\Phi_I{\cal O}_B)-\tr(\Phi_I{\cal O}_W),
\end{equation}
and the F-term condition of $\Phi_I$ gives
\begin{equation}
{\cal O}_B={\cal O}_W.
\end{equation}
Therefore, the superpotential (\ref{WQOQ}) does not depend on the
choice between ${\cal O}_B$ and ${\cal O}_W$.

Let us compare the two superpotentials obtained by the minor and major
flavor branes.
\begin{equation}
W_{\rm minor}=\wt q_i\Phi_I^i{}_{j'}q^{j'},\quad
W_{\rm major}=\wt Q_{i'}{\cal O}^{i'}{}_jQ^j.
\label{wmwm}
\end{equation}
These two are represented as cycles made of orange and green arrows
in Figure \ref{fundmass0.eps}.
In (\ref{wmwm}) color indices are explicitly written.
Notice that the existence of these terms requires
the chirality of the quark fields should be opposite
between minor and major flavor branes.
If we have a bi-fundamental field in the representation $(\fund,\fundbar)$
at the edge $I$,
minor branes give quarks in the representation
$(\fundbar,1)$ and $(1,\fund)$ while
major branes give ones in
$(\fund,1)$ and $(1,\fundbar)$
(Table \ref{quarks}).
\begin{table}[htbp]
\caption{Two types of flavor branes and representations of
quark fields}
\label{quarks}
\begin{center}
\begin{tabular}{cl}
\hline
\hline
& $SU(N)\times SU(N)'$ \\
\hline
bi-fundamental & $\Phi_I^i{}_{j'}(\fund,\fundbar)$ \\
minor brane & $q^{i'}(1,\fund)$, $\wt q_i(\fundbar,1)$ \\
major brane & $Q^i(\fund,1)$, $\wt Q_{i'}(1,\fundbar)$ \\
\hline
\end{tabular}
\end{center}
\end{table}

\begin{figure}[htbp]
\centerline{\includegraphics{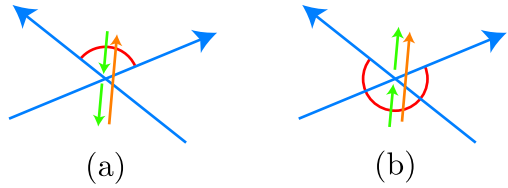}}
\caption{Graphical representation of minor branes (a) and major branes (b),
and corresponding fields.}
\label{minormajor.eps}
\end{figure}

In Figure \ref{minormajor.eps}, the difference of quark representations
is expressed by the orientation of arrows.

\subsection{BPS conditions} \label{BPS.subsec}

\begin{figure}[htbp]
\centering{\includegraphics[scale=0.6]{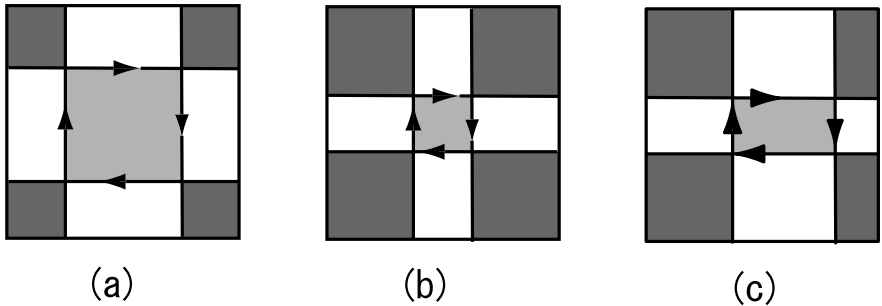}}
\caption{Changing the position of cycles of NS5-brane does not change the quiver diagram and superpotential. Are these deformations preserve BPS conditions? Does this deformation have any physical implications in the quiver gauge theory side? These are exactly the question we are going to ask in this subsection and the next. As we will see, the answer is that (a) and (b) preserves $\scN=1$ SUSY, whereas (c) does not.}
\label{cycledeform.eps}
\end{figure}

Throughout this review we have emphasized the fact that brane tilings have physical interpretations. But some readers might still be a bit skeptical about the meaning of the word `physical', because so far we have only used only topological/graphical aspects of brane configurations. Namely, we can use the bipartite graph instead of the fivebrane diagram, and then essentially everything we have discussed so far is contained in a single bipartite graph. If brane tilings genuinely represent physical fivebrane systems, then the deformation of the position of 1-cycles of NS5-branes on the torus (as shown in Figure \ref{cycledeform.eps} should have some implications on the gauge theory side. This is an important problem, since it goes beyond topological/graphical information for the first time in the study of brane tilings.

In this section, we investigate BPS conditions of fivebrane systems both in the weak coupling limit and the strong coupling limit, following \cite{IIKY}. We also determine the moduli parameters of brane configurations, which will be identified in \S\ref{marginal.subsec} with (part of) exactly marginal deformations of corresponding superconformal quiver gauge theories .

\subsubsection{The weak coupling limit} \label{weak.subsubsec}
We first study BPS conditions for the brane configuration in the case of weak coupling limit $g_{\rm str}\to 0$.
%in \S\ref{weak.sec}, \ref{bps.sec} and \ref{sec.gc}.
Then we determine the number of moduli parameters
of brane systems for general tilings.

In some simple cases, we can explicitly solve BPS conditions 
and determine the constraints imposed on the coefficients of
the equation defining the NS5-brane worldvolume.
It is, however, difficult to solve the BPS conditions in general cases.
This difficulty can be avoided by considering
the opposite limit, $g_{\rm str}\rightarrow\infty$,
which we discuss in \S\ref{strong.subsec}.
In this limit, we can easily solve the BPS conditions,
and we obtain results similar to those in the weak coupling case.

For our purpose, it is convenient to define
a function $Q(x^5,x^7)$ on the torus
as follows.
Let $\Pi(x^5,x^7)$ be the two-dimensional plane along the $46$ directions
with fixed $x^5$ and $x^7$ coordinates
in the $4567$ space.
In a four-dimensional space, two two-dimensional surfaces generically
intersect at isolated points, and we can thus define the intersection number of them.
We define the function $Q(x^5,x^7)$ as the intersection number
of the holomorphic curve $\Sigma$ and the plane $\Pi(x^5,x^7)$.
By definition, this function takes integer values at generic points
and jumps by one along cycles which are
defined as the projections of boundaries (punctures) of $\Sigma$
at infinity.
We can use the function $Q(x^5,x^7)$ to describe
the asymptotic structure of the surface $\Sigma$.
If the function $Q(x^5,x^7)$ takes only the values $0$ and $\pm1$,
we can construct the corresponding bipartite graph in the manner mentioned in \S\ref{quiver.subsubsec}.
But, in general, $Q(x^5,x^7)$ may not satisfy this condition.
Because $Q(x^5,x^7)$ is defined as a projection,
the condition $|Q(x^5,x^7)|\geq2$ does not imply the emergence of $(N,k)$-branes with $|k|\geq2$.
In the weak coupling limit, the worldvolume of an NS5-brane is a holomorphic curve,
and such branes never appear.

To clarify the relation between the structure of the surface $\Sigma$
and the function $Q(x^5,x^7)$,
let us focus on the asymptotic form of $\Sigma$.
Let us label the edges of the toric diagram
by $\mu=1,2,\ldots,d$ as we go around the sides in a counterclockwise manner.
We also define the primitive integral normal vector
$(m_\mu,n_\mu)$ of the $\mu$-th edge.
For each edge $\mu$, we have an external line
with direction $(m_\mu,n_\mu)$ in the web-diagram
constructed on the $46$ plane.
If a side $e$ of a toric diagram consists of $n_e$ edges,
we have $n_e$ parallel external lines associated with this side.

As discussed in \S\ref{NS5shape.subsubsec} \eqref{leading}, the corresponding asymptotic parts of $\Sigma$ are given by
\begin{equation}
P_e(x,y)=0,
\label{pazero}
\end{equation}
where $P_e(x,y)$ is the `restriction' of
the Newton polynomial to the side $A$
defined by
\begin{equation}
P_e(x,y)=\sum_{(i,j)\in e}c_{i,j}x^iy^j.
\label{newtona}
\end{equation}
Here, the summation is taken over only the points
on the side $e$.
Thus, the parameters responsible for the asymptotic behavior
of $\Sigma$ are those assigned to the points on the
perimeter of the toric diagram.
These variables
are actually redundant.
For example, we can use the freedom of rescaling $x$ and $y$
to eliminate two parameters.
Moreover, one coefficient can be eliminated through an overall
rescaling of $P(x,y)$.
Thus, we have $d-3$ complex parameters
associated with the asymptotic form of $\Sigma$.

\paragraph{another parameterization}
Instead of the coefficients in the Newton polynomial,
it is more convenient to introduce another set of
variables which is directly related to the
asymptotic structure of the surface $\Sigma$.
The surface $\Sigma$ has $d$ punctures
corresponding to each external line in the web-diagram.
The asymptotic form of $\Sigma$ around the puncture $\mu$
is a cylinder.
Its projection onto the $4$-$6$ plane
gives an external line of the web-diagram,
and its projection onto the $5$-$7$ plane
gives a cycle $\balpha_\mu$ in the torus.
We define the parameters $M_\mu$ and $\zeta_\mu$,
representing the positions of the lines and the cycles, respectively, by
\begin{equation}
M_\mu=\lim_{\rightarrow\mu}(n_\mu x^4-m_\mu x^6),\quad
\zeta_\mu=\lim_{\rightarrow\mu}\left(n_\mu x^5-m_\mu x^7+\frac{1}{2}\right),
\label{mzeta2}
\end{equation}
or equivalently,
\begin{equation}
-e^{2\pi(M_\mu+i\zeta_\mu)}=\lim_{\rightarrow\mu}
x^{n_\mu}y^{-m_\mu}.\label{mzeta}
\end{equation}
These are defined as the limits in which we approach the puncture $\mu$
in the curve $\Sigma$.
At this point, only the fractional part of $\zeta_\mu$ is defined,
because the $x^5$ and $x^7$ coordinates are periodic.
The parameter $M_\mu$ determines the position in the $4$-$6$-plane
of the external line of the web-diagram corresponding to
a puncture $\mu$.
This is also regarded as the `moment' generated by the
tension of the $\mu$-th external line.
The parameter $\zeta_\mu$ determines the position of
the cycle $\balpha_\mu$ in the torus.
We have a set of $2d$ real parameters $\{M_\mu,\zeta_\mu\}$.
These parameters must be related to the $d-3$ complex parameters
in the Newton polynomial $P(x,y)$ associated with the perimeter
of the toric diagram.
Due to the translational symmetry, only $2d-4$ parameters
in $\{M_\mu,\zeta_\mu\}$
are relevant to the shape of the surface $\Sigma$.
We still have two more real parameters in $\{M_\mu,\zeta_\mu\}$.

The discrepancy described above is resolved if we take account of
constraints imposed on $\{M_\mu,\zeta_\mu\}$.
As mentioned above, $M_\mu$ can be regarded as
the moments acting on the brane system.
Because the brane configurations
we are considering here are BPS and stable,
the total moment must vanishes:
\begin{equation}
\sum_{\mu=1}^d M_\mu=0.
\label{momentzero}
\end{equation}
This can be directly proved as follows.
Because $\Sigma$ is holomorphic,
the pull-back of the $(2,0)$-form $ds\wedge dt$ onto
$\Sigma$ vanishes:
\begin{equation}
ds\wedge dt|_\Sigma=0.
\end{equation}
We can decompose this into the following two equations:
\begin{eqnarray}
(dx^4\wedge dx^7-dx^6\wedge dx^5)|_\Sigma&=&0,\label{4765}\\
(dx^4\wedge dx^6-dx^5\wedge dx^7)|_\Sigma&=&0.\label{4657}
\end{eqnarray}
With Stokes' theorem,
we can rewrite the integration of (\ref{4765}) over $\Sigma$
as
\begin{equation}
0
=\int_\Sigma(dx^4\wedge dx^7-dx^6\wedge dx^5)
=\int_{\partial\Sigma}(x^4dx^7-x^6dx^5)
=\sum_{\mu=1}^d \lim_{\rightarrow\mu} (n_\mu x^4-m_\mu x^6),
\label{momentcancel}
\end{equation}
and this is identical to the relation (\ref{momentzero}).
In the final step
of (\ref{momentcancel}),
we have used the fact that
the boundary of the surface $\Sigma$ is the union of
cycles $\balpha_\mu$,
and the integral of $(dx^5,dx^7)$ over a cycle $\balpha_\mu$
gives the integral vector $(m_\mu,n_\mu)$.

We also obtain a similar constraint on the parameters $\zeta_\mu$.
Using the relation (\ref{4657}), we can show
\begin{equation}
\int_{{\mathbb T}^2} dx^5dx^7Q(x^5,x^7)
=\int_\Sigma dx^5\wedge dx^7
=\int_\Sigma dx^4\wedge dx^6
=\int_{\partial\Sigma}x^4dx^6=0.
\label{intqcond}
\end{equation}
In other words, the average of the function $Q(x^5,x^7)$
on the torus vanishes.
In the final step, we have used the fact that the
punctures of $\Sigma$ are points in the $x^4$-$x^6$ plane,
and the integral vanishes.
Because the function $Q(x^5,x^7)$ is defined as
a step function that is discontinuous along
the cycles $\balpha_\mu$, whose positions are
determined by the parameters $\zeta_\mu$,
(\ref{intqcond}) imposes one constraint on the
parameters $\zeta_\mu$.

As mentioned above, the function $Q(x^5,x^7)$
is a step function which jumps by one
along the cycles $\balpha_\mu$.
We can decompose this function into step functions $q_\mu(x^5,x^7)$
associated with each cycle $\balpha_\mu$;
\begin{equation}
Q(x^5,x^7)=\sum_{\mu=1}^d q_\mu(x^5,x^7).
\label{qbyq}
\end{equation}
The function $q_\mu(x^5,x^7)$
jumps by one on the cycle $\balpha_\mu$,
whose position on the torus is specified by the
parameter $\zeta_\mu$.
The explicit form of the function $q_\mu(x^5,x^7)$ is
\begin{equation}
q_\mu(x^5,x^7)=[[-n_\mu x^5+m_\mu x^7+\zeta_\mu]],
\label{qdef}
\end{equation}
where $[[\cdots]]$ is defined by
\begin{equation}
-1/2\leq [[x]]-x\leq 1/2,\quad
[[x]]\in{\mathbb Z},
\end{equation}
for a real variable $x$.
The function $q_\mu(x^5,x^7)$ is not periodic but satisfies
\begin{equation}
q_\mu(x^5+p,x^7+q)=q_\mu(x^5,x^7)
+qm_\mu-pn_\mu,
\end{equation}
for an arbitrary integral vector $(p,q)$.
Note that the definition of $q_\mu(x^5,x^7)$ depends on
not only the fractional part of $\zeta_\mu$ but also on its
integral part.
In \eqref{mzeta2} we defined only the fractional part of
the parameters $\zeta_\mu$.
Let us define the parameter $\zeta_\mu$ including
the integral part by
\begin{equation}
\zeta_\mu=\int_{{\cal F}_0}q_\mu(x^5,x^7)dx^5dx^7,
\label{zteaq}
\end{equation}
where ${\cal F}_0$ is the specific fundamental region
${\cal F}_0=\{(x^5,x^7)|-1/2\leq x^5,x^7\leq 1/2\}$.
We can easily check that this definition gives
the same fractional part as the previous definition (\ref{mzeta2})
of the parameters.
The integral parts of $\zeta_\mu$ defined in this way
contribute to the function $Q(x^5,x^7)$ through
(\ref{qbyq}).

Now we can rewrite the constraint (\ref{intqcond})
imposed on %the function 
$Q(x^5,x^7)$
in terms of the parameters $\zeta_\mu$.
Substituting the relation (\ref{qbyq}) into (\ref{intqcond}),
and using (\ref{zteaq}),
we obtain
\begin{equation}
\sum_{\mu=1}^d\zeta_\mu=0,
\label{t2cond}
\end{equation}
and this decreases the number of independent parameters
by one.
As a result, we have $d-3$ physical degrees of freedom
associated with the parameters $\zeta_\mu$.

The curve is specified by the parameters $M_\mu$
and $\zeta_\mu$ and the coefficients $c_{k,l}$ for the internal points
of the toric diagram.
In the next section,
we show that the BPS conditions of the D5-branes
fix some of them and leave $d-3$ parameters unfixed.
Later we explicitly show in simple examples that
the free parameters are $\zeta_\mu$, subject to the
constraint (\ref{intqcond}).

%%%%%%%%%%%%%%%%%%%%%%%%%%%%%%%%%%%%%%%%%%%%%%%%%%%%%%
\paragraph{BPS conditions and deformation of branes}\label{bps.sec}
To this point, we have considered only the NS5-brane
in a system.
To realize the gauge theory,
we need to introduce D5-branes,
and these, in fact, impose extra constraints
on the parameters of the surface $\Sigma$.

As we mentioned above,
an NS5-brane wrapped on $\Sigma$ preserves the ${\cal N}=2$ supersymmetry.
Let $\epsilon_1$ and $\epsilon_2$ be the parameters of two supersymmetries.
If we introduce D5-branes into this system,
each of them breaks the supersymmetry down to ${\cal N}=1$,
which is specified by
$\epsilon_2=(Z/|Z|)\epsilon_1$,
where $Z$ is the central charge of the D5-brane,
given by
\begin{equation}
Z=\int_{\rm D5} ds\wedge dt
 =\oint_{\partial\rm D5} \log x\frac{dy}{y}.
\label{zintdsdt}
\end{equation}
Note that the topology of each D5-brane is a disk,
and its boundary $\partial\rm D5$ is a 1-cycle on the NS5-brane $\Sigma$.

In the context of the mirror Calabi-Yau
geometry,
this integral is simply the period.
By taking the T-duality
along one of the directions transverse to the brane system,
say the $x^9$ direction,
the NS5-brane is mapped to the Calabi-Yau $P(x,y)=uv$,
where $u,v \in \mathbb{C}$, and we have a new $\mathbb{S}^1$-fiber
from T-duality (see \S\ref{D6.subsec}).
For the Calabi-Yau manifold, we have a holomorphic 3-cycle $\Omega$. In our coordinates, $\Omega$ can be written as
\begin{equation}
\Omega=\frac{dx}{x}\wedge\frac{dy}{y}\wedge\frac{du}{u}.
\end{equation}
We can trivially integrate this $\Omega$ along the $uv$-fiber direction,
and we are left with
\begin{equation}
\int_{{\rm D6}}\Omega
=\int_{{\rm D6}} \frac{dx}{x}\wedge\frac{dy}{y}\wedge\frac{du}{u}
\propto \int_{\rm D5} \frac{dx}{x}\wedge\frac{dy}{y}
=\oint_{\partial {\rm D5}} \log(x) \frac{dy}{y}.
\end{equation}
This is the central charge in (\ref{zintdsdt}).%
\footnote{The argument that
the period reduces to the integral
of a differential on 1-cycle of torus appeared long ago in \cite{Klemm:1996bj},
although in a slightly different context.}

As we see below, for complicated examples,
it is difficult to directly analyze the shape of the parameter space.
It is still possible, however, to count the number of dimensions of that space.
Let us define the following quantities:
$d$ is the number of lattice points on the boundary of a toric diagram;
$I$ is the number of lattice points inside a toric diagram;
and $S$ is the area of a toric diagram.
Since we have one complex coefficient for each term
of a Newton polynomial, i.e., for each lattice point of a toric diagram,
we have $I+d$ complex parameters.
These variables
are actually redundant.
For example, we can use the freedom of rescaling $x$ and $y$
to eliminate two parameters.
Moreover, one coefficient can be eliminated through the overall
rescaling of $P(x,y)$.
Thus we have $I+d-3$ complex parameters,
or $2(I+d-3)$ real parameters.

Let $n_g$ be the number of $SU(N)$ gauge groups.
For the brane system to be BPS,
all the central charges $Z_i$ ($i=1,\ldots, n_g$) of the D5-branes must have the same argument:
\begin{equation}
\arg Z_1=\arg Z_2=\cdots =\arg Z_{n_g}.
\label{zreal}
\end{equation}
Since the number of D5-branes is twice the area of the toric diagram,
we have $2S-1$ conditions.
Thus we compute
\begin{equation}
2(I+d-3)-(2S-1)
=d-3,
\end{equation}
where we have used 
Pick's theorem,
\begin{equation}
S=I+\frac{d}{2}-1.
\end{equation}

In order to use (\ref{zintdsdt})
to compute the central charges $Z_i$ of the D5-branes,
we need to know the homotopy
classes of the D5-boundaries on the surface $\Sigma$.
We have already explained in \S\ref{untwist.subsec} how to know this.

%%%%%%%%%%%%%%%%%%%%%%%%%%%%%%%%%%%%%%%%%%%%%%%%

As simple examples, we determine the parameter spaces
of brane configurations corresponding to generalized conifolds,
which include ${\mathbb C}^3$, the conifold, SPP,
as special cases.
First, we consider the SPP case, and then, we
discuss the generalization.

\begin{exa}[SPP]
The case of Suspended Pinched Point (SPP) is already discussed in Figure \ref{quiverread} and Figure \ref{untwistspp.eps}, 
The Newton polynomial for SPP is
\begin{equation}
P(x,y)=y(x-x_\alpha)+(x-x_\gamma)(x-x_\delta).
\end{equation}
We have used the rescaling of $y$ and $P(x,y)$ to set the coefficients
of $xy$ and $x^2$ terms to $1$.
We can also set one of the quantities $x_\alpha$, $x_\gamma$, and $x_\delta$ to
an arbitrary value
through a rescaling of the variable $x$.
Here we set $x_\alpha=-1$.
The surface $\Sigma$ is a sphere with five punctures
[see Figure \ref{untwistspp.eps}(b)].
Let us use $x$ as a holomorphic coordinate of the Riemann sphere.
If we regard $x^4$ and $x^5$ as the latitude and the longitude
on the sphere,
we have two punctures, one at the north pole ($x=\infty$) and one at the south pole
($x=0$), which correspond to the cycles $\beta$ and $\epsilon$, respectively.
The three other punctures are at $x=x_\alpha=-1$, $x=x_\gamma$, and $x=x_\delta$.
The parameters $M_\mu$ and $\zeta_\mu$
are given by
\begin{eqnarray}
&&
e^{2\pi(M_\alpha+i\zeta_\alpha)}=
e^{2\pi(M_\beta+i\zeta_\beta)}=1,
\nonumber\\&&
e^{2\pi(M_\gamma+i\zeta_\gamma)}=-\frac{1}{x_\gamma},\quad
e^{2\pi(M_\delta+i\zeta_\delta)}=-\frac{1}{x_\delta},\quad
e^{2\pi(M_\epsilon+i\zeta_\epsilon)}=x_\gamma x_\delta.
\end{eqnarray}
We can easily see that the constraints (\ref{momentzero})
and (\ref{t2cond}) actually hold.

By tracing the boundaries of the three $(N,0)$ faces
[the white faces in Figure \ref{untwistspp.eps}(a)]
in the untwisting procedure,
we obtain three contours on the Riemann sphere
(see Figure \ref{sppcontour.eps}).
We should evaluate the integral (\ref{zintdsdt})
along these contours.
The function $\log x$ has a cut along a meridian
(the wavy line in Figure \ref{sppcontour.eps}),
and the differential
\begin{equation}
\frac{dy}{y}=
\frac{dx}{x-x_\gamma}
+\frac{dx}{x-x_\delta}
-\frac{dx}{x-x_\alpha}
\label{yrational}
\end{equation}
has poles at three punctures
$\alpha$, $\delta$, and $\gamma$.
\begin{figure}[t]
\centerline{\includegraphics[scale=1]{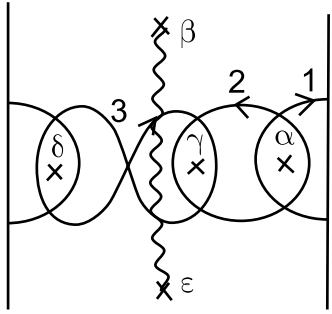}}
\caption{Three contours obtained from the boundaries
of the three $(N,0)$ faces, `1', `2', and `3',
in Figure \ref{untwistspp.eps}(a).
The symbols $\times$ indicate punctures corresponding to
the cycles in the tiling, and
the wavy line between the two punctures $\beta$ and $\epsilon$
represents a branch cut of $\log x$.
}
\label{sppcontour.eps}
\end{figure}
\begin{figure}[b]
\centering{\includegraphics[scale=0.4]{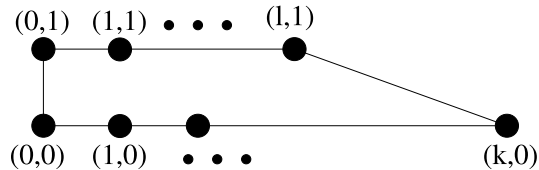}}
\caption{Toric diagram of a generalized conifold.}
\label{gconifoldTD}
\end{figure}
As shown in Figure \ref{sppcontour.eps},
each contour goes around two of these three punctures, 
and the integrals along them pick up the
residues of the differential $dy/y$.
We easily obtain
\begin{subequations}
\begin{eqnarray}
Z_1=\oint_1\log x\frac{dy}{y}
&=&2\pi i(\Log x_\alpha-\Log x_\delta),\\
Z_2=\oint_2\log x\frac{dy}{y}
&=&2\pi i(\Log x_\gamma-\Log x_\alpha),\\
Z_3=\oint_3\log x\frac{dy}{y}
&=&2\pi i(\Log x_\delta-\Log x_\gamma+2\pi i),
\end{eqnarray}
\end{subequations}
where $\Log$ is the principal value of the logarithm
defined with the branch cut in Figure \ref{sppcontour.eps}.
We have the extra $2\pi i$ in the third equation because
the contour crosses the cut.
For these central charges to have the same arguments,
the three punctures $\alpha$, $\gamma$ and $\delta$
must be at the same latitude:
\begin{equation}
1=|x_\alpha|=|x_\delta|=|x_\gamma|.
\end{equation}
(Note that we set $x_\alpha=-1$.)
This implies that all the parameters $M_\mu$ vanish,
and the unfixed parameters are
$\zeta_\gamma$ and $\zeta_\delta$.

\qed
\end{exa}

\begin{exa}[Generalized Conifolds] \label{gc.exa}
We can easily generalize this analysis to the case of a generalized conifold
$x^ky^l=uv$ with arbitrary $k$ and $l$.
The toric diagram of a generalized conifold 
is displayed in Figure \ref{gconifoldTD}. 
The toric diagram has $k$ edges on the bottom and $l$ edges on the top,
and we denote the corresponding cycles by $\alpha_i$ and $\beta_i$, respectively.
We also have two cycles, $\gamma$ and $\delta$,
corresponding to the other two edges of the toric diagram.
The Newton polynomial is then
\begin{equation}
P(x,y)
=\prod_{i=1}^k(x-x_{\alpha_i})+y\prod_{i=1}^l(x-x_{\beta_i})=0,
\end{equation}
where $x_{\alpha_i}$ and $x_{\beta_i}$ are the positions of the punctures $\alpha_i$ and $\beta_i$.
We set the coefficients of $x^k$ and $yx^l$ terms to $1$
by rescaling $y$ and $P(x,y)$.
We can also set either $x_{\alpha_i}$ or $x_{\beta_i}$ to an arbitrary value.
We here set $x_{\alpha_1}=-1$.
The normal vectors $(m_\mu,n_\mu)$ are given for each cycle by
\begin{equation}
\alpha_i:(0,-1),\quad
\beta_i:(0,1),\quad
\gamma:(-1,0),\quad
\delta:(1,k-l).
\end{equation}
The parameters $M_\mu$ and $\zeta_\mu$ are given by
\begin{eqnarray}
&&
e^{2\pi(M_{\alpha_i}+i\zeta_{\alpha_i})}=-\frac{1}{x_{\alpha_i}},\quad
e^{2\pi(M_{\beta_i}+i\zeta_{\beta_i})}=-x_{\beta_i},\nonumber\\
&&
e^{2\pi(M_\gamma+i\zeta_\gamma)}=\frac{\prod_{i=1}^k(-x_{\alpha_i})}{\prod_{i=1}^l(-x_{\beta_i})},\quad
e^{2\pi(M_\delta+i\zeta_\delta)}=1.
\end{eqnarray}
We use $x$ as a coordinate on the Riemann sphere $\Sigma$.
The two punctures $\delta$ and $\gamma$ are at the north ($x=\infty$) and south poles ($x=0$),
and the function $\log x$ has a cut along a meridian connecting them.
The other $k+l$ punctures are poles of
the differential
\begin{equation}
\frac{dy}{y}
=\sum_{i=1}^k\frac{dx}{x-x_{\alpha_i}}
-\sum_{i=1}^l\frac{dx}{x-x_{\beta_i}}.
\end{equation}

The tiling for the generalized conifold has $k$ down-going cycles $\alpha_i$ and $l$ up-going cycles $\beta_i$.
Before performing the untwisting operation,
we need to fix the order of these vertical cycles in the tiling.
The choice of the ordering corresponds to the choice of
the toric phase.
Let $\eta_i$ ($i=1,\ldots,k+l$) be the set of two
kinds of cycles $\alpha_i$ ($i=1,\ldots,k$) and
$\beta_i$ ($i=1,\ldots,l$) ordered as
\begin{equation}
\arg x_{\eta_1}\leq \arg x_{\eta_2}\leq\cdots\leq\arg x_{\eta_{k+l}}.
\end{equation}
The cycles $\eta_i$ divide the tiling into $k+l$ strips.
\begin{figure}[htbp]
\centerline{\includegraphics[scale=1]{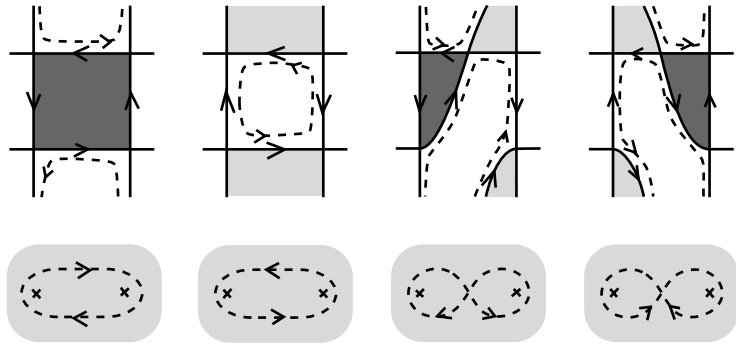}}
\caption{The upper four diagrams depict pieces of tilings of generalized conifolds.
The tiling of the generalized conifold $x^ky^k=uv$ consists of
$k+l$ of them.
Each strip has one $(N,0)$ face.
The dashed lines represent the boundaries of $(N,0)$ faces.
The lower four diagrams depict the contours obtained through the untwisting
operation from the boundaries.
Each of them encloses two punctures (indicated by $\times$)
corresponding to the two sides of the strip.
}
\label{gc.eps}
\end{figure}
Let us focus on the strip between the cycles $\eta_i$ and $\eta_{i+1}$.
There are four different types of strips, depending on the orientation of
the two sides of the strips (Figure \ref{gc.eps}).
In any case, the strip includes one $(N,0)$ face,
and the untwisting operation maps its boundary to
a contour on $\Sigma$ enclosing two punctures, $\eta_i$ and $\eta_{i+1}$
(see Figure \ref{gc.eps}).
We can easily show that the contour integral is given by
\begin{equation}
Z_i=\oint_{\eta_i}\log x\frac{dy}{y}=
\left\{\begin{array}{l}
2\pi i(\Log x_{\eta_{i+1}}-\Log x_{\eta_i})\quad\mbox{or}\\
2\pi i(\Log x_{\eta_{i+1}}-\Log x_{\eta_i}+2\pi i),
\end{array}\right.
\end{equation}
where the additional term $2\pi i$ is included when the contour crosses
the branch cut of $\log x$.
For the $k+l$ central charges $Z_i$ to satisfy the BPS conditions
(\ref{zreal}),
the following equations must hold:
\begin{equation}
1=|x_{\alpha_1}|=|x_{\alpha_2}|=\cdots=|x_{\alpha_k}|
=|x_{\beta_1}|=|x_{\beta_2}|=\cdots=|x_{\beta_l}|.
\end{equation}
(Note that we set $x_{\alpha_1}=-1$.)
This relation means that all the parameters $M_\mu$ vanish.
Therefore, the unfixed parameters are
$\zeta_{\alpha_i}$ ($i=2,\ldots,k$)
and $\zeta_{\beta_i}$ ($i=1,\ldots,l$),
which describe the positions of the poles $\alpha_i$ and $\beta_i$
aligned on the equator.
Because we first fixed the order of these poles,
the moduli space is a part of $\mathbb T^{d-3}$.
Different orders of the poles on the equator can be interpreted
as different toric phases related by the Seiberg duality.
The total moduli space defined as the union of all the phases
is $\mathbb T^{d-3}$.

\qed
\end{exa}

%%%%%%%%%%%%%%%%%%%%%%%%%%%%%%%%%%%%%%%%%%%%%%%
\subsubsection{The strong coupling limit}\label{strong.subsubsec}
In this section, we determine the parameter spaces of brane configurations
in the large $g_{\rm str}$ limit.
The reason we study this limit even though the relation to gauge theories is not clear is that
the analysis of the parameter space in this case is quite simple.
We can still use the parameters $M_\mu$ and $\zeta_\mu$ defined in (\ref{mzeta2})
to describe the asymptotic forms of brane configurations,
and we find below that all the $M_\mu$ are fixed, and
BPS configurations are parameterized only by $\zeta_\mu$
with the condition (\ref{t2cond}) imposed.

As discussed in \S\ref{strong.subsec} and in \S\ref{note.subsec}, in the strong coupling limit,
the tension of D5-branes is much greater than that of NS5-branes,
and the D5 worldvolume wrapped on the ${\mathbb T}^2$ is almost flat (Figure \ref{NS5divide}).
This means that the D5 worldvolume is a point in the $x^4$-$x^6$
plane, and
all the external lines (the projection of the NS5-branes onto the $46$-plane)
of the web diagram meet at this point (Figure \ref{NS5junction}).
Therefore, we have no degrees of freedom to deform the web diagram.
This implies that $M_\mu=0$.
The parameters of the brane configuration are only
the positions of the NS5-branes in the internal space.
In other words, a brane configuration is
completely determined by $\zeta_\mu$, the
positions of the cycles in the tiling.
The constraint imposed on these parameters
coming from the requiring of preserving supersymmetry is analyzed in \cite{IIKY}. Here we only summarize their result.

Define the function $Q(z)$ which gives the NS5 charge
on the $57$-plane, where $z=x^5+ix^7$, the complexified coordinate in $\bT^2$ directions. This is equal to the function given in (\ref{qbyq}).
In \S\ref{weak.subsubsec}, this function is defined as the `projection'
of NS5-branes, while here it represents the NS5 charge of the worldvolume.

Then, the BPS condition is stated as, when the vanishing of axion is assumed,
\begin{eqnarray}\label{ch-avr}
\int_{\cal F} Q(z)d^2z=0.
\end{eqnarray}
This is the same as (\ref{intqcond}).

This means that 
the brane configurations in the strong coupling limit are parameterized
by only the parameters $\zeta_\mu$,
with the condition (\ref{ch-avr}) imposed.
This is identical to the case of the weak coupling
limit.
We emphasize that in the analysis in the strong coupling limit
we can consider general bipartite graphs.
Although we cannot explicitly determine the moduli space of general tilings
in the weak coupling limit, the result obtained here strongly indicates
that the moduli space is always parameterized by only the parameters $\zeta_\mu$,
and the other parameters, $M_\mu$ and $c_{i,j}$ associated with
the internal points of the toric diagrams are fixed by
the BPS conditions.

An important difference between the strong coupling and weak coupling limits is that
the function $Q(x^5,x^7)$ in the strong coupling limit is not the projection but
the actual NS5 charge.
Therefore we have $(N,k)$-branes with $|k|\geq2$
if $|Q(x^5,x^7)|\geq2$ at some points on the torus.
If such branes appear, we cannot use bipartite graphs to
determine the gauge groups and matter content.
In the following,
we solve the condition (\ref{ch-avr}) explicitly in two cases, 
the conifold and the SPP,
and we show that such tilings do appear in the latter example.

\begin{exa}[conifold]
First, we consider the conifold.
The toric diagram of the conifold and corresponding brane configuration
are given in Figure \ref{conima.eps}.
We can fix the positions of one horizontal and one vertical cycle
by using the translational symmetry,
and hence there are only two physical parameters, $x$ and $y$,
satisfying $0\leq x,y\leq 1$,
which represent the separations of parallel cycles.
\begin{figure}[htbp]
\centerline{\includegraphics[scale=1]{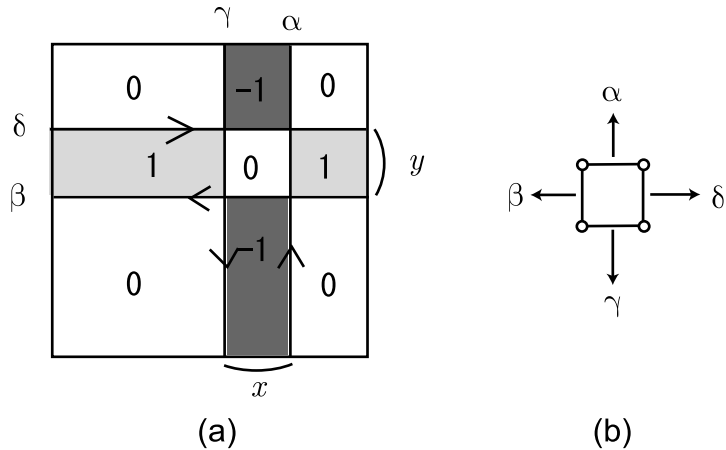}}
\caption{The tiling of the conifold (a) and the corresponding toric diagram (b).}
\label{conima.eps}
\end{figure}
In order for the average charge to vanish,
we need both positive and negative charges,
and the unique charge assignment is given in Figure \ref{conima.eps}.
Furthermore,
(\ref{ch-avr}) demands that two shaded faces have the same area.
This is the case only when $x=y$.
Therefore, the parameter space is $\mathbb S^1$.
Note that this is consistent with our previous analysis in the weak coupling limit.
(The conifold represents the $k=l=1$ case of the generalized conifold in Example \ref{gc.exa}.)
\end{exa}

\begin{exa}[SPP]
The next example is the SPP, whose toric diagram is given in Figure \ref{untwistspp.eps}.
This toric diagram has 5 edges, and thus the number of physical parameters should be 2.
The corresponding tiling has 5 cycles.
We fix the positions of the vertical cycle $\alpha$ at $x^5=0$
and the horizontal cycle $\epsilon$ at $x^7=0$.
Let $x$, $y$ and $z$ be the $x^5$ coordinates of the intersections of the remaining cycles,
$\beta$, $\gamma$ and $\delta$, and the $x^5$ axis, respectively.
We choose $x$, $y$ and $z$ such that $0<x,y,z<1$ and $y<z$.
We can classify the configuration of tilings into three cases (See Figure \ref{spp-iso.eps}),
according to the order of $x$, $y$ and $z$.
\begin{figure}[htbp]
\centerline{\includegraphics[scale=0.8]{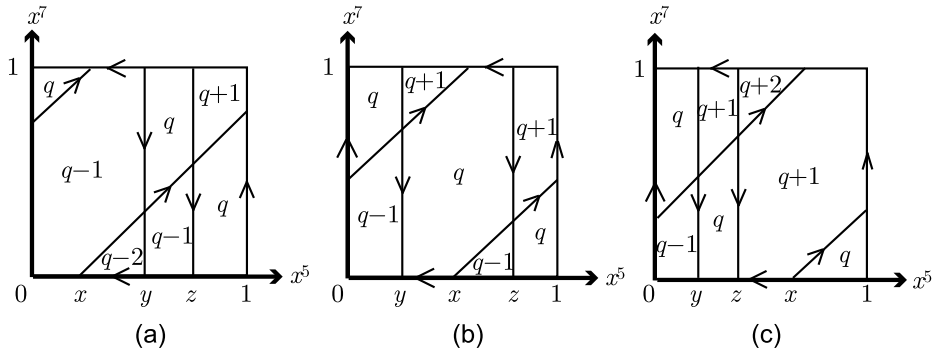}}
\caption{Three configurations for the brane tiling of the SPP case.}
\label{spp-iso.eps}
\end{figure}
By setting the charge on one face in the tiling, 
the charges on the other faces are automatically determined.
If we assign the charges as in Figure \ref{spp-iso.eps},
the BPS condition (\ref{ch-avr}) becomes
\begin{equation}\label{chvan-spp}
0=\int_{\cal F}Q(z)d^2z=\frac{1}{2}+q+x-y-z.
\end{equation}
Let us take $y$ and $z$ as independent variables.
Then, through the condition (\ref{chvan-spp}), 
we can uniquely determine $q$ and $x$ as functions of $y$ and $z$.
The result is depicted in the ``phase diagram'' appearing in Figure \ref{spp-phase-2.eps}.
\begin{figure}[t]
\centerline{\includegraphics[scale=0.5]{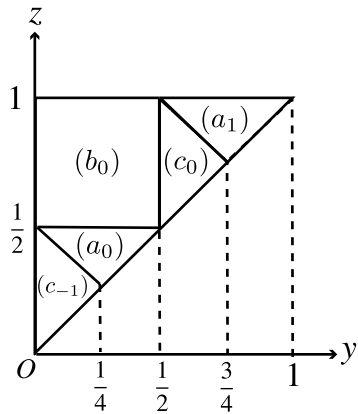}}
\caption{``Phase diagram'' for the SPP case. 
(a$_q$), (b$_q$) and (c$_q$) refer to the configurations 
(a), (b) and (c), respectively, with charge $q$ 
in Figure \ref{spp-iso.eps}.}
\label{spp-phase-2.eps}
\end{figure}
The lines dividing the triangle in Figure \ref{spp-phase-2.eps}
represent the transition points
at which the parameter $x$ crosses $x=y$, $x=z$, and $x=0$
(or, $x=1$, which is identified with $x=0$).
The subscripts of the labels represent the charges $Q$.
In the region (b$_0$), the tiling has faces with NS5 charges $-1$, $0$ and $1$,
while in the regions (a$_q$) and (c$_q$), the tilings have faces with NS5 charge $+2$ or $-2$, in addition to $0$ and $\pm1$.
Such configurations cannot be translated into bipartite graphs.
Generally, in the large $g_{\rm str}$ limit,
such types of tilings are allowed.
\end{exa}

%%%%%%%%%%%%%%%%%%%%%%%%%%%%%%%%%%%%%%%%%%%%%%%%%%%%%%%%%%%
\subsubsection{Wilson lines} \label{Wilson.subsubsec}
Up to now, we have only discussed geometric deformations of
the brane configurations, and we have
obtained $d-3$ independent degrees of freedom, both in the weak and the strong coupling limit.
These are positions $\zeta_\mu$ $(\mu=1,\ldots,d)$
of $d$ cycles with one constraint imposed,
and two of them are unphysical due to the translational invariance.
Because the brane configurations preserve
${\cal N}=1$ supersymmetry,
all scalar parameters must belong to chiral multiplets.
This implies that every real parameter
should be combined with its `superpartner' to
form a complex parameter.

What is the partner of the parameter $\zeta_\mu$? It was proposed in \cite{IIKY} that the answer is given by the Wilson line associated with the cycle $\balpha_\mu$. Since each cycle is the projection of NS5-brane worldvolume wrapped on $1$-cycle
on the torus, we can make gauge invariant quantity by
integrating the $U(1)$ gauge field $A$ on the NS5-brane along the cycle $\balpha_\mu$;
\begin{equation}
W_\mu=\oint_{\balpha_\mu} A.
\end{equation}

A direct way to prove this is to perform the supersymmetry transformation
of the fields on NS5-brane. Another argument is to use S and T dualities to transform the brane configuration to other ones in which multiplet structure
is clearer. See \cite{IIKY} for more details.

There is one constraint (\ref{t2cond}) imposed on the parameters $\zeta_\mu$.
The multiplet structure requires that this should be the case for the Wilson lines
$W_\mu$, too.
Actually, with the Stokes' theorem, we can show that,
\begin{equation}
\sum_{\mu=1}^dW_\mu
=\sum_{\mu=1}^d\oint_{\balpha_\mu} A
=\int_\Sigma dA=0.
\end{equation}
At the final step, we used $dA=0$.

We can also show that
two of $d-1$ degrees of freedom of Wilson lines
are redundant.
Let us consider gauge transformations
of the RR $2$-form field in the bulk.
They also change the gauge field on the NS5-brane as
\begin{equation}
\delta C_2=d\Lambda,\quad
\delta A=\Lambda.
\end{equation}
If the parameter $\Lambda$ is a closed $1$-form,
this transformation changes only $A$.
The Wilson lines are transformed by
\begin{equation}
W_\mu\rightarrow W_\mu'=W_\mu+\oint_{\balpha_\mu}\Lambda.
\label{wtow}
\end{equation}
We have two independent closed $1$-form on
the torus, and the gauge transformation
(\ref{wtow}) decreases the number of physical degrees of
freedom by two.

Summarizing this subsection, we have $d-3$ complex parameters in the fivebrane system, corresponding to geometric deformation of branes and Wilson lines. In the next subsection, we will see that they actually correspond to exactly marginal deformation in quiver gauge theories.

%%%%%%%%%%%%%%%%%%%%%%%%%%%%%%%%%%%%%%%%%%%%%%%%%%%%%%%%%%%%%%%%%%%%%%%%%%%%%%%%%
\subsection{Marginal deformations of quiver gauge theories} \label{marginal.subsec}
In previous subsection we have studied the BPS condition of moduli parameters, and determined the moduli parameters of fivebrane systems. Here we study exactly marginal deformations of quiver gauge theories (\S\ref{marginal.subsubsec}) along the lines of \cite{Leigh:1995ep}, and will see that (part of) exactly marginal deformations in quiver gauge theory side correspond to geometrical deformation of branes in fivebrane side (\S\ref{comparison.subsubsec}).

\subsubsection{Exactly marginal deformation of quiver gauge theories} \label{marginal.subsubsec}

In our previous discussion of superpotential, we chose the superpotential coupling $h_k$ to be $\pm$ 1 for each $(N,\pm 1)$-brane labeled by $k$ (see \eqref{hkpm1}). This is quite unnatural physically and we are motivated to study more general coefficients. Another parameter we have is the gauge coupling $g_a$ (together with $\theta$-angle $\theta_a$, which we will comment on in later sections) for each $(N,0)$-brane labeled by $a$.

In this parameter space spanned by $g_a$ and $h_k$, we search for the space of exactly marginal deformations, or the conformal manifold, which is defined by the vanishing of $\beta$-functions:

\beq
\{ \beta_a=\beta_k=0 \} \subset \{g_a, h_k \}. \label{betaak0}
\eeq

One should stress that this analysis is purely field theoretical, but interestingly enough, the description of quiver gauge theories by brane tilings and fivebrane diagrams still turns out to be useful. %, even though our analysis here is just purely field-theoretical.

As already shown in \eqref{NSVZ}, the beta function for $g_a$ is related to the
anomalous dimensions $\gamma_I$ of the
bi-fundamental fields $\Phi_I$
by the NSVZ exact beta-function formula\cite{Novikov:1983uc}

\begin{equation}
\beta_a\equiv\mu\frac{d}{d\mu}\frac{1}{g_a^2}
=\frac{N}{1-g_a^2 N/8\pi^2}
\left[3-\frac{1}{2}\sum_{I\in a}(1-\gamma_I)\right],
\label{betagauge}
\end{equation}
where the summation is taken over the fields
coupled to the gauge group $SU(N)_a$.

Because the conformal dimension of the field $\Phi_I$
is $1+(1/2)\gamma_I$,
the beta-function of the coefficient $h_k$
in (\ref{hkterm}) is given by
\begin{equation}
\beta_k\equiv\mu\frac{d}{d\mu}h_k=-h_k\left[3-\sum_{I\in k}\left(1+\frac{1}{2}\gamma_I\right)\right],
\label{betapot}
\end{equation}
where $3-\sum_{I\in k}\left(1+\frac{1}{2}\gamma_I\right)$ is equal to the dimension of the superpotential coupling $h_k$.

In general, conformal manifold is dimension zero, that is, we expect isolated conformal fixed points. This is because the number of equation in \eqref{betaak0} is equal to the number of parameters. In the supersymmetric case, however, both the gauge and
superpotential couplings
(\ref{betagauge}) and
(\ref{betapot}) are given in terms of the anomalous dimensions $\gamma_I$,
and there may be identically vanishing linear combinations of these
$\beta$-functions of the form
\beq
\begin{split}
\beta[S_A]
&\equiv
\sum_a S_a \frac{1}{N}\beta'_a-\sum_k S_k \frac{\beta_k}{h_k} \\
&=\sum_a S_a\left[3-\frac{1}{2}\sum_{I\in a}(1-\gamma_I)\right] \\
& \ \ \ \ \ \ \ \  +\sum_k S_k\left[3-\sum_{I\in k}\left(1+\frac{1}{2}\gamma_I\right)\right] ,
\label{sumsbeta}
\end{split}
\eeq
with some numerical coefficients $S_a$ and $S_k$.
Instead of $\beta_a$, here we have used
\begin{equation}
\beta'_a=\left(1-\frac{g_a^2N}{8\pi^2}\right)\beta_a
=\mu\frac{d}{d\mu}\frac{1}{g_a^{'2}},
\end{equation}
which is the $\beta$-function for the coupling $1/g^{'2}_a$ defined by \footnote{Although this redefinition is not essential in our discussion, we remark that this definition of $g^{'2}_a$
is not artificial. The coupling $g_a$ is the canonical coupling (for which kinetic term has canonical normalization), whereas the coupling $g'_a$ is the holomorphic coupling. The difference between them is the origin of the factor $\frac{1}{1-g_a^2 N/8 \pi^2}$, in the derivation of NSVZ $\beta$-function in \cite{ArkaniHamed:1997mj}.}

\begin{equation}
d\left(\frac{1}{g^{'2}_a}\right)
=\left(1-\frac{g_a^2N}{8\pi^2}\right)d\left(\frac{1}{g_a^2}\right).
\label{gprime}
\end{equation}

These linear combinations of $\beta$-functions
correspond to RG invariant couplings parameterizing
the orbits of RG flow in the coupling space.
We can assume, at least in the vicinity of the IR fixed manifold,
a one-to-one correspondence between these orbits and points
in the fixed manifold.
Thus, in order to determine what marginal deformations exist
in gauge theories,
we should look for vanishing linear combinations of the form (\ref{sumsbeta}).
This analysis was carried out in \cite{Uranga:1998vf} for orbifolded conifolds,
and in \cite{Benvenuti:2005wi} for $Y^{p,q}$, including the conifold.
In this work, we generalize these analysis to general quiver gauge theories described by
brane tilings (cf. \cite{Wijnholt:2005mp} for analysis from the point of view of exceptional collections). See also \cite{Erlich:1999rb} and \cite{Lunin:2005jy}.

\begin{figure}[htbp]
\centering{\includegraphics[scale=0.4]{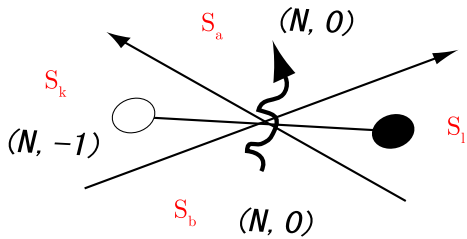}}
\caption{The condition \eqref{sumsasumsk} say we have $S_a+S_b=S_k+S_l$. Here $S_a, S_b$ are numbers assigned to $(N,0)$-brane faces, and $S_k, S_l$ are assigned to $(N,\pm 1)$-faces.}
\label{sumsasumsk.eps}
\end{figure}

We now search for coefficients $S_A$ such that
the linear combination of $\beta$ in (\ref{sumsbeta})
vanishes for any $\gamma_I$.
For the cancellation of the coefficients of $\gamma_I$,
the condition
\begin{equation}
\sum_{a\in I}S_a=\sum_{k\in I}S_k,
\label{sumsasumsk}
\end{equation}
must hold, where the summation on the left-hand side is taken over
gauge groups coupling to the chiral field $I$,
and that on the right-hand side is taken over
terms in the superpotential including the fields $I$.
On the brane tiling, chiral fields correspond to the
intersections of cycles,
and $S_a$ and $S_k$ are numbers assigned to
the $(N,0)$ and $(N,\pm1)$ faces, respectively.
For each intersection, we have two pairs of
faces contacting each other at the intersection,
and the relation (\ref{sumsasumsk}) implies that
the sum of $S_A$ for the two faces in one pair
is the same as that for the two faces in the other pair (Figure \ref{sumsasumsk.eps}).

\begin{figure}[htbp]
\centering{\includegraphics[scale=0.4]{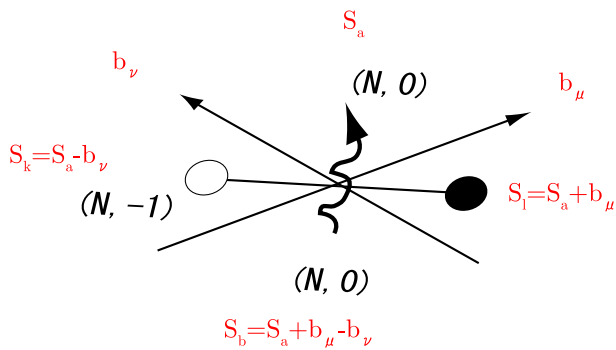}}
\caption{The condition \eqref{ssb} ensures \eqref{sumsasumsk}. $b_{\mu}$ and $b_{\nu}$ are numbers assigned to cycles of NS5-branes.}
\label{ssb.eps}
\end{figure}

This condition is equivalent to
the existence of numbers $b_\mu$ assigned to cycles
which satisfy the relation
\begin{equation}
S_A-S_B=b_\mu
\label{ssb}
\end{equation}
for two faces $A$ and $B$ adjacent on a cycle $\balpha_\mu$ (Figure \ref{ssb.eps}). 
In \eqref{ssb}, we assume that when the cycle $\balpha_{\mu}$ is up-going,
the faces $A$ and $B$ are on the right and left sides of the cycle, respectively.

The set of numbers $\{S_A,b_\mu\}$ satisfying the relation
(\ref{ssb})  are called ``baryonic number assignments'' in \cite{IIKY}.
We can easily show that
the quantities $b_\mu$ must satisfy the relation
\begin{equation}
\sum_{\mu=1}^d b_\mu\balpha_\mu=0
\label{balpha}
\end{equation}
in order for there to exist $S_A$ satisfying the relation (\ref{ssb}).
Because of this relation, the number of independent degrees of freedom
of $b_\mu$ is $d-2$.
If we fix all the $b_\mu$ and one $S_A$, the relation (\ref{ssb}) determines
all the other $S_A$.
Therefore, we have $d-1$ linearly independent baryonic number assignments.

The baryonic number assignments defined above have already appeared in the discussion of fractional branes in \S\ref{fractional.subsec}. There the numbers are used to give anomaly-free rank distributions
of gauge groups\cite{Benvenuti:2004wx,Butti:2006hc}, and the number $S_a$ in this section is denoted by $N_a$, which is the rank of the gauge group corresponding to gauge group $a$ \footnote{Of course, in the discussion of fractional branes, $N_a$ must be a positive integer, which is different from our present discussion.}.
 Another use of baryonic number assignments is
to give anomaly-free baryonic $U(1)$ charges of chiral multiplets.\cite{Benvenuti:2004wx,Butti:2006hc}, as will be discussed in \S\ref{versus.subsec}.
We can identify $S_a$ assigned to $(N,0)$ faces as
baryonic $U(1)$ charges of open string endpoints on the faces\cite{Imamura:2006ie},
and the baryonic $U(1)$ charge of a chiral multiplet $I$
is given by the difference $S_a-S_b$, where $a$ and $b$ are
two $(N,0)$ faces contacting each other at the intersection $I$.
This is why we call sets of numbers $\{S_A,b_\mu\}$
baryonic number assignments. We will come back to this point later in \S\ref{versus.subsec}

Now we obtain the condition (\ref{ssb})
for the cancellation of the $\gamma_I$ terms in
(\ref{sumsbeta}),
and if this condition is satisfied, we are left with
\begin{equation}
\beta[S_A]
=3\sum_{\rm faces} S_A -3\sum_{\textrm{intersections}}\bar{S}_I,
\label{sum_beta_general2}
\end{equation}
where we have replaced
the sum over both indices $a$ and $k$
by a sum over all faces in the brane tilings,
and the matter contribution is given by the sum over intersections.
We have also introduced $\bar{S}_I$, defined as half of (\ref{sumsasumsk}):
\begin{equation}
\bar{S}_I=\frac{1}{2}\sum_{a\in I}S_a=\frac{1}{2}\sum_{k\in I}S_k.
\end{equation}

\begin{figure}[t]
\centering
{\small
\begin{tabular}{cc}
\includegraphics[width=60mm]{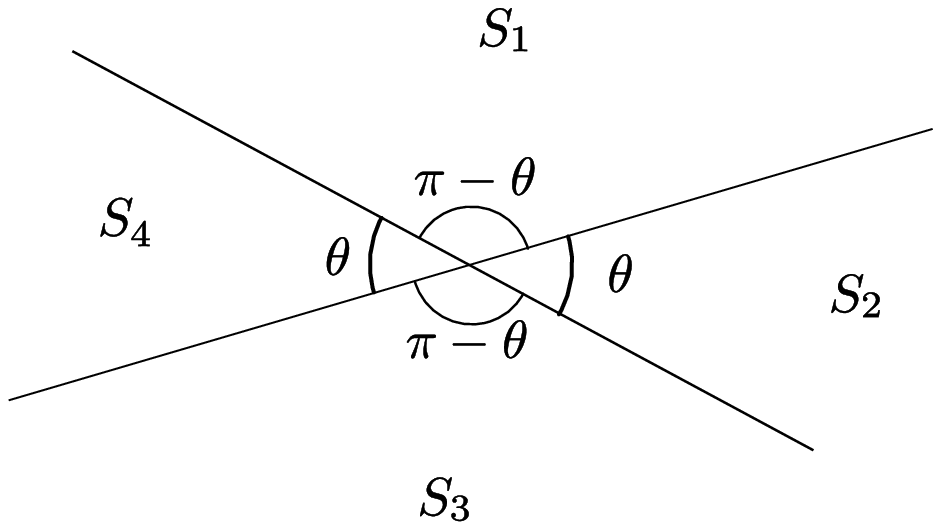}&
\includegraphics[width=60mm]{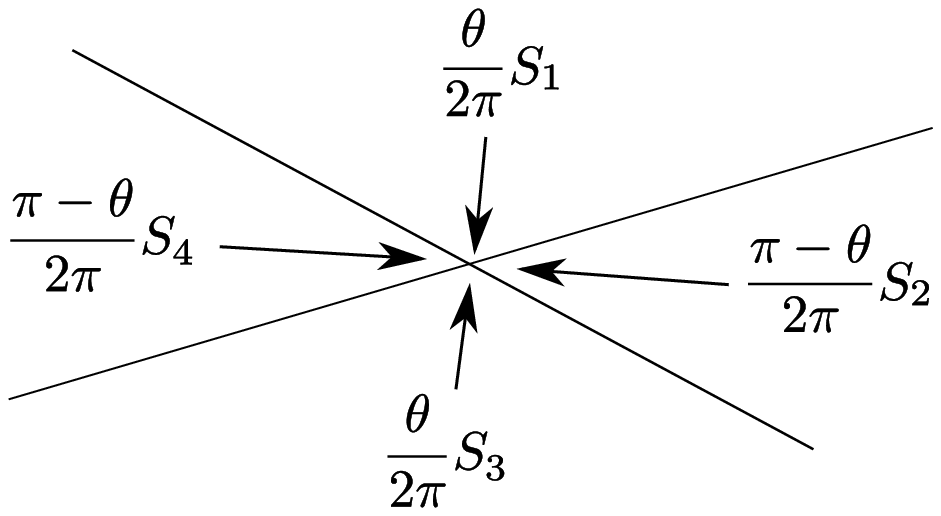}\\
(a) Baryonic charges and angles&
\multicolumn{1}{p{60mm}}{(b) The charges of faces moved to the intersection}
\end{tabular}
}
\caption{Changing the charge distribution of faces to cancel the charges of the intersection.}
\label{s_mean}
\end{figure}

Interestingly, we can show that
if the coefficients $S_A$ satisfy the condition (\ref{ssb}),
the right-hand side of (\ref{sum_beta_general2}) automatically vanishes 
as well as the $\gamma_I$-dependent terms.
To prove this, 
we move a fraction of the total baryonic charge on a given face
equal to the external angle of that face divided by $2\pi$
to the corners of that face.
For example, of
the charge $S_2$ in Figure \ref{s_mean},
the amount $[(\pi-\theta)/2\pi]S_2$
is moved to the corner $\theta$.
The sum of the baryonic charges moved to the
corners of one face is
the original baryonic charge of the face,
because the sum of the external angles of a polygon is identically $2\pi$.
The sum of the baryonic charges of the corner around one intersection point is
the mean value of the baryonic charges of the faces surrounding that point.
In the case of Figure \ref{s_mean},
the sum is
\begin{equation}
\begin{split}
\frac{\theta}{2\pi}S_1+\frac{\pi-\theta}{2\pi}S_2+
\frac{\theta}{2\pi}S_3+\frac{\pi-\theta}{2\pi}S_4
&=\frac{\theta}{2\pi}(S_1+S_3)+\frac{\pi-\theta}{2\pi}(S_2+S_4)\\
&=\frac{\theta}{2\pi}(S_1+S_3)+\frac{\pi-\theta}{2\pi}(S_1+S_3)\\
&=\bar{S}_I.
\end{split}
\end{equation}
Hence $\sum S_A =\sum \bar{S}_I$ holds, 
and \eqref{sum_beta_general2} is zero.

Thus we have $d-1$ vanishing linear combinations of $\beta$ functions,
and this implies that there are $d-1$ RG invariant
parameters
\begin{equation}
f^{(I)}=\sum_aS^{(I)}_a\frac{1}{Ng^{'2}_a}
-\sum_kS^{(I)}_k\log h_k,
\label{rginvariant2}
\end{equation}
where $I=1,\ldots,d-1$ labels
linearly independent baryonic number assignments.
The coupling $g_a'$ is obtained as follows by integrating (\ref{gprime}):
\begin{equation}
\frac{1}{g^{'2}_a}=\frac{1}{g_a^2}-\frac{N}{8\pi^2}\log\left(\frac{1}{g_a^2}\right).
\end{equation}

\begin{figure}[htbp]
\centering{\includegraphics[scale=0.4]{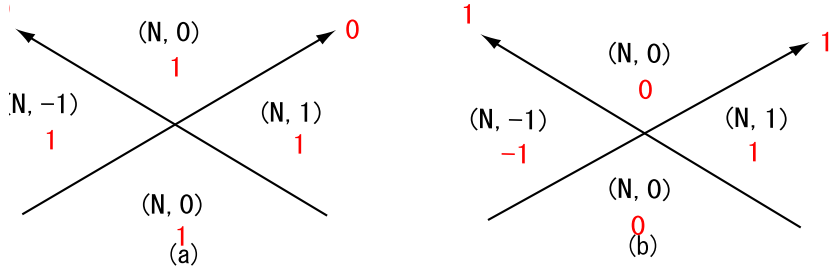}}
\caption{The two types of baryonic number assignments which exist for all bipartite graphs. In \S\ref{comparison.subsubsec}, the marginal deformations corresponding to these baryonic number assignments will be identified with background supergravity fields.}
\label{chargeassignment}
\end{figure}

We now comment on two special number assignments.
The first one is (see Figure \ref{chargeassignment} (a))
\begin{equation}
S^{(1)}_A=1
\quad\forall A,\quad
b^{(1)}_\mu=0
\quad\forall \mu.
\label{allsone}
\end{equation}
This gives the RG invariant coupling
\begin{equation}
f^{(1)}=\sum_a\frac{1}{Ng^{'2}_a}
-\sum_k\log h_k.
\label{diag}
\end{equation}
Roughly speaking, this is related to the gauge coupling $g_{\rm diag}$
of the diagonal $SU(N)$ subgroup by $f^{(1)}\sim 1/(Ng_{\rm diag}^2)$.
The second one is (see Figure \ref{chargeassignment} (b))
\begin{equation}
S^{(2)}_A=Q\quad
\mbox{for $(N,Q)$ face $A$},\quad
b^{(2)}_\mu=1\quad\forall\mu,
\label{allbone}
\end{equation}
which gives the RG invariant parameter
\begin{equation}\label{f2}
f^{(2)}=\sum_k\pm \log h_k.
\end{equation}
Because this depends only on superpotential couplings
and does not include gauge couplings,
we can regard this
as a generalization of the $\beta$ deformation\cite{Leigh:1995ep}.

In ${\cal N}=1$ gauge theory, all couplings belong to
chiral multiplets and are complex.
The superpotential couplings are
complex, and the gauge couplings are also combined
with the $\theta$-angles to form the complex
couplings
\begin{equation}
\tau_a=\frac{\theta_a}{2\pi}+\frac{4\pi i}{g_a^2}.
\end{equation}
Thus, we have in total $d-1$ complex exactly marginal deformations.

%%%%%%%%%%%%%%%%%%%%%%%%%%%%%%%%%%%%%%%%%%%%%%%%%%%%%%%%%
\subsubsection{Comparison of parameters}\label{comparison.subsubsec}
Up to now, we analyzed the $\beta$-functions
of gauge and superpotential couplings, and
we found $d-1$ complex marginal deformations.
We also found that the corresponding brane configuration
has $d-3$ complex degrees of freedom
corresponding to changing the shape and Wilson lines.
In this section, we give an argument aimed at determining
the relations among these deformations and degrees of freedom.

To begin with, let us consider how the couplings are obtained in string theory.
We first focus on the real parameters.
The gauge coupling for each $SU(N)_a$
can be read off of the Born-Infeld action
of D5-branes as
\begin{equation}
\frac{1}{Ng^{'2}_a}
=
\frac{1}{Ng_a^2}-\frac{1}{8\pi^2}\log\frac{1}{g_a^2}
\sim\frac{A'_a}{4\pi Ng_{\rm str}\alpha'}
-\frac{1}{8\pi^2}\log\frac{A'_a}{4\pi g_{\rm str}\alpha'}
,
\label{oong}
\end{equation}
where $A_a'$ is the area of the face $a$
evaluated with the real shape of the D5-brane worldvolume.
Then,
each term in the superpotential is
induced by the string worldsheet wrapped on the
corresponding $(N,\pm1)$ face and is roughly given by
\begin{equation}
-\log|h_k|
\sim
-\log\left(g_{\rm str}^{(n/2)-1}
e^{-A'_k/(2\pi\alpha')}\right)
=
\frac{A'_k}{2\pi\alpha'}
+\left(\frac{n}{2}-1\right)\log\frac{1}{g_{\rm str}},
\label{loghk}
\end{equation}
where $n$ is the number of fields included in the interaction
term.

Because of the difficulty involved in quantizing strings
in the background with D5- and NS5-branes,
we cannot obtain precise relations between
the parameters in gauge theories and
those in string theory.
The purpose of this section is to derive only rough
relations among them.
For this purpose,
we focus only on the first terms in the final expressions
in (\ref{oong}) and (\ref{loghk}).
Furthermore, we ignore the $N$ and $g_{\rm str}$ dependences
of these terms, and we simplify the relations
as
\begin{equation}
\frac{1}{N{g_a'}^2}\sim A_a,\quad
-\log|h_k|\sim A_k.
\label{coupling2area}
\end{equation}
In these relations,
in addition to the simplification
mentioned above,
we have replaced the areas $A_A'$,
which are evaluated with the real shapes of the D5-brane worldvolumes,
by the areas $A_A$ on a flat torus
divided by straight cycles.
This combination of simplification results in
an approximate treatment that is so rough
that we cannot obtain any quantitative results
from the analysis in the following.
On the other hand, these simplifications make the equations below quite simple
and clarify the qualitative relations among
the parameters.

Replacing the terms in
the RG invariant parameters (\ref{rginvariant2})
by the corresponding areas according to (\ref{coupling2area}),
we obtain
\begin{equation}
f^{(I)}=\sum_AS_A^{(I)}A_A
=\int_{\cal F} S^{(I)}(x^5,x^7)dx^5dx^7,
\label{saaa}
\end{equation}
where the function $S^{(I)}(x^5,x^7)$ is defined so that
$S^{(I)}(x^5,x^7)=S_A^{(I)}$ on the face $A$.
The integration region ${\cal F}$ is an arbitrary fundamental region.
Then, using the parameters $b_\mu$ in (\ref{ssb}) and
the functions $q_\mu(x^5,x^7)$ defined in (\ref{qdef}),
we obtain the function $S(x^5,x^7)$ as
\begin{equation}
S^{(I)}(x^5,x^7)=\sum_{\mu=1}^d b_\mu^{(I)} q_\mu(x^5,x^7)+c,
\end{equation}
where $c$ is a new parameter determining the constant part of $S^{(I)}$.
The periodicity of $S(x^5,x^7)$ is guaranteed by the condition
(\ref{balpha}).
If we substitute this into
(\ref{saaa}), we obtain
\begin{equation}
f^{(I)}=\sum_{\mu=1}^d b_\mu^{(I)}\zeta_\mu+c.
\label{sabz}
\end{equation}
This relation shows how the RG invariant parameters $f^{(I)}$
in the gauge theory
are related to the parameters $\zeta_\mu$ and $c$ in the
string theory.

Among the parameters $\zeta_\mu$ and $c$ in the string theory,
we know that $\zeta_\mu$ represent the positions of the cycles $\mu$
on the torus and describe the shape of the brane configuration.
What is the other parameter, $c$?
We can identify this degree of freedom
with the expectation value of the dilaton field in the following way.

If we substitute the assignment (\ref{allsone})
into (\ref{sabz}),
we obtain
\begin{equation}
c=f^{(1)}
\sim \frac{1}{Ng_{\rm diag}^2}.
\label{sabz1}
\end{equation}
Hence, we find that the parameter $c$ is the diagonal gauge coupling (\ref{diag}).
It is basically the gauge coupling of the theory
on D5-branes wrapped on the torus without NS5-branes attached.
We can read off the coupling
from the action of the D5-branes, and we find
\begin{equation}
c=\frac{A_{\rm tot}}{\alpha'e^\phi}.
\end{equation}
With this equation, we can identify the parameter $c$
with the expectation value of $e^{-\phi}$.
(More precisely, $c$ depends not only on the dilaton but also
on the size of ${\mathbb T}^2$.)
This correspondence can easily be extended to the
correspondence between complex parameters.
The diagonal gauge coupling $1/g_{\rm diag}^2$ is combined with
the theta angle $\theta_{\rm diag}$ of the diagonal gauge group,
and we can read off the relation
$\theta_{\rm diag}\sim C_{57}$
from the D5-brane action.
We thus obtain the relation
\begin{equation}
\tau_{\rm diag}\sim ic\sim
C_{57}+\frac{i}{e^\phi}.
\end{equation}
We can show that
the right-hand side of this relation is actually the
scalar component of one chiral multiplet
by checking the transformation law of fermions
in type IIB supergravity.

Now we have $d-1$ RG invariant parameters
in the gauge theory
and $d-2$ parameters in the string theory.
There is still one more parameter in the gauge theory,
namely the $\beta$-like deformation
given by the baryonic charges (\ref{allbone}).
Substituting (\ref{allbone}) into (\ref{sabz}),
we obtain
\begin{equation}
f^{(2)}
=\sum_{\mu=1}^d \zeta_\mu+c.
\label{f2b}
\end{equation}
If the constraint (\ref{t2cond}) is imposed on
the parameters $\zeta_\mu$,
the first term on the right-hand side of (\ref{f2b}) vanishes,
and this does not give an independent degree of freedom.
To realize the $\beta$-like deformation $f^{(2)}$,
we need to relax the constraint
imposed on the parameters $\zeta_\mu$.
In other words, the marginal deformation $f^{(2)}$
corresponds to a supergravity
field which modifies the constraint (\ref{t2cond}).
We can easily see that the axion field $C$ is such a field.

In \S\ref{strong.subsubsec}, we assumed a vanishing axion field
when we obtained the BPS conditions.
If, instead, we consider a non-vanishing axion field, the constraint (\ref{t2cond}) is modified and it corresponds to the $\beta$-like deformation (\ref{f2b}).

The correspondence between $f^{(2)}$ and the axion field
can be extended to the correspondence between complex parameters.
With the supersymmetry transformation law of type IIB supergravity,
we can show that the partner of the axion field
is $B_{57}$.
The non-vanishing expectation value of this component of $B_2$
contributes to the phase of the coupling $h_k$
as $h_k\sim e^{-(C+iB_{57})}$ through the coupling of $B_2$ and the string world sheet,
and we obtain the correspondence
\begin{equation}
f^{(2)}\sim C+iB_{57}.
\end{equation}

Having obtained the relations between
background supergravity fields and
the two marginal deformations $f^{(1)}$ and $f^{(2)}$,
the $d-3$ other marginal deformations $f^{(I)}$ ($I=3,\ldots,d-1$)
are naturally matched with the parameters $\zeta_\mu+iM_\mu$.
If we change the parameter $\zeta_\mu$,
the cycle $\mu$ moves on the torus,
and the areas of the faces touching the cycle
change.
This changes the corresponding coupling constants.
If we change the Wilson line $W_\mu$,
the $\theta$-angles associated with the $(N,0)$ faces
touching the cycle are changed
by the interaction term in the brane action
\begin{equation}
\int_{\rm junctions}A^{\rm (NS5)}\wedge F^{\rm (D5)}\wedge F^{\rm (D5)}.
\end{equation}

Summarizing, the relations between exactly marginal deformations in the gauge theory and the degrees of freedom in the brane tiling are given by the following:

\begin{subequations}
\begin{eqnarray}
\mbox{diagonal gauge coupling} & \leftrightarrow & C_{57}+ie^{-\phi}, \label{dilaton}\\
\mbox{$\beta$-like deformation} & \leftrightarrow & C+iB_{57}, \label{axion}\\
\mbox{other $d-3$ deformations} & \leftrightarrow & \zeta_\mu+iW_\mu. \label{zeta}
\end{eqnarray}
\end{subequations}
The combination $C_{57}+ie^{-\phi}$ might look strange, but by taking T-duality along $57$, we are back to the familiar combination $C+ie^{-\phi}$.
We should note that the correspondence proposed above is only the zero-th order approximation, and there may be a mixing
among these parameters that cannot be captured with the rough analysis given above.

This concludes our somewhat long discussion of BPS conditions and marginal deformations. We have seen that fivebrane systems can go beyond bipartite graphs to uncover interesting facts about gauge theories, confirming the interpretation of bipartite graphs as fivebrane systems.

%%%%%%%%%%%%%%%%%%%%%%%%%%%%%%%%%%%%%%%%%%%%%%%%%%%%%%%%%%%%%%%%%%%%%
\subsection{Kasteleyn matrix and another fast forward algorithm} \label{another.subsec}

In \S\ref{strong.subsec} we have already explained one ``forward algorithm''. 
In this subsection we explain another way of extracting toric data from bipartite graphs. Along the way we introduce the Kasteleyn matrix, which is a standard tool in dimer models.

\paragraph{Kasteleyn matrix}

As said in Introduction, the problem of counting the number of perfect matchings is solved for any planar graph by Kasteleyn. There he proposed to use the technique of the so-called Kasteleyn matrix and its determinant, the Kasteleyn determinant.

One way of formulating the Kasteleyn determinant is to use height function. Out of all perfect matchings, choose a reference perfect matching, which we denote $D_0$. The choice of $D_0$ is arbitrary. In order to define height function $h(D,D_0)$ for another perfect matching $D$ , we superimpose $D$ onto $D_0$.
The height change $h_1(D,D_0)$ and $h_2(D,D_0)$ is given by the difference of the height change if we go around $\balpha$ and $\bbeta$-cycles, respectively:
\beq
(h_1(D,D_0),h_2(D,D_0))=(\langle D-D_0, \balpha \rangle, \langle D-D_0, \bbeta \rangle).
\eeq

The characteristic polynomial $P(x,y)$ is defined by
\beq
P(x,y)=\sum_D (-1)^{ h_1(D) h_2(D)+h_1(D)+h_2(D)} x^{h_1(D)} y^{h_2(D)}, \label{characteristic}
\eeq
where summation is taken over all perfect matchings. In a sense, this is an example of partition function for dimer model defined in \eqref{partition}. We need to specify a reference perfect matching $D_0$ in this definition, but this polynomial is independent of $D_0$ up to the choice of overall multiplicative powers of $x$ and $y$.

\begin{exa}
As a simple example, we consider the bipartite graph corresponding to SPP. The six perfect matchings are shown in Figure \ref{bipartiteconnection}.

\begin{figure}[htbp]
\centering{\includegraphics[scale=0.3]{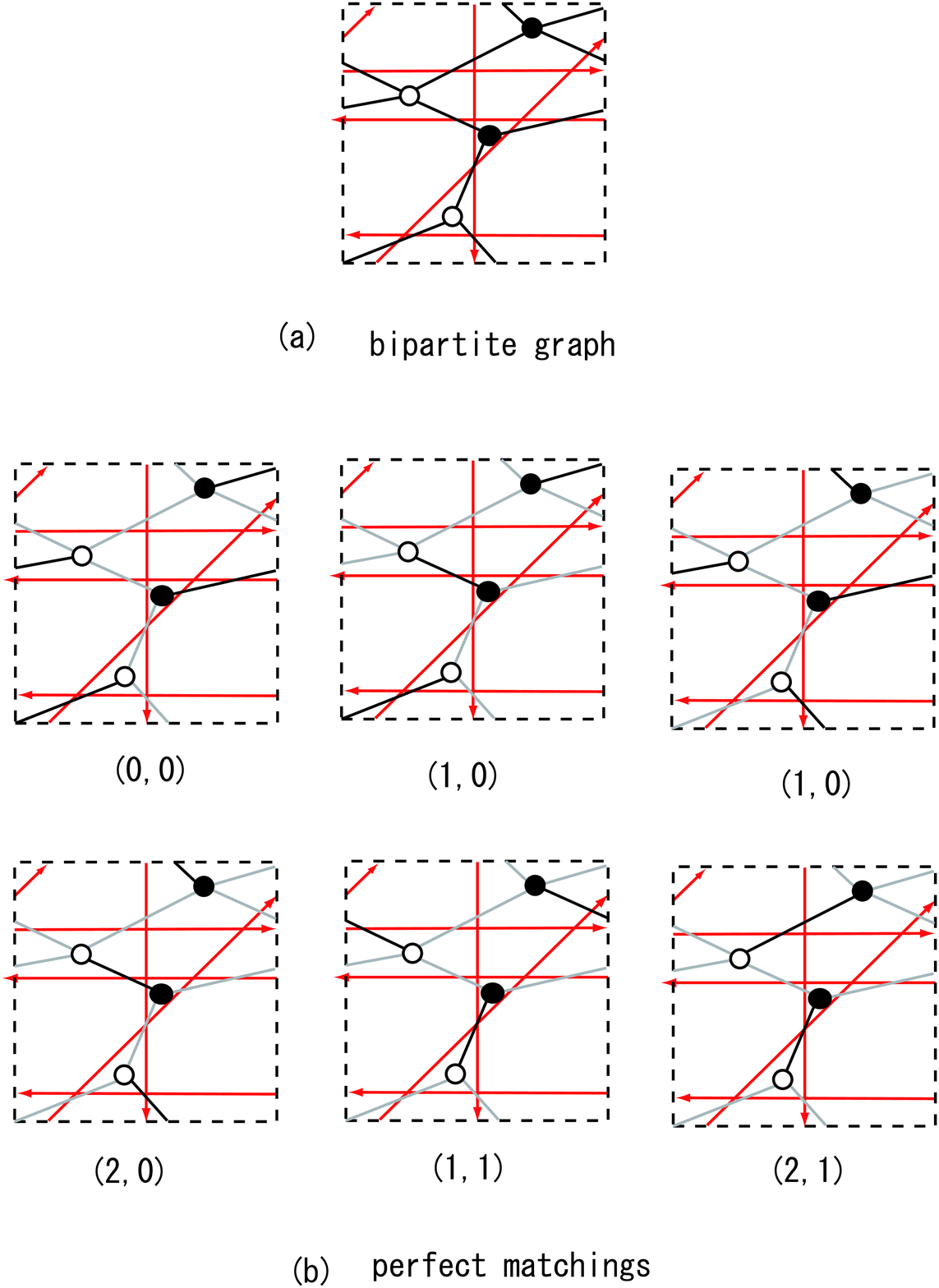}}
\caption{The 6 perfect matchings for SPP bipartite graph (Figure \ref{bipartiteconnection}). The numbers assigned to each perfect matching shows the height function.}
\label{SPPPM}
\end{figure}

The characteristic polynomial of this graph is given by
\beq
P(x,y)=1-2x+x^2-xy-x^2 y.\label{SPPK1}
\eeq
\qed
\end{exa}

This definition, although simple, requires the knowledge of all perfect matchings, which is not easy in general (after all, we do not the number of perfect matchings in advance because that is what we want to know!). In practical computations, it is sometimes better to use another method, i.e. to use the so-called Kasteleyn matrix.

In order to make things simple, we first consider bipartite graphs on two-dimensional plane, not torus. We also assume that the number of edges connecting arbitrary pair of vertices is at most one. Let us denote the set of black (white) vertices by $B$ and $W$, respectively. We define the matrix $A=(A_{bw})_{b\in B, w\in W}$ as the adjacency matrix 
\beq
A_{bw}=
\begin{cases}
1 & \textrm{($b$ and $w$ are connected by an edge )}\\
0 & \textrm{(otherwise)} \label{defAbw}
\end{cases}.
\eeq

Consider the permanent of the matrix $A$ defined by
\beq
\textrm{perm}(A)=\sum_{\sigma\in S_n} A_{1 \sigma(1)} A_{2 \sigma(2)}\ldots A_{n \sigma(n)}. \label{perm}
\eeq

Then from its definition \eqref{defAbw} it follows that each term in the RHS of \eqref{perm} is non-zero (and takes value $1$) only when each pair $(i, \sigma(i))$ is connected by an edge. This means that the RHS is non-zero precisely when the set $\{ \left(1,\sigma(1)\right),\left(2,\sigma(2)\right),\ldots, \left(n,\sigma(n)\right) \}$ is a perfect matching of the bipartite graph. In other words, we have one-to-one correspondence between non-zero terms in the permanent and the perfect matchings of the bipartite graph. Thus the partition function $Z$, which counts the number of perfect matchings, is given by the permanent of $A$:
\beq
Z=\textrm{perm}(A).
\eeq

Unfortunately, this expression is not useful for practical computations, since permanent does not have good properties as determinant does\footnote{For example, we do not have 
\beq
\mbox{perm}(AB)\ne \mbox{perm}(A)\ \mbox{perm}(B).
\eeq
}. We therefore modify the matrix by
\beq
B_{bw}=s_{bw} A_{bw},
\eeq
and choose appropriate signs $s_I$ so that the sign of each term in the expansion of determinant 
\beq
\textrm{det}(B)=\sum_{\sigma\in S_n} \textrm{sgn}(\sigma) B_{1 \sigma(1)} B_{2 \sigma(2)} \ldots B_{n \sigma(n)}
\eeq
have the same sign. Then we have $Z=\pm {\rm det}B$. \footnote{In a sense, this is `bosonization' of fermions.}

The condition for this to occur was analyzed in \cite{KasteleynGraph}. The result is that for each face in the bipartite graph with $2m$ edges of polygon, the product of $s_{bw}$ around the face should be given by $(-1)^{m+1}$. For example, the product is $-1$ for square, and $+1$ for hexagon.
It was also shown in \cite{KasteleynGraph} that we can always choose $s_I$ to satisfy this condition \cite{KasteleynGraph}. %(In fact, if you look at the proof it suffices that the condition is satisfied at the boundary).

\begin{figure}[htbp]
\centering{\includegraphics[scale=0.5]{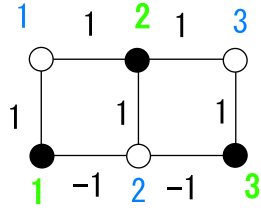}}
\caption{The example of sign choice for the bipartite graph of Figure \ref{bipartite}. Since we have squares, the signs are chosen so that the product of all weights is -1 for each square. The black numbers represent weight assigned to edges, and the green and blue numbers are labels of black and white vertices, respectively.}
\label{KasteleynSigneg}
\end{figure}

Example of such a sign choice for the dimer in Figure \ref{PM} is shown in Figure \ref{KasteleynSigneg}.
In this case, the Kasteleyn matrix is given by

\beq
K=\left(
\begin{array}{ccc}
1 & -1& 0 \\
1 &1 &1 \\
0 &-1  & 1\\
\end{array}
\right),
\eeq
whose determinant gives 
\beq
\textrm{det}K=3,
\eeq
 which is the correct number as we have verified in Figure \ref{PM}.

We now consider a bipartite graph on $\bT^2$. In this case, the Kasteleyn matrix is modified to be a Laurent polynomial in two variables $x$ and $y$:

\beq
K_{bw}(x,y)=\sum_{   \genfrac{}{}{0pt}{}{I} { I\ni b, I\ni w}     }  s_I A_I x^{\langle I, \balpha \rangle} y^{\langle I, \bbeta \rangle}. \label{Kbw}
\eeq

%% \genfrac{開き括弧}{閉じ括弧}{分数の横棒の太さ}{数式スタイル}{上}{下}
where $x$ and $y$ are variables, and $s_I$ is chosen as in the previous section to compensate the sign coming from the determinant. The summation is over all edges $I$ connecting $b$ and $w$. Previously we have assumed we have at most one such $I$, but we are now considering more general case.

The determinant of this Kasteleyn matrix is called characteristic polynomial, and is written $P(x,y)$. We have ambiguities in the choice of $\alpha$ and $\beta$-cycles, but you can show $P(x,y)$ is independent of such choice, up to overall multiplication by $x$ and $y$. %You can also show that this definition coincides %with the previous definition \cite{characteristic}

From the characteristic polynomial, the total number of perfect matchings is computed to be
\beq
Z=\frac{1}{2} \left( -P(1,1)+P(1,-1)+P(-1,1)+P(-1,-1) \right).
\eeq
From this expression, you will immediately notice similarity with one-loop amplitude in string theory. That is, $z$ and $w$ correspond to the choice of spin structures in $\balpha$ and $\bbeta$-directions, respectively.

\paragraph{Fast forward algorithm}

Now the forward algorithm is easy to state. Start from a bipartite graph, write down Kasteleyn matrix, and obtain the characteristic polynomial $P(x,y)$. Then the toric diagram is the Newton polygon of $P(x,y)$\cite{Hanany:2005ve,Franco:2006gc}, where the Newton polygon $\Delta(P)$ for Laurent polynomial $P(x,y)=\sum_{(k,l)}c_{k,l}x^k y^l$ is given by
\beq
\Delta(P)=\textrm{convex hull of } \{(k,l) \in \bZ^2 \left| c_{k,l}\ne 0 \right.\}.
\eeq

We have given two different definitions of characteristic polynomial, one given in \eqref{characteristic} and the other as the determinant of Kasteleyn matrix \eqref{Kbw}. Moreover, we have certain ambiguities in the definition of Kasteleyn matrix. It can be shown that the corresponding Newton polygon is unique up to $SL(2,\bZ)$-transformation and translation.

\begin{exa}
The example of the choice of signs and $\alpha, \beta$-cycles is shown in Figure \ref{Kasteleyneg.eps}.

\begin{figure}[htbp]
\centering{\includegraphics[scale=0.4]{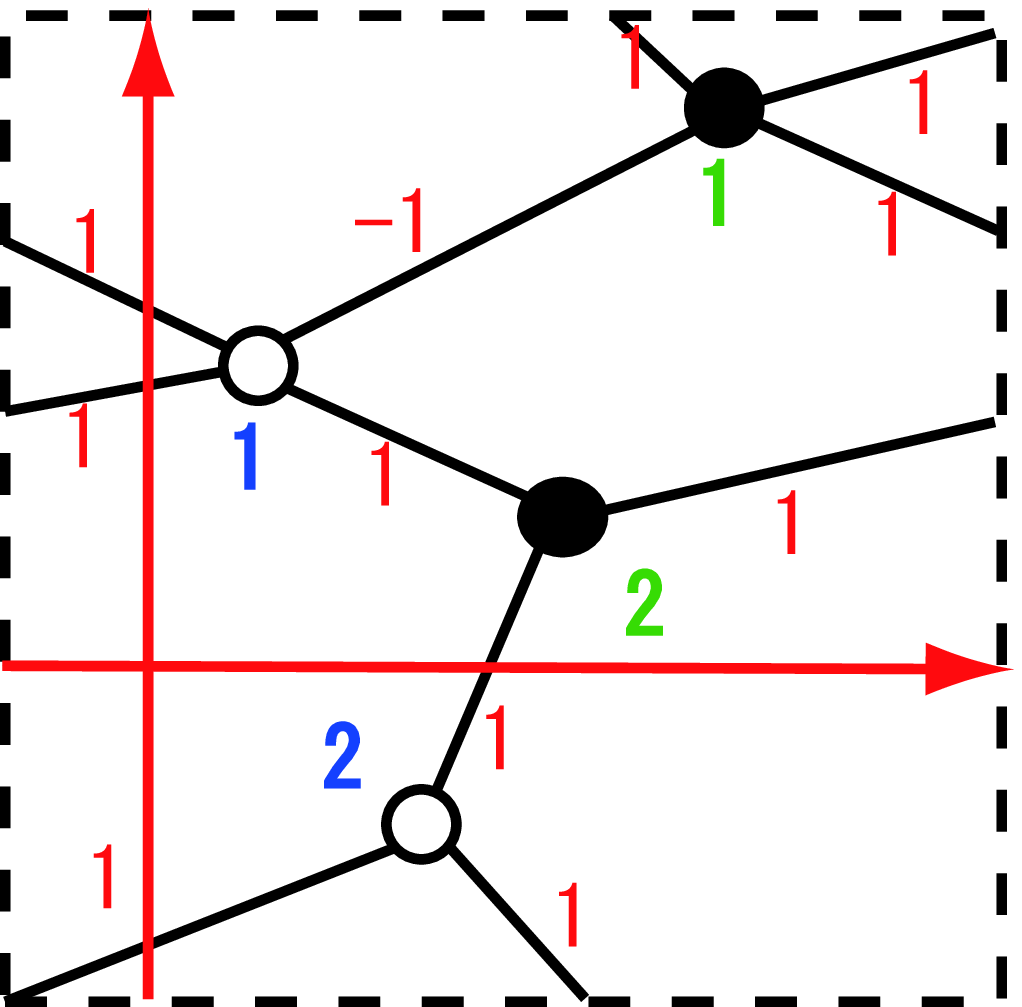}}
\caption{Example of the choice of signs and $\alpha, \beta$-cycles for a bipartite graph on torus. The red lines represent the choice of $\alpha$ and $\beta$-cycles,}
\label{Kasteleyneg.eps}
\end{figure}

In this example the Kasteleyn matrix is given by
\beq
P(x,y)={\rm det }
\left(\begin{array}{cc}
-1+w  & 1+w \\
1+w & z 
\end{array}
\right)
=-z+zw-1-2w-w^2. \label{SPPK2}
\eeq

You can directly check that the Newton polygon for \eqref{SPPK1} and \eqref{SPPK2} both give the same lattice polygon, which coincide with the SPP toric diagram given for example in Figure \ref{untwistspp.eps}.
\qed
\end{exa}

\begin{figure}[htbp]
\centering{\includegraphics[scale=0.4]{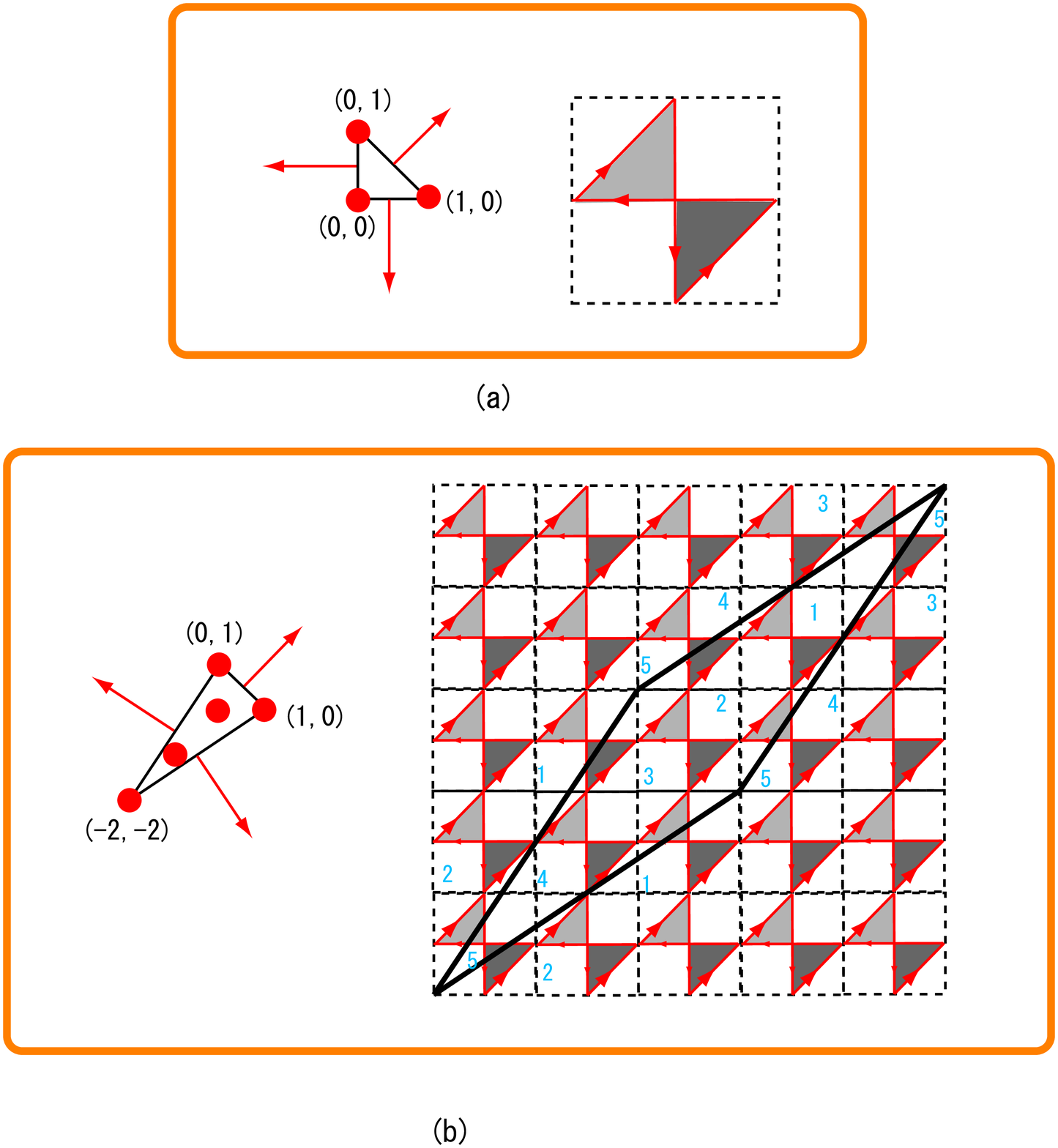}}
\caption{Taking orbifold by finite subgroups of torus corresponds to changing the fundamental region of torus. In this example, (a) shows an example of $\bC^3$, and (b) shows an example of $\bC^3/\bZ_5$ with $\bZ_5\simeq (1,2,2)/5$. Note the fivebrane diagram shown in (b) is the same as that of Figure \ref{SU3eg}.}
\label{equivariant.eps}
\end{figure}

We now give some remarks.
\begin{itemize}
\item In some literature the Pfaffian of the antisymmetric matrix $\tilde{K}$ is used instead of det$K$:
\beq
\tilde{K}=
\left(
\begin{array}{cc}
0 & K \\
 -^{t}K&  0
\end{array}
\right).
\label{Kasteleynform}
\eeq
Since
\beq
%\textrm{Pf}(\tilde{K})=(-1)^{N_W/2 (N_W/2-1)/2}
\textrm{Pf}(\tilde{K})=\pm \textrm{det}(K)
\eeq
%(where $N_W$ is the number of terms in a superpotential, or the number of vertices in bipartite graphs), 
we can use $\textrm{Pf}\tilde{K}$ instead of $\textrm{det}(K)$.

\item The form of the $\tilde{K}$ shown in \eqref{Kasteleynform} is reminiscent of the Dirac operator. Indeed, Kasteleyn matrix is considered to be a discrete analogue of the Dir ac operator on the graph. Similarly, we can define discrete analogue of Green functions and holomorphic functions on the graph. See \cite{KenyonLaplacian,Mercat}.

\item Since Abelian orbifold of toric Calabi-Yaus corresponds to enlarging the fundamental domain (see Figure \ref{equivariant.eps} for an example), once we know the characteristic polynomial of the parent theory it is easy to compute characteristic polynomial for orbifold theory \cite{Hanany:2005ve}.
\end{itemize}

%%%%%%%%%%%%%%%%%%%%%%%%%%%%%%%%%%%%%%%%%%%%%%%%%%%%%%%%%%%%%%%%%%%%%%%%%%%%%%
\subsection{Perfect matchings and F-term constraints} \label{GLSM.subsec}

In Introduction, we introduced the concept of perfect matchings. Perfect matchings have already appeared in the discussion of Kasteleyn matrix (\S\ref{another.subsec}). In this subsection, we explain another use of perfect matchings\footnote{Perfect matchings also appear in the discussion of orientifolds \cite{IKY}.}.
%(\S\ref{orientifold.sec}. }.

Let us begin by defining a natural product between an edge $I$ (corresponding to bifundamental field $\Phi_I$) and a perfect matching $m_{D}$:

\beq
\langle I, m_{D} \rangle=
\begin{cases}
1 & \mathrm{ if }~ I \in m_{D} \\
0 & \mathrm{ otherwise }
\end{cases}
.
\eeq

For each perfect matching, prepare a complex variable $\rho_{D}$. Then the claim is that
\beq
X_I=\prod_{D} \rho_{D}^{\langle  I , m_{D} \rangle} \label{Xp}
\eeq
solves the F-term constraint $\frac{\partial W}{\partial X_I}=0$.

The proof of this fact proceeds as follows. Suppose we are going to consider the F-term relation for the bifundamental field $X_I$, whose corresponding arrow begins at vertex $a$ and ends at vertex $b$. Then superpotential is

\beq
W= \pm \left(X_I \prod_{J\in a, J\ne I} X_J -X_I \prod_{J\in b, J\ne I} X_J  \right) +\ldots,
\eeq
where $I\in a$ means that $a$ is one of the endpoints of the edge $I$, and $\ldots$ is independent of $X_I$. As discussed in \S\ref{quiver.subsubsec}, the superpotential obtained from bipartite graphs and corresponding to toric Calabi-Yau cones always takes this form.

Then the F-term constraint for $\Phi_I$ is 
\beq
X_I \frac{\partial W}{\partial X_I}= \prod_{J\in a} X_J - \prod_{J\in b} X_J =0,\eeq
or, written in variables $\tilde{p}_{D}$, 
\beq
\prod_{D} \prod_{J\in a}  \rho_{D}^{\langle  J , m_{D} \rangle}
=\prod_{D} \prod_{J\in b} \rho_{D}^{\langle  J , m_{D} \rangle},
\eeq
or 
\beq
\prod_{J\in a}  \rho_{D}^{\langle  J , m_{D} \rangle}
=\prod_{J\in b} \rho_{D}^{\langle  J , m_{D} \rangle} \textrm{ ($\forall D$)},
\eeq
or
\beq
  \rho_{D}^{\sum_{J\in a} \langle  J , m_{D} \rangle}
=\rho_{D}^{\sum_{J\in b} \langle  J , m_{D} \rangle} \textrm{ ($\forall D$)}.
\eeq
But since $D$ is a perfect matching, both LHS and RHS of this equation is $\rho_{D}$, and thus the statement if proven.

The formula \eqref{Xp} is useful in the discussion of flavor branes in \S\ref{flavor.subsec}. Historically, the motivation to introduce $\rho_{D}$ comes from the relation with gauged linear sigma model (GLSM). The problem is to compute the vacuum moduli space of $\scN=1$ supersymmetric quiver gauge theories. When we consider the Calabi-Yau setup as in \S\ref{D3.subsec}, the Calabi-Yau manifold in geometry side should be seen as a vacuum moduli space of gauge theory. Therefore, by solving F-term conditions and D-term conditions, we are expected to obtain toric Calabi-Yau cone. 

In simple example such as conifold, it is not difficult to carry this out, but in general, this is not so easy. Calabi-Yau manifolds can be written in the form of K\"ahler quotient, or more physically as a gauged linear sigma model \cite{Witten:1993yc}. In gauge theory language, this amounts to use D-term constraints.  In gauge theory, however, we have F-term constraints as well. Thus first, we have to solve F-term constraints by good variables, and only later we can write everything in terms of D-term, or in the language of GLSM. The field transformation \eqref{Xp} is first studied in such a context in \cite{Feng:2001xr}, and $\rho_D$ are interpreted as fields in GLSM\footnote{See \cite{Sarkar:2007iq} for recent discussions related to these points.}. For details, see \cite{Feng:2001xr} or the review\cite{He:2004rn}.

%%%%%%%%%%%%%%%%%%%%%%%%%%%%%%%%%%%%%%%%%%%%%%%%%%%%%%%%%%%%%%%%%%%%%%%%%%%%%%

\subsection{Seiberg duality} \label{Seiberg.subsec}

In Figure \ref{manytoone} of \S\ref{D3.subsec}, we have mentioned that the relation between toric diagram and the quiver diagram is in general many-to-one. The reason (or at least part of the reason) for this phenomena is that we have Seiberg duality. In this subsection, we will discuss this point in more detail.

Let us very briefly review Seiberg duality \cite{Seiberg:1994pq}. It is a duality between ``electric theory'' and ``magnetic theory''.
As an ``electric theory'', consider $\scN=1$ supersymmetric $SU(N_c)$ gauge theory with $N_f$ fundamentals (quarks) $Q^i$ and antifundamentals (antiquarks) $\bar{Q}_i$ without superpotential, and with $N_f$ in the ``conformal window'' ($\frac{3}{2}N_c < N_f < 3 N_c$) \footnote{We can also consider Seiberg duality outside conformal window, but we restrict ourselves to conformal window in this paper.}. The matter contents of this theory is summarized in Table \ref{beforeSD}.

\begin{table}[htbp]
\caption{The matter contents of electric theory.}
\begin{center}
\begin{tabular}{|c|c|c|c|}
\hline
 & $SU(N_c)$ & $SU(N_f)_L$ & $SU(N_f)_R$ \\
\hline
\hline
$Q$ & $\fund$ & $\fund$ & $1$ \\
\hline
$\bar{Q}$ & $\fundbar$ & $1$ & $\fundbar$ \\
\hline
\end{tabular}
\label{beforeSD}
\end{center}
\end{table}

The dual theory, ``magnetic theory'', is the $\scN=1$ supersymmetric gauge theory with $SU(N_f-N_c)$ gauge groups and $N_f$ fundamentals (quarks) $q_i$ and antifundamentals (antiquarks) $\bar{q}^i$. We have, in addition, meson fields ${M^i}_j$. The superpotential is given by
\beq
W=q_i {M^i}_j \bar{q}^j.
\eeq
The mesons ${M^i}_j$ are written in fields of electric theory as ${M^i}_j=Q^i \bar{Q}_j$, but in magnetic theory counted as independent parameters. The matter content of this theory is summarized in Table \ref{afterSD}.
\begin{table}[htbp]
\caption{The matter content of the magnetic theory, which is dual to electric theory shown in Table \ref{beforeSD}.}
\begin{center}
\begin{tabular}{|c|c|c|c|}
\hline
 & $SU(N_f-N_c)$ & $SU(N_f)_L$ & $SU(N_f)_R$ \\
\hline
\hline
$q$ & $\fund$ & $\fundbar$ & $1$ \\
\hline
$\bar{q}$ & $\fundbar$ & $1$ & $\fund$ \\
\hline
$M$ & $1$ & $\fund$ & $\fundbar$ \\
\hline
\end{tabular}
\label{afterSD}
\end{center}
\end{table}
The claim of Seiberg duality is that these two theories flow in IR to the same fixed point under RG flow. In other words, magnetic theories and electric theories are UV description of the same physics in IR. We have many non-trivial checks of this proposal, such as the matching of  't Hooft anomalies and matching of gauge invariant operators.

%%%%%%%%%%%%%%%%%%%%%%%%%%%%%%

\begin{figure}[htbp]
\centering{\includegraphics[scale=0.8]{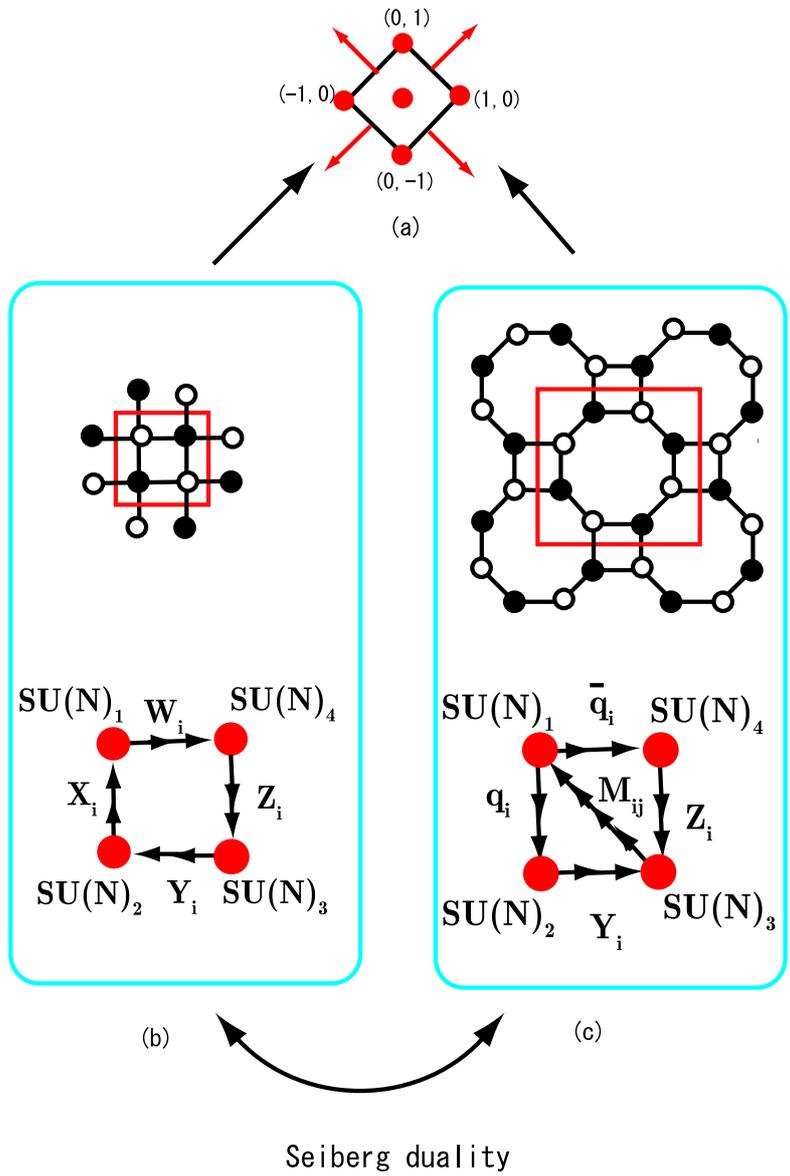}}
\caption{The quiver corresponding to $K_{\bP^1\times \bP^1}$. The two phases shown in Figure \ref{manytoone} are related by Seiberg duality.}
\label{F0phases.eps}
\end{figure}

Now we move onto more complicated example from quiver gauge theories. We take dual of the quiver shown in Figure \ref{F0phases.eps} (b), which corresponds to $K_{\bP^1\times \bP^1}$. The matter contents are summarized in Table \ref{beforeSDF0}.

%\begin{figure}[htbp]
%\centering{\includegraphics[scale=0.8]{F0Iquiver.eps}}
%\caption{The quiver corresponding to $K_{\bP^1\times \bP^1}$.}
%\label{FOIquiver.eps}
%\end{figure}

\begin{table}[htbp]
\caption{The matter content of one quiver of $\bF_0$. Here $i=1,2$.}
\begin{center}
\begin{tabular}{|c|c|c|c|c|}
\hline
 & $SU(N)_1$ & $SU(N)_2$ & $SU(N)_3$ & $SU(N)_4$ \\
\hline
\hline
$X_i$ & $\fund$ & $\fundbar$ & & \\
\hline
$Y_i$ &  &$\fund$ & $\fundbar$ &  \\
\hline
$Z_i$ & & & $\fund$ & $\fundbar$ \\
\hline
$W_i$ & $\fundbar$ & & & $\fund$ \\
\hline
\end{tabular}
\label{beforeSDF0}
\end{center}
\end{table}

The superpotential read off from bipartite graph in Figure \ref{F0phases.eps} (b) is given by
\beq
\begin{split}
W&=\epsilon^{ij}\epsilon^{kl} \tr (X_i Y_k Z_j W_l) \\
&=X_1 Y_1 Z_2 W_2 -X_1 Y_2 Z_2 W_1 -X_2 Y_1 Z_1 W_2 +X_2 Y_2 Z_1 W_1.
\end{split}
\eeq

Now we take Seiberg duality with respect to the gauge group $SU(N)_1$. For $SU(N)_1$, $N_c=N$ and $N_f=2N$ (we have two arrows stating from or ending at that node), so we are certainly in the conformal window, and the rank of the dual gauge group is again $N_f-N_c=N$. From the previously discussed rules, the matter contents of the dual theory should be as in Table \ref{afterSDF0}, and the quiver diagram is shown in Figure \ref{F0phases.eps} (c).
\begin{table}[htbp]
\begin{center}
\caption{The matter contents of the theory obtained by taking Seiberg duality with respect to the node $1$.}
\begin{tabular}{|c|c|c|c|c|}
\hline
 & $SU(N)_1$ & $SU(N)_2$ & $SU(N)_3$ & $SU(N)_4$ \\
\hline
\hline
$q$ & $\fundbar$ & $\fund$ & & \\
\hline
$Y_i$ &  &$\fund$ & $\fundbar$ &  \\
\hline
$Z_i$ & & & $\fund$ & $\fundbar$ \\
\hline
$\bar{q}_i$ & $\fund$ & & & $\fundbar$ \\
\hline

$M_{ij}$ & & $\fundbar$ &  & $\fund$ \\
\hline
\end{tabular}
\label{afterSDF0}
\end{center}
\end{table}

Here the meson field $M_{ij}$ is, written in original variables,
\beq
M_{ij}=X_i Z_j.
\eeq
and the new superpotential is given by
\beq
\begin{split}
W_{I}^{\textrm{dual}}&=W_{I}+M_{ij}q_i \bar{q}_j \\
&= M_{21}Y_1 Z_2-M_{11}Y_2 Z_2-M_{22}Y_1 Z_1 +M_{12}Y_2 Z_1+M_{ij}q_i \bar{q}_j.
\end{split}
\eeq
But you can directly verify that this superpotential and quiver diagram  can be derived from the superpotential obtained from the bipartite graph show in Figure \ref{F0phases.eps} (c).

\begin{figure}[htbp]
\centering{\includegraphics[scale=0.8]{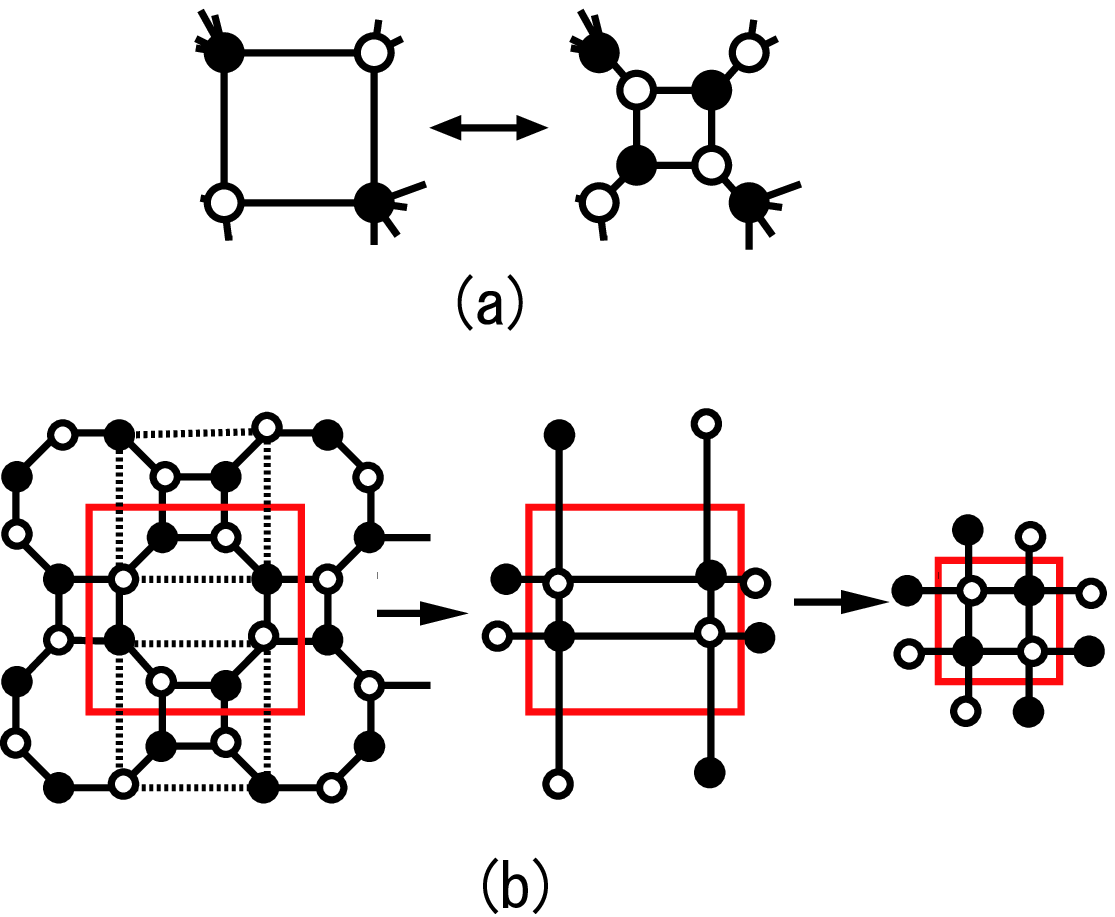}}
\caption{Graphical representation of Seiberg duality by bipartite graph (a). The example of $K_{\bP^1\times \bP^1}$ is shown in (b).}
\label{SeibergDual.eps}
\end{figure}

\begin{figure}[htbp]
\centering{\includegraphics[scale=0.4]{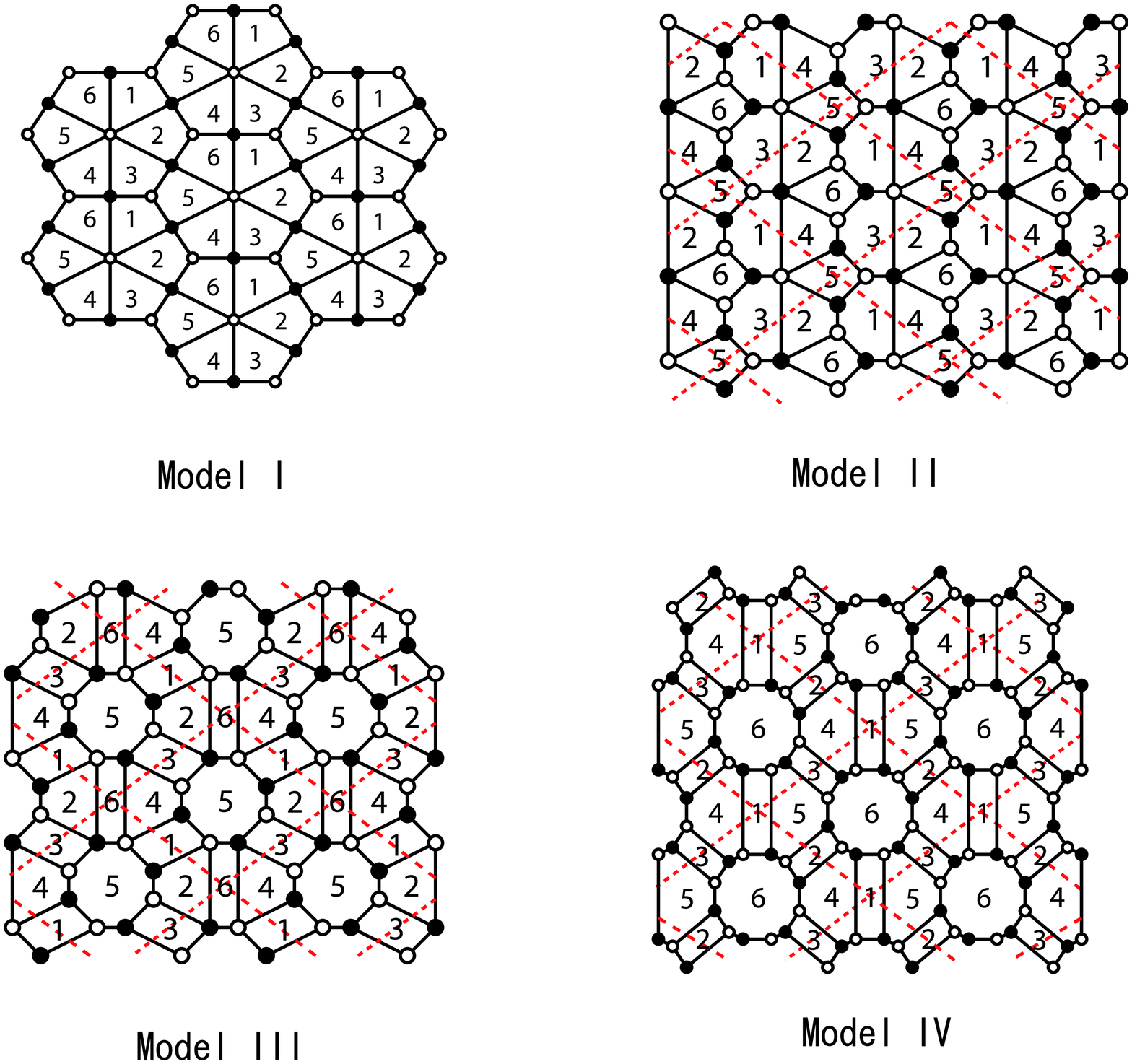}}
\caption{Four bipartite graphs, or ``toric phases'' of del Pezzo 3. In this case we have four bipartite graphs, which are called Model I, II, III and IV. The bipartite graphs are all related by Seiberg duality (Figure \ref{SeibergDual.eps}) and integrating out massive fields (Figure \ref{Massive}). Each region surrounded by red lines represents a fundamental region of torus. Figure adapted and modified from \cite{Franco:2005rj}.}
\label{dP3phases}
\end{figure}

\begin{figure}
\centering{\includegraphics[scale=0.8]{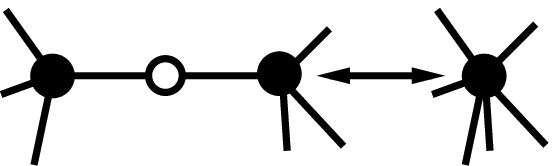}}
\caption{Integrating out massive fields corresponds to replacing (a) by (b).}
\label{Massive}
\end{figure}

From this example, we can learn two facts. First, Seiberg duality can be represented graphically as in Figure \ref{SeibergDual.eps}.
Second, we can guess that all quivers corresponding to the same toric diagram is related by chain of Seiberg dualities. This conjecture is sometimes phrased as ``toric duality is Seiberg duality''\cite{Beasley:2001zp}.
The examples of del Pezzo 3 is shown in Figure \ref{dP3phases}. In this case four bipartite graphs are known and you can explicitly verify that all the bipartite graphs are related by graphical operations as in Figure \ref{SeibergDual.eps}. In verifying this, you will sometimes need to integrate out/in massive fields, whose graphical representation is shown Figure \ref{Massive}.

%%%

To prove or disprove this conjecture, probably we need a better understanding of Seiberg duality. In the past, there are many proposals for understanding Seiberg duality, such as Picard-Lefschetz monodromy \cite{Feng:2001bn,Cachazo:2001sg,Ooguri:1997ih}, mutation \cite{Cachazo:2001sg,Wijnholt:2002qz}, and tilting equivalence of derived categories \cite{Berenstein:2002fi,Braun:2002sb,Mukhopadhyay:2003ky}. The relation between all of them is not clear at the moment (see \cite{Herzog:2003zc}, however, for an interesting attempt), and awaits further study.

Another possible clue might come from the idea of isoradial embedding, which we discussed in \S\ref{R-charge.subsubsec}. As pointed out in \cite{Hanany:2005ss}, this might open the way to single out one `canonical' brane tilings. For example, in Figure \ref{F0phases.eps} we have shown two bipartite graphs corresponding to the same toric diagram, but only one of them (b) can be isoradially embedded. In this connection, we mention the recent work of \cite{IshiiUeda} which relates the existence of R-charges satisfying \eqref{Rcond} to the smoothness of the moduli space.

\begin{figure}
\centering{\includegraphics[scale=0.5]{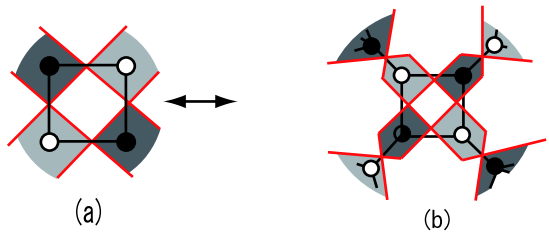}}
\caption{Seiberg duality (Figure \ref{SeibergDual.eps}) rewritten as fivebrane diagrams.}
\label{FivebraneSD.eps}
\end{figure}

Finally would also like to re-look at the contents of this section from fivebrane viewpoint. The graphical representation of Seiberg duality now becomes Figure \ref{FivebraneSD.eps} in fivebrane diagram. It is fair to say that such complicated rearrangement is not well-understood as of this writing. In NS5/D4 brane setup, Seiberg duality is interpreted as Hanany-Witten type of brane crossing \cite{Giveon:1998sr,Elitzur:1997fh}, and the rearrangement of fivebranes in NS5/D5 setup as in Figure \ref{FivebraneSD.eps} should be similarly interpreted, although much more involved.

We also remark that in the strong coupling limit, cycles of NS5-brane become straight in order to preserve SUSY and therefore we do not have the situation as in Figure \ref{FivebraneSD.eps} (b). This means we have fewer possible bipartite graphs (and therefore fewer quiver gauge theories) in the strong coupling limit than in the weak coupling limit. In fact, it is verified for all toric Fano case that we have only one quiver diagram in the strong coupling limit \cite{UY3}. This means that several quiver diagrams in the weak coupling limit should merge into one in the strong coupling limit, but how that actually happens is not known, due to the difficulty of analyzing in general $g_{\mbox{str}}$. We need to investigate brane configurations in general string coupling constant in order to answer the question.

%%%%%%%%%%%%%%%%%%%%%%%%%%%%%%
\subsection{A look at mirror D6-brane picture}\label{D6.subsec}

The discussion up to now mostly uses fivebrane setup, with D5-branes and NS5-branes.
We also commented on the Calabi-Yau setup with D3-branes in \S\ref{D3.subsec}.
These are two different descriptions of the same physics.
In fact, we have one more description, using mirror Calabi-Yau with D6-branes \cite{Feng:2005gw}. This viewpoint is also important, especially in connection with mirror symmetry (for application to homological mirror symmetry, see \S\ref{HMS.sec}).

Let us again begin with the fivebrane system as in Table \ref{configre.tbl}. Take T-duality along 9-directions (or $\phi_3$-direction in the notation of \S\ref{D3.subsubsec}). Then, NS5-brane is turned into a Calabi-Yau manifold $\scW$ and D5-branes are turned into D6-branes wrapping 3-cycles of Calabi-Yau $\scW$.

\begin{table}[htbp]
\caption{By taking T-duality along 9-directions from Table \ref{configre.tbl}, we have a brane configuration as shown here. NS5-brane is turned into a Calabi-Yau manifold $\scW$, and D5-branes are turned into D6-branes wrapping 3-cycles of $\scW$.}
\label{configmirror.tbl}
\begin{center}
\begin{tabular}{c||cccc|cccccc}
\hline
& 0 & 1 & 2 & 3 & 4 & 5 & 6 & 7 & 8 & 9 \\
\hline
CY3 & &  & &  &$\circ$ & $\circ$&$\circ$ & $\circ$&$\circ$ & $\circ$\\
D6  & $\circ$ & $\circ$ & $\circ$ & $\circ$ && $\circ$ && $\circ$ & & $\circ$\\
\hline
\end{tabular}
\end{center}
\end{table}

The geometry of $\scW$ is given by
\beq
P(x,y)=uv,
\eeq
with $x,y\in \bCx$ and $u,v\in\bC$.
This is the mirror Calabi-Yau of the toric Calabi-Yau three-fold $\scM$. This form of mirror Calabi-Yau geometry $\scW$ is known since long ago \cite{Hori:2000kt,Hori:2000ck}. The $uv$ directions represent the 9-direction of T-duality.

In order to identify the gauge theories on D6-branes, we first need to know which 3-cycles D6-branes wrap. 

\begin{figure}[htbp]
\centering{\includegraphics{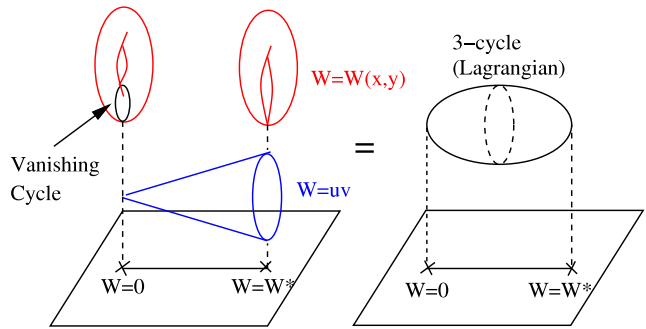}}
\caption{We have a 3-cycle for each critical point of Newton polynomial.}
\label{mirrorfiber}
\end{figure}

Represent $\scW$ as double fibration over $W$-plane;
\beq
uv=W, ~~ P(x,y)=W.
\eeq
For each point of $W$-plane, the fiber is generically $\bP^1\times \Sigma_g$. Here $\bP^1$ comes from $u,v$-fiber, and $\Sigma_g$ comes from $x,y$-fiber. At some special points, however, some cycles of fiber degenerates. For example, at $z=0$, the $uv$-fiber degenerates. Also, at critical values of $P(x,y)$, 1-cycle of Riemann surface $P(x,y)$ shrinks. Therefore, if we start from $W=0$ and go to $W=W_*$ ($W_*$ is a critical point of $P(x,y)$), we have a 3-cycle (Figure \ref{mirrorfiber}).  In fact, these 3-cycles span a basis of $H^3(\scW,\bZ)$, and D6-branes wrap these 3-cycles. The number of such 3-cycles is given by the number of critical points, and as proven in Appendix of \cite{Feng:2005gw}\footnote{the situation is subtle in genus 0 case. See \cite{Feng:2005gw}.} is given by twice the area of the toric diagram. 

Let us label critical points of $P(x,y)$ by $a$ (this is in accord with our previous convention that $a$ is the label for gauge groups, since the number of critical points is the same with the number of gauge groups, as we will see). We denote the corresponding 3-cycle by $S_a$. To be precise, to specify $S_a$ we need to specify a set of paths from $W=0$ to $W=W_*$, which is called the distinguished set of vanishing paths.

More formally, an ordered set $(c_a)_{a=1}^{N_G}$ of smooth paths $c_a: [0,1] \to \bC$ is called a {\em distinguished set of vanishing paths} if 

\begin{enumerate}
  \item the base point $c_a(0)$ is a regular value of $W$
        independent of $a$,
  \item $\{ c_a(1) \}_{a=1}^{N_G}$ is the set of critical values of $W$,
  \item $c_a$ has no self-intersection,
  \item images of $c_a$ and $c_b$ intersects only at the base point.
  \item $c_a'(0) \neq 0$ for $a = 1, \dots, N_G$.
  \item $\arg c_{a+1}'(0) < \arg c_a'(0)$, $a=1,\ldots N_G-1$,
         for a choice of a branch of the argument map.
 \end{enumerate}

As distinguished set of vanishing paths, we simply take straight paths from $W=0$ to $W=W_*$.
Given a distinguished set
$(c_a)_{a=1}^{N_G}$ of vanishing paths,
we lift $c_a$ to $\tilde{c}_a: [0,1]\to (\bC^{\times})^2$ on $(\bC^{\times})^2$ starting from a point $p\in W^{-1}(c(0))$.\footnote{For this to be possible, we need to assume that $W_{\Delta}:(\bC^{\times})^2\to \bC$ is an {\em exact Lefschetz fibration}, namely all the critical values are distinct and all the critical points are non-degenerate. This is the case as long as $W_{\Delta}$ is a generic Newton polynomial of $\Delta$.}
Then corresponding
{\em distinguished basis}
$(C_a)_{a=1}^{N_G}$
{\em of vanishing cycles}
is defined by
\beq
 C_a = \{p \in W^{-1}(c_a(0)) \mid
           \lim_{t \rightarrow 1} \widetilde{c}_p(t) = p_a\},
 \qquad 1 \le a \le N.
\eeq
They are 
%exacg
Lagrangian submanifolds
of $W^{-1}(c_a(0))$, and coincide with the intersection of 3-cycle $S_a$ and $W^{-1}(0)$. These Lagrangian submanifolds will will play an important role in our discussion of directed Fukaya category in \S\ref{HMS.sec}.

Finally we summarize the gauge theory content, but the discussion is almost the same with \S\ref{D5cycle.subsubsec}. We have a gauge group, or a node of  the quiver diagram, for each cycle $S_a$. We have an open string, or a bifundamental field for each intersection of $S_a$, and superpotential terms correspond to whether we can span a disc. This is to be expected, since T-duality along 9-direction does not change complicated geometry spanning 4567-directions. The vanishing cycle $C_a$ coincides with the 1-cycle discussed in \S\ref{D5cycle.subsubsec}, for which we used the same expression. In \S\ref{HMS.sec} we will see that this simple picture is directly translated into the language of Fukaya category.

\subsection{Summary}
In this section, we discussed various aspects of brane tilings, which is not discussed in previous section. In \S\ref{fractional.subsec}, we discussed inclusion of fractional branes, which corresponds to general anomaly-free rank assignments. It was shown there that condition of gauge anomaly cancellation comes from the condition of D5-brane charge conservation. 

In \S\ref{flavor.subsec} we discussed inclusion of flavor branes. By including these brane it is possible to include fundamental and anti-fundamental representations (quarks and anti-quarks).
Flavor branes we consider in this paper are D7-branes in Calabi-Yau setup, and become D5-branes in our fivebrane systems. The D5-branes are represented as an intersection point of two zig-zag paths, and  for each intersection point two types of flavor branes, i.e. major and minor flavor branes, are possible. We proposed superpotential terms for these fields, and show that the massless loci of these quarks coincide with the worldvolume of D7-branes.

As an another topic, we discussed deformation of fivebrane systems. The conventional bipartite graph does not care about such deformation, but it is natural to ask whether such deformation of branes have any physical significance. We answered this question in \S\ref{BPS.subsec} and \S\ref{marginal.subsec}. In \S\ref{BPS.subsec}, we discussed BPS conditions imposed on fivebrane systems, both in the weak and the strong coupling limit. In the weak coupling limit or decoupling limit, deformation corresponds to changing coefficients of Newton polynomial, and the dimension of moduli of deformations can be computed to be $d-3$, where $d$ is the number of lattice points on the boundary of the toric diagram. Knowing the precise shape of the moduli, however, is difficult in general, except in simple cases like generalized conifolds. In the strong coupling limit, the analysis of BPS conditions is much simpler and corresponds to change of the positions of NS5 cycles. In this limit, the dimension of deformation moduli space is again given by $d-3$.
These geometric deformations are combined with Wilson line to make $d-3$ complex parameters.

In \S\ref{marginal.subsec} we discussed exactly marginal deformations of $\scN=1$ superconformal quiver gauge theories. Exactly marginal deformations are parameterized by ``baryonic number assignments'', and the complex dimension of conformal manifold is $d-1$. In \S\ref{comparison.subsubsec}, these marginal deformations are compared with deformation with fivebranes as studied in \S\ref{BPS.subsec}.
Out of these $d-1$ complex parameters, $d-3$ are identified with geometric deformation of branes and Wilson lines. The other two, diagonal gauge coupling and $\beta$-like deformation, are identified with background supergravity fields.

In \S\ref{another.subsec} and \S\ref{GLSM.subsec}, we discussed more combinatorial aspects of brane tilings. In \S\ref{another.subsec}, after introduction of Kasteleyn matrix, we saw that Kasteleyn matrix can be used to obtain toric diagram from bipartite graphs. In \S\ref{GLSM.subsec}, we discussed an interesting fact that perfect matchings solve F-term conditions, and also the relation with gauged linear sigma model.

\S\ref{Seiberg.subsec} is a discussion on Seiberg duality. Using a simple example, we explained Seiberg duality in quiver gauge theories. Interestingly, Seiberg duality can be understood as a simple graphical operation. We also described the conjecture that all quivers corresponding to the same toric diagram are related by chain of dualities.

In the final subsection (\S\ref{D6.subsec}), we quickly explained another Calabi-Yau setup with intersecting D6-branes. In this description, D6-branes are wrapping 3-cycles of Calabi-Yau, and intersecting D6-branes give rise to quiver gauge theories. Some concepts, such as vanishing cycles, will play crucial roles in the discussion of mirror symmetry in \S\ref{HMS.sec}.

\newpage

%%%%%%%%%%%%%%%%%%%%%%%% Part II %%%%%%%%%%%%%%%%%%%%%%%%%%%%%%%%%%%
\part{Applications of brane tilings}\label{part2}
%%%%%%%%%%%%%%%%%%%%%%%%%%%%%%%%%%%%%%%%%%%%%%%%%%%%%%%%%%%%%%%%%%

\section{Application to AdS/CFT correspondence}\label{AdSCFT.sec}

\subsection{Description of the problem}
AdS/CFT correspondence\cite{Maldacena:1997re,Gubser:1998bc,Witten:1998qj,Aharony:1999ti} is definitely one of the most important developments in string theory in recent years. Although still a conjecture, intensive checks have been performed since then, especially in the case of the duality between type IIB string theory on $AdS_5\times S^5$ and $\mathcal{N}=4$ super Yang-Mills theory. In this case, an integrable structure is found both on gauge and on gravity side (at least up to certain orders in perturbation theory) and it is still an quite active area of research (for example, see \cite{Tseytlin:2003ii,Beisert:2004ry,Tseytlin:2004xa,Zarembo:2004hp,Swanson:2005wz,Plefka:2005bk} for reviews).

All these developments are very important and exciting, but at the same time we should always keep in mind that $\mathcal{N}=4$ theory is very special and far different from our real QCD \footnote{However, recent study on quark-gluon plasma and RHIC physics suggests that they are perhaps not that different when we consider finite temperature effects.}. For example, they are conformal invariant with vanishing $\beta$-functions with no mass gap, no confinement, no gaugino condensation and no chiral symmetry breaking. Also, since $AdS_5\times S^5$ has very high symmetry group $PSU(2,2|4)$, we are not sure from this example whether AdS/CFT holds because of such high symmetry or because of dynamics.

Of course, studying (even large $N_c$) QCD is very difficult\footnote{See, however, very interesting proposal about holographic QCD \cite{Sakai:2004cn,Sakai:2005yt}.}; without SUSY we lose control. Still, we can study $\mathcal{N}=1$ SYM. Although certainly different from QCD, $\scN=1$ SYM is believed to share many qualitative features with QCD such as gaugino condensation, chiral symmetry breaking, confinement, mass gap. Thus it is extremely important to try to extend AdS/CFT to $\mathcal{N}=1$ case.

We already have many works on $\mathcal{N}=1$ AdS/CFT. One of the simple ways of reducing supersymmetry is to take orbifolds \cite{Kachru:1998ys}. On the gauge theory side we consider quiver gauge theories as discussed in \S\ref{D3.subsec}, and these are claimed to be dual to type IIB string theory on $AdS_5\times (S^5/\Gamma)$. Later, generalization to more complicated geometry is done by Klebanov and Witten\cite{Klebanov:1998hh}. They found a quiver gauge theory in UV, which they argued in the IR flows to a strongly coupled superconformal field theory perturbed by a certain superpotential. They further claimed that this IR superconformal field theory is dual to type IIB string theory on $AdS_5\times T^{1,1}$, where $T^{1,1}$ is a manifold whose cone is the conifold which is discussed in detail in Appendix \ref{T11.subsec}.

It is natural to  consider more general situation of 
$AdS_5\times S$ where $S$ is a five-dimensional manifold. $S$ must have Killing spinors in order for the dual gauge theory to have $\mathcal{N}=1$ superconformal symmetry, and must be Einstein since $AdS_5\times S$ must be a solution to (super-)gravity. Thus, $S$ must be a Sasaki-Einstein manifold\cite{Morrison:1998cs,Acharya:1998db}. Here the condition that $S$ is Sasaki ensures the existence of Killing spinors, which is needed for supersymmetry. Proper definition of Sasaki manifolds will be given in \S\ref{basics.subsubsec}.

%is that an odd-dimensional manifold $S$ is called Sasaki iff its metric cone is K\"ahler. Here metric cone of manifold $S$ (together with its metric $g$) is defied by $C(S)=S\times \mathbb{R}_{+}$.

Unfortunately, the study of AdS/CFT in this case is quite limited for some time.
Part of the reasons is that the gauge theory is strongly coupled and perturbation is not applicable. One possible check of the correspondence is through the relation proposed by Gubser\cite{Gubser:1998vd}:
\beq
\mathrm{Vol}(S)=\frac{\pi^3}{4} \frac{1}{a}, \label{vol=a}
\eeq
where Vol$(S)$ is the volume of Sasaki-Einstein manifold $S$, and $a$ is the ``central charge'' of four-dimensional $\scN=1$ superconformal field theories\footnote{Actually, we have formulae for the volume of divisors as well. If we denote by $\Sigma_A$ the pull-back of toric divisors of toric Calabi-Yau to Sasaki-Einstein manifold, then its volume is related to the R-charge $R[X_A]$ of the corresponding bifundamental field $X_A$ to be 
\beq
R[X_A]=\frac{\pi}{3}\frac{\textrm{Vol}(\Sigma_A)}{\textrm{Vol}(S)}.
\eeq
We do not discuss the verification of this formula; the method is almost the same with method of \S\ref{versus.subsec}. See \cite{Berenstein:2002ke,Intriligator:2003wr,Herzog:2003wt,Herzog:2003dj} for discussions.
}. 

Let us explain the definition of the central charge. In two-dimensional conformal field theories, we have a beautiful theorem by Zamolodchikov\cite{Zamolodchikov:1986gt} which states that the central charge is monotonically decreasing along the renormalization group flow. In the four-dimensional conformal field theory, the central charge $a$ is the candidate for such central charge \cite{Cardy:1988cw}. In fact, the conjectured $a$-theorem states that $a_{UV}>a_{IR}$ along RG-flow.

In order to define $a$, consider curved background, and then we have the trace anomaly:
\beqa
\langle T_{\alpha}^{\alpha}\rangle =\frac{c}{16\pi^2} W_{\mu\nu\rho\lambda}^2
-\frac{a}{16\pi^2}R^2_{\mu\nu\rho\lambda},
%\langle \partial_{\mu} R_{\mu}\rangle =\frac{c-a}{24\pi^2}\epsilon_{\mu\nu\rho\%lambda}
%{R^{\mu\nu}}_{\beta\gamma} R^{\rho\lambda\beta\gamma}+\frac{5a-3c}{9\pi^2}B_{\m%u\nu}\tilde{B}^{\mu\nu} \label{divR}
\eeqa
where $W_{\mu\nu\rho\lambda}$ is a Weyl tensor, and $R^2$ term is the so-called Euler density. and  The central charges $a$ and $c$ are defined as coefficients of this expression. These central charges $a$, $c$ are related to the 't Hooft anomaly $\Tr R^3$ and $\Tr R$ as in \cite{Anselmi:1997am, Anselmi:1997ys}
\beq
a=\frac{3}{32}\left(3 \Tr R^3-\Tr R \right), ~~ c=\frac{1}{32}\left(9 \Tr R^3-5\Tr R \right). \label{ac}
\eeq
The important point is that, through 't Hooft anomaly matching condition \cite{tHooft:1979bh}, the value of these central charges can be computed in UV\footnote{We assume here that the global symmetry is the same in IR and in UV and no accidental symmetry appears. This is believed to be true for all the cases we consider here. }, where perturbative calculation is applicable. 

This means that $a$ can be computed in UV, which is a good news, and we have a hope of verifying \eqref{vol=a} for many examples. However, Gubser was able to verify this formula only for $S^5$ and $T^{1,1}$. There are again several reasons for this.
First, in order to know the volume of Sasaki-Einstein manifolds, (at least naively) one has to know the explicitly form of the metric, which was known only for these two cases. Another reason is that we have multiple global $U(1)$s in UV, and superconformal $U(1)_R$ is a mixture of UV R-symmetry with these global $U(1)$ symmetries. Third, the precise relation between (toric) Calabi-Yau geometry and (quiver) gauge theory was not known.

All of these problems are now solved. First, recently a new infinite class of explicit metrics has been constructed, which are called $Y^{p,q}$ and $L^{a,b,c}$ (\S\ref{explicit.subsubsec}). Using these explicit metrics we can of course compute their volume. In more general case of toric Sasaki-Einstein manifold, a general formula for volumes of toric Sasaki-Einstein manifolds is later given (\S\ref{volume.subsec}). 
Also, since we are considering toric case, we can find their dual gauge theories by brane tiling techniques as previously explained in \S\ref{strong.subsec}. With regard to $U(1)_R$-symmetry, we have $a$-maximization (\S\ref{a-max.subsec}), and we can compute the central charge.

In this section, we describe these Sasaki-Einstein manifolds and their role in AdS/CFT correspondence. %Strictly speaking, not all of the materials in this section is directly relevant to brane tilings.
We begin with the discussion of gauge theory side, namely $a$-maximization (\S\ref{a-max.subsec}). We next explain the geometry of Sasaki-Einstein manifolds (\S\ref{Sasaki.subsec}), and the volume-minimization procedure (\S\ref{volume.subsec}). All of these discussions are combined in \S\ref{versus.subsec}, where we discuss the matching of $a$-maximization and volume minimization.

Some reviews on this topic (mainly from Sasaki-Einstein side) include \cite{Martelli:2004wu}, which summarizes early developments and mostly concentrates on $Y^{p,q}$. More recent review \cite{Sparks:2007us}, although intended mainly for mathematicians, should also be useful.

\subsection{Gauge theory side: $a$-maximization}\label{a-max.subsec}

In this subsection we briefly explain how to compute the central charge $a$. As we discussed, the problem is the mixing between various $U(1)$ symmetries. When we have a single $U(1)$ (this is R-symmetry) in UV, that $U(1)$ is directly identified with the superconformal $U(1)_R$. In general, however,  we have global flavor symmetries in IR, and symmetry argument alone does not determine superconformal $U(1)_R$ uniquely. Therefore, superconformal $U(1)_R$ is given as a mixture of all these $U(1)$s:
\beq
R_t=R_0+\sum_M t_M F_M. \label{trial}
\eeq
This $R_t$ is a possible candidate for superconformal $U(1)_R$-symmetry, and called ``trial R-symmetry''.
What Intriligator and Wecht has shown \cite{Intriligator:2003jj} is that superconformal $U(1)_R$ satisfies 
\beq
9 \Tr(R^2 F_M)=\Tr F_M, \label{trRF}
\eeq
\beq
\Tr R F_M F_N <0. \label{trRFF}
\eeq
Interestingly, these two conditions are rewritten as the maximization of trial $a$-function for trial R-charge $R_t$ in \eqref{trial}:
\beq
a(t)=\frac{3}{32}(3\Tr R_t^3-\Tr R_t). \label{at}
\eeq
%Note that this takes the same form as the usual $a$-function \eqref{ac}, with $a$ replaced by $a(t)$.
In other words, the claim is that superconformal $U(1)_R$ maximizes ``trial $a$-function $a(t)$''. \footnote{Of course, this statement is equivalent to the statement that \eqref{trRF} and \eqref{trRFF} holds, so we can use these equations instead anyway. The reason we are stating in this form is that \eqref{at} takes the same for as \eqref{ac}, with R-charge replaced by ``trial R-charge''. The real implications of this fact, however, are far from being understood.}

More practically, suppose we are given a quiver diagram together with a superpotential. Then we want to parameterize possible $U(1)_R$-symmetries. In order to do that, we assign trial R-charge $x_I$ to each bifundamental field. In order to parameterize $U(1)_R$-symmetry, $x_I$ have to be chosen so that the theory is conformal, or
\beq
\beta_a=\frac{d}{d\log \mu}\frac{1}{g_a^2}=0, \ \  \\
\beta_k=\frac{d}{d\log \mu}h_k=0.
\eeq

These conditions are already spelled out in \eqref{Rcond} (see also \eqref{betagauge} and \eqref{betapot}), which we reproduce here for convenience:
\beq
\sum_{I\in k} x_I=2,\ \ \sum_{I\in a}(1-x_I)=0 \label{Rcond2}
\eeq
We have $N_G+N_W$ conditions imposed on $N_F$ parameters, where $N_G$ is the number of gauge groups, $N_W$ is the number of superpotential terms, and $N_F$ is the number of bifundamental fields. Since bipartite graph is written on torus, we have $N_G-N_F+N_W=0$ and we have as many conditions as many parameters.
But the conditions \eqref{Rcond2} are not independent. In fact, we expect $d-1$ parameter space of solutions to these equations\footnote{
In gravity side, this corresponds to the following geometrical fact:
\beq
\textrm{dim }H_3(S,\bZ)=d-3.
\eeq
Ramond-Ramond four-form $C_4$ in type IIB string theory can be dimensionally reduced to obtain one-form gauge fields $A_I$:
\beq
C_4=\sum_{I=1}^{d-3} A_I\wedge \mathcal{H}_I,
\eeq
where $\mathcal{H}_I$ is a three-form Poincar\"e dual to $C_I$. These $d-3$ gauge symmetries are interpreted as baryonic global symmetries in quiver gauge theory side. In addition to baryonic symmetries, we have $U(1)^3$ isometry of toric variety, which correspond to two mesonic global symmetries and one R-symmetry. Therefore, we have $(d-1)$ non-R global symmetries. 
}.

After solving these equations, write down $a$-function in ultraviolet:
\beq
a=\frac{3}{32}\left(
2N_G +\sum_I \left[
3(x_I-1)^3-(x_I-1)
\right]
\right).
\eeq
Here $2N_G$ corresponds to contribution from gauginos (recall that $N_G$ is the number of $SU(N)$ gauge groups, or twice the area of the toric diagram $\Delta$), and $-1$ of $r_I-1$ comes from the difference of R-charge between scalar and fermion components. The summation is over all elementary fields, and $x_I$ denotes the R-charge of the field $I$, as we discussed. Also, in the case we want to consider, the relation $a=c$ holds \cite{Henningson:1998gx,Freedman:1999gp} and thus $\Tr R=0$, which means $a$ simplifies to
\beq
a=\frac{9}{32}\left(
N_G +\sum_I 
(x_I-1)^3
\right).
\eeq

\begin{exa}[$T^{1,1}$]
Let us first try the example of $T^{1,1}$. The toric diagram, fivebrane diagram and the quiver diagram is shown in Figure \ref{conifoldcycles.eps}, and the superpotential is shown in \eqref{conifoldW}.
In this case, we prepare four variables $x_1, x_2, y_1, y_2$, corresponding to four bifundamentals $A_1, A_2, B_1, B_2$, respectively. Then the trial $a$-function is given by
\beq
a(x_1,x_2,x_3,x_4)=\frac{9}{32}\left[
2+(x_1-1)^3+(x_2-1)^3+(y_1-1)^3+(y_2-1)^3
\right].
\eeq

The superpotential is given by \eqref{conifoldW}, and although we have two conditions from superpotential, both of them is the same and is given by 
\beq
x_1+x_2+y_1+y_2=2. \label{sum=2}
\eeq
The second condition of (\ref{Rcond2}) is again the same for two vertices, and is given by
\beq
(1-x_1)+(1-x_2)+(1-y_1)+(1-y_2)=2.
\eeq
This is the same as (\ref{sum=2}). 
We therefore learn that $x_i, y_i$ should satisfy
\beq
x_1+ x_2+ y_1+y_2=2.
\eeq
We have three parameters left, which is consistent with the above result since $d-1=3$. The extremization of $a$-function is easy, and the result is
\beq
{x_i}_*={y_i}_*=\frac{1}{2} ~(\text{for all i}),
\eeq
with the extremal value given by
\beq
a(x_1,x_2,y_1,y_2)=\frac{9}{32}\left[
2\cdot 1+4(\frac{1}{2}-1)^3
\right]=\frac{27}{64}.
\eeq
The AdS/CFT relation \eqref{vol=a} predicts that the volume of corresponding Sasaki-Einstein manifold (which is denoted $T^{1,1}$, as we will see) is given by

\beq
\textrm{Vol}(T^{1,1})=\frac{\pi^3}{4}\frac{64}{27}=\frac{16\pi^3}{27}. \label{T11volfroma}
\eeq

\qed
\end{exa}

\begin{exa}[del Pezzo 2] \label{dP2amax}
Now we are going to treat the case of del Pezzo 2. In this case, we do not have the explicit metric.

The toric diagram, fivebrane diagram and quiver diagram is shown in Figure \ref{dP2cycles.eps}.
\begin{figure}
\centering{\includegraphics[scale=0.45]{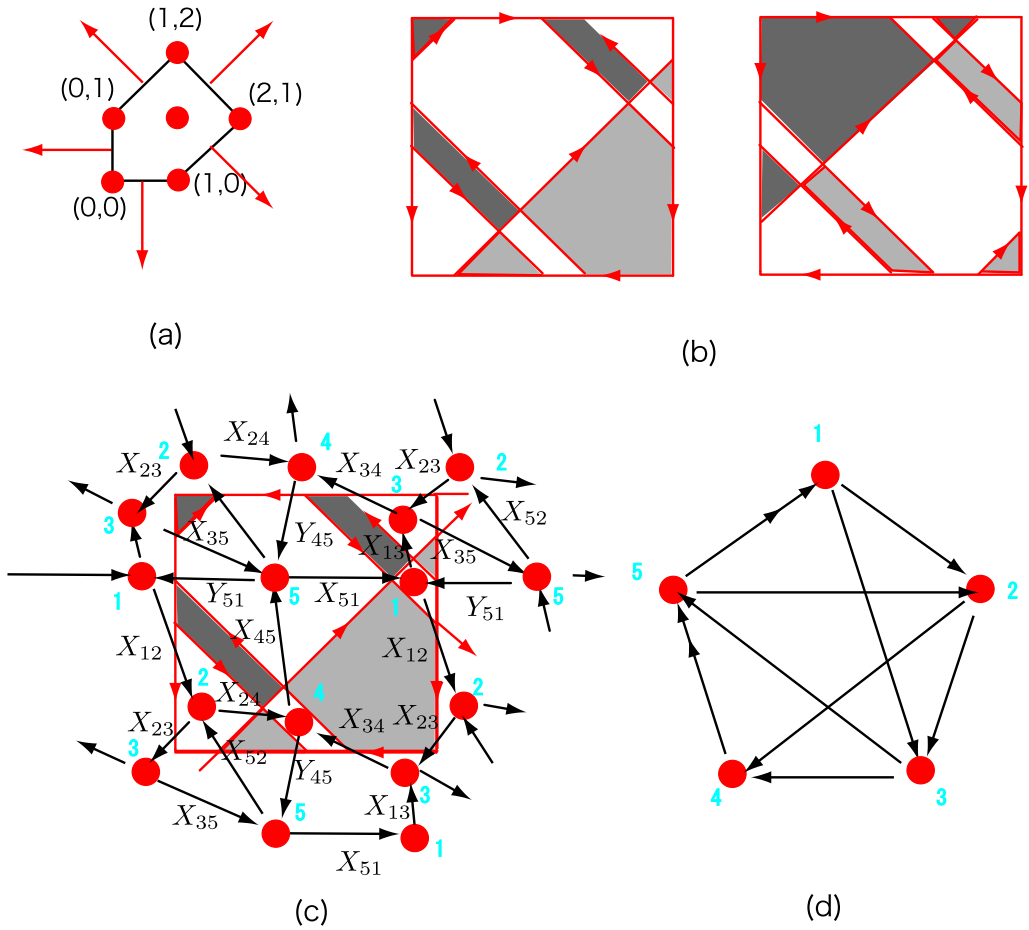}}
\caption{The toric diagram (a), fivebrane diagram (b), periodic quiver (c) and the quiver diagram (d). In (b), we have two seemingly different fivebrane diagrams, but you can verify that they give the same quiver and the same superpotential.}
\label{dP2cycles.eps}
\end{figure}
In Figure \ref{dP2cycles.eps} the arrow which starts from node $i$ and ends at $j$ is denoted by $X_{ij}$. Sometimes we have two arrows beginning and ending at the same vertex, and in that case we use $X$ and $Y$ to denote them, such as $X_{51}$ and $Y_{51}$.
In this notation, the superpotential is given by
\beq
\begin{split}
W=&+\tr (X_{51} X_{12} X_{23} X_{34} X_{45}) %B
-\tr (X_{51} X_{13} X_{34} Y_{45} ) %W
-\tr (Y_{15} X_{12} X_{24} X_{45} X_{51}) %W 
\\
&+\tr (Y_{51} X_{13} X_{35}) %B
-\tr(X_{35} X_{52} X_{23}) %W
+\tr(Y_{45} X_{52} X_{24}) %B
%
%W=&-(a_{15}, a_{54}, a_{43}, a_{32}, a_{21}) %B
%+(a_{15},b_{54},a_{43},a_{31}) %W
%+(b_{15},a_{54},a_{42},a_{21}) %W 
%\\
%&-(b_{15},a_{53},a_{31}) %B
%+(a_{53},a_{32},a_{25}) %W
%-(b_{54},a_{42},a_{25}) %B
%
%W=&-(a_{15}, a_{54}, a_{43}, a_{32}, a_{21}) %B
%+(a_{15},b_{54},a_{43},a_{31}) %W
%+(b_{15},a_{54},a_{42},a_{21}) %W 
%\\
%&-(b_{15},a_{53},a_{31}) %B
%+(a_{53},a_{32},a_{25}) %W
%-(b_{54},a_{42},a_{25}) %B
\end{split}
\eeq

Therefore, the conditions coming from superpotential are
\beq
\begin{split}
x_{51}&+x_{12}+x_{23}+x_{34}+x_{45}=x_{51}+x_{13}+x_{34}+y_{45}\\
&=y_{51}+x_{12}+x_{24}+x_{45}=y_{51}+x_{13}+x_{35}\\
&=x_{35}+x_{52}+x_{23}=y_{45}+x_{52}+x_{24}=2,
\end{split}
\eeq
where $x_{ij}$ and $y_{i,j}$ are variables corresponding to bifundamental fields $X_{ij}$ and $Y_{i,j}$.

The condition coming from vertices, i.e. the second condition of (\ref{Rcond2}), are written as, (beginning from vertex $1$ to vertex $5$)
\beq
\begin{split}
(1-x_{12})&+(1-x_{13})+(1-x_{51})+(1-y_{51})\\
=(1-x_{12})&+(1-x_{52})+(1-x_{24})+(1-x_{23})\\
=(1-x_{13})&+(1-x_{23})+(1-x_{34})+(1-x_{35})\\
=(1-x_{24})&+(1-x_{34})+(1-x_{45})+(1-y_{45})\\
=(1-x_{51})&+(1-y_{51})+(1-x_{52})+(1-x_{35})+(1-x_{45})+(1-y_{45})
=2
\end{split}
\eeq
We have, $6+5=11$ conditions, but they are not independent and can be solved as
\beq
\begin{split}
x_{23}&:=x,~~ x_{52}:=y,~~ x_{12}=z,~~ y_{51}=w,~~ x_{34}=w-x,~~ \\
x_{45}&=x_{13}=x+y-w,~~y_{45}=x+z,\\
 x_{51}&=x_{24}=2-x-y-z,~~ x_{45}=2-x-y.
\end{split} \label{dP2parameters}
\eeq
Note that we have 3 remaining variables, and this is consistent with the expectation since $d-1=3$.

The trial $a$-function is given by
\beq
\begin{split}
a=\frac{9}{32}&\left[
5+ (x-1)^3+(y-1)^3+(z-1)^3+(w-1)^3+(w-x-1)^3 \right.\\
&+ 2(x-y-w-1)^3+(x+z-1)^3+2(1-x-y-z)^3\\
&\left.+(1-x-y)^3
\right].
\end{split}
\eeq

Extremization of this function gives
\beq
\begin{split}
x_{*}&=\frac{1}{2}(-5+\sqrt{\mathstrut{33}}),~~
y_{*}=\frac{1}{4}(9-\sqrt{\mathstrut{33}}),~~ \\
w_{*}&=\frac{1}{16}(17-\sqrt{\mathstrut{33}}),~~
z_{*}=\frac{2}{16}(19-3\sqrt{\mathstrut{33}}), 
\end{split}
\eeq
and the extremal value of $a$ is given by
\beq
a=\frac{243}{1024}(-59+11\sqrt{\mathstrut{33}})
\eeq

When we know the value of $a$, we can predict the volume of Sasaki-Einstein manifolds by using the relation (\ref{vol=a}). The result is 
\beq
\textrm{Vol}(S_{dP_2})=\frac{\pi^3}{4}\frac{1}{a}=\frac{(59+11\sqrt{\mathstrut{33}})\pi^3}{486}. \label{dP2vol}
\eeq

We have thus succeeded in obtaining the volume of $S^1$ bundle over del Pezzo 2, whose metric is not known!\footnote{In fact, historically this was first derived from the field theory computation in the pioneering work of \cite{Bertolini:2004xf}. The gravity computation, namely volume minimization, was later developed and field theory prediction was confirmed.} We will see below that this value is reproduced by volume minimization procedure.

\end{exa}

\subsection{Sasaki-Einstein geometry} \label{Sasaki.subsec}
\subsubsection{Basics} \label{basics.subsubsec}
We begin this subsection with a quick introduction to Sasaki-Einstein geometry. See the book \cite{BGbook} for complete discussions. 

%First, as we already learned in section ***, we have

Since we want to study $AdS_5/CFT_4$ correspondence, for the most part we concentrate on the case of five-dimensional Sasaki-Einstein manifolds, although some parts (especially the discussion on volume minimization in \S\ref{volume.subsec})  apply to general odd-dimensional Sasaki manifolds.

 A manifold $S$ with metric $g$ is called Sasaki (resp. Sasaki-Einstein) if and only if 
its metric cone $(C(S),\overline{g})=(\mathbb{R}_{+}\times S, dr^2+r^2 g)$ is K\"ahler (resp. K\"ahler and Ricci-flat) (see Figure \ref{cone.eps} in \S\ref{D3.subsec} for figure of metric cone.). Here, $r$ is a coordinate of $\mathbb{R}_{+}$. Note also $r=0$ is not included, since we have used $\mathbb{R}_{+}$. In all examples expect $S^5$, $r=0$ is a singular point. 

By explicit computation of Ricci tensor, the condition that metric cone being Ricci flat is equivalent to the condition that $S$ is Einstein\footnote{Usually, when we say Einstein, the proportionality constant of the Ricci tensor and the metric is arbitrary, but it is fixed to be $2(n-1)$ in Sasaki-Einstein geometry. The reason is that metric is canonically normalized.}:
\beq
\mbox{Ric}=2(n-1)g.
\eeq

Sasaki-Einstein manifolds have several important concepts, but for physics applications, the Reeb vector field is the most important. The Reeb vector field $\xi$ is defined by \footnote{We do not bother about the difference between $\xi$ on Calabi-Yau cone and its pull-back to $S$, and use the same symbol for both.}
\begin{align}
\xi=J\left(r\frac{\partial}{\partial r}\right),
\end{align}
where $J$ is the complex structure on $C(S)$, and $r$ is the radial coordinate

A closely related concept is the contact one-form $\eta$,
\beq
\eta=J\left(\frac{dr}{r} \right). \label{eta}
\eeq 
This $\eta$ satisfies $\eta\wedge (d\eta)^{n-1}\ne 0$ and defines a contact structure\footnote{A contact structure is an odd-dimensional analogue of symplectic structure.} on $S$. From its definition, you will see that this $\eta$ is dual of $\xi$ with respect to the metric.

%%%%
\paragraph{Classification of Sasaki-Einstein manifolds} 
Sasaki manifolds are divided into the following three classes, according to the orbits of Reeb vector field. The classification is shown below\footnote{In some references, what is called quasi-regular here is called non-regular, and quasi-regular have broader meaning, including regular and non-regular. in the following we assume that quasi-regular means quasi-regular in the narrower sense as defined above. This terminology in accordance with physics literatures, for example \cite{Martelli:2006yb}.}.

\begin{equation}
\begin{cases}
\mbox{the orbit of }\xi \mbox{ closes} & 
\begin{cases}
U(1) \mbox{ action of } \xi \textrm{ free } & \textrm{regular} \\
U(1) \mbox{ action of } \xi \textrm{ not free } &  \textrm{quasi-regular}\\
\end{cases}

\\

\mbox{the orbit of } \xi \mbox{ does not close} & \textrm{irregular} \\
\end{cases}
.
\label{Sasakicl}
\end{equation}
%\qed
%\end{dfn}

Now we are going to explain the meaning of this classification.

\begin{itemize}
\item When $S$ is regular, the lengths of orbits of $\xi$ are all equal, and $S$ is a $U(1)$ principle bundle over a one-dimensional lower ($(2n-2)$-dimensional) K\"ahler-Einstein manifold. This means the theory of regular Sasaki-Einstein manifold boils down to the theory of K\"ahler-Einstein manifolds in the base.
 
\item When $S$ is quasi-regular, the orbit of $\xi$ closes, but there exists at least one point $x$ on $S$, whose stabilizer $\Gamma_x$ is non-trivial. In this case, since $\Gamma_x$ is a non-trivial subgroup of $U(1)$ (Note that $\xi$ cannot vanish since it has norm $1$), there exists certain integer $m$ such that $\Gamma_x$ is isomorphic to $\mathbb{Z}_m$. Then the length of orbit passing through $x$ is $1/m$ times the length of generic orbit (Figure \ref{quasiregular}). In this case, the quotient space is an orbifold, and $S$ is a $U(1)$-principle  orbibundle over $(2n-2)$-dimensional K\"ahler-Einstein orbifold.

\begin{figure}[htbp]
\centering{\includegraphics[scale=0.5]{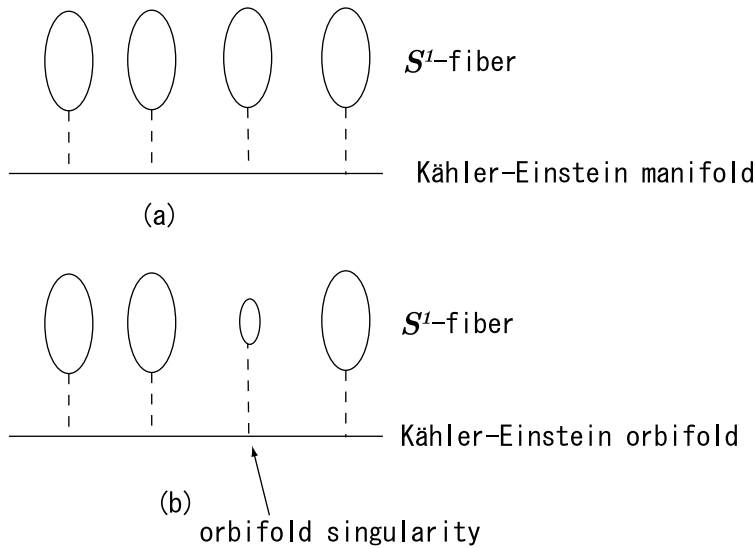}}
\caption{In regular case (a), the length of the $S^1$ fiber is the same for all points in the base. In quasi-regular case (b), however, we have an orbifold singularity and the length of the fiber is smaller at that point.}
\label{quasiregular}
\end{figure}

\item 
When irregular, the quotient space is not well-defined and exists only as a transverse structure.
\end{itemize}

We now learn that  $(2n-1)$-dimensional Sasaki-Einstein manifold is sandwiched between two K\"ahler structures. The metric cone, which is $2n$-dimensional, is by definition K\"ahler. We also have K\"ahler-Einstein structure in 
$(2n-2)$-dimensions, although that exists only as a transverse structure in general.

\subsubsection{Explicit metrics of Sasaki-Einstein manifolds} \label{explicit.subsubsec}
%\subsubsection{Explicit metrics of five-dimensional Sasaki-Einstein manifolds}

In order to check AdS/CFT relation \eqref{vol=a}, we need to compute volume of Sasaki-Einstein manifolds and apparently we need explicit form of the metric.

%Recently, there has been important developments in this field, but before stating them, let us mention what we know about the metric of five-dimensional Sasaki-Einstein manifolds, before 2004.

Surprisingly enough, the examples of explicit metrics are limited for a long time to only two examples: five-dimensional sphere $S^5$ and homogeneous space $T^{1,1}$.
These are both regular, whose associated four and six-dimensional K\"ahler-Einstein are summarized in \ref{table:reg}. See Appendix \S\ref{T11.subsec} for detailed discussion of $T^{1,1}$.

\begin{table}[htbp]
\begin{center}
\begin{tabular}{|c|c|c|}
\hline
4d &5d& 6d\\
\hline
\hline
 $\bb{CP}^2$ & $S^5$ & $\bb{C}^3$ \\
\hline
 $\bb{CP}^1\times \bb{CP}^1 $ & $T^{1,1}$& conifold\\
\hline
\end{tabular}
\caption{For a long time, $S^5$ and $T^{1,1}$ are the only examples  of five-dimensional Sasaki-Einstein manifolds which we know explicit metrics. In the Table we also show associated 4- and 6-dimensional K\"ahler-Einstein manifolds.}
\label{table:reg}
\end{center}
\end{table}

\paragraph{Regular and quasi-regular case}
We are now going to treat each of the classification \eqref{Sasakicl} in detail. First, if you consider regular case, the Sasaki-Einstein manifold is a $S^1$-bundle over (real) four-dimensional K\"ahler-Einstein
manifold, so the problem is to classify complex two-dimensional 
K\"ahler-Einstein manifold with $c_1>0$.

 These manifolds are classified in mathematics \cite{GH}, and they are either  $\mathbb{P}^1 \times \mathbb{P}^1$, or blow-up of $\mathbb{P}^2$, $\mathbb{P}^2$ up to eight points (at generic points) . 
 Let us call $k$-point blow-up of $\mathbb{P}^2$ the del Pezzo $k$ or $k$-th del Pezzo surface, and denote them by  $dP_k$. Of all the del Pezzo surfaces, only $\mathbb{P}^1 \times \mathbb{P}^1$, $\mathbb{P}^2$ and $dP_k (k=1,2,3)$ are toric.

Of the surfaces in this classification, we have already discussed the case of $\mathbb{P}^1 \times \mathbb{P}^1$ and $\mathbb{P}^2$, whose corresponding Sasaki-Einstein manifold is $S^5$ and $T^{1,1}$. The question is whether we have a metric on the remaining $dP_k$.

In fact, it is known that we do not have any K\"ahler-Einstein metric on $dP_1$ and $dP_2$ \footnote{As written here, in the case of $c_1 >0$, we have in general obstruction to the existence of K\"ahler-Einstein metrics. This is in sharp contrast with $c_1=0$ or $c_1<0$ case. In fact, the famous Calabi conjecture asks whether we have K\"ahler-Einstein metrics in these cases, and the positive answer is given by Aubin\cite{Aubin} and Yau\cite{Yau} for $c_1<0$ and by Yau\cite{Yau} for $c_1=0$. }. The Matsushima theorem \cite{Matsushima} states that the set of holomorphic vector fields on compact K\"ahler-Einstein manifold is reductive (namely a direct sum of Abelian Lie algebras and semisimple Lie algebras), but that is not the case for $dP_1$ and $dP_2$.

The remaining case is $dP_k (k=3,4,\ldots, 8)$. 
In this case, the existence of metric is known by the works of Tian and Yau \cite{Tian-Yau}
for all $k$ with $3 \le k \le 8$ \footnote{In the case $n\ge 5$, we have a non-trivial moduli space of complex structures, whose complex dimension is $\ge n-4$.}. Therefore, if we consider $S^1$ bundle over $dP_k (k=3,4,\ldots ,8)$, that is a five-dimensional Sasaki-Einstein manifold. We will denote these $S_{dP_k}$ for the remainder of the paper.  Unfortunately, the explicit form of the metric is not known\footnote{Explicit K\"ahler-Einstein metric is known when the blow-up points are in a special symmetric configuration in $dP_6$ \cite{Calabi}. }
. 

For the quasi-regular case, Boyer-Galicki has shown many existence theorems, using existence of metrics of Fano orbifolds, but the physical significance of their metrics is not clear at the moment. 
See the papers \cite{BG1,BG2,BG3,Coevering} and the review \cite{BGreview}.

\paragraph{Explicit construction of irregular metrics: $Y^{p,q}$ and $L^{a,b,c}$}

Finally, we discuss the irregular case, which is the most difficult.
For irregular case, explicit construction of metric, or even the existence proof of metric was not known for a long time. In fact, Cheeger and Tian conjectured in 1994 that irregular Sasaki-Einstein metrics do not exist \cite{Cheeger}. But in 2004, Gauntlett-Martelli-Sparks-Waldrum\cite{Gauntlett:2004yd} has shown (motivated from study of supergravity \cite{Gauntlett:2004zh}) that we have countably infinite Sasaki-Einstein metrics on $S^2\times S^3$, which are called $Y^{p,q}$. Here $p$, $q$ are integers such that $\textrm{hcf}(p,q)=1$ and  $q<p$.
This metric $Y^{p,q}$ is quasi-regular if and only if $4p^2-3q^2$ can be written as a square of some integer, and irregular otherwise. In particular, these are counterexamples to Cheeger-Tian conjecture. 
Also, $Y^{p,q}$ is topologically $S^2\times S^3$ (see Appendix A of \cite{Gauntlett:2004yd}). 
We can apply similar construction for higher-dimensional odd Sasaki-Einstein manifolds as well \cite{Chen:2004nq,Gauntlett:2004hh,Gauntlett:2004hs}.

\begin{figure}[htbp]
\begin{center}
\includegraphics[scale=0.7]{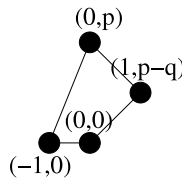}
\caption{The toric diagram of $Y^{p,q}$. }
\label{fig:YpqTD}
\end{center}
\end{figure}

It was later shown the metric $Y^{p,q}$ is toric. The toric diagram is shown in \ref{fig:YpqTD}. 
This metric has cohomogeneity $1$, where cohomogeneity refers to (real) dimension of generic orbit of isometry group. In the case of $Y^{p,q}$, the Lie algebra of isometry group is $\mathfrak{su}(2)\times\mathfrak{u}(1)\times \mathfrak{u}(1)$. We have $\mathfrak{u}(1)^3$-isometry, since it is toric, but in this case symmetry is enhanced to $\mathfrak{su}(2)\times\mathfrak{u}(1)\times \mathfrak{u}(1)$.

The explicit form of the metric is given as follows:
\beq
\begin{split}
ds_{Y^{p,q}}^2&=\frac{1-cy}{6}(d\theta^2+\sin^2\theta d\phi^2)+\frac{1}{w(y) q(y)}dy^2\\
&+\frac{q(y)}{9}(d\psi-\cos\theta d\phi)^2+w(y)\left[
d\alpha+f(y)(d\psi-\cos\theta d\phi)
\right]^2, \label{Ypqmetric}
\end{split}
\eeq
where
\beq
w(y)=\frac{2(a-y^2)}{1-cy}, ~~ q(y)=\frac{a-3y^2+2cy^3}{a-y^2},
~~ f(y)=\frac{ac-2y+cy^2}{6(a-y^2)}.
\eeq

The relation between $a,c$ and $p,q$ is slightly complicated, and explained in \S\ref{Ypq.subsec}.

If you take $p=2,q=1$ in Figure \ref{fig:YpqTD}, the toric diagram coincides with the toric diagram of $dP_1$. As discussed above, we have obstruction to $dP_1$ and we do not have K\"ahler-Einstein metric. But this explicit metric tells us that if you take a $S^1$ bundle and change Sasaki structures appropriately, we have Einstein metric and thus Sasaki-Einstein manifold! This point will be clarified further by ``volume minimization'' which will be explained in \S\ref{volume.subsec}.

We noted above that $Y^{p,q}$ has cohomogeneity $1$. Recently, the converse statement is proven by \cite{Conti}, which states that arbitrary cohomogeneity $1$ five-dimensional Sasaki-Einstein metric coincides with one of $Y^{p,q}$. Then the remaining case is cohomogeneity $2$ case. 

Explicit form of five-dimensional Sasaki-Einstein manifold with cohomogeneity $2$ is subsequently given by Cvetic-Lu-Page-Pope\cite{Cvetic1,Cvetic2,Martelli:2005wy}, and the metric is denoted by $L^{a,b,c}$. 
%This is a generalization of $Y^{p,q}$.

Let $a,b,c,d$ be integers satisfying  
\beq 
a\le b, c\le b, d=a+b-c, \text{hcf}(a,b,c,d)=1, \text{hcf}(\{a,b\},\{c,d\})=1,
\eeq
where the last of these equations says that any one of $a,b$ is relatively prime with any one of $c,d$. Then we have infinitely many Sasaki-Einstein metrics on $S^2\times S^3$, labeled by $a,b,c$ \footnote{In some literature, this is written as $L^{p,q,r}$, but this sometimes causes confusion with $p$, $q$ of $Y^{p,q}$. In fact, $a,b,c$ of $L^{a,b,c}$ and $p,q$ of $Y^{p,q}$ are related by $a=p-q$, $b=p+q$, $c=p$, as we will explain. Also, more symmetric expression should be  $L^{\left.a,b\right| c,d}$, but this expression is not much used.}.  These metrics have generically cohomogeneity $2$, and generically irregular. 
These $L^{a,b,c}$ metrics are also toric, and contain $Y^{p,q}$ as a special case of $a=p-q$, $b=p+q$, $c=p$.

This metric is constructed from Kerr Black hole solution by taking some scaling limit. The rough idea is that since Kerr Black hole is a solution to Einstein equation, if we can take certain limit and make the solution, we have Sasaki condition as well\footnote{Actually, before \cite{Cvetic1,Cvetic2}	, \cite{HSY} has used similar methods to construct infinite family of Einstein metrics on $S^2\times S^3$. These metrics are not Sasaki-Einstein, however.}. 

Again we do not explain these metrics in detail, but let us write down the explicit form of the metric to give you some feeling.
First, the metric of five-dimensional Kerr-AdS black hole \cite{HHTR} is given by 
\beq
\begin{split}
ds_5^2=&-\frac{\Delta}{\rho^2}\left[dt-\frac{a\sin^2\theta}{\Xi_a}d\phi-\frac{b\cos^2\theta}{\Xi_b}d\psi 
\right]^2
+\frac{\rho^2 dr^2}{\Delta}+\frac{\rho^2d\theta^2}{\Delta_{\theta}} \\
+&\frac{\Delta_{\theta}\sin^2\theta}{\rho^2}\left[adt-\frac{r^2+a^2}{\Xi_a}d\phi\right]^2
+\frac{\Delta_{\theta}\cos^2\theta}{\rho^2}\left[bdt-\frac{r^2+b^2}{\Xi_b}d\psi
\right]^2\\
+&\frac{1+g^2 r^2}{r^2 \rho^2}\left[abdt-\frac{b(r^2+a^2)\sin^2\theta}{\Xi_a}d\phi-\frac{a(r^2+b^2)\cos^2\theta}{\Xi_b}d\psi
\right]^2,
\end{split}
\eeq
where
\begin{subequations}
\begin{eqnarray}
\Delta&=&\frac{1}{r^2}(r^2+a^2)(r^2+b^2)(1+g^2 r^2)-2m, \\
\Delta_{\theta}&=&1-g^2 a^2\cos^2\theta-g^2 b^2 \sin^2\theta, \\
\rho^2&=&r^2+a^2\cos^2\theta+b^2\sin^2\theta, \\
\Xi_a &=&1-g^2 a^2, ~~\Xi_b=1-g^2 b^2 .
\end{eqnarray}
\end{subequations}

Euclideanize this metric by
\beq
t\to i\tau,~~ g\to \frac{i}{\sqrt{\mathstrut{\lambda}}},~~a\to ia,~~b\to ib,
\eeq
and take the scaling limit
\beq
\begin{split}
a=&\frac{1}{\sqrt{\mathstrut{\lambda}}}\left(1-\frac{1}{2}\alpha\epsilon \right), ~~
b=\frac{1}{\sqrt{\mathstrut{\lambda}}}\left(1-\frac{1}{2}\beta\epsilon \right), \\
r^2=&\frac{1}{\lambda}(1-x\epsilon),~~m=\frac{1}{2\lambda}\mu \epsilon^3,
\end{split}
\eeq
with $\epsilon\to 0$. Then what we have is the metric of $L^{a,b,c}$. 
\beq
\lambda ds_5^2=(d\tau+\sigma)^2+ds_4^2,
\eeq
with
\beq
\begin{split}
ds_4^2&=\frac{\rho^2}{4\Delta_x}dx^2+\frac{\rho}{\Delta_{\theta}}d\theta^2   \\
&+\frac{\Delta_x}{\rho^2}\left(\frac{\sin^2\theta}{\alpha}d\phi+\frac{\cos^22\theta}{\beta}d\psi \right)^2  \\
&+\frac{\Delta_{\theta}\sin^2\theta \cos^2\theta }{\rho^2}
\left(
\frac{\alpha-x}{\alpha}d\phi-\frac{\beta-x}{\beta}d\psi
\right)^2 ,
\end{split}
\eeq

where
\begin{subequations}
\beq
\sigma=\frac{(\alpha-x) \sin^2\theta}{\alpha} d\phi+\frac{(\beta-x) \cos^2\theta}{\beta} d\psi,
\eeq
\beq
\Delta_x=x(\alpha-x)(\beta-x)-\mu,
\eeq
\beq
\rho^2=\Delta_{\theta}-x,
\eeq
\beq
\Delta_{\theta}=\alpha\cos^2\theta+\beta\sin^2\theta.
\eeq
\end{subequations}
Also in this case, the integers $a,b,c$ do not appear in the explicit form of the metric. As in $Y^{p,q}$ case, they are obtained from the condition that the local form of the metric shown above can be extended globally. This metric is also toric.
%\begin{figure}[htbp]
%\begin{center}
%\includegraphics[scale=0.5]{toric_generic.eps}
%\caption{$L^{a,b,c}$のトーリック図. 但し, ここで$k$と$l$は$ck+dl=1$を満たす二つの整数である. }
%\label{LabcTD}
%\end{center}
%\end{figure}

\paragraph{Existence and uniqueness of Sasaki-Einstein metrics}
Apart from the cases discussed above, no explicit metric is known for toric Sasaki-Einstein manifolds. However, we have existence and uniqueness theorems, proved recently, for all toric Sasaki-Einstein metrics. We now briefly comment on this.

The existence statement is as follows. \footnote{This statement is the combined versions of theorems in \cite{FOW} and \cite{CFO}. In their paper they consider the slightly more general case of toric diagram with height $l$. In this case, not the canonical bundle itself, but their $l$-th power, becomes trivial.}. 

\begin{thm}[Futaki-Ono-Wang\cite{FOW,CFO}]
For any toric Sasaki-Einstein manifold, by deforming the Sasaki structure varying the Reeb vector field, we get a Sasaki-Einstein metric.
\qed
\end{thm}

Here ``by deforming the Sasaki structure varying the Reeb vector field'' means the procedure of volume minimization as will be explained in \S\ref{volume.subsec}.
It follows from this theorem that we have Sasaki-Einstein metrics on $S_{dP1}, S_{dP_2}$. for example. Recall that we do not have K\"ahler-Einstein metrics on four-dimensional manifold $dP_1, dP_2$. These two facts are not in contradiction.
Since we do not have K\"ahler-Einstein metrics, we do not have regular Sasaki-Einstein metric on its $S^1$ bundle. By changing Reeb vector field, however, we have irregular Sasaki-Einstein metric.

The uniqueness is also shown in \cite{CFO}.
\begin{thm}[Uniqueness of Sasaki-Einstein metrics, \cite{CFO}]
For arbitrary toric Sasaki-Einstein manifold. the identity component of automorphism group acts transitively on the space of all Sasaki-Einstein metrics whose Reeb vector is the same as that of $g$.
\qed
\end{thm}
 The proof is similar to the case of uniqueness of K\"ahler-Einstein metric with $c_1>0$ \cite{Bando-Mabuchi}.

Before closing this subsection, let us comment on some related topics. First, he metric we have described so far is singular at $r=0$, and some authors have discussed the construction of complex Ricci-flat metrics which extends smoothly up to $r=0$\cite{Oota:2006pm, Futaki:2007ip,Martelli:2007pv}, and their gauge theory interpretation is also discussed. Second, recently some works have been done on the numerical construction of metrics of Calabi-Yau two- and three-folds \cite{Headrick:2005ch,Donaldson_Numerical,Douglas:2006hz,Douglas:2006rr,Doran:2007zn}, although so far they are no applied to AdS/CFT as far as the author is aware of.

\subsection{Gravity side: volume minimization}\label{volume.subsec}
After finishing brief review of Sasaki-Einstein geometry,
 we now discuss the computation of volume by volume minimization, following Martelli-Sparks-Yau (for the toric case \cite{Martelli:2005tp} and generalization given in\cite{Martelli:2006yb}). We first discuss general case, and only later specify the discussion to toric case. Since brane tilings are so far limited to the toric case, we can limit ourselves to toric case from the outset. The reason we choose to present more general story first is that it makes transparent the derivation of volume minimization.

In this section we consider arbitrary odd dimensional Sasaki(-Einstein) manifold $S$ with dimension $2n-1$. In $AdS_5/CFT_4$, we have of course $n=3$, but volume minimization itself can be considered in arbitrary $n$. 

Let us denote by $\mathbb{T}^s$ the maximal torus of the automorphism group Aut($S$, and its Lie algebra $\mathfrak{t}_s$. In the toric case $s=n$, but for a while we also consider the case $s<n$. Also, note that we have $s>1$ in the irregular case since the orbits of Reeb vector do not close. 

Now, in order to obtain volumes of Sasaki-Einstein manifolds, at least naively we need metrics, so first we ask ourselves how to obtain the metric. Of course, the answer is to extremize the Einstein-Hilbert action
\beq
S_{EH}[g]=\int_S \left[ R+2(n-1)(3-2n)\right]d\mu.
\eeq
The result of extremization is
\beq
\textrm{Ric}=(2n-2) g,
\eeq
which says metric is Einstein and its metric cone being Ricci-flat.

Now the interesting thing is simple calculation (and variation with respect to radial directions) shows that this Einstein-Hilbert action is proportional to the volume of Sasaki-Einstein manifold:

\beq
S[g]=4(n-1) \text{Vol}(S)[g].
\eeq

Now the problem reduces to finding the extremal value of $\text{Vol}(S)[g]$. Still, this problem is a variational problem in infinite-dimensional space. However, the  next fact shows that the problem again reduces to a finite-dimensional problem:

\begin{prop}
$\text{Vol}(S)[g]$ depends only on Reeb vector $\xi$, and is independent of deformation of transverse K\"ahler structures.
\qed
\end{prop}

This might look surprising, but similar statement, that the volume depends only on K\"ahler class, is known in K\"ahler-Einstein case as well and the proof is similar.

Now we consider extremization of  
\beq
\text{Vol}: \mathcal{R}(C(S)) \to \mathbb{R}_{+},
\eeq
which is a finite-dimensional variational problem.
Here the domain $\mathcal{R}(C(S))$ is defined by
\beq
\begin{split}
\mathcal{R}(C(S))=\{
\xi \in \mathfrak{t}_s \left|  \right.& C(S) \text{ has some metric, } \\
 & \mathbb{T}^s\text { is a Hamiltonian action on the metric, }\\
& \xi\text{ is a Reeb vector field for the metric } \},
\end{split}
\eeq
and the condition $\xi \in \mathfrak{t}_s$ ensures that the action of $\mathbb{T}^s$ is of Reeb type \cite{Moraes}. 

The first and second variation of function Vol is computed to be
\beqa
d\text{Vol}(Y)&=&-n\int_S\eta(Y)d\mu, \\
d^2\text{Vol}(Y,Z)&=&n(n+1)\int_S\eta(Y)\eta(Z)d\mu, 
\eeqa
where $Y, Z$ are holomorphic Killing vectors of $\mathfrak{t}_s$, $\eta$ a contact one-form defined in \eqref{eta}, and $d\mu$ is a Riemannian measure.

Now the RHS of the first equation actually coincides with the Futaki invariant\cite{Futaki1,Futaki2}, which is the well-known obstruction to the existence of K\"ahler-Einstein metrics. Thus minimizing volume dynamically sets the Futaki invariant to zero, thereby eliminating an obstruction to the existence of K\"ahler-Einstein metrics. Also, the second equation shows the convexity of $\text{Vol}(S)$, from which the uniqueness of critical point automatically follows.

Now let us explain how to compute volume Vol($S$) using localization.
Write the volume in the form
\beq
\text{Vol}(S)=\frac{1}{2^{n-1} (n-1)!}\int_{C(S)}e^{-r^2/2} \frac{\omega^n}{n!}.
\label{vol1}
\eeq

Since $r^2/2$ is known to be a Hamiltonian function of Reeb vector field, let us write $r^2/2$ by $H$. Then (\ref{vol1}) becomes
\beq
\text{Vol}(S)=\frac{1}{2^{n-1} (n-1)!}\int_{C(S)} e^{-H} e^{\omega}.
\eeq
The RHS of this equation can be computed by Duistermaat-Heckman formula, and the result is the localization formula for volume.

There exist certain subtleties, however. Since Reeb vector field does not have any fixed point at $r>0$, all contributions come from $r=0$, which is a singular point in metric cone. To apply Duistermaat-Heckman, we therefore need to take resolutions of singular manifold and apply the formula to non-singular blown-up manifold. It can be shown that the result is independent of the choice of resolution.

Anyway, the result is given by
%\begin{thm}[Martelli-Sparks-Yau]\label{DHresult}
\beq
\frac{\text{Vol}(g)}{\text{Vol}(S^{2n-1})}=
\sum_{\{F \} }\frac{1}{d_F}\int_F \prod_{m=1}^R
\frac{1}{\langle \xi, u_m \rangle^{n_m}}
\left[
\sum_{a\ge 0} \frac{c_a(\mathcal{E}_m)}{\langle \xi, u_m\rangle^a}
\right]^{-1},
\label{DHresult}
\eeq
where 
\begin{itemize}
\item $\{F\}$ is the connected component of fixed point set
\item For each connected component $F$, the $\mathbb{T}^s$ action on the normal bundle $\mathcal{E}$ is determined by weights $u_1,\ldots, u_R\in \mathfrak{t}_s^*$, and correspondingly we have the decomposition$\mathcal{E}=\oplus_{m=1}^R \mathcal{E}_m$.
\item $c_a(\mathcal{E}_m)$ is the $a$th Chern class of $\mathcal{E}_m$.
\item In general, we take partial resolution of $C(S)$ when we apply Duistermaat-Heckman theorem. This means that fiber of $\mathcal{E}$ has in general orbifold singularities, and takes the form $\mathbb{C}^l/\Gamma$. Here I have written $d_F=|\Gamma|$. 
\end{itemize}
%\qed
%\end{thm}

In this way, we know the form of $\textrm{Vol}(S)$, at least in principle, regardless of whether $S$ is toric or not. In practice, however, it is difficult to use \eqref{DHresult}. We therefore specialize to the toric case, and in this case the formula simplifies considerably.

\paragraph{The toric case}\label{subsec:toric}
We next impose the toric condition. In this case, the domain of volume $\mathcal{R}(C(S))$ is shown to be $\mathcal{C}$, by the following fact: 

\begin{prop}[\cite{Martelli:2005tp}]
The space of K\"ahler metrics is  
\beq
\mathcal{C}_{\text{int}}\times \mathcal{H}^1(\mathcal{C}^*),
\eeq
where $\mathcal{C}_{\text{int}}$ denotes the interior or $\mathcal{C}$, which is precisely the domain of Reeb vector. Also, $\mathcal{H}^1(\mathcal{C}^*)$ denotes the set of  smooth degree one homogeneous functions on $\mathcal{C}^*$.
\qed
\end{prop}
To write down $\textrm{Vol}[b]$ more explicitly, let us remind ourselves that $C(S)$ is a $\mathbb{T}^n$-fibration\footnote{Of course, this is not a genuine $\mathbb{T}^n$-fibration since cycles of $\mathbb{T}^n$ shrink at the boundaries of $\mathcal{C}^*$.} over $\mathcal{C}^*=\mu(C(S))$
, and write their symplectic coordinates $(y^1,\ldots y^n,\phi^1,\ldots,\phi^n)$, where $(y_1,\ldots y_n)$ are coordinates of $\mathcal{C}^*=\mu(C(S))$, and $\phi^1,\ldots,\phi^n)$ are coordinates of $\mathbb{T}^n$. In this case, the symplectic potential, which is obtained from K\"ahler potential by Legendre transformation, is written as 
\footnote{This $G(y)$ satisfies Monge-Amp\`ere equation, which is nonlinear. Finding explicit solution is thus very difficult, and for $L^{a,b,c}$, the form of the symplectic potential obtained by integrating metric takes very complicated form\cite{Oota-Yasui-G}. }

\beq
G=G_{\text{can}}+G_{\xi}(y)+h(y) \label{Gstd}.
\eeq
Here $G_{\text{can}}$ id the canonical symplectic potential obtained by Guillemin \cite{Guillemin}:
\beq
G_{\text{can}}(y)=\frac{1}{2}\sum_{a=1}^d \langle y,v_a \rangle \log \langle y,v_a \rangle , 
\eeq
$G_{\xi}$ is determined by Reeb vector $\xi$:
\beq
G_{\xi}(y)=\langle \xi,y \rangle \log \langle \xi,y \rangle-\frac{1}{2}\left(\sum_{a=1}^d\langle y,v_a \rangle\right)\log\left(\sum_{a=1}^d\langle y,v_a \rangle\right),
\eeq
and $h(y)$ is a smooth function at the boundary of $\mathcal{C}$.

From this symplectic potential, the metric is given by
\beq
g=G_{ij}dy^i dy^j+G^{ij}d\phi^i d\phi^j,
\eeq
where $G_{ij}=\frac{\partial^2 G}{\partial y^i \partial y^j}$ and $G^{ij}$ is its inverse matrix. 

Also, computing det($G_{ij}$) from (\ref{Gstd}) gives the following form:
\beq
\text{det}(G_{ij})=f(y)\prod_a \frac{1}{\langle y, v_a\rangle},
\eeq
where $f(y)$ is a smooth function. As shown by Abreu (\cite{Abreu}, Theorem 2.8), this is the condition that the corresponding K\"ahler metric is smooth. This ensures that we always have a metric for arbitrary $\xi$ taken $\mathcal{C}_{\text{int}}$,  We thus have $\mathcal{R}(C(S))=\mathcal{C}_{\text{int}}$ as announced.

If we write the Reeb vector field as
\beq
\xi=\sum_{i=1}^n b_i \frac{\partial}{\partial \phi_i},
\eeq
$\textrm{Vol}[b]$ can be written down explicitly, with the result\footnote{This formula is originally derived in \cite{Martelli:2005tp}, and derives from the fact that the volume can explicitly be written down in toric case. The toric version of (\ref{DHresult}) gives seemingly different expression, but they are shown to be equivalent.}

\beq
\text{Vol}[b]=V[b_1=n,b_2,\ldots,b_n]=\frac{(2\pi)^n}{24}\sum_{a=1}^d \frac{(v_{a-1},v_a,v_{a+1})}{(b,v_{a-1},v_a)(b,v_a,v_{a+1})},
\eeq
where $v_a$ stands for the generator of the fan $\mathcal{C}$, and are numbered counterclockwise manner. Also, Vol$[b]$ can always be minimized for $b_1$, with the result $b_1=n$. In the expression given above, we have already set $b_1=n$.

In Martelli-Sparks-Yau, 
\beq
Z[b]\equiv \frac{1}{4(n-1)(2\pi)^n}S_{EH}=\frac{1}{(2\pi)^n}\text{Vol}[b]
=\frac{1}{24}\sum_{a=1}^d \frac{(v_{a-1},v_a,v_{a+1})}{(b,v_{a-1},v_a)(b,v_a,v_{a+1})}
\eeq
is denoted $Z$, and the minimization of Vol$[b]$ or $Z[b]$ for $b=(n,b_2,\ldots,b_n) \in \mathcal{C}_{\text{int}}$ is called `Z-minimization' or ``volume minimization''. 

Now let us give you some examples.
\begin{exa}[conifold]
For the conifold case,
\beq
v_1=(1,1,1), v_2=(1,0,1), v_3=(1,0,0), v_4=(1,1,0).
\eeq
Therefore, if we write Reeb vector as $b=(x,y,t)$, 
\beq
Z[x,y,t]=\frac{(x-2)x}{8yt(x-t)(x-y)}.
\eeq
Minimizing this gives, 
\beq
b_{\text{min}}=(3,\frac{3}{2},\frac{3}{2}),
\eeq
and 
\beq
\textrm{Vol}(T^{1,1})=\frac{16 \pi^3}{27}.
\eeq

This coincides with the explicit formula for volume obtained from explicit metric in \eqref{T11vol}. It also coincides with the prediction from AdS/CFT \eqref{T11volfroma}, which is a non-trivial check of AdS/CFT.
\qed
\label{T11.exa}
\end{exa}

\begin{exa}[del Pezzo 2]
We next consider the volume of the Sasaki-Einstein manifold $S_{dP_2}$. 

The vectors specifying the toric diagram is, for example, 
\beq
v_1=(1,0,0), v_2=(1,0,1), v_3=(1,1,2), v_4=(1,2,1), v_5=(1,1,0).
\eeq

The $Z$-function is, again for $b=(x,y,t)$,
\beq
Z[x,y,t]=\frac{(x-2)(-t^2+2t(x+y)+(3x-y)(x+y))}{8yt(t-x-y)(t+x-y)(t-3x+y)}.
\eeq
Minimizing this function, we easily find
\beq
b_{\text{min}}=\left(3,\frac{9}{16}(-1+\sqrt{\mathstrut{33}}),\frac{9}{16}(-1+\sqrt{\mathstrut{33}}) \right),
\eeq
and corresponding minimum of $Z$ gives
\beq
\textrm{Vol}(S_{dP2})=\frac{(59+11\sqrt{\mathstrut{33}})\pi^3}{486}.
\eeq

This is consistent with the AdS/CFT prediction, as given in \ref{dP2vol}.
\qed
\label{dP2.exa}
\end{exa}

\paragraph{Relation with equivariant index}

Here we explain the relation of the volume $Vol(S)$ of Sasaki-Einstein manifold  $S$ with the equivariant index of Cauchy-Riemann operator on the metric cone $C(S)$. This argument applies to non-toric case as well. 

Consider Cauchy-Riemann operator $\overline{\partial}$ on $C(S)$, and consider elliptic complex
\beq
0\to \Omega^{0,0}(C(S))\to \Omega^{0,1}(C(S)) \to \ldots 
 \to \Omega^{0,n}(C(S))\to 0.
\eeq
Let us denote the cohomology of this complex by $\mathcal{H}^p \equiv H^{0,p}(C(S),\mathbb{C})$. In this case, $\mathcal{H}^0$ is an infinite-dimensional space, contrary to the compact case. 

Since the action of  $\mathbb{T}^r$ commutes with $\overline{\partial}$, the action of  $\mathbb{T}^r$ on $\mathcal{H}^0$ is induced.
The equivariant index for $q\in \mathbb{T}^r$ is defined by %\footnote{Here $L$ stands for Lefshetz number.}
\beq
L(q,C(S))=\sum_{p=1}^n (-1)^p \text{Tr}\{ q| \mathcal{H}^p(C(S))\}. \label{Lefschetz}
\eeq

From equivariant index theorem\cite{Atiyah-Singer},
\beq
L(q,C(S))=\sum_{\{F \}}\int_F 
\frac{\text{Todd}(F)}
{\prod_{\lambda=1}^R \prod_j (1-q^{u_{\lambda}}e^{-x_j})}. \label{Lresult}
\eeq
Here the symbols $F, \mathcal{E}_{\lambda}, u_{\lambda}$ are the same as in (\ref{DHresult}), and $x_j$ stands for the first Chern-class of the bundle $L_i$ when we decompose $\mathcal{E}_{\lambda}\to F$ into the direct sum of line bundles $\mathcal{L}_j$.
\beq 
\mathcal{E}_{\lambda}=\oplus_{j=1}^{n_{\lambda}}\mathcal{L}_{j},
\eeq

\beq
x_j=c_1(\mathcal{L}_j).
\eeq

If you compare (\ref{DHresult}) and (\ref{Lresult}), you will notice some similarities between them. In fact, it turns out that  $C(q,C(S))$ and volume $\text{Vol}(S)$ are related by the following relation:
\beq
\mbox{Vol}(S)=\mbox{Vol}[b]=\text{lim}_{t\to 0} t^n L(\exp(-t b),C(S)).
\eeq
That is, although $L(q,C(S))$ diverges in the limit $q\to 1$ (this simply means we have infinitely many holomorphic functions on $C(S)$), the asymptotic coefficient of the divergent part gives you the volume $\text{Vol}(S)$. In this sense, $L(q,C(S))$ contains more information than $\text{Vol}(S)$. $L(q,C(S))$ simply counts holomorphic functions on cone, and in toric case, we have one-to-one correspondence with lattice points of $\B$\cite{Fulton}.

Physically speaking, $L(q,C(S))$ corresponds to counting of mesonic operators in gauge theory; see \S\ref{counting.subsec} for more discussion.

\subsection{$A$-maximization versus volume minimization} \label{versus.subsec}

We now explain the proof of the formula \eqref{vol=a} for toric Sasaki-Einstein manifolds. We now know the relation between toric Sasaki-Einstein manifolds and quiver gauge theories by the method of brane tilings.  Since central charge in quiver gauge theory side is computed by $a$-maximization \S\ref{a-max.subsec}, and since volume in gravity side is computed by volume minimization \S\ref{volume.subsec}, in principle we have no difficulty. We have already discuss such a check  in Examples \ref{T11.exa} and \ref{dP2.exa}, taking the conifold and del Pezzo 2 as examples. We now move onto the more complicated example of $Y^{p,q}$.

\paragraph{$Y^{p,q}$ quivers}
The quiver gauge theory corresponding to $Y^{p,q}$ can be obtained by the methods explained in \S\ref{D3.subsubsec}. Historically, it was first obtained by more indirect methods in \cite{Benvenuti:2004dy}. 

\begin{figure}
\centering{\includegraphics[scale=0.5]{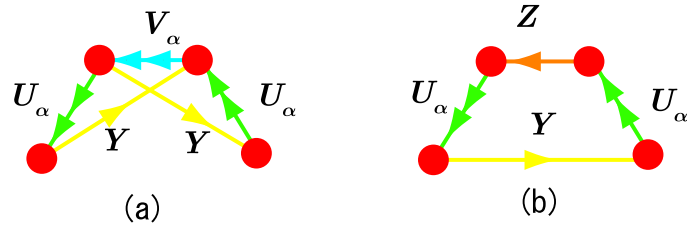}}
\caption{The quiver corresponding to $Y^{p,q}$ can be constructed by using $p-q$ pieces of (a) and $q$ pieces of (b).}
\label{Ypqpieces.eps}
\end{figure}

\begin{figure}
\centering{\includegraphics[scale=0.35]{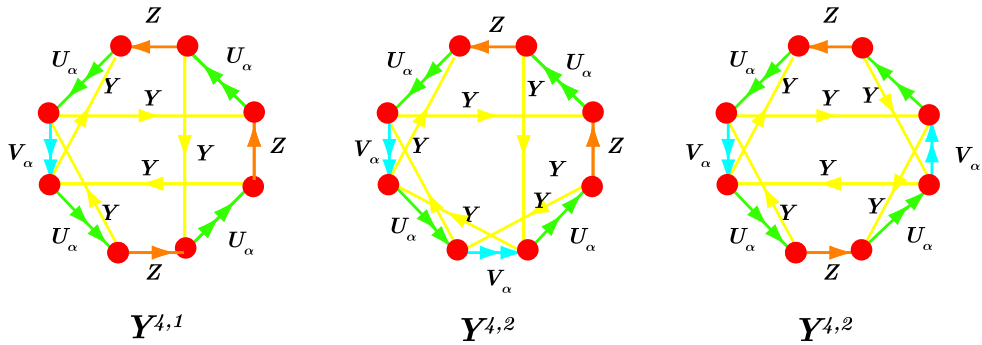}}
\caption{The quiver for $Y^{4,1}$ and $Y^{4,2}$. We have shown two quivers for $Y^{4,2}$, which are related by Seiberg duality. One of the quivers have previously appeared in Figure \ref{theproblem} in \pageref{theproblem}.}
\label{Ypqquiver.eps}
\end{figure}

The quiver corresponding to $Y^{p,q}$ can be constructed using $p-q$ pieces of Figure \ref{Ypqpieces.eps} (a), and $q$ pieces of Figure \ref{Ypqpieces.eps} (b). We have shown the example of $Y^{4,1}$ and $Y^{4,2}$ in Figure \ref{Ypqquiver.eps}. In Figure \ref{Ypqquiver.eps}, two quivers corresponding to $Y^{4,2}$ are shown. In general, we have several different quivers corresponding to the order of $q$ and $p-q$ pieces, but they are connected by Seiberg dualities.

We have four types of fields, $U_{\alpha=1,2},V_{\alpha=1,2},Y$ and $Z$. For $Y^{p,q}$, we prepare $2p$ gauge groups connected by $4p+2q$ bifundamental fields and we have $p$ $SU(2)$ doublets $U_{\alpha=1,2}$, $q$ $SU(2)$ doublets $V_{\alpha=1,2}$, $p+q$ singlets $Y$, and $p-q$ singlets $Z$.

The superpotential is build from various cubic and quartic terms that are represented by $\Tr UVY$ and $\Tr UZUY$:
\beq
W=\sum_k \epsilon^{\alpha\beta} \left(U^k_{\alpha} V^k_{\beta} Y^{2k+2} +V^k_{\alpha} U^{k+1}_{\beta} Y^{2k+3}\right)+\epsilon_{\alpha\beta} \sum_k Z^k U^{k+1}_{\alpha} Y^{2k+3} U^k_{\beta},
\eeq
where the first term corresponds to $2q$ triangles of Figure \ref{Ypqpieces.eps} (a), and the second corresponds to $p-q$ rectangles of Figure \ref{Ypqpieces.eps} (b). The meaning of index $k$ should be clear from examples above. The corresponding bipartite graph can be found in \cite{Benvenuti:2004dy}. The case of $L^{p,q,r}$ is more complicated but essentially similar, so we do not discuss it here.

Now you can carry out $a$-maximization for this quiver and check the prediction with volume minimization. That is possible, but at the same time things are already complicated for this $Y^{p,q}$ example. In this method it seems to difficult to extend the result to arbitrary toric Calabi-Yau manifold.

\paragraph{Parametrization of global symmetries}
In order to systematically perform $a$-maximization to quiver gauge theories corresponding to arbitrary toric Calabi-Yau, we need to solve conditions \eqref{Rcond2}. If you recall the discussion in \S\ref{R-charge.subsubsec}, you will recognize that the best way is to solve these equations is to use an interpretation as angles. To state it more formally, assign number $a_\alpha$ to each vertex $\alpha$ on the boundary of the toric diagram. If $\alpha$ is sandwiched between two zig-zag paths $\mu$ and $\mu+1$, this $a_\alpha$ represents the angle between two paths $\mu$ and $\mu+1$. For a bifundamental field lying at the intersection $I$ of two zig-zag paths $\mu$ and $\nu$, its charge is given by the angle formed by $\mu$ and $\nu$:
\beq
R_\alpha=\sum_{\mu < \alpha <\nu} a_\alpha, \label{RchargeBZ}
\eeq
where the summation is over all vertices in the minor angle formed by two zig-zag paths $\mu$ and $\nu$.

%The condition for $U(1)_R$-charge %is given by
Moreover, since they are angles (divided by $\pi$), they should sum up to 2:
\beq
\sum_\alpha a_\alpha=2. \label{suma2}
\eeq
These numbers parametrize $d-1$ possible candidates for R-symmetry. The number $d-1$ is exactly the same with that expected from gravity side calculation. These numbers can also be used for flavor symmetries \cite{Butti:2005vn}. In that case, we replace \eqref{suma2} by
\beq
\sum_\alpha a_\alpha=0, \label{suma2-2}
\eeq
and out of these, $d-3$ baryonic symmetries are constrained further by 
\beq
\sum_\alpha w_\alpha a_\alpha=0,\label{sumva}
\eeq
where $w_\alpha (\alpha=1,2,\ldots, d)$ are two-dimensional vectors spanning the toric diagram. In our previous notation, $v_{\alpha}=(w_\alpha,1)=(p_{\alpha}, q_\alpha,1)$.

You can rewrite these formulae by another set of variables $b_{\mu}$ assigned to each primitive normal $\mu$ to the toric diagram. For a vertex $\alpha$ between two normals $\mu$ and $\mu+1$, $b_{\mu},b_{\mu+1}$ and $a_\alpha$ are related by
\beq
b_{\mu+1}-b_{\mu}=a_\alpha,
\eeq
and (\ref{RchargeBZ}) are now translated into 
\beq
R_I=b_{\nu}-b_{\mu}\label{RchargeBZ-2},
\eeq
whereas in the R-charge case, \eqref{suma2} means that $b_{\mu}$ is multi-valued as we go around the perimeter of the toric diagram. For non-R flavor symmetries, $b_{\mu}$ is single valued. Out of $d-1$ flavor symmetries, we can further pick out $d-3$ baryonic symmetries by requiring $b_{\mu}$ to satisfy
\beq
\sum_{\mu} b_{\mu} \balpha_{\mu}=0. \label{balpha2}
\eeq
This is exactly the same with \eqref{balpha} and therefore $b_{\mu}$ is precisely the ``baryonic number assignments'' as discussed previously in \S\ref{marginal.subsec} in the discussion of fractional branes.

Let us explain why the condition \eqref{balpha} picks up baryonic symmetries. Recall that mesonic operators are gauge invariant operators of the form $\mbox{Tr}(X_1X_2\ldots )$, and therefore represented as closed path $\scC$ in the bipartite graph \footnote{In this sense, mesonic operators are ``closed strings''. By contrast, baryonic operators are identified as D-strings connecting two $(N,0)$-regions \cite{Imamura:2006ie}, and therefore ``open strings''.}. Since mesonic operators are not charged under baryonic symmetries. To represent this condition, introduce a one-form $\scB$ by
\beq
\scB \equiv \sum_I B_I \delta(I),
\eeq
where $\delta(I)$ is a one-form delta-function supported on an edge $I$, and $Q_I$ is the baryonic charge of the edge $I$. Then we should have 
\beq
\int_{\scC} \scB=0
\eeq
for all closed paths $\scC$. This means 1-form $\scB$ is exact, and can be written as
\beq
\scB=d \scS,
\eeq
for
\beq
\scS=\sum_a S_a \delta(a),
\eeq
with $a$ labels the face of the $(N,0)$-brane, as in previous sections.
In other words, there exist set of integers $S_a$ for each $(N,0)$ face such that for each edge $I$ between two faces $a$ and $b$, we have
\beq
B_I=\mbox{sign}(a,b)(S_a-S_b).
\eeq
You can easily verify (we simply have to take $S_a$ to be $S_A$ of the corresponding face in \eqref{ssb}; see Figure \ref{ssb.eps}) that this condition is satisfied if we impose \eqref{ssb} or equivalently if $b_{\mu}$ satisfies \eqref{balpha2}.

\begin{exa}
As an example, let us take del Pezzo 2 discussed in Example \ref{dP2amax}.
In this case, in order to parametrize possible R-symmetry we prepare five variables $a_1,\ldots a_5$ corresponding to vertices in Figure \ref{dP2vertices}, with the constraint $\sum_{\alpha=1}^5 a_\alpha=2$. The charge assignment is given in Table \ref{dP2chargeassign}. If you compare this with previous parameterization in \eqref{dP2parameters}, we the relation between $a_1,\ldots a_5$ and $x,y,z,w$ is as given in Table \ref{dP2chargeassign}. Therefore, the two parameterizations are equivalent.

\begin{figure}[htbp]
\centering{\includegraphics[scale=0.6]{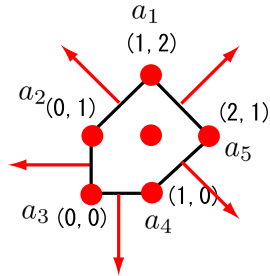}}
\caption{We assign $a_1,\ldots a_5$ to each lattice point on the boundary of the toric diagram.}
\label{dP2vertices}
\end{figure}

\begin{table}[htbp]
\begin{center}
\caption{The charge assignment \eqref{RchargeBZ} for del Pezzo 2. Comparison with previous parameterization in Example \ref{dP2amax} is also indicated.}
\begin{tabular}{|c|c|c|c|}
\hline
charge  & multiplicity & bifundamental field & previous parameterization \\
\hline
\hline
$a_1$ & 2 & $X_{13},X_{45}$ & $x+y-w$ \\
\hline
$a_2$ & 1 & $X_{34}$ & $w-x$ \\
\hline
$a_3$ & 1 & $X_{23}$ & $x$ \\
\hline
$a_4$ & 1 & $X_{12}$ & $z$ \\
\hline
$a_5$ & 2 & $X_{51}, X_{24}$ & $2-x-y-z$ \\
\hline
$a_1+a_2$ & 1& $X_{52}$ & $y$ \\
\hline
$a_2+a_3$ & 1& $Y_{51}$ & $w$ \\
\hline
$a_3+a_4$ & 1& $Y_{45}$ & $x+z$ \\
\hline
$a_4+a_5$ & 1& $X_{35}$ & $2-x-y$ \\
\hline
\end{tabular}
\label{dP2chargeassign}
\end{center}
\end{table}

\qed
\end{exa}

You can try more examples by using $Y^{p,q}$ quivers shown in Figure \ref{Ypqpieces.eps} and Figure \ref{Ypqquiver.eps}.

\paragraph{Equivalence of a-maximization and volume minimization}
Using this parameterization, the formula for trial $a$-function is given by
%We have AdS/CFT relation:
\beq
a=\frac{9}{32}\Tr R^3
=\frac{9}{32}\left(
F+\sum_{\mu < \nu} |\langle w_{\mu}, w_{\nu}\rangle | (\sum_{\mu<A<\nu} a_A-1)^3
\right), \label{BZa}
\eeq
where $v_{\mu} (\mu=1,2,\ldots,d)$ denotes primitive normals of the toric diagram labeled in counterclockwise manner. Also, we have $\langle w_{\mu}, w_{\nu} \rangle $ in the sum because the number of intersection points of two path with winding $w_{\mu}=(p_{\mu},q_{\mu})$ and $w_{\nu}=(p_{\nu},q_{\nu})$ is given by 
\beq
\left| \langle w_{\mu}, w_{\mu} \rangle \right| =\textrm{det}\left(\begin{array}{cc} p_{\mu} & p_{\nu} \\ q_{\mu} & q_{\nu} \end{array}\right).
\eeq
Also, in the summation $\mu<\nu$ means that the angle from $\mu$ to $\nu$ measured in a counterclockwise manner is smaller than $\pi$, and is present in order to prevent double counting of bifundamental fields.

%This is a function of $d-1$ parameters.

Now the problem is to compare the function \eqref{BZa} to Volume functional $\mbox{Vol}[b]$. The function \eqref{BZa} has $d-1$ parameters, whereas $\mbox{Vol}[b]$ is a function of only two parameters (recall $b_1=3$). This means that we have to eliminate  $d-3$ more parameters in gauge theory side, which correspond to baryonic symmetries. As shown in Appendix of \cite{Butti:2005vn}, we can prove $\Tr U(1)_B^3$=0, and \eqref{BZa} becomes a quadratic function for $d-3$ baryonic symmetries, . This means that the derivative becomes a linear function, and we can delete all such parameters, to obtain a function of remaining two parameters. Butti and Zaffaroni has shown \footnote{
The original proof by Butti and Zaffaroni is quite complicated, 
the expression \eqref{BZa} is simplified \cite{Benvenuti:2006xg,Lee:2006ru} 
and the problem reduces to the maximization of 
\beq
a[a_1,a_2,\ldots a_d]=\frac{9}{32} \sum_{\genfrac{}{}{0pt}{}{\alpha,\beta,\gamma=1}{\alpha<\beta<\gamma}}^d d_\alpha d_\beta d_\gamma \left|\mathrm{det}(v_\alpha,v_\beta,v_\gamma)\right| \label{aphiphiphi}
\eeq
under the constraint
\beq
\sum_\alpha d_\alpha=2,
\eeq
where $v_\alpha=(1,w_\alpha)$ is the 3-vectors spanning the fan.
It was shown later in \cite{Kato:2006vx} that this function has unique critical point, under the constraint $d_\alpha \ge 0$.
} that this function of two variables coincide with volume functional.

In this way the problem of proving \eqref{vol=a} is done for all toric Sasaki-Einstein manifolds.
This agreement is quite impressive, but we would like to stress here that this problem is not yet solved in complete generality. For example, as we have discussed above, the parameterization we used applies only to `minimal' quivers. For example, the proof in \cite{Butti:2005vn} assumes\footnote{Even if the number of bifundamental fields is larger than this, we sometimes have a cancellation and the formula for $a$-functions reduces to the `minimial' case. However, this seems not to occur in general.} that the number of bifundamental fields is given by
\beq
N_f=\frac{1}{2}\sum_{i,j}|\langle v_i, v_j\rangle |.
\eeq
This formula applies to all quivers which appear in the strong coupling limit,
but not for quiver gauge theories obtained after Seiberg duality (see \S\ref{Seiberg.subsec}). In general, we have several different quivers corresponding to the same toric Calabi-Yau, as shown in Figure \ref{dP3phases}. In this case, the central charges of this quiver gauge theories should be the same, and we point out it is still open to prove this fact, at least to the best of the author's knowledge. In fact, if you look at specific examples, say del Pezzo 3 (see examples described in \cite{Butti:2005vn}), then the $a$-function for one phase seems to be different for other three phases, and the result seems to agree only after maximization.

Finally, to check \eqref{vol=a} for non-toric case should be quite interesting but a difficult at the same time since we do not have brane-tiling techniques for non-toric case. See \cite{Butti:2006nk} for discussion in the non-toric del Pezzo case.

\subsection{Summary}

In this section, we discussed the duality between type IIB string theory on $AdS_5\times S$ ($S$ is a five-dimensional Sasaki-Einstein manifold) and four-dimensional $\scN=1$ superconformal quiver gauge theories. The check of this correspondence was done through the holographic relation between the volume of Sasaki-Einstein manifolds and the central charge of superconformal quiver gauge theories:
\beq
{\rm Vol}(S)=\frac{\pi^3}{4}\frac{1}{a}. \label{vol=a-2}
\eeq

 The gauge theory dual of a Sasaki-Einstein manifold can be found by brane tiling techniques as discussed in \S\ref{D5NS5.sec}. On the gauge theory side, the central charge of quiver gauge theories was computed by $a$-maximization (\S\ref{a-max.subsec}). On gravity side, the volume of Sasaki-Einstein manifold can be directly computed for $Y^{p,q}$ and $L^{a,b,c}$ whose explicit form of the metric is known. More generally, volume of arbitrary toric Sasaki-Einstein manifold was computed by volume minimization (\S\ref{volume.subsec}). Finally, in \S\ref{versus.subsec} we discussed the verification of the formula \eqref{vol=a-2}.
We also discussed related topics, such as some Sasaki-Einstein, and the relation of volume to equivariant index.

%%%%%%%%%%%%%%%%%%%%%%%%%%%%%%%%%%%%%%%%%%%%%%%%%%%%%%
\section{Application to homological mirror symmetry}\label{HMS.sec}

In this section, we discuss homological mirror symmetry (HMS), which is an another topic where brane tilings play an important role.
%application of brane tilings, .

As we have seen, brane tilings provide powerful techniques to study $\mathcal{N}=1$ supersymmetric quiver gauge theories. But that is not the only use of brane tilings. As we have seen, they are directly connected with the geometry of Calabi-Yau manifold, which means that brane tilings are useful not only for quiver gauge theories, but also for studying toric Calabi-Yau geometry. Since mirror symmetry is the statement about Calabi-Yau manifolds, it is natural to expect that brane tilings are of use in mirror symmetry as well.

\begin{figure}[htbp]
\centering{\includegraphics[scale=0.37]{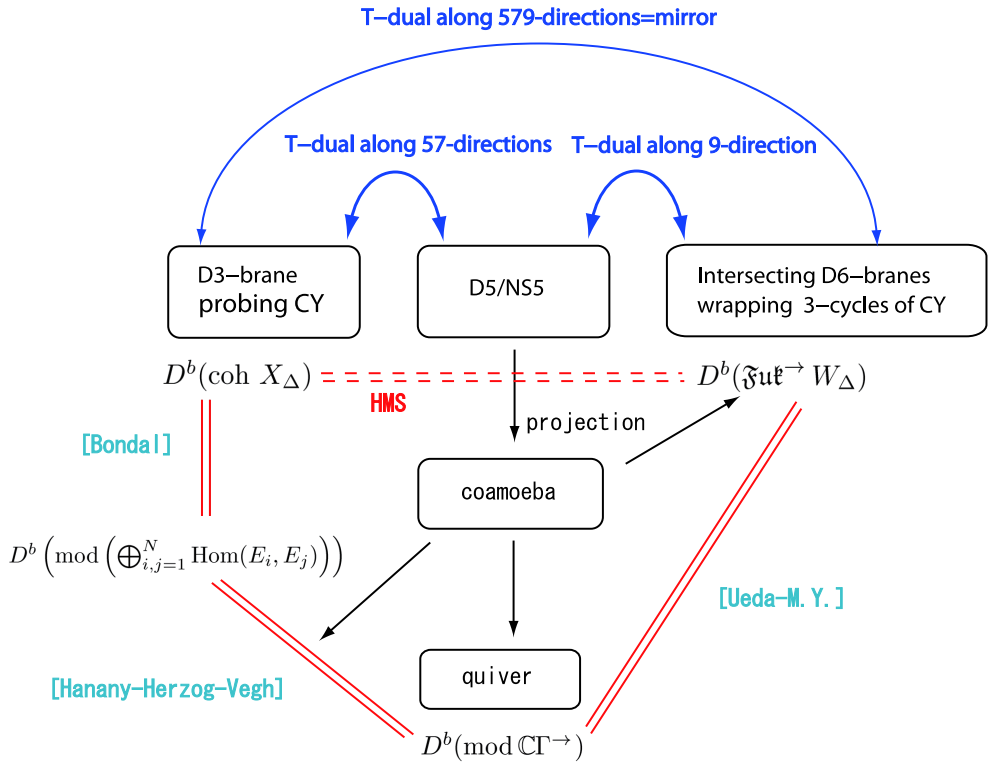}}
\caption{These chain of T-dualities show that we can use brane tilings to study mirror symmetry, The blue arrows in the upper half of this figure represent various T-dualities. Brane tiling represents D5/NS5 configuration, which is T-dual to D3-branes with Calabi-Yau $\scM$, and to D6-branes with mirror Calabi-Yau $\scW$. The lower half of this figure represents the strategy to prove HMS, as will be explained later in this section.}
\label{Tdualex}
\end{figure}

Let us tell you the story in greater detail. We have seen so far that quiver gauge theories have three different realizations in string theory: D3-branes probing Calabi-Yau $\cal{M}$ (\S\ref{D3.subsec}), D5-branes with NS5-brane (\S\ref{D5NS5.sec}), and intersecting D6-branes wrapping three-cycles of Calabi-Yau $\scW$ (\S\ref{D6.subsec}). The  first two are related by T-duality along two-cycle of Calabi-Yau, as explained previously in \S\ref{D3.subsubsec}. 
%Of course, you can take other T-dualities. For example, if you take T-duality along 1-cycle of Calabi-Yau, then you have D4-branes with NS5-brane. This type of configuration is known long before the advent of brane tilings and known as ``elliptic models''. 
Also, taking further T-duality takes the second into the third (see Figure \ref{Tdualex}).
%in D5/NS5 systems to make them into D6-branes and Calabi-Yau $\scW$.

In this way, brane tilings know not only about quiver gauge theories, but also about the geometry of toric Calabi-Yau manifold $\scM$ and its mirror $\scW$. It is thus quite natural to use brane tilings and its physical interpretation to study mirror symmetry for toric Calabi-Yau manifolds. This is exactly what we are going to do in this section. We are going to  apply the technology of brane tiling to prove homological mirror symmetry, based on our papers \cite{UY1,UY2,UY3,UY4}. The surprising fact is that complicated mathematical problem can be understood quite intuitively from brane perspective. 

\subsection{Homological mirror symmetry} \label{HMS.subsec}

Mirror symmetry has been studied for almost twenty years. %, since the seminal work by Candelas et al\cite{Candelas:1990rm}. 
Although many more interesting dualities are later found, it is still an important example of duality because (at least part of) mirror symmetry can be  rigorously formulated in mathematics. 

There are several mathematical formulations of mirror symmetry. For example, the first formulation is the so-called topological mirror symmetry, which states that for any Calabi-Yau $\scM$, there exists another Calabi-Yau $\scM$ with Betti numbers $b^{1,1}$ and $b^{2,1}$ interchanged.
%\beq
%b^{1,1}(\scM)=b^{2,1}(\scW), ~~
%\eeq
Another formulation is classical mirror symmetry\footnote{Perhaps this name is a misnomer, and the reader should be careful with the name `classical'. It should not be taken `classical' as opposed to `quantum'.}, which states surprising relation between the K\"ahler moduli of Calabi-Yau $\scM$ and the complex structure moduli of Calabi-Yau $\scW$, and vice versa. 

Homological mirror symmetry (see \cite{Ballard} for recent review) is yet another mathematical formulation of mirror symmetry, which is  proposed by Kontsevich in 1994\cite{Kontsevich_HAMS}, which is inspired by the preprint by Kenji Fukaya\cite{Fukaya93}. It is believed to be one of the most powerful formulations of mirror symmetry. For example, it is believed (although not proved completely) that classical mirror symmetry is obtained as the corollary to this homological mirror symmetry. HMS is also believed to be equivalent to the geometric mirror symmetry of Strominger-Yau-Zaslow\cite{SYZ}, which roughly states that mirror symmetry is the T-duality along the 3-cycle of Calabi-Yau. See 

The statement of mirror symmetry is, for Calabi-Yau three-fold $\scM$ and its mirror $\scW$, given as follows:
\begin{conj}[Homological Mirror Symmetry]
\beq
D^b (\coh \mathcal{M}) \cong D^b (\Fuk \mathcal{W}).
\label{HMS}
\eeq
\end{conj}
Here $D^b (\coh \mathcal{M})$ denotes the ``derived category of coherent sheaves'' on $\mathcal{M}$, and $D^b (\Fuk \mathcal{W})$ denotes the ``derived Fukaya category'' of $\scW$. The (rough) definitions of these mathematical terminologies will be explained in later sections.

In physics terms, HMS states that the `category' of A-branes is the same as the `category' of B-branes. We are going to explain below what category is, but for the moment think of it as some generalized notion of set. Therefore, roughly speaking \eqref{HMS} states
\beq
\textrm{(category of B-branes)} \cong \textrm{(category of A-branes)}.
\label{HMS2}
\eeq

Historically, when homological mirror symmetry was first proposed, physicists had no ideas how these categories could come into string theory. It was only several years later that several authors proposed \cite{Douglas_DCN,Sharpe:1999qz,Diaconescu:2001ze,Aspinwall:2001pu,Lazaroiu:2001rv} that the category of B-branes is in fact the derived categories of coherent sheaves $D^b (\coh \cal{M})$, and the category of A-branes is the derived Fukaya category $D^b (\Fuk \cal{W})$.

The reader might wonder at this point what the advantage of using all these complicated mathematical terminologies is. First, as we have already mentioned, HMS is the most powerful mathematical formulation of mirror symmetry known so far.
The second point is that this formulation of mirror symmetry will help to understand the equivalence of underlying physical setup. In usual formulation of mirror symmetry, we compute some numbers on both sides (e.g compute period in B-model side and Gromov-Witten invariant in A-model side), and compare them. Of course, this is a non-trivial check of mirror symmetry, but the real reason for such a match is not clearly seen. HMS basically says that we get the same answer because we are considering equivalent theories.
This is the reason we are interested in HMS.

%This formulation is more powerful than comparing numbers. The general argument is that we obtain the same number simply because we are in a sense considering the same setting.

More broadly speaking, HMS is a part of `categorification' program. Categorification refers to the process of  replacing set-theoretic statements by categorical statements (see \cite{Baez_Cat} for reviews). 
They have come to play an important role in several areas in string theory, for example in topological string theory\cite{Gukov:2004hz,Gukov:2005qp,Gukov:2007tf} (in relation with Khovanov homology\cite{Khovanov_Jones,Khovanov_sl3,KhovanovRozansky1,KhovanovRozansky2}) and in geometric Langlands program\cite{Kapustin:2006pk,Gukov:2006jk,Witten:2007td,Gukov:2007ck,Gukov:2007zz}. 

Actually, what we are going to prove is the so-called Fano version of HMS \footnote{Physically speaking original HMS \eqref{HMS} basically follows from the corresponding Fano version \eqref{HMSFano}, although we have no rigorous mathematical proof as far as the author knows.}. It states 
\beq
D^b (\coh X_\Delta) \cong D^b (\dirFuk W_{\Delta}),
\label{HMSFano}
\eeq
where $X_\Delta$ is the complex two-dimensional toric Fano variety (or their orbifolds (or stacks)) obtained from toric diagram $\Delta$. In our setup, Calabi-Yau $\scM$ is the so-called local Calabi-Yau, and is actually a canonical bundle over a complex two-dimensional toric variety, and that toric variety is denoted by $X_{\Delta}$. When $X_{\Delta}$ is a smooth manifold, it is one of the K\"ahler-Einstein manifolds $\bP^2, \bP_1\times \bP_1$ or their blow-ups which are discussed in \S\ref{explicit.subsubsec} in the discussion of regular Sasaki-Einstein manifolds.
Also, $\dirFuk W_{\Delta}$ refers to the directed Fukaya category obtained from Newton polynomial $W_{\Delta}$ of $\Delta$. Of course, Newton polynomials have ambiguities of coefficients (as shown in \eqref{newton}), but $\dirFuk W_{\Delta}$ is independent of such coefficients as long as they are generic \footnote{The reason for this is as follows. $\dirFuk W_{\Delta}$ is defined by symplectic geometry, and thus independent of complex structure deformations. Since changing coefficients of $W_{\Delta}$ corresponds to complex structure deformations, we see that $\dirFuk W_{\Delta}$ is independent of coefficients}.
%In our setup, toric Fano varieties are specified is a toric variety whose fan is specified directly from the toric diagram, and toric Calabi-Yau is a canonical bundle over the corresponding toric Fano variety.
 Note in previous sections we have used $P(x,y)$ to denote Newton polynomials, but we use the expression $W_{\Delta}$ (or sometimes $W$ for simplicity) so that it is in accord with mirror symmetry literature.

The homological mirror symmetry \eqref{HMS} is demonstrated for two-tori in \cite{Polishchuk:1998db} and quartic K3 surface in \cite{Seidel:2003kb}. The Fano version \eqref{HMS2} was proved by Seidel \cite{Seidel_VC2}
for $\bP^2$ and $\bP^1 \times \bP^1$,
Auroux, Katzarkov and Orlov
\cite{Auroux-Katzarkov-Orlov_WPP} for Hirzebruch surfaces
and Ueda \cite{Ueda_HMSTdPS} for $\bP^2$ blown-up
at two or three points. In this section we prove \eqref{HMS2} for all toric Fano varieties, using brane tilings and coamoebae. In particular, all known results of Fano version of HMS \eqref{HMS2} are reproduced by our method. We also extend our relation to the case of orbifolds  by Abelian subgroups of torus (\S\ref{equivariant.subsubsec}), which is a new result in mathematics.

\subsection{The category of A-branes and B-branes} \label{category.subsec}

In this subsection, we briefly review the category of A- and B-branes, so that we have a better understanding of \eqref{HMS} and \eqref{HMS2}. The discussion here is brief and the interested readers are referred to reviews
 \cite{Aspinwall:2004jr,Sharpe:2003dr,Lazaroiu:2003md}. %For HMS, understanding of A-brane category is more important.

\subsubsection{The definition of category} \label{category.subsubsec}
Category is an abstract concept in mathematics. Although most physicists might be unfamiliar with such a theoretical framework, its definition is quite simple. A category $\mathcal{C}$ consists of 

\begin{itemize}
\item
The set $\Ob(\mathcal{C})$ of objects.
\item
The set $\Homfrak_{\scC}(\mathcal{C})$ of morphisms. Each morphism $c$ has a source object $a_1$ and target object $a_2$, and in this case we write $c: a_1\to a_2$. The set of objects from $a_1$ to $a_2$ is written $\hom_{\scC}(a_1,a_2)$ or simply $\hom (a_1,a_2)$.
\item
Composition of morphisms for two morphisms $c_1: a_1\to a_2$ and $c_2: a_2\to a_3$. This is denoted $c_1 \cdot c_2$ or simply $c_1 c_2$, and is an another morphism from $a_1$ to $a_3$ (i.e. $c_1 \cdot c_2: a_1\to a_3$).
\end{itemize}

In addition, they have to satisfy some axioms, or consistency conditions (such as existence of identify and associativity), which we do not mention here. Physically speaking, objects correspond to D-branes and morphisms correspond to open strings connecting two D-branes. We will discuss more about this in the discussion of A-brane category in \S\ref{Fukaya.subsec}, but before that we move onto the discussion of the category of B-branes.
%and composition of morphisms correspond to junction of two open strings, and we will see this fact later in the computation of A-brane category.

%We now move onto the discussion of more detailed discussion of homological mirror symmetry.

\subsubsection{The category of B-branes} \label{Bbrane.subsubsec}
We first describe the category of B-branes, although much of what I have to say here is not needed in understanding the proof of homological mirror symmetry itself. Also, the readers should be careful because the explanation of this section is quite sketchy and not so accurate.

As said before, it is believed now that D-branes in B-models on Calabi-Yau three-fold $\scM$ are represented by ``derived category of coherent sheaves on $\scM$``, $D^b (\coh \mathcal{M})$.\footnote{Sometimes this is simply denoted by $D^b(\scM)$.}

Coherent sheaves are, roughly speaking, generalization of vector bundles. It is natural that vector bundles appear since D-branes have gauge fields on them. We use coherent sheaves because they are more general and they have better mathematical properties than vector bundles.

To obtain $D^b (\coh \mathcal{M})$, we next consider the complex\footnote{More accurately, we consider bounded complexes. Namely, we only consider complexes such that $E_i=0$ when the absolute value of $i$ is sufficiently large. This is the reason for the expression $D^b$, where $b$ stands for `bounded'.} of vector bundles
\beq
E: \ \ \ldots  \to E_{i-2 } \to E_{i-1} \to E_{i} \to E_{i+1} \to E_{i+2} \to \ldots.
\eeq
By a complex we mean that composition of two subsequent morphisms is zero.
We define morphisms between complexes as chain maps between complexes. 

Roughly speaking, we consider complexes in order to take anti D-branes into account. Namely, if the complex
\beq
E: \ \ \ldots  \to 0 \to 0 \to \overbrace{E_i}^i \to 0 \to 0 \to \ldots
\eeq
(with $E$ in the $i$ the position)
describes the D-brane, the complex
\beq
E[1]: \ \ \ldots  \to 0 \to 0 \to 0 \to  \overbrace{E_i}^{i+1} \to 0  \to \ldots\label{shiftone}
\eeq
describes anti D-brane \footnote{Strictly speaking, this slightly misleading, since if this explanation is take literally then you shift the position $E$ once more to obtain $E[2]$, you should be back to the D-brane, but $E[2]$ is a different complex from the original one $E$.}.

To obtain derived categories, we introduce the concept of quasi-isomorphism. Since the composition of two subsequent morphisms are zero and we can consider cohomology of the complex. A morphism between complexes is called quasi-isomorphism if the morphism induces isomorphism on the cohomologies. The idea of derived category is to identify all objects which are quasi-isomorphic, or said more formally to add formal `inverses' of quasi-isomorphisms. For example, the complex
\beq
 \ldots  \to 0 \to 0 \to E \overbrace{\to}^{\textrm{id}} E \to 0 \to 0 \to \ldots,
\eeq
with one D-brane and anti D-brane, is unstable and is equivalent by tachyon condensation to 
\beq
\ldots  \to 0 \to 0 \to 0 \to 0 \to 0 \to 0 \to \ldots ,
\eeq
and thus we should identify all such objects. Objects in the derived categories of coherent sheaves are (bounded) complexes of coherent sheaves, divided by the equivalence relation that two quasi-isomorphic complexes are identified. As morphisms, we include chain maps, and inverses of quasi-isomorphisms.

If you are familiar with K-theory in the context of tachyon condensation \cite{Witten:1998cd,Witten:2000cn}, then you can think of derived categories as an more elaborate version of K-theory.

We do not discuss derived categories in detail here, since we have ample mathematical references on these topics. For leisurely introduction to derived categories, see the review \cite{Thomas}. We also have many mathematics introductions \cite{GelfandManin,Residues,KashiwaraShapira,Keller} and physics reviews \cite{Aspinwall:2004jr,Sharpe:2003dr}.

\paragraph{Exceptional collections}

We comment on the concept of exceptional collections, since they are useful in computing B-brane category. See \cite{Herzog:2005sy} for more detailed discussions.
A object $\mathcal{E}$ in $D^b(\coh \scM)$ is called an {\em exceptional object} if and only if
\beq
\Ext^q(\mathcal{E},\mathcal{E})=
\begin{cases}
\bC & q=0 \\
0 & q\ne 0
\end{cases}
.
\eeq

A collection of exceptional objects $\mathcal{E}=(\mathcal{E}_1,\mathcal{E}_2,\ldots, \mathcal{E}_n)$ is called an {\em exceptional collection} if
\beq
\Ext^q(\mathcal{E}_i,\mathcal{E}_j)=0 \textrm{ for } i>j \textrm{ and } \forall q,
\eeq
and an exceptional collection is called a {\em strongly exceptional collection} if 
\beq 
\Ext^q(\mathcal{E}_i,\mathcal{E}_j)=0 \textrm{ for } i\ne j \textrm{ and } \forall q .
\eeq

An exceptional collection must be strong to generate a physical quiver gauge theory\cite{Herzog:2004qw,Aspinwall:2004vm}. Also, exceptional collections are called full (or complete) if they generate $D^b( \coh \scW)$. Roughly speaking, they are analogues of basis of vector spaces. Physically, full strong exceptional collections are fractional branes, from which all topological D-branes can be constructed.
For the toric case we are now going to consider, the existence of full exceptional collections was shown by Kawamata \cite{Kawamata}. 

Suppose we have a full strong exceptional collection (the set of fractional branes) $\scE=(E_1,E_2,\ldots, E_n)$. Then it is natural to guess that we can compute category of B-branes from that information. In fact, that is actually the case.
Thanks to the theorem of Bondal\cite{Bondal}, $D^b(\coh X_{\Delta})$ can be computed to be
\beq
D^b (\coh X_{\Delta})  \cong D^b \left(\mod \left(\bigoplus_{i,j=1}^N \Hom (E_i,E_j) \right) \right).   \label{Bondal}
\eeq

%.... \footnote{Here we simply borrow known results in mathematics, 
%but as we will seen in section **, there is also proposal for computing B-model side from coamoeba and brane tilings. In this sense, }

\subsubsection{The path algebra of quiver with superpotential} \label{pathalg.subsubsec}
The RHS of \cite{Bondal} might still seem unfamiliar, but as we will see, they are directly related to the path algebra of the quiver, an algebra defined from the quiver diagram.

Let us define the path algebra of quiver (for more formal definition, see \cite{UY3}). As explained in \S\ref{quiver.subsubsec}, a quiver $Q$ is an oriented graph. A path of a quiver is a sequence of arrows such that the endpoint of a path is the same is the startpoint of the next path. A path algebra $\bC Q$ (over $\bC$) is spanned by all such paths, and the product of two such paths are defined by concatenation. We define the product to be zero when we cannot concatenate two paths.

Suppose we are given a quiver $Q$ with a superpotential $W$ (this is called quiver with relations). Let us denote them collectively as $\Gamma=(Q,W)$. Its path algebra $\Gamma$ is obtained from $\bC\Gamma$ by identifying all paths which are F-term equivalent. 

\begin{exa}
As a simple example, for $\bC^3$, the quiver is a single node with three arrows  which we label $X,Y,Z$ (Figure \ref{C3cycles.eps}). The path algebra of the quiver is generated by three letters $X,Y,Z$. The F-term relation says that $X,Y,Z$ all commute, and thus the path algebra for quiver with superpotential is given by $\bC\Gamma=\bC[X,Y,Z]$.
\qed
\end{exa}

To go to the Fano version of homological mirror symmetry, we need a path algebra for subquiver of $Q$. Suppose we are given a bipartite graph, and choose an arbitrary perfect matching. Given a perfect matching, we can delete all arrows on the edges of the perfect matching. The path algebra for the resulting quiver with superpotential is denoted $\bC\Gamma^{\to}$. The path algebra for this quiver is dependent on the choice of the perfect matching, but it can be verified (at least for all toric Fano varieties \cite{UY3}) that its derived category $D^b (\mod\bC\Gamma^{\to})$ is independent of such a choice.

\begin{exa}
For the example of $\bC^3$, perfect matching corresponds to the choice of one edge from $X,Y,Z$. If we choose $X$, then the resulting $\bC\Gamma^{\to}$ is given by $\bC\Gamma^{\to}=\bC\Gamma/(X)=\bC[Y,Z]$.
\qed
\end{exa}

% superpotential (or more often cal%led quiver with relations in mathematics literature). The path algebra (over $\%bC$)of a quiver is spanned by all paths on the quiver, As a path, we also inclu%ded identity for each node. The composition of two paths are defined by concate%nation, and 0 if we cannot concatenate them. When we consider a path algebra of a quiver with superpotential, F-term equivalent paths are identifies (see more %formal definition, see \cite{UY3}). 

%Choose an arbitrary perfect matching, and delete all the edges in the perfect m%atching from the superpotential

Now we go back to the problem of computing the category of B-branes.
According to the conjecture by Hanany-Herzog-Vegh\cite{Hanany:2006nm}, we can construct full strong exceptional collection $(E_1, E_2,\ldots E_N)$ from brane tilings such that
\beq
\bigoplus_{i,j=1}^N \Hom (E_i,E_j) \cong \bC \Gamma^{\to}, \label{HHV}
\eeq
where $\bC \Gamma^{\to}$ is the path algebra for subquiver $\Gamma^{\to}$, as we have defined above. We do not explain explicit way of reading off exceptional corrections from bipartite graphs: see \cite{Hanany:2006nm,UY3} for details. 
 This conjecture is proven for toric Fano varieties \cite{Hanany:2006nm,UY3}, but for more general toric Fano stacks specified from arbitrary toric diagram, the proof is still lacking \footnote{\eqref{HHV} suggests that exceptional collections and quiver diagrams are somehow related. In fact, we can write down the corresponding quiver if we know exceptional collections of some Calabi-Yau, regardless of whether it is toric or not. Unfortunately, in practice it is difficult to know exceptional collections explicitly for completely general Calabi-Yau manifolds, and this method is limited so far to toric case and non-toric del Pezzo \cite{Herzog:2003zc} ($dP_k$ with $k \ge 4$) . For toric case, brane tiling method is more powerful, but for non-toric case, exceptional collections is the only method known so far.}.

The importance of this fact is that, by
combining \eqref{Bondal} and \eqref{HHV}, we have
\beq
D^b (\coh X_{\Delta})  \cong D^b \left(\mod \left( \bigoplus_{i,j=1}^N \Hom (E_i,E_j) \right)\right) \cong D^b (\mod \bC \Gamma^{\to}). \label{XGamma}
\eeq
Thus, in order to prove HMS, all we have to do is to check that 
\beq
D^b (\mod \bC \Gamma^{\to}) \cong D^b(\dirFuk W_{\Delta}).
\eeq
See Figure \ref{Tdualex} for the flowchart of this proof.
In the next subsection we are going to compute the RHS, namely the Fukaya category.

\subsubsection{The category of A-branes} \label{Abrane.subsubsec}

In order to prove homological mirror symmetry \eqref{HMS}, we need to compute both sides and check the equality. The usual lesson of mirror symmetry is A-model side is much more difficult than in the B-model side, and that is also the case in our situation. The B-brane category is already computed by Bondal's theorem \eqref{XGamma}.%The left hand side (B-model side) is already known for toric del Pezzo varieties.

So, we can concentrate on the right hand side, which is known as Fukaya category. Actually, computing Fukaya category is very difficult, and even giving precise mathematical definition to Fukaya category in full generality is in itself a very difficult task\cite{FOOO}. But fortunately, for our case, the relevant definitions are given in \cite{Seidel_VC1}.

Before defining Fukaya category, we mention that Fukaya category belongs to a special class of category known as $A^{\infty}$-category. 
An $A_\infty$-category $\scA$ is a special kind of category consisting of
\begin{itemize}
 \item the set $\Ob(\scA)$ of objects,
 \item for $c_1,\; c_2 \in \Ob(\scA)$,
       a $\bZ$-graded vector space $\hom_\scA(c_1, c_2)$
       called the space of morphisms,
 \item operations (composition of morphisms)
$$\
 \m_l : \hom_\scA (c_{l-1},c_l) [1] \otimes \dots
          \otimes \hom_\scA (c_0,c_1) [1]
 \longrightarrow \hom_\scA (c_0,c_l) [1]
$$
of degree $+1$ for $l=1,2,\ldots$
and $c_i \in \Ob(\scA)$,
$i=0,\dots, l$,
satisfying
the {\em $A_\infty$-relations}
\begin{eqnarray}
 \sum_{i=0}^{l-1} \sum_{j=i+1}^l
  (-1)^{\deg a_1 + \cdots + \deg a_i}
  \m_{l+i-j+1}\left(a_l \otimes \cdots \otimes a_{j+1} \right. \\
   \left. \otimes 
    \m_{j-i}(a_j \otimes \cdots \otimes a_{i+1})
     \nonumber
   \otimes
    a_i \otimes \cdots \otimes a_1 \right) = 0,
  \label{A_infty.eq}
\end{eqnarray}
for any positive integer $l$,
any sequence $c_0, \dots, c_l$ of objects of $\scA$,
and any sequence of morphisms
$a_i \in \hom_{\scA}(c_{i-1}, c_i)$
for $i = 1, \dots, l$.
Here, degrees are counted after shifts;
if $a_i \in \hom^p_{\scA}(c_{i-1}, c_i)$,
then $\deg a_i = p - 1$ in $\hom_{\scA}(c_{i-1}, c_i)[1]$. 
Also, $[1]$ means the shift of $\bZ$-grading by one, as in \eqref{shiftone}.
\end{itemize}

The first and second is almost the same with the general definition of categories, except for grading. What characterizes $A_{\infty}$-categories is the existence of $\m_l$ and $A_{\infty}$-relation \eqref{A_infty.eq}. This looks very complicated, but this comes from the nilpotency of certain operator on graded vector space.
We do not go into detailed explanation of this relation since they are not needed from what we are going to explain. We also do not bother about the $\bZ$-grading in this review. See \cite{Keller1,Keller2} for more details. Here we only mention that this kind of structure appears also in open string field theory \cite{Gaberdiel:1997ia,Zwiebach:1997fe,Nakatsu:2001da}
 (For closed string theory, similar structure known as $L_{\infty}$-structure appears).

\begin{exa}
As an example of an $A_\infty$-category comes from
a {\em differential graded category},
i.e., a category whose spaces of morphisms are complexes
such that the differential $d$ satisfies the Leibniz rule
with respect to the composition $\circ$;
\beq
 d (x \circ y) = (d x) \circ y + (-1)^{\deg x} x \circ (d y).
\eeq
It gives rise to an $A_\infty$-category by
\begin{align*}
 \m_1(x) &= (-1)^{\deg x} d x,\\
 \m_2(x, y) &= (-1)^{(\deg x + 1) \deg y} x \circ y, \\
 \m_k &= 0 \qquad \text{for} \ k > 2.
\end{align*}
\qed
\end{exa}

Let us proceed to define Fukaya category. Fix a distinguished basis of vanishing cycles, as explained in \S\ref{D6.subsec}. From the ordering of the distinguished set of vanishing paths, we also have an ordering for vanishing cycles $C_a$. Although we are choosing a specific choice of the distinguished basis of vanishing cycles, you can verify that the final category $D^b(\Fuk \scW)$ is independent of such a choice\cite{Seidel_VC1,Seidel_VC2}. We can also verify that $D^b(\dirFuk W_{\Delta})$ is independent of the choice of the Newton polynomial $W_{\Delta}$ as long as it is generic. This is because $D^b(\dirFuk W_{\Delta})$ is defined by the language of symplectic geometry, and thus independent of complex structure deformations.

Now the  directed Fukaya category $\dirFuk W$
is an $A_\infty$-category, with objects, morphisms and composition of morphisms given by
\begin{itemize}
\item The set of objects is given by the set of vanishing cycles $(C_a)_{a=1}^{N_G}$. As explained in \S\ref{D6.subsec}, this corresponds to the 3-cycles of D6-branes projected onto the Riemann surface $W(x,y)=0$, or in fivebrane setup, the boundary $C_a$ of D5-brane discs $D_a$ in \S\ref{D5cycle.subsubsec}.

\item
Morphisms are given by
\beq
 \hom_{\dirFuk W}(C_a, C_b) = 
  \begin{cases}
    0 & a > b, \\
   \bC \cdot \id_{C_a} & a = b, \\
   \bigoplus_{p \in C_a \cap C_b} \vspan_{\bC}\{ p \}
      & a < b. \\
  \end{cases}
\eeq
This definition says morphisms are basically intersection points of vanishing cycles. Since we have massless open strings at each intersection point of D-branes, morphisms correspond to open strings. The inequality $a<b$ means the ordering of vanishing cycles as we explained above, and the distinction of $a>b$ and $a<b$ corresponds to the fact we are now considering Fano case, which is the origin of the word `directed' Fukaya category.

\item 
For a positive integer $k$,
a sequence
%$(C_{i_0}^\flat,\ldots,C_{i_k}^\flat)$
$(C_{a_0},\ldots,C_{a_k})$
of objects,
and morphisms $p_l \in C_{a_{l-1}} \cap C_{a_l}$
for $l = 1,\ldots,k$,
the $A_\infty$-operation $\m_k$ is given by
counting with signs
the number of holomorphic disks
with Lagrangian boundary conditions;
\beq
 \m_k(p_k,\ldots,p_1) = \sum_{p_0 \in C_{a_0} \cap C_{a_k}}
     \# \overline{M}_{k+1}(C_{a_0},\ldots,C_{a_k};p_0,\ldots,p_k) p_0.
\eeq
Here,
$\overline{M}_{k+1}(C_{a_0},\ldots,C_{a_k};p_0,\ldots,p_k)$
is moduli space of holomorphic maps $\phi : D^2 \to W^{-1}(0)$
from the unit disk $D^2$
with $k+1$ marked points
$(z_0,\ldots,z_k)$
on the boundary
respecting the cyclic order,
with the following boundary condition:
Let $\partial_l D^2 \in \partial D^2$
be the interval
between $z_l$ and $z_{l+1}$,
where we set $z_{k+1} = z_0$.
Then
$\phi(\partial_l D^2) \subset C_{a_l}$ and $\phi(z_l) = p_l$
for $l = 0, \dots, k$.
The sign should be chosen such that is $A_{\infty}$-relation \eqref{A_infty.eq} is satisfied. See \cite{UY3} for details \footnote{In all toric Fano case, such as toric del Pezzo, it is not difficult to find such a choice. It is in general unknown at present, however, whether such choice of signs can be read off from coamoebae and bipartite graphs. We thank Hiroshi Ohta for discussion on this point.}.

\begin{figure}[htbp]
\centering{\scalebox{0.3}{\input{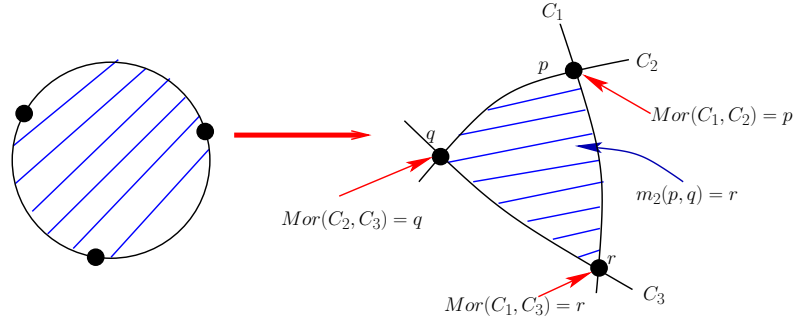}}}
\caption{We have composition of morphism $m_2$.}
\label{m3}
\end{figure}

\begin{figure}[htbp]
\centering{\scalebox{0.3}{\input{mkex.pstex_t}}}
\caption{We have composition of morphism $m_k$.}
\label{mk}
\end{figure}

Physically speaking, this corresponds to disk instantons in the topological A-model. In the quiver gauge theory language, it corresponds to a term in the superpotential (Figure \ref{m3} corresponds to Figure \ref{discamp} in \ref{quiver.subsubsec}). We use discs because we are considering tree-level amplitudes in string theory \footnote{We should be able to define higher-genus analogue of Fukaya category, but corresponding mathematical structure seems to be unknown.}. 
 \end{itemize}

Now we can finally compute Fukaya category\footnote{Actually, in order to compute Fukaya category, we need to fix spin structures and grading. See \cite{UY2,UY3} for such details}. First, we know the objects of Fukaya category, or the cycles D6-branes wrap, by the untwisting operation as we discussed in \S\ref{untwist.subsec}. Before untwisting, these cycles are simply the face of the bipartite graph, or region of $(N,0)$-brane in the fivebrane diagram. Of course, physically speaking this should be the case, but in the next section we verify this claim explicitly. Second, it is immediate to read off morphisms of Fukaya category, since they are simply intersection points of cycles. Finally, it is also immediate to know whether we can see from the untwisted diagram whether we can span a disk bounding intersection points. If we can span a disk, then we have a composition of morphism, and this corresponds to a term in the superpotential.

In closing this section we are going to make a technical remark on what we mean by `derived' \footnote{The reader should be careful at this point because $D^b$ of $D^b (\Fuk \scW)$ does not denote derived category in the usual sense, as in the case of $D^b (\coh \mathcal{M})$.} Fukaya category, since in the statement of HMS \eqref{HMS}, we have used $D^b(\Fuk \scW)$ instead of $\Fuk \scW$. $D^b (\Fuk \scW)$ is the triangulated category obtained from $\Fuk \scW$ by the methods of Bondal and Kapranov\cite{BondalKapranov}\footnote{See Definition 3 and Lemma 4 of \cite{UY3} for how to define this.}. The reason we are using  $D^b(\Fuk \scW)$ instead of  $\Fuk \scW$ is that the former is triangulated, whereas the latter is not. Since  $D^b(\coh \scM)$ is triangulated, we have to make  $\Fuk \scW$ into a triangulated category in order for the equality \eqref{HMS} to hold, and the result is precisely given by $D^b(\Fuk \scW)$.

\subsection{A-brane category from brane tilings: an example} \label{Fukaya.subsec} 

We are now going to explain the computation of Fukaya category, taking del Pezzo 1 as an example. The basic story is the same for all toric Fano varieties. For more examples, see original papers \cite{UY1,UY2,UY3}.

\subsubsection{The coamoeba and vanishing cycles for del Pezzo 1} \label{dP1.subsubsec} \label{sc:dp1}

\begin{figure}[htbp]
\centering{\includegraphics[scale=0.4]{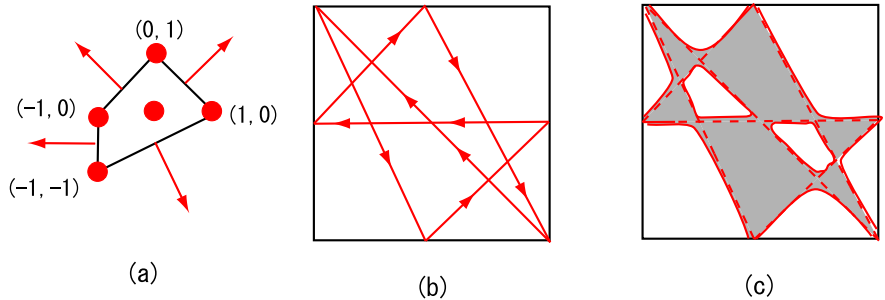}}
\caption{The toric diagram (a), asymptotic boundaries (b) and coamoeba (c) for del Pezzo 1. In this case, asymptotic boundary does not coincide with the real boundary of coamoeba.
}
\label{dP1_lines.eps}
\end{figure}

Here we study the vanishing cycles of the mirror of del Pezzo 1
and their images by the argument map. The toric diagram for del Pezzo 1 is shown in Figure \ref{dP1_lines.eps} (a).

As a Newton polynomial, we take
\beq
 W_{\Delta}(x, y) = x + y - \frac{1}{x} + \frac{1}{x y}.\label{Wform}
\eeq

The asymptotic boundary of the coamoeba of $W_{\Delta}^{-1}(0)$
coincides with the unique admissible configuration of lines
in Figure \ref{dP1_lines.eps} (b). In verifying this we have used the equation for asymptotic boundary \eqref{asymbound} and the form of $W_\Delta$ \eqref{Wform}. We therefore obtain a schematic picture
of the coamoeba shown in the Figure \ref{dP1_lines.eps} (c).

\begin{figure}[htbp]
\centering{\includegraphics[scale=0.5]{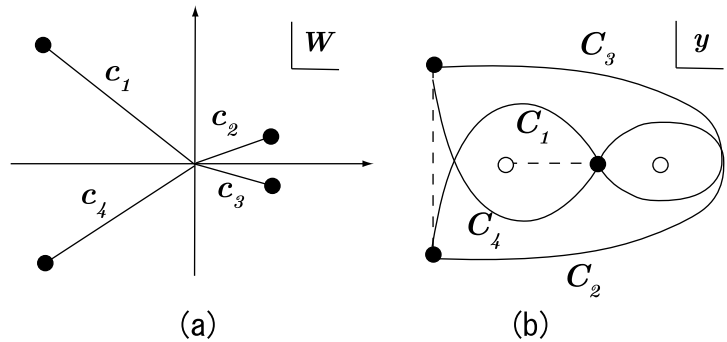}}
\caption{Distinguished basis of vanishing cycles (a) and their images by the map $\pi$ (b). In (b) the black dots are branched points, and white dots correspond to $y=0$ and $y=1$.}
\label{dP1_vc.eps}
\end{figure}

\begin{figure}[htbp]
\centering{\includegraphics[scale=0.5]{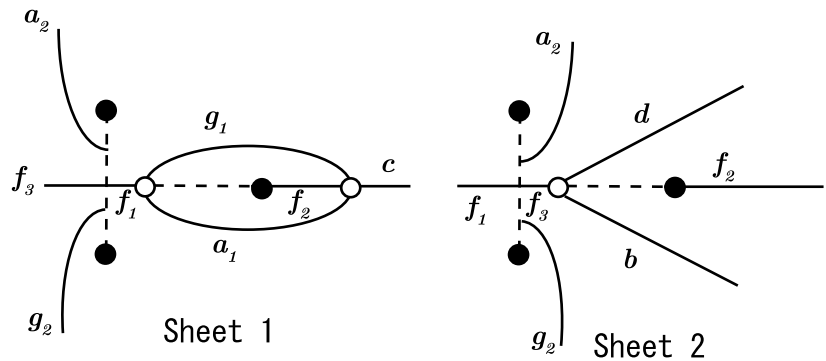}}
\caption{The surface $W_{\Delta}^{-1}(0)$ is divided into two sheets. Here again, the black dots are branched points, and the black dotted lines represent branch cuts. The white dots are $y=0$ and $y=1$. The bold lines correspond to black bold lines in Figure \ref{dP1_coamoeba_cut.eps}.}
\label{dP1_sheet.eps}
\end{figure}

The holomorphic map $W_{\Delta}$ has four critical points.
Take the origin as the base point and
let $(c_a)_{a=1}^4$ be the distinguished set of vanishing paths,
defined as the straight line segments from the origin
to the critical values as in Figure \ref{dP1_vc.eps} (a) (see \S\ref{D6.subsec} for discussion of vanishing cycles from D6-brane viewpoint).
In order to study the geometry of $W_{\Delta}^{-1}(0)$, consider the second projection
\beq
\begin{array}{cccc}
 \pi : & W_{\Delta}^{-1}(0) & \to & \bCx \\
 & \rotatebox{90}{$\in$} & & \rotatebox{90}{$\in$} \\
 & (x, y) & \mapsto & y.
\end{array}
\label{pidef}
\eeq
The fiber $\pi^{-1}(y)$ consists of two points
for $y \in \bCx \setminus \{ 1 \}$,
whereas $\pi^{-1}(1)$
consists of only one point
(the other point goes to $x = 0$).
Figure \ref{dP1_vc.eps} (b)
shows the image by $\pi$
of the distinguished basis $(C_a)_{a=1}^4$
of vanishing cycles
along the paths $(c_a)_{a=1}^4$. We have used numerical computation to plot this figure. In this figure, 
the black dots are the branch points,
the white dots are $y=0$ and $y=1$,
the solid lines are the images of the vanishing cycles
and the dotted lines are cuts
introduced artificially to divide $W_{\Delta}^{-1}(0)$
into two sheets
as shown in Figure \ref{dP1_sheet.eps}. 

\begin{figure}[htbp]
\centering{\includegraphics[scale=0.5]{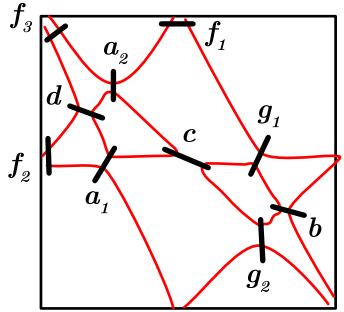}}
\caption{We cut the coamoeba into pieces. The black bold lines correspond to dotted lines in Figure \ref{dP1_sheet.eps}.}
\label{dP1_coamoeba_cut.eps}
\end{figure}

\begin{figure}[htbp]
\centering{\includegraphics[scale=0.35]{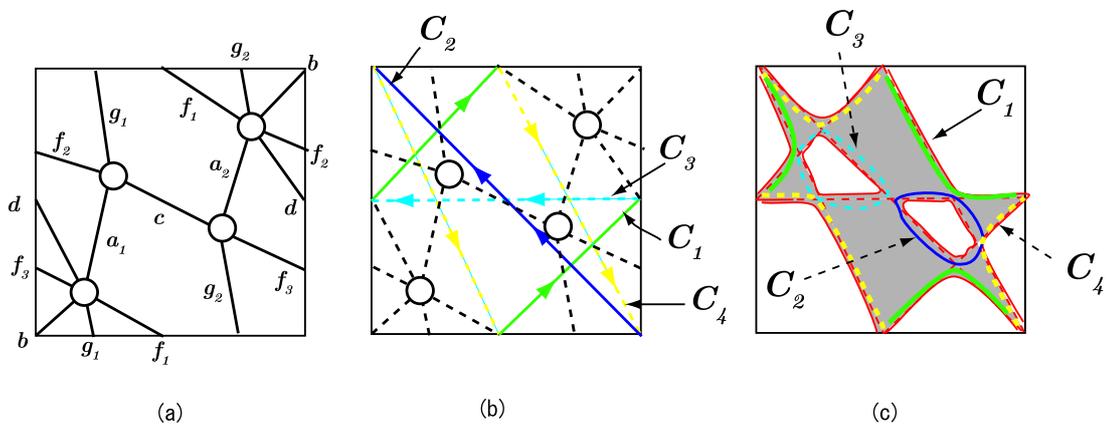}}
\caption{(a) is the surface $W_{\Delta}^{-1}(0)$ obtained by untwisting. The vanishing cycles $C_i$ are represented in (b). Represented on original torus (c), these vanishing cycles represent holes of coamoeba, or regions of D5-branes, as expected from physics intuitions.}
\label{dP1_glued.eps}
\end{figure}

To study the image of the vanishing cycles
by the argument map,
we cut the coamoeba into pieces
along the bold lines in Figure \ref{dP1_coamoeba_cut.eps} (b).

These lines look as in Figure \ref{dP1_sheet.eps}
on the two sheets.
They cut $W_{\Delta}^{-1}(0)$ into the union of
two quadrilaterals and four triangles,
glued along ten edges.
By gluing these pieces,
one obtains an elliptic curve minus four points
in Figure \ref{dP1_glued.eps} (a).
One can see that the vanishing cycles on $W_{\Delta}^{-1}(0)$
looks as in Figure \ref{dP1_glued.eps} (b)
using Figures \ref{dP1_vc.eps},
\ref{dP1_sheet.eps}, and
\ref{dP1_glued.eps} (a). 
One can also see that
the images of the vanishing cycles by the argument map
encircles the holes in the coamoeba
as shown in Figure \ref{dP1_glued.eps} (c).
This is consistent with our physics intuition that vanishing cycles are D-branes, and therefore should correspond to such holes in the coamoeba.

Now that we have computed  $\dirFuk \scW$, and thus $D^b (\dirFuk \scW)$. We have explained the case of $dP_1$, but they are similar for toric Fano varieties. Then, by explicit comparison we can show that\footnote{In this comparison we make use of minimal models of $A_{\infty}$-category, which is standard in homological perturbation theory \cite{Kadeishvili,Merkulov,KontsevichSoibelman}. See \cite{UY3} for details.}
\beq
D^b (\dirFuk \scW) \cong D^b (\mod \bC \Gamma^{\to}).
\eeq 
Combining this with \eqref{XGamma}, we finally arrive the desired HMS
\beq
D^b(\coh X_{\Delta})\cong D^b (\dirFuk \scW_{\Delta}).
\eeq

\subsubsection{Torus equivariant homological mirror symmetry} \label{equivariant.subsubsec}

So far, we have explained how to prove homological mirror symmetry using brane tilings. This is quite interesting both physically and mathematically, but skeptics might still want some new results. Actually, as a bonus of this new proof, we can almost trivially generalize our discussion to orbifold case, which is a new result in mathematics.

Let us consider integer matrix 
\beq
P=
\left(
\begin{array}{cc}
p &q \\
r& s
\end{array}
\right)
\eeq
with det$\, P\ne 0$. For toric diagram $\Delta$, $P^T(\Delta)$ becomes an another lattice polygon, and we can consider the corresponding Newton polynomials $W_{\Delta}$ and $W_{P^T({\Delta})}$. 
It is natural to expect homological mirror symmetry $\Delta$
\beq
D^b(\coh X_{\Delta}) \cong D^b (\dirFuk W_{\Delta})  \label{HMSDelta}
\eeq
and homological mirror symmetry for $P^T(\Delta)$
\beq
D^b (\coh X_{P^T(\Delta)}) \cong D^b (\dirFuk W_{P^T(\Delta)}) \label{HMSPDelta}
\eeq
are related. In fact, by the method of brane tiling and coamoeba, this can almost trivially seen, since considering $P^T(\Delta)$ instead of $\Delta$ corresponds to the change of fundamental region of bipartite graph (as an example, compare Figure \ref{C3cycles.eps} and \ref{SU3egquiver}).

This means that if we can prove \eqref{HMSDelta}, then \eqref{HMSPDelta} follows almost immediately! We call this ``torus equivariant homological mirror symmetry'' \cite{UY4}, which is a new result in mathematics. In fact, the study of homological mirror symmetry for toric Fano orbifolds is quite limited in mathematics literature.

\subsection{Summary}

In this section, we discussed the application of brane tilings to homological mirror symmetry. After a brief description of the statement and implications of homological mirror symmetry (\S\ref{HMS.subsec}), We explained in \S\ref{category.subsec} the rough idea for the mathematical formulation of category of A-branes and B-branes, respectively. The category of B-branes is computed by combining Bondal's theorem and Hanany-Herzog-Vegh proposal. The category of A-branes, or Fukaya category, is computed in \S\ref{Fukaya.subsec} using brane tilings and coamoebae. The object of Fukaya category, a vanishing cycle, is the cycle of D5-branes obtained after untwisting, and a morphism between them is an intersection point of these cycles. Composition of morphisms correspond to counting of holomorphic discs, or a term in the superpotential.

By using coamoebae and bipartite graphs, we have given completely new proof of homological mirror symmetry, applicable to all toric Fano varieties and their orbifolds. This proof is not only mathematically rigorous, but also quite intuitive from physics viewpoint. We have also succeeded in obtaining new mathematical result, namely torus equivariant homological mirror symmetry, as discussed in \S\ref{Fukaya.subsec}.

%%%%%%%%%%%%%%%%%%%%%%%%%%%%%%%%%%%%%%%%%%%%%%%%%%%%%%%%%
%%%%%%%%%%%%%%%%%%%%%%%%%%%%%%%%%%%%%%%%%%%%%%%%%%%%%%%%%

\section{Other topics}\label{others.sec}
In this section, we comment on application to some more topics. The discussion of this section is brief, and the reader is referred to literature for details.

\subsection{Orientifolds of brane tilings}\label{orientifold.subsec}
Recently, orientifolds of brane tilings was studied in \cite{Franco:2007ii,IKY}. There are several possibilities of orientifolds preserving $\scN=1$ supersymmetry, which are shown in Table \ref{oplanes.tbl}.\footnote{As another option, we can put O5-branes in 012357-directions parallel to D5-branes. But probably we then lose the connection with conventional quiver gauge theories.} In the bipartite graph O5-planes (resp. O7-planes) are represented as fixed points (resp. fixed lines). In this paper we restrict ourselves to the case of O5-planes.
% (Figure \ref{tttt.eps}).

\begin{table}[htbp]
\caption{The structure of fivebrane systems with orientifolds. As shown in this table, both O5-planes and O7-planes preserve $\mathcal{N}=1$ supersymmetry. In this paper we only consider the case of O5-planes.}
\label{oplanes.tbl}
\begin{center}
\begin{tabular}{c|cccc|cccc|cc}
\hline
\hline
& $0$ &$1$& $2$& $3$& $4$& $5$& $6$& $7$& $8$& $9$ \\
\hline
D5 & $\circ$ & $\circ$ & $\circ$ & $\circ$ & & $\circ$ && $\circ$ \\
NS5 & $\circ$ & $\circ$ & $\circ$ & $\circ$ & \multicolumn{4}{c|}{$\Sigma$} \\
O5 & $\circ$ & $\circ$ & $\circ$ & $\circ$ & $\circ$ && $\circ$ & \\
O7 & $\circ$ & $\circ$ & $\circ$ & $\circ$ & $\circ$ &&& $\circ$ & $\circ$ & $\circ$ \\
\hline
\end{tabular}
\end{center}
\end{table}

If we construct the gauge theory for an orientifolded fivebrane system
by the naive orientifold projection,
the daughter theory, the theory obtained by the projection
from the parent theory, in general possesses
gauge anomalies.

For example, let us consider the case of a fixed point on a edge of the bipartite graph. The two gauge groups on both sides of the edge is projected to a single gauge group $SU(N)$. The bifundamental field corresponding to the edge is projected down to symmetric or anti-symmetric representations of the $SU(N)$ gauge group (depending on the RR-charge of the O5-plane). The anomaly coefficient for these representations are then given by
\begin{equation}
d_\asymm=(N-4)d_\fund,\quad
d_\symm=(N+4)d_\fund. \label{dasymsym}
\end{equation}
These are different from the contribution $Nd_\fund$ of
the bi-fundamental field in the parent theory,
and we need extra ingredient in order to cancel
the gauge anomaly.

This anomaly can be cured by introducing an appropriate number of
fundamental or anti-fundamental representation fields (quarks and anti-quarks),
By analogy with the relation between the gauge anomaly cancellation
and the D5-brane charge conservation in the un-orientifolded case (see \S\ref{fractional.subsec}),
it is natural to expect the
emergence of the fundamental representation in the orientifolded
brane tilings is also guaranteed by the D5-brane charge conservation. This is in fact verified in \cite{IKY}.

\begin{figure}[htbp]
\centering{\scalebox{0.5}{\includegraphics{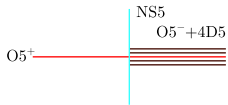}}}
\caption{When O5-brane intersects with NS5-brane, its RR-charge changes its sign. Moreover, in order to conserve RR-charge, inclusion of four flavor D5-branes (including their mirrors) on top of O5-plane are required.}
\label{ns5o5.eps}
\end{figure}

The key for verifying this is the fact that O5-planes carry the D5-brane charge, and its signature
changes when it intersects with NS5-branes\cite{Evans:1997hk,Hanany:2000fq,Hyakutake:2000mr,Bergman:2001rp}. If you take account of the RR-charge conservation,
 the simplest way to satisfy the conservation law is
to introduce four (including mirror images)
D5-branes on top of O5$^-$-plane compensating the
change of O5-plane's RR-charge at the intersection of
O5 and NS5 (Figure \ref{ns5o5.eps}).
These flavor D5-branes indeed cancels the extra contribution of anomaly coming from orientifolding \eqref{dasymsym}. By this way we can indeed show that that D5-brane charge conservation implies gauge anomaly cancellation.

This looks good, but at the same time we have invented an another problem. In \cite{Franco:2007ii}, a set of rules to associate $\mathbb{Z}_2$-parities to mesonic operators are proposed, and there $\bZ_2$-parities of four O5-planes, which are called ``transposition parity'' or `T-parity' in \cite{IKY}, are used. These $\bZ_2$-parities tell us the action of $\bZ_2$ orientifold projections onto gauge invariant operators.

Since RR-charge of O5-plane change its sign in crossing NS5, RR-charges of O5-planes cannot be directly identified with T-parities assigned to four O5-planes. In \cite{IKY} we have clarified the relation between RR-charges and T-parities, and proposed several rules, for example the ``angle rule''.

We can further generalize the discussion. For example, so far we have required the existence of flavor branes coinciding with D5-branes, but we can loosen the condition and can discuss more general flow of D5-brane charge along NS5-brane. We can also extend the argument to include flavor branes. We do not give full details here and refer the interested readers to \cite{IKY}.

\subsection{Application to phenomenological model building} \label{pheno.sec}

Another possible application of brane tilings is to string phenomenology.
The are some proposals of constructing MSSM (or their modifications)  or GUT on D3-brane probing Calabi-Yau singularity \cite{Verlinde:2005jr,Buican:2006sn}. Although the original proposal \cite{Verlinde:2005jr} uses del Pezzo 8, which is non-toric, we can also use toric Calabi-Yaus. See  \cite{Berenstein:2006pk,Wijnholt:2007vn,Heckman:2007zp} for related discussions. 

\begin{figure}[htbp]
\centerline{\includegraphics{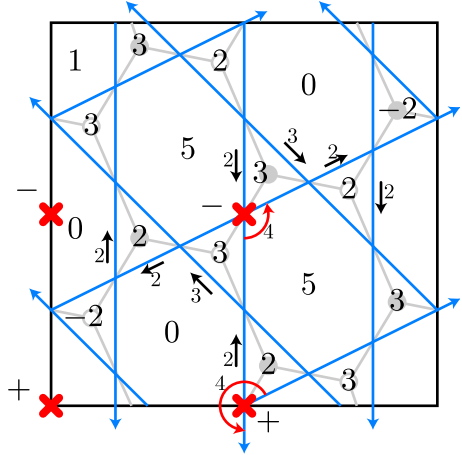}}
\caption{A fivebrane realization of supersymmetric Georgi-Glashow model,
which is a famous example of models of dynamical supersymmetry breaking. The red crosses represent the position of O5-planes, and small black arrows represent the flow of NS5 charge.  Red lines represent flavor branes (Figure \ref{fundmass0.eps}). See \cite{IKY} for full explanation.}
\label{ex1.eps}
\end{figure}

Brane tilings can also be used for supersymmetry breaking sector, and have also opened up a new possibility of constructing models of dynamical supersymmetry breaking in concrete string theory setup. For example, it is possible to construct supersymmetry Georgi-Glashow model from orientifolded  brane tilings  \cite{Franco:2007ii,IKY}. Figure \ref{ex1.eps} shows fivebrane diagrams
of the model $\bC^3/\bZ_6'$ geometry, whose corresponding gauge theory is supersymmetric Georgi-Glashow model.
This fivebrane diagram includes six hexagonal faces
with integers $0$, $0$, $0$, $1$, $5$, and $5$ assigned to each.
These faces give the gauge group $Sp(0)\times SU(0)\times SO(1)\times SU(5)\sim SU(5)$.
We also have $\ol{\bf 10}$ from the fixed point at the center
and $\bf5$ from the contacting point of the $SU(5)$ face and the $SO(1)$ face.
The D5-brane charge is conserved without introducing any flavor branes,
and we have no more extra fields. This verified that we have indeed supersymmetric Georgi-Glashow model. 
%We can also use toric Calabi-Yaus to make models of gauge mediation as in \cite{GarciaEtxebarria:2006rw}.

%See also related work \cite{Antebi:2007xw}.

We can also discuss metastable dynamical supersymmetry breaking. It has recently been discovered \cite{Intriligator:2006dd} that the SQCD with massive quarks has metastable vacua, which can be made to have arbitrary long lifetime. This metastable SUSY breaking is a generic phenomena, and has opened up a new way of making realistic models \cite{Murayama:2006yf,Murayama:2007fe}. They are also ubiquitous in string theory \cite{Ooguri:2006pj,Franco:2006ht,Kawano:2007ru}, and their string theory realization is already known, including some of its moduli stabilization \cite{Nakayama:2007du}. 

So far, the analysis is mostly done using Type IIA theory with NS5 and D4, but the same kind of analysis can be done  for quiver gauge theories. It was verified that, with the addition of fractional branes and flavor branes, we have metastable vacua  in quiver gauge theories corresponding to brane tilings \cite{Franco:2006es,GarciaEtxebarria:2007vh}.

Finally, brane tilings are useful in understanding D-brane instantons, which are interesting because they can  produce non-perturbative superpotential and possibly can explain the hierarchy of Yukawa coupling. See \cite{Ibanez:2007tu,Ibanez:2007rs,Ibanez:2006da,Argurio:2007vqa,Florea:2006si,Bianchi:2007wy}, for example, for related discussions.

\subsection{Similarities with intersection of BPS solitons}\label{soliton.sec}
In previous sections we have emphasized the fact that brane tiling business is basically projection of $(\bCx)^2$ onto angular torus directions $\bT^2$. This idea should not be related to the specific setup we have been considering so far.  We expect similar structures for any system with $\bT^2$ isometry, at least in principle. In fact, our recent analysis \cite{FNOSY} has shown that similar structures appear in the consideration of intersection (or web) of solitons in the Higgs phase of the (bosonic part of) five-dimensional $\scN=1$ supersymmetric $U(N_c)$ gauge theory on $\bR\times (\bC^{\times})^2\sim \bR^{2,1}\times \bT^2$ with $N_f=N_c$ Higgs scalars in the fundamental representation. The moduli matrix formalism \cite{Eto:2006pg,Eto:2006ng} tells us that the most general $1/4$-BPS states are parametrized (in the infinite gauge coupling limit) by a Laurent polynomial $H_0(z_1,z_2)$ of two complex variables $z_1,z_2$. The vortex sheet in 1234 directions are represented in the infinite gauge coupling limit by the equation $H_0(z_1,z_2)=0$, which is exactly the same equation as that of the NS5-brane of the brane tiling. Motivated (partly) by these facts, we found a direct correspondence between amoeba/tropical geometry and solitons/gauge theories. The full detail will appear in our paper \cite{FNOSY}.

Despite many similarities, we should warn the reader at the same time that this system is physically not equivalent to the fivebrane system represented by brane tilings. For example, as shown in Table \ref{soliton.tbl}, we have no analogue of D5-branes and we have instead instanton charge spread along ``vortex sheet'' $\Sigma$. The number of supersymmetries is eight in this case, whereas we are mostly considering fivebrane configuration with four supercharges. Thus the similarity is only at the technical level, but these two systems might possibly have more direct connection by considering more general setup. Unfortunately, D-brane realization of this vortex-instanton system is currently not known (see \cite{Lambert:1999ix,Eto:2004vy}, however, where D-brane configuration for D-branes on $\bC^{\times}$ is discussed).

\begin{table}[htbp]
\caption{The configuration of $1/4$-BPS solitons considered in \cite{FNOSY}. Compare it with the fivebrane configuration shown in Table \ref{config_fst.tbl}. Two directions 2 and 4 are compactified, and $\Sigma$ is a two-dimensional surface in 1234 space. In the infinite gauge coupling limit, the surface $\Sigma$ is given by $H_0(z_1,z_2)=0$, where $H_0$ is a Laurent polynomial of two variables $z_1,z_2 \in \bC^{\times}$. This is exactly the same situation as in the NS5-brane surface of the brane tiling in the decoupling limit.  The counterpart of D5-brane, however, does not exist and we have instead an instanton charge spread along the vortex sheet.}
\label{config_soliton.tbl}
\begin{center}
\begin{tabular}{c|c|cccc}
\hline\hline
& 0 & 1 & 2 & 3 & 4 \\
\hline
vortex & $\circ$ & \multicolumn{4}{c}{$\Sigma$}\\
instanton  & $\circ$ & &  & &  \\
\hline
\end{tabular}
\end{center}
\label{soliton.tbl}
\end{table}

\subsection{Counting of gauge invariant operators} \label{counting.subsec}

Recently, the problem of enumerating gauge invariant operators has been discussed \cite{Benvenuti:2006qr,Hanany:2006uc,Noma:2006pe,Feng:2007ur,Forcella:2008bb}. For example, we count mesonic operators according to charges of $U(1)^3$-isometry, and make the generating function of the number of such operators. In the toric case, this is basically the generating function defined in \eqref{Lefschetz}. In this counting, combinatorial operation known as plethystics plays an important role. For the toric case brane tiling are of great use, although their discussion can be applicable to non-toric case as well.
Counting of baryonic operators and fermionic operators is also discussed \cite{Forcella:2007wk,Butti:2007jv}.
This generating function should be part of the index of four-dimensional $\scN=1$ superconformal field theories \cite{Kinney:2005ej,Romelsberger:2005eg,Nakayama:2005mf,Nakayama:2006ur,Nakayama:2007jy,Biswas:2006tj,Mandal:2006tk}.

\subsection{Three-dimensional generalization of dimer models}
So far, we have mainly talked about dimer models on two-dimensional torus, which corresponds to Calabi-Yau three-fold. The natural question is whether we can generalize these to higher dimensions. The discussion is almost parallel up to some part. For example, in toric Calabi-Yau four-fold case, the fan is spanned by four-vectors and the toric diagram $\Delta$ becomes a convex lattice polytope in $\bR^3$. In this case the Newton polynomial becomes a Laurent polynomial of three variables
\beq
W(x,y,z)=\sum_{(i,j,k)\in \Delta \cap \bZ^3} c_{i,j,k}x^i y^j z^i.
\eeq
Of course, we can also consider their coamoeba $\tilde{\scA}(W)$, as the image of $W^{-1}(0)$ by the map $\Arg$:

\begin{equation}
\begin{array}{ccccc}
 \Arg & : & (\bCx)^3 & \to & (\bR / \bZ )^3 \\
 & & \rotatebox{90}{$\in$} & & \rotatebox{90}{$\in$} \\
 & & (x, y, z) & \mapsto &
  {\displaystyle \frac{1}{2 \pi}(\arg(x), \arg(y), \arg(z))}.
\end{array}
\end{equation}

The important difference arises here. In Calabi-Yau three-fold case, the dimension of $W^{-1}(0)$ and coamoeba $\tilde{\scA}(W)$ is both (real dimension) two. Namely, although $\Arg$ itself a projection from $(\bCx)^2$ to $(\bR / \bZ )^2$, the dimension of $W^{-1}(0)$ and coamoeba $\tilde{\scA}(W)$ is the same. This is the reason why coamoebae are so powerful in studying $W^{-1}(0)$ itself. In Calabi-Yau four-fold case, however, the dimension of $W^{-1}(0)$ is (real dimension) four, which is greater than that of dimension three coamoeba $\tilde{\scA}(W)$. Therefore, our expectation is that although coamoeba still contains some important information about coamoeba $\tilde{\scA}(W)$, they are not so directly related as in our case. See however, the recent proposal of \cite{Lee:2006hw,Lee:2007kv,Kim:2007ic}, in which three-dimensional generalization of brane tiling, called ``crystal model'' is proposed. 

If you look at the problem from the viewpoint of AdS/CFT correspondence, you can also see the difficulty in more physical terms. The Calabi-Yau is four-fold, so the Sasaki-Einstein manifold is seven-dimensional. Therefore, we are considering M-theory on $AdS_3\times X_7$ with $N$ coincident M5-branes placed at the tip, where $X_7$ is a seven-dimensional Sasaki-Einstein manifold, and this theory is believed to be dual to some three-dimensional boundary theory, which is probably not a conventional gauge theory. In fact, the number of degrees of freedom of the theory on $N$ coincident M5-branes should scale as $N^3$ \cite{Klebanov:1996un,Klebanov:1997kc,Harvey:1998bx,Henningson:1998gx}, not as $N^2$ as in conventional gauge theories. 
It might still be possible to study AdS side using eleven-dimensional supergravity \footnote{As  discussed in \S\ref{volume.subsec}, the volume minimization procedure works in this as as well. The counterpart of $a$-maximization, however, is unfortunately not known. It seems that naive extension of the formula to four-dimensional Calabi-Yau does not seem to work.}. The related, though different question is that of tri-Sasakian manifolds. Here, the message of tri-Sasakian manifolds are odd-dimensional manifolds whose metric cone is hyperK\"ahler. In this case, the dual gauge theory is argued to be three-dimensional $\mathcal{N}=3$ SCFT. See \cite{Oh:1998qi,Ahn:1999ec,Fre':1999xp,Fabbri:1999hw,Billo:2000zr,Gauntlett:2005jb,Lee:2006ys,Yee} for discussions related to these points.

\section{Conclusions and discussions} \label{final.sec}
%\section{Conclusions and discussions}\label{final.sec}
Brane tilings are bipartite graphs on $\bT^2$. They are powerful in analyzing arbitrary $\mathcal{N}=1$ supersymmetric quiver gauge theories (both conformal and non-conformal) realized on D3-branes probing toric Calabi-Yau. They are more powerful than conventional quiver diagrams, and capture the information about the superpotential. The beauty of brane tilings is that we can treat such a huge class of quiver gauge theories by writing simple bipartite graphs.

Throughout this paper we have emphasized the fivebrane picture underlying all these developments. In our viewpoint, what is more essential than bipartite graph is the fivebrane diagram, which is the projection of fivebranes (D5-branes and NS5-branes) onto $\bT^2$. In fact, this explains many of previously mysterious `rules' or `algorithms' which are often stated in the literature without proper physical explanation. We have explained the fivebrane systems both in the weak (\S\ref{weak.subsec}) and the strong (\S\ref{strong.subsec}) coupling limit, and they are related by untwisting (\S\ref{untwist.subsec}).
Also, we have seen many more developments on this topic in \S\ref{more.sec}, such as inclusion of fractional branes and flavor branes, analysis of BPS conditions, and relation with mirror D6-brane setup.

Contrary to the common belief, brane tilings are not limited to the study of quiver gauge theories and have many applications to various topics, such as AdS/CFT correspondence ($\mathcal{N}=1$ case) (\S\ref{AdSCFT.sec}), homological mirror symmetry (\S\ref{HMS.sec}) %,junction of solitons in supersymmetric gauge theories 
and string-phenomenological model building. (\S\ref{others.sec})

In each of these developments, brane tilings shed new light on these topics. The appearance of similar structures in many seemingly different contexts is not a coincidence, and we have toric Calabi-Yau behind them. In principle, brane tiling can appear anywhere if you have toric Calabi-Yau three-fold in scene.

%%%
\bigskip

Despite these impressive success, I would like to point out that many important open questions are left, some of which we list below.

\paragraph{The physical significance of the dimer model}

The real physical significance of dimer model in fivebrane systems is far from being clarified. In our viewpoint, fivebrane diagrams are more essential than bipartite graphs. It is true, at the same time, that some concepts of dimer model, such as perfect matchings, are at least technically essential in understanding many aspect of brane tilings. Whether they are only technicalities or have wider implications we do not know. As an example, it would be interesting to understand the physical significance of perfect matching from fivebrane systems. Another clue might come from partition function of the dimer model. The characteristic polynomial (the determinant of Kasteleyn matrix) is a kind of partition function and is useful, as we have explained in \S\ref{another.subsec}, but they are again used in some technical way. But if the dimer model has real physical significance, the partition function of the dimer model should have some role to play. In this connection, some literature of dimer models might be of interest\cite{Kenyon,KenyonLaplacian}. We also point out that the Ronkin function, which is the thermodynamic limit of the partition function of the dimer model, have also been discussed in different contexts \cite{Okounkov:2003sp,Maeda:2006we,FNOSY}.

\paragraph{Dynamics of $\mathcal{N}=1$ supersymmetric quiver gauge theories}
It would be an extremely interesting question to ask whether brane tiling actually knows about rich dynamics of $\mathcal{N}=1$ supersymmetric quiver gauge theories, such as chiral symmetry breaking, gaugino condensation, confinement, mass gap.

Since we now know that brane tilings are physical fivebrane systems, in principle we should be able to do that. 
%One of the possible difficulties is that contrary to the conifold case \cite{Klebanov:2000hb}, we do not have explicit metrics in more general toric Calabi-Yau, which makes the analysis difficult. 
Probably for this purpose the bipartite graph is not enough and we will need to analyze physical fivebrane systems in more detail. 
A more ambitious goal is to try to find new phenomena in $\mathcal{N}=1$ supersymmetric gauge theories.

\paragraph{$A$-maximization}
How to understand $a$-maximization (\S\ref{a-max.subsec}) and R-charges (\S\ref{R-charge.subsubsec}) from fivebrane viewpoint is still an open question. Despite some proposals in the literature, they are not necessarily satisfactory as explained in \S\ref{R-charge.subsubsec}.

\paragraph{More on $\mathcal{N}=1$ case of AdS5/CFT4}
Application to $\mathcal{N}=1$ case of AdS/CFT correspondence is quite important, and as we emphasized in \S\ref{AdSCFT.sec}, this topic is far from being settled, as emphasized there. For example, the check of AdS/CFT in this case is limited mainly to the verification of the formula \eqref{vol=a}, and we definitely need more to understand physics. 
 One big difference from the conventional $\mathcal{N}=4$ case is that the gauge theory side is strongly coupled, and we cannot apply perturbation. The gravity side, however, still seems to be tractable. See \cite{Schvellinger:2003vz,Kim:2003vn,Benvenuti:2005cz,Benvenuti:2005ja,Benvenuti:2007qt,Berenstein:2007wi,Berenstein:2007kq} for related discussions. More speculatively, since dimer models are in a sense exactly solvable, the appearance of dimer model might be suggesting some sort of integrability in $\scN=1$ AdS/CFT.

\paragraph{Sasaki-Einstein geometry}
 A related topic is the relation with Sasaki-Einstein geometry. As we have discussed, Sasaki-Einstein geometry is closely related to K\"ahler-Einstein geometry, and we have much mathematical study on the existence of K\"ahler-Einstein metrics. In \cite{Gauntlett:2006vf}, it is clarified that some obstructions in K\"ahler-Einstein metrics can be interpreted in the gauge theory side. There is in general a conjecture that existence of K\"ahler-Einstein metric is related to stability in geometric invariant theory\cite{Yauopen}, and one formulation is to use the concept of K-stability\cite{Donaldsonscalar}. Whether these deep mathematical structures have any role to play in quiver gauge theories is not clear, but they should if AdS/CFT is somehow correct.

\paragraph{Mirror Symmetry} 
The connection with mirror symmetry is already fruitful, and it is interesting to pursue further this connection. See \cite{Hanany:2001py,Stienstra:2005wy,Stienstra:2005wz} for different approach from ours.

The version of homological mirror symmetry we discussed in the text concerns only the topological information about D-branes. In order to truly understand the spectrum of physical D-branes, we have to consider the stability of D-branes, and homological mirror symmetry should also be extended. In this connection, tachyon condensation is also expected to play an important role. For the discussion of stability, see \cite{Douglas:2000ah,Aspinwall:2001dz,Douglas:2002fj,Aspinwall:2004mb,Bridgeland:2005my}.

Orientifold version of homological mirror symmetry would also be interest, since we do not have much literature. We expect several differences, such as $A_{\infty}$-structures are replaced by $L_{\infty}$-structures in orientifold case\cite{Diaconescu:2006id}, but the basic line of argument should be similar.

Finally, the computation of Fukaya category as shown in \S\ref{Fukaya.subsec} corresponds to computation of Yukawa couplings in intersecting D-brane models \cite{Aldazabal:2000cn,Cremades:2003qj}. Therefore, homological mirror symmetry is strongly connected with the seemingly different topic of string phenomenology.

\paragraph{BPS Solitons}
 To understand solitons in supersymmetric quiver gauge theories from the fivebrane configuration is an interesting problem.  By realizing gauge theory in string theory, it sometimes becomes easier to understand BPS states in gauge theories. Since it seems that not much is known about solitons in supersymmetric quiver gauge theories (see \cite{Lechtenfeld:2006wu}, however), fivebrane configuration with some extra D-branes (for example, D1-branes and D3-branes) will provide new information in solitons in supersymmetric quiver gauge theories.

\paragraph{dessin d'enfants}
In certain area of mathematics, the bipartite graphs are called dessin d'enfants, and has an interesting relation with number theory and algebraic geometry (see \cite{Schneps,Zvonkin}). Whether this viewpoint gives new physics or not is an interesting question. See the recent work \cite{Stienstra:2007} for related discussion. Note that dessins appear recently in physics in different context \cite{Ashok:2006br}.

\paragraph{Lattice gauge theory}
 It is notoriously difficult to construct supersymmetric lattice gauge theory. Some of these works use supersymmetric orbifolds, which arise naturally from string theory constructions \cite{Kaplan:2002wv,Cohen:2003xe,Cohen:2003qw,Ohta:2006qz}. The basic idea behind these works is the deconstruction\cite{ArkaniHamed:2001ca}. Generalization to more general toric Calabi-Yau case is interesting, although it is not clear how to deconstruct general quiver diagram. 

\paragraph{Topological string theory}
In Introduction, we have mentioned that dimer models appear also in topological string theory. In a sense, the work \cite{Feng:2005gw} has clarified the connection, albeit in an indirect way. So far, nobody has found direction connection between the dimer models in brane tilings and those in topological string theory (Figure \ref{3dYoung}), although there are already several interesting attempts (see \cite{Heckman:2006sk}). Part of the reasons for the difficulty is that the two dimer models are slightly different. For example, dimers in brane tilings are on $\bT^2$, but in topological string theory dimers are on two-dimensional plane (Figure \ref{3dYoung}).

The connection with Donaldson-Thomas invariants \cite{DonaldsonThomas,ThomasCasson} is also interesting (see recent works of \cite{Szendroi:2007nu}). Since their generating function coincide with Gromov-Witten invariants in the toric case \cite{Maulik1,Maulik2}, we expect many interesting connection with topological string theory, such as topological vertex \cite{Aganagic:2003db}.

\paragraph{String cosmology}
Brane tilings can possibly be applied to (string) cosmology. The now-famous KKLMMT scenario \cite{Kachru:2003sx} uses a pair of D3-brane and anti D3-brane separated along a warped throat of compact Calabi-Yau, whose local geometry near the throat is well-approximated by the conifold. Although we do not have explicit form of smooth metric, the similar story should also apply to more general toric Calabi-Yau cones. Whether this generalization gives something new we do not know, but at least should allow more flexibility in more realistic model building.

\bigskip

The list can go on and on, but let us stop here. 
As we have seen, brane tilings are certainly linked with various topics, and the whole richness of the subject only recently begins to be explored. We expect more tantalizing developments, and for that we should await further study. 

\section*{Acknowledgments}

First and foremost, it is a pleasure to express my sincere gratitude to my advisor Tohru Eguchi for his guidance and advice. The author would also like to thank Yosuke Imamura and Kazushi Ueda for intensive discussions on this topic. Without them it would be almost impossible to write this review. 
%We are also grateful to thank Kazushi Ueda for educating the author on many mathematical aspects of mirror symmetry, part of which is reflected in \S\ref{HMS.sec}. 
We are also grateful to Toshiaki Fujimori, Hiroshi Isono, Keisuke Kimura, Muneto Nitta, Kazutoshi Ohta and Norisuke Sakai for fruitful collaborations, from which this paper has grown up.
We would also like to acknowledge inspiring discussions with Minoru Eto, Akito Futaki, Akishi Kato, Sang-min Lee, Takayuki Nakashima, Keisuke Ohashi, Hajime Ono, Takeshi Oota and Yukinori Yasui. My thanks also go to Hiroshi Ooguri for comments on the manuscript.
Finally, this paper is based on talks I gave at numerous universities and workshops. I would like to thank all the audience for their interests, questions, comments, and criticisms, which are of great help in clarifying many points in this review.

\appendix

\section{T-duality and Buscher's rule} \label{Buscher.sec}

In this appendix, we derive Buscher's rule\cite{Buscher:1987qj,Buscher:1987sk}, which is used in \S\ref{strong.subsec}. The contents of this section is based on \cite{ImamuraNote}. 

We consider T-duality of either type IIA or IIB theory. The difference does not matter as far as we are considering NS-NS sector.
We are going to take T-duality along $x^9$-direction, which we assume to be compactified. Decompose 10-dimensional metric in the form
\beq
ds_{10}^2=ds_9^2 +g_{99}(dx^9+v_1)^2.
\eeq
Here $v_1$ is a gauge field. Under the translation in $x^9$-directions
\beq
x^9\to x^9+\lambda,
\eeq
with $\lambda$ a parameter independent of $x^9$, $v_1$ transforms as 
\beq
v_1\to v_1-d\lambda.
\eeq
It is well-known that this $v_1$ couples to Kaluza-Klein mode.

Next, decompose NS-NS 2-form as
\beq
B=b_2+b_1\wedge(dx^9+v_1).
\eeq
By looking at the action $S\propto \int B_2$, it is easy to see that $b_1$ couples to winding mode.

Since T-duality interchanges winding mode and Kaluza-Klein mode, the corresponding gauge fields should also be interchanged and we have 
\beq
v_1'=b_1, ~~b_1'=v_1.
\eeq

Let us look at this problem from the viewpoint of supergravity.
The NS-NS part of 10-dimensional type II supergravity (whether IIA or IIB) is given by 
\beq
S=\int d^{10} \sqrt{\mathstrut{-g}}e^{-2\phi} \left[
\left(R+4(\partial \phi)^2\right)-\frac{1}{12}H_3^2 \label{SUGRAaction}
\right].
\eeq

We are going to use the smeared solution, namely assume that all fields are independent of $x^9$-directions.

Now again decompose the metric
\beq
ds_{10}^2=ds_9^2 +g_{99}(dx^9+v_1)^2 \label{dsdecomp},
\eeq
and the denote the field strength of $v_1$ by $f_2$.
Decompose
\beq 
H_3=h_3+h_2\wedge(dx^9+v_1),~~ B_2=b_2+b_1\wedge (dx^9+v_1). \label{HBdecomp}
\eeq
From the relation $H_3=dB_2$, we have
\beq
h_3=db_2-b_1\wedge f_2, ~~h_2=db_1.
\eeq

If you plug \eqref{dsdecomp}, \eqref{HBdecomp} into \eqref{SUGRAaction}, we have 9-dimensional action

\beq
S=\int dx^9 \sqrt{-g^{(9)}}e^{-2\varphi} \left[
\left(R^{(9)}-(\partial \sigma)^2+4(\partial \varphi)^2 \right) 
-\left(\frac{e^{2\sigma}}{4}f_2^2+ \frac{e^{-2\sigma}}{4}h_2^2 \right)
-\frac{1}{12}h_3^2
\right], \label{9daction}
\eeq
where 9-dimensional dilaton $\varphi$ is defined by
\beq
\varphi=\phi-\frac{\sigma}{2}.
\eeq

Then the form of the action in \eqref{9daction} is invariant under the following transposition:
\beq
h_3'=h_3,~~h_2'=f_2,~~ f_2'=h_2,~~\sigma'=-\sigma, ~~\varphi'=-\varphi. \label{Buscher} 
\eeq
For the the potential, we have
\beq
v_1'=b_1,~~b_1'=v_1,~~b_2'=b_2+b_1\wedge v_1. \label{Buscher2}
\eeq
This is the Buscher rule we are going to use in the main text. By applying \eqref{Buscher2} twice along 5 and 7 directions, we have a formula as shown in \eqref{Bjump}.

\section{Some details on $T^{1,1}$ and $Y^{p,q}$}
In this Appendix we explain $T^{1,1}$ and $Y^{p,q}$ in more detail.
\subsection{Notes on $T^{1,1}$} \label{T11.subsec}
Here we explain five-dimensional Sasaki-Einstein manifolds $T^{1,1}$. I have included this because $T^{1,1}$ is the most typical example of five-dimensional Sasaki-Einstein manifolds whose metric is explicitly known. Of course, we have explicit metric $Y^{p,q}$ and $L^{a,b,c}$, as explained in the main text, but their metric is very complicated and it is sometimes good to think of $T^{1,1}$ example first.

The metric cone of $T^{1,1}$ is the famous conifold
\beq
z_1^2+z_2^2+z_3^2+z_4^2=0. \label{conifolddef}
\eeq
You can actually see that this is a cone. If $(z_1,z_2,z_3,z_4)$ satisfies (\ref{conifolddef}), then $(r z_1,r z_2,r z_3,r z_4)$ with $r\in \mathbf{R}_{+}$ also satisfies (\ref{conifolddef}).
This means that $T^{1,1}$ is extracted from the conifold as the intersection of (\ref{conifolddef}) with
\beq
|z_1|^2+|z_2|^2+|z_3|^2+|z_4|^2=\rho^2. \label{conicond}
\eeq
Here $\rho \in \mathbb{R}_{> 0}$ is a new coordinate which corresponds to radial direction of the metric cone. If you write $z_i=x_i+i y_i ~ (i=1,2,3,4)$, we have from (\ref{conifolddef}) and (\ref{conicond}) that
\begin{subequations}
\beq
x_1^2+x_2^2+x_3^2+x_4^2  =\frac{\rho^2}{2}, \label{sub1}
\eeq

\beq
y_1^2+y_2^2+y_3^2+y_4^2 =\frac{\rho^2}{2}, \label{sub2}
\eeq
\beq
x_1 y_1+x_2 y_2+x_3 y_3+x_4 y_4=0. \label{sub3}
\eeq
\end{subequations}
(\ref{sub1}) says $(x_1,x_2,x_3,x_4)$ spans $S^3$, and \eqref{sub2}, \eqref{sub3} says $(y_1,y_2,y_3,y_4)$ move on the intersection of $S^3$ with a plane, which is $S^2$.
This means that $T^{1,1}$ is a $S^2$-fibration over $S^3$.
Since all $S^2$ bundle over $S^3$ are known to be trivial, $T^{1,1}$ is topologically equivalent to $S^2\times S^3$.

Another way of understanding $T^{1,1}$ is to write it as a homogeneous metric on $S^2\times S^3$. First, (\ref{conifolddef}) and (\ref{conicond}) says that $SO(4)\sim SU(2)\times SU(2)$
acts transitively on $T^{1,1}$. Second, the stabilizer of arbitrary point, say, $(\frac{1}{\sqrt{2}},\frac{i}{\sqrt{2}},0,0)$, is $SO(2)\sim U(1)$ which rotates $z_3$ and  $z_4$. This means that 
\beq
T^{1,1}=\frac{SU(2)\times SU(2)}{U(1)_{\textrm{diag}}}. \label{T11AsCoset}
\eeq
The reason for the suffix diag in $U(1)$ will be explained in a moment.

\eqref{T11AsCoset} can also be derived in slightly different manner.
Let us define 
\beq
Z=\frac{1}{\sqrt{\mathstrut{2}}}z^{i}\sigma_i=\frac{1}{\sqrt{\mathstrut{2}}}
\left(
\begin{array}{cc}
z_3+iz_4 & z_1-i z_2 \\
z_1+i z_2 & -z_3+i z_4
\end{array}
\right).
\eeq
The (\ref{conifolddef}) and (\ref{conicond}) becomes
\beq
\textrm{det}Z=0, ~\textrm{tr}Z Z^{\dagger}=\rho^2.
\eeq
If you write one solution of these equations, say
\beq
Z_0=\frac{1}{2}(\sigma_1+i \sigma_2),
\eeq
the general solution is given by
\beq
Z=L Z_0 R, ~L\in \
SU(2)_L, ~R\in SU(2)_R. \label{Zexpr}
\eeq

However, the expression \eqref{Zexpr} has some redundancy.
 If we write $(L,R)=(\Theta, \Theta^{\dagger})$ with
\beq
\Theta=
\left(
\begin{array}{cc}
e^{i\theta} & 0\\
0& e^{-i\theta}\\
\end{array}
\right),
\eeq
($\theta$ a real parameter), we have
\beq
L Z_0 R=Z_0.
\eeq
This shows $T^{1,1}$ is of the form \eqref{T11AsCoset}, and $U(1)_{\text{diag}}$ is generated by $\Theta$. This verifies (\ref{T11AsCoset}). 

The explicit form of the metric is (see the Appendix of \cite{Candelas:1989js})

\beq
\begin{split}
ds^2_{T^{1,1}}=&\frac{1}{9}(d\psi+\cos\theta_1d\phi_1+\cos\theta_2d\phi_2)^2 \\
+&\frac{1}{6}(d\theta_1^2+\sin^2\theta_1d\phi_1^2) 
+\frac{1}{6}(d\theta_2^2+\sin^2\theta_2d\phi_2^2) ,
\end{split}
\label{T11met} 
\eeq
where $0\le \psi <4\pi, 0\le \theta_i <\pi, 0\le \phi_i<2\pi$.

The metric of the metric cone is, by defining $r=\sqrt{3/2}\ \rho^{2/3}$ from previously given $\rho$, 
\beq
ds_{\text{conifold}}^2=dr^2+r^2 ds_{T^{1,1}}^2 .
\eeq
The volume computed from this metric is
\beqa
\text{Vol}(T^{1,1})=\int_{0}^{4\pi} d\psi \int_{0}^{\pi} d\theta_1 d\theta_2 \int_0^{2\pi}d\phi_1 d\phi_2\sqrt{\mathstrut{g}}=\frac{16\pi^3}{27}. \label{T11vol}
\eeqa

Historically, this Sasaki-Einstein manifold is studied in the context of Kaluza-Klein supergravity. In \cite{Romans:1984an} Einstein manifolds $T^{p,q}$ ($p$ and $q$ relatively prime integers) are discovered and their metric obtained.
The explicit form of their metric is given by
\beq
\begin{split}
ds^2_{T^{p,q}}&=\lambda^2(d\psi+p\cos\theta_1d\phi_1+q\cos\theta_2d\phi_2)^2 \\
&+\Lambda_1^{-1}(d\theta_1^2+\sin^2\theta_1d\phi_1^2)+\Lambda_2^{-1}(d\theta_2^2+\sin^2\theta_2d\phi_2^2), 
\end{split}
\eeq
and the condition for Einstein manifold $R_{\mu\nu}=4g_{\mu\nu}$ amounts to 
\beq
\begin{split}
&\lambda^2=p^2\Lambda_1^2+q^2\Lambda_2^2=2, \\
&\Lambda_1\left(1-\frac{p^2\Lambda^2 \Lambda_1}{2}\right)=4,~~
\Lambda_2\left(1-\frac{q^2\Lambda^2 \Lambda_2}{2}\right)=4.
\end{split}
\eeq
These Einstein manifolds $T^{p,q}$ are given by the identification\beq
(L,R)\sim (R \Theta^p, L {\Theta^{\dagger}}^q)
\eeq
in our previous explanation.

Out of these infinite number of Einstein manifolds, only the case with $p=q=1$ is the Sasaki-Einstein manifold. This is the reason for the name $T^{1,1}$.

\subsection{Details of $Y^{p,q}$ metric}\label{Ypq.subsec}

Here we explain the relation between parameters $a$ and $c$ of the $Y^{p,q}$ metric \eqref{Ypqmetric} and the integers $p, q$.more detail of the metric of $Y^{p,q}$ shown in \eqref{Ypqmetric}. In particular, the relation of the parameter $a$, $c$ with integers $p$, $q$ is discussed.

First, if $c=0$, we can set $a=3$ by changing variables appropriately.  
In this case, if you rewrite the metric using new variables $\cos\omega=y, \nu=6\alpha$,
\beq
\begin{split}
ds^2&=\frac{1}{9}(d\psi-\cos\theta d\phi-\cos\omega d\nu)^2 \\
&+\frac{1}{6}(d\theta^2+\sin^2\theta d\phi^2)+\frac{1}{6}(d\omega^2+\sin^2\omega d\nu^2).
\end{split}
\eeq
If you take the period of $\nu$ to be $2\pi$, and that of $\psi$ $4\pi$, then this coincides with the metric of $T^{1,1}$ shown previously (\ref{T11met}).

We therefore consider the case $c\ne 0$. In this case, by rescaling $y$ appropriately, we can set $c=1$, and the only remaining parameter is $a$. Explicit computation verifies the Einstein condition $\text{Ric}=4g$. You can also verify that it is Sasaki.

Now the problem is whether this local form of the metric can be extended globally without any singularities. This is analyzed in detail in the original paper \cite{Gauntlett:2004yd} and the review \cite{Sparks:2007us}, so let us tell you only the rough story. 

First, let us denote by $B$ the four-dimensional base space specified by
$(\theta, \phi, y, \psi)$, and make this $B$ topologically $S^2\times S^2$.
For that purpose, we first set 
\beq
0\le \theta \le \pi, 0\le \phi \le 2\pi,
\eeq
so that $(\theta, \phi)$-part becomes standard metric of $S^2$.
The two-dimensional space spanned by remaining variables $(y, \psi)$
are fibered over this $S^2$. Now the problem is what this fiber is.

We assume $1-y > 0, a-y^2>0$ so that $1-y > 0, a-y^2>0$ becomes positive, 
and choose 
\beq
0<a<1\label{a01}
\eeq 
so that $y$ is between two zeros $y_1,y_2$ of $q(y)$. Here, of the solution to $a-3y^2+2y^3=0$,   one is  negative and two are positive when $0<a<1$. The negative solution is denoted by  $y_1$, and the smaller one of the two positive solutions are denoted $y_2$. $\psi$-direction is $S^1$ when $y_1\le y \le y_2$, and shrinks to one point at $y=y_1,y_2$. This means $(y,\psi)$-direction is diffeomorphic to $S^2$. One mighty worry about apparent singularity at $y=y_1,y_2$, but by changing coordinates we can verify that the metric is non-singular there. In this way, we have verified that  $B$ is $S^2\times S^2$, exactly as in the conifold case.

Next we make $\alpha$-direction a $S^1$-fibration over $B=S^2\times S^2$. Take the period of $\alpha$ to be 
\beq
0\le \alpha \le 2\pi l,
\eeq
then $l^{-1}A$ becomes a $U(1)$ connection  over $B=S^2\times S^2$. Such $U(1)$-bundle is determined by how to patch two bundles on each $S^2$ of $B=S^2\times S^2$, which in turn is determined by Chern number $H^2(S^2;\mathbb{Z})=\mathbb{Z}$. Let us write these $p$, $q$. Then
\beq
P_1 \equiv\frac{1}{2\pi} \int_{C_1}dA=pl, ~~
P_2 \equiv\frac{1}{2\pi} \int_{C_2}dA=ql. \label{P1P2old}
\eeq

From explicit computation, 
\beq
P_1=\frac{y_1-y_2}{6y_1 y_2},~~
P_2=-\frac{(y_1-y_2)^2}{9y_1 y_2}. \label{P1P2}
\eeq

Combining \eqref{P1P2old} and \eqref{P1P2}, we finally arrive at the conclusion that  $a$ and $p$, $q$ are related by
\beq
\frac{3}{2 (y_1(a)-y_2(a))}=\frac{p}{q},
\eeq
where $y_1(a)$, $y_2(a)$ are  $y_1$, $y_2$ as a function of $a$. 

By changing $a$ in the region (\ref{a01}), it is shown that this equation has solution for arbitrary relatively prime integers $p$, $q$ with $q<p$. Thus we have metric for all values of all integers $p$, $q$ with $q<p$, which is denoted by $Y^{p,q}$.

(\ref{P1P2}) determines $l$ as 
\beq
l=\frac{q}{3q^2-2p^2+p\sqrt{4p^2-3q^2}}. \label{leq}
\eeq

Summarizing previous discussions, the parameter region of variables are given by
\beq
0\le \theta \le \pi, 0\le \phi \le 2\pi, y_1\le y\le y_2, 0\le \psi \le 2\pi, 
0\le \alpha \le 2\pi l,
\eeq
with $l$ given by \eqref{leq}.

Finally, the volume of $Y^{p,q}$ is computed to be
\beq
\textrm{Vol}(Y^{p,q})=\frac{q^2[2p+\sqrt{4p^2-3q^2}]}{3p^3 [3p^2-2q^2 +\sqrt{4p^2-3q^2}}.
\label{Ypqvol}
\eeq

\bibliographystyle{utcaps}
\bibliography{mthesis_arXiv_v2}

\end{document}